\documentclass[prc,aps,final,twocolumn,nofootinbib]{revtex4-1}
\usepackage{graphicx}
\usepackage{mathrsfs}
\usepackage{bm}
\usepackage{amssymb}

\newcommand{\bel}[1]{\begin{equation}\label{#1}}

\def\ramaSM{\vadjust{\vbox to 0pt{\vss \hbox to \hsize
{\hskip\hsize \quad $\Leftarrow$\quad {\it SM}\hss}\vskip3.5pt}}}
\def\ramaKA{\vadjust{\vbox to 0pt{\vss \hbox to \hsize
{\hskip\hsize \quad $\Leftarrow$\quad {\it KA}\hss}\vskip3.5pt}}}
\def\ramaJB{\vadjust{\vbox to 0pt{\vss \hbox to \hsize
{\hskip\hsize \quad $\Leftarrow$\quad {\it JB}\hss}\vskip3.5pt}}}
\def\ramaDG{\vadjust{\vbox to 0pt{\vss \hbox to \hsize
{\hskip\hsize \quad $\Leftarrow$\quad {\it PR}\hss}\vskip3.5pt}}}
\def\rama{\vadjust{\vbox to 0pt{\vss \hbox to \hsize
{\hskip\hsize \quad $\Leftarrow$\quad
{$\Longleftarrow$}\hss}\vskip3.5pt}}}

\def\r{{\bf r}}
\def\p{{\bf p}}
\def\q{{\bf q}}

\def\d{\hbox{d}}
\def\be{\begin{equation}}
\def\ee{\end{equation}}
\def\bea{\begin{eqnarray}}
\def\eea{\end{eqnarray}}
\def\l{\label}
\def\s{{\bf s}}
\def\om{\omega}
\def\Om{\Omega}

\def\hahat{\hat{H}}

\def\hahat0{\hat{H}_0}

\def\bs{\bigskip}
\def\ms{\medskip}

\def\cos{\hbox{cos}}
\def\sin{\hbox{sin}}

\def\sinh{\hbox{sinh}}

\def\exp{\hbox{exp}}

\def\sinh{\hbox{sinh}}
\def\Im{{\mbox {\rm Im}}}
\def\Re{{\mbox {\rm Re}}}

\def\eps{\epsilon}

\def\vareps{\varepsilon}
\def\e{e}
\def\epsi{\mathcal{E}}
\def\siml{\hbox{\kern.1em \lower.6ex \hbox{$\sim$} \kern-1.12em
          \raise.6ex \hbox{$<$} \kern.1em }}
\def\simg{\hbox{\kern.1em \lower.6ex \hbox{$\sim$} \kern-1.12em
          \raise.6ex \hbox{$>$} \kern.1em }}
\def\s{{\bf s}}

\def\bs{\bigskip}
\def\ms{\medskip}

\def\siml{\hbox{\kern.1em \lower.6ex \hbox{$\sim$} \kern-1.12em
 \raise.6ex \hbox{$<$} \kern.1em}}
\def\simg{\hbox{\kern.1em \lower.6ex \hbox{$\sim$} \kern-1.12em
 \raise.6ex \hbox{$>$} \kern.1em}}

\newcommand{\beqar}{\begin{eqnarray}}
\newcommand{\eeqar}[1]{\label{#1} \end{eqnarray}}

\def\isospin{\rm q}

\begin{document}

\title{SEMICLASSICAL APPROACHES TO NUCLEAR DYNAMICS}
\author{A.G. Magner}
\email{Email: magner@kinr.kiev.ua}
\author{D.V. Gorpinchenko}
\affiliation{\it  Institute for Nuclear Research, 03680 Kyiv, Ukraine} 
\author{J.\ Bartel}
\affiliation{\it Institut Pluridisciplinaire Hubert Curien, CNRS/IN2P3,
Universit\'e de Strasbourg,  F-67000 Strasbourg, France} 
\bigskip
\date{April 22, 2016}

\bigskip

\begin{abstract}

The extended  Gutzwiller trajectory approach is presented for the
semiclassical description of nuclear collective dynamics, in line with
the main topics of the fruitful activity of V.G.\ Solovjov.
Within the Fermi-liquid droplet model, the leptodermous effective-surface
approximation was applied to calculations of energies, sum rules and
transition densities for the neutron-proton asymmetry of the isovector
giant-dipole resonance and found to be in good agreement with the
experimental data.
By using the Strutinsky shell-correction method, the semiclassical
collective transport coefficients such as nuclear inertia, friction,
stiffness, and moments of inertia can be derived beyond the quantum 
perturbation approximation of the response-function theory and the
cranking model.
The averaged particle-number dependence of the low-lying collective
vibrational states are described in good agreement with basic experimental
data, mainly due to an enhancement of the collective inertia as compared
to its irrotational flow value.
Shell components of the moment of inertia are derived in terms of the
periodic-orbit free-energy shell corrections.
A good agreement between the semiclassical extended Thomas-Fermi moments
of inertia with shell corrections and the quantum results is obtained for
different nuclear deformations and particle numbers.
Shell effects are shown to be exponentially dampted out with increasing
temperature in all the transport coefficients.
\end{abstract}
\maketitle

\section{INTRODUCTION}
\label{introd}

Many experimental data on the fundamental properties  of
fission, and  
collective excitations 
  in
nuclei,  were 
successfully explained 
by using the  
macroscopic-microscopic approaches to
the description of 
finite Fermi systems with strongly interacting
nucleons \cite{migdal,strut,myswann69,fuhi,fraupash,bormot,mix,belyaevzel,SOLbook1981,SOLbook1992,hofbook}. 
One of the transparent and, at the same time, fruitful 
  ways
to study 
collective excitations in complex nuclei was suggested by
V.G. Solovjov and his collaborators within the 
semi-microscopic Quasiparticle-Phonon Model (QPM) 
\cite{SOLbook1981,SOLbook1992,SOLQPMI-EPNP1978,SOLQPMII-EPNP1980,SOLQPMIII-EPNP1983,SOLQPMIV-EPNP1983,SOLQPMV-EPNP1985}. 
We 
should also 
mention the 
theoretical mean-field approaches, based in particular
on the cranking model
\cite{bormot,inglis,bohrmotpr,valat}. 
By using the cranking model and the Strutinsky shell-correction method
(SCM)
\cite{strut,fuhi,fraupash}, 
shell effects in the vibrational and rotational bands 
were intensively studied
\cite{migdal,bormot,mix}. 
  These 
approaches are 
  rooted in
the self-consistent finite-Fermi-system theory 
\cite{migdal,khodsap} and its
mean-field Hartree-Fock (HF) \cite{brink} and 
HF-Bogoliubov (HFB) \cite{solovjov1,ringschuck,brackquenNPA1981} 
approximations. For small-amplitude
collective excitations
they are equivalent to the Random Phase Approximation (RPA) of the QPM.
However, the HF and HFB approaches go beyond the RPA since
they can be applied
for large-amplitude collective motion as well.
Concerning all these approaches, we should also
mention  
pairing correlations
\cite{SOLsuperfluidity-NPA1958,belyaev61,SOLbook1963,belzel,belsmitolfay87},  
  and high spins physics (see  
  \cite{sfraurev} for a review).
The problem of the coexistence of  
oblate-prolate shapes in relation to 
rotational bands was discussed in
\cite{matmatnak2010_12,hinsatnak2010_11}.

For the  description of 
nuclear collective excitations 
within the general response-function theory
\cite{bormot,hofbook}, which can be associated with the RPA of the QPM, 
the basic idea is to parametrize
the complex dynamical problem of the collective motion of many
strongly interacting particles in terms of a few main collective
variables found from the physical
meaning of the considered dynamical problem, for example 
the nuclear surface itself \cite{strtyap,strmagbr,strmagden}
or its multipole deformations \cite{bormot,solovjov1}. 
We can then study the response of the dynamical quantities, describing
    the nuclear collective motion in terms of these variables, 
to an external field.
Thus, we get important information on the transport
properties of nuclei. 
For such a theoretical description of the 
collective motion it is often very important to take 
into account the 
temperature dependence of the
dissipative nuclear characteristics as the friction coefficient, as
shown in \cite{hofbook,hofivyam,hofmann,ivhofpasyam}.

However, a precise quantum description 
of dissipative phenomena in
nuclei is rather complicated because  one has
to take into account
the residual interaction beyond the mean-field approximation.
Therefore, a  simpler
Fermi-liquid drop  model (FLDM) 
\cite{kolmagpl,magkohofsh,kolmagsh,belyaev}
accounting for some
macroscopic properties of the many-body Fermi system can be helpful
to understand the global average properties of the collective motion. 
Such a model is based on the Landau Fermi-liquid theory
\cite{landau,abrikha,pinenoz,baympeth},  applied for the nuclear interior 
and  some simple
macroscopic boundary conditions on the nuclear surface
\cite{strmagbr,strmagden,kolmagsh,magstr,magboundcond}.  An
extension of the effective-surface method  
\cite{strtyap,strmagbr,strmagden} to 
neutron-proton
 systems,  that is accounting for the asymmetry, 
spin-orbit and
Swiatecki's derivative terms in the
local energy-density approach 
\cite{rungegrossPRL1984,marquesgrossARPC2004}
  is given in  
\cite{magsangzh,BMRV,BMRPhysScr2014,BMRps2015,BMRprc2015}.
 A more extensive discussion of other 
macroscopic approaches, in particular with different boundary conditions 
can be found
in the review  \cite{belyaev}.  
In \cite{magkohofsh}, the response-function theory was applied to 
describe collective nuclear excitations as the isoscalar quadrupole 
mode. The
transport coefficients, such as friction and inertia, are simply 
calculated within
the macroscopic FLDM, and
their temperature dependence  can be easily 
discussed
\cite{magkohofsh,kolmagsh,belyaev}.
The isospin asymmetry of 
heavy nuclei near their stability line 
and the structure of the isovector dipole resonance
are studied within the FLDM
\cite{kolmagsh,belyaev,BMRV,BMRPhysScr2014,BMRps2015,BMRprc2015,kolmag}.
In this way,  
the giant multipole resonances were described, in particular,  
by taking into account
 a gradual transition with increasing
  temperature from a zero sound mode to the hydrodynamic first sound
\cite{kolmagpl,magkohofsh,belyaev}. 
The friction 
  phenomenon is described 
in \cite{kolmagpl,magkohofsh} 
  as being 
due to  nucleon-nucleon collisions, which were taken into
account in the relaxation-time approximation
  (see e.g.\ \cite{abrikha,pinenoz,baympeth} 
for a general description, 
and \cite{kolmagpl,belyaev} for a specific account of a temperature and
frequency dependence (retardation effects))
\cite{kolmagpl,belyaev,landau}. 
Relations to some general problems of the 
response-function theory \cite{hofbook}
and their understanding,
taking the example of an analytically solved
model based on a 
nontrivial temperature-dependent Fermi-liquid
theory, can be found in \cite{belyaev}. 
One of the most important questions which was clarified there 
is the temperature
dependence of friction and interaction-coupling constants.

Within this extended macroscopic (FLDM) theory, one can determine the
structure of the isovector 
dipole resonance (IVDR) as a splitting 
of the collective states due to the nuclear
symmetry interaction between neutrons and protons
 near the stability line 
\cite{kolmagsh,belyaev,BMRV,BMRPhysScr2014,BMRps2015,BMRprc2015,kolmag}.
Also, the neutron skin of exotic nuclei with a large neutron excess 
is still one of the exciting subjects of nuclear physics 
and nuclear astrophysics
\cite{myswann69,myswnp80pr96,myswprc77,myswiat85,danielewicz1,pearson,danielewicz2,vinas1,vinas2,vinas3,vinas4,vinas5}.
Simple and accurate solutions for the isovector  particle density
distributions were obtained within the nuclear effective-surface (ES)
approximation
\cite{strtyap,strmagbr,strmagden,magsangzh,BMRV,BMRPhysScr2014,BMRps2015,BMRprc2015}
which exploits the saturation property
of nuclear matter and a narrow diffuse-edge region in finite heavy
     nuclei.
In particular,  in the 
  Extended 
Thomas--Fermi (ETF) approach 
\cite{brguehak,sclbook}
(with Skyrme forces
\cite{chaban,reinhard,reinhardSV,jmeyer,bender,revstonerein,ehnazarrrein,pastore})
this can be done for any  
  deformation 
by using an expansion  in a small
leptodermic parameter. 
The latter can be defined as the diffuse surface thickness of a heavy
    nucleus relative to its mean curvature radius, proportional to 
$A^{-1/3}$,
   where $A$ is the nuclear particle number.  
 For deformed nuclear shapes
  such an approach can be carried through under the constraint on some 
  multipole moments. 
The accuracy of the ES  approximation in the ETF
approach without the spin-orbit (SO) and asymmetry terms was checked
\cite{strmagden} by comparing with the results of  the HF
\cite{brink,ringschuck} and ETF calculations  
\cite{brguehak,sclbook} for different
Skyrme forces. The ES approach \cite{strtyap,strmagbr,strmagden}
was then extended by taking the SO interaction,
 and asymmetry effects into account
\cite{magsangzh,BMRV,BMRPhysScr2014,BMRps2015,BMRprc2015}.
Solutions for the
isoscalar and isovector particle densities and
energies at the quasi-equilibrium
in the ES approach of the ETF approach were applied to
analytical calculations of the neutron skin and isovector stiffness
coefficients in   
leading order of the leptodermic parameter
and to the derivations of the macroscopic boundary conditions 
\cite{strtyap,strmagbr,strmagden,magsangzh,BMRV,BMRprc2015}
  and compared 
  with those 
obtained in the liquid droplet model (LDM) 
\cite{myswann69,myswnp80pr96,myswprc77,myswiat85}.  
These analytical expressions for the
surface-energy constants can also be used
for IVDR calculations within the FLDM (see \cite{belyaev} and  
references therein).

A further interesting application
of the semiclassical response-function 
theory would consist in the study of the
properties of collective transport phenomena, in particular the
low-lying excitations and rotational 
bands in heavy deformed nuclei.
One may consider nuclear collective rotations
within the cranking model
as a response of the nuclear system to the 
Coriolis external-field perturbation.
The moment of inertia (MI)
can be calculated as a kind of susceptibility 
with respect to this external field.
The rotation frequency of the rotating Fermi system 
is determined 
  in the cranking model 
for a given nuclear
angular momentum through a constraint, as
for any other integral of motion, as in particular the particle number
conservation \cite{ringschuck}. In order to simplify
such a rather complicated problem, the Strutinsky shell correction method (SCM)
\cite{strut,fuhi} was adjusted to the collective nuclear rotations in
\cite{mix,fraupash}. The collective MI
is expressed as function of the particle number and temperature
in terms of a smooth part
and an oscillating shell correction. The smooth component can be described
by a suitable semiclassical 
macroscopic model, like the dynamical ETF approach
\cite{brguehak,sclbook,bloch,amadobruekner,rockmore,jenningsbhbr,bartelnp,bartelpl}
which has proven to be both simple and precise.
For the definition of the MI shell correction, one can apply the Strutinsky
averaging procedure to the single-particle (s.p.) MI, in the same way as for
the well-known free-energy shell correction.

For a deeper understanding of the quantum results and the correspondence
between classical and quantum physics of the MI shell components, it
is worth to analyze these shell components in terms of periodic orbits
(POs), what is now well established as the semiclassical periodic-orbit
theory (POT)
\cite{sclbook,gutz,bablo,strumag,bt76,creagh,migdalrev,MKApanSOL1}
(see also its
extension to a given angular-momentum projection along with the
energy of the particle \cite{magkolstr},
to the particle densities
\cite{strmagvvizv1986,brackrocciaIJMPE2010} and
pairing correlations
\cite{brackrocciaIJMPE2010,friskguhr1993,AB_jphys2002,BAPZ_ijmpe2004}).
Gutzwiller was the first who developed the POT for
completely chaotic Hamiltonians with only one integral of motion
 (the particle energy) \cite{gutz}.
The Gutzwiller approach of the POT, extended
to potentials with continuous symmetries, for the description of the
nuclear shell structure can be found in 
\cite{sclbook,strumag,creagh,migdalrev,MKApanSOL1,smod}.
The semiclassical shell-structure corrections to the level density
and energy have been tested for a large number of s.p.\ Hamiltonians
  in two and three dimensions (see, for instance, 
\cite{sclbook,migdalrev,MKApanSOL1,magosc,ellipseptp,spheroidpre,spheroidptp,MAFptp2006,magNPAE2010,MVApre2013,kkmabPS2015}).
For a Fermi gas the
 entropy shell corrections of the POT as a sum of periodic orbits were
derived in 
\cite{strumag}, and with its help,
simple analytical expressions for the
shell-structure energies in cold nuclei were obtained 
 in \cite{sclbook}.
These shell-correction energies are in good agreement with the
quantum SCM results, for
instance for the elliptic and spheroidal cavities, including the
 superdeformed bifurcation region within the improved 
stationary-phase method (improved SPM or shortly ISPM)
\cite{migdalrev,MKApanSOL1,ellipseptp,spheroidptp,MAFptp2006,MVApre2013,kkmabPS2015}.
In particular in three dimensions, the superdeformed bifurcation
nanostructure leads, as function of deformation, to the
double-humped shell-structure energy with the first and second
potential wells in heavy enough nuclei
\cite{sclbook,migdalrev,MKApanSOL1,smod,spheroidptp,magNPAE2010},
which is well known as the double-humped fission barriers in the region
of actinide nuclei.
At large deformations the second well can be understood semiclassically,
for spheroidal type shapes, through the bifurcation of equatorial orbits
into equatorial and the shortest three-dimensional periodic orbits.
For finite heated fermionic systems, it was also shown
\cite{sclbook,strumag,brackrocciaIJMPE2010,kolmagstr,magkolstrizv1979,richter}
within the POT that the
shell-structure of the entropy, the thermodynamical (grand-canonical) potential
 and the free-energy shell
corrections can be obtained by multiplying the terms of the POT expansion
by a temperature-dependent factor, which is exponentially decreasing
with temperature. For the case of the so called
``{\it classical rotations}$\,$'' around the symmetry $z$ axis of the nucleus,
the MI shell correction is obtained
  at finite temperature
for any rotational frequency 
within the extended Gutzwiller approach (EGA) to the
 POT through the averaging of the
individual angular momenta aligned along this symmetry axis
\cite{magkolstr,kolmagstr,magkolstrizv1979}.
A similar POT problem, dealing with the magnetic susceptibility of fermionic
systems, like metallic clusters and quantum dots, was worked out in
\cite{richter,fraukolmagsan}.

It was suggested \cite{dfcpuprc2004} to use the spheroidal cavity and
the classical perturbation approach to the POT by Creagh 
\cite{sclbook,creagh1996}
to describe the collective rotation of deformed nuclei around an axis ($x$
axis) perpendicular to the symmetry $z$ axis.
The small parameter of the POT perturbation approximation turns out to be
proportional to the rotational frequency, but also to the classical action
(in units of $\hbar$), which causes an additional restriction to Fermi
systems (or particle numbers) of small enough size, in contrast to the usual
semiclassical POT.

In \cite{mskbg,mskbPRC2010,GMBBps2015,GMBBprc2016}, 
the nonperturbative EGA POT
was used for the calculation of the MI shell corrections within the
mean-field cranking model for both the collective and the alignment
rotations. In these studies of the statistical equilibrium  
nuclear rotations,
the semiclassical MI shell corrections were obtained in good agreement with the
quantum results
in the case of the harmonic-oscillator potential.
We extended this approach for collective rotations
perpendicular to the   
symmetry axis to the analytical calculations of
the MI shell corrections for the  case of different mean fields,
in particular with spheroidal
shapes and sharp edges in the phase space representation, also
taking into account the ETF surface corrections to the
  MI shell components 
\cite{GMBBprc2016}.
The main purpose was here  to study semiclassically, 
within the ISPM
\cite{migdalrev,MKApanSOL1,ellipseptp,spheroidptp,MAFptp2006,MVApre2013,kkmabPS2015},
the enhancement effects in the MI, 
due to the bifurcations of
periodic orbits in the superdeformed region.

In the present review
in Section II  
we give a general presentation of the periodic-orbit theory 
  in the EGA 
using phase-space variables, a theory
  that is valid for any mean-field potential.
In Section III we present
the ETF local-density approach within the
ES approximation and apply it to the 
  study 
of the 
isovector dipole-resonance structure by using the FLDM.
  We show in Section IV how transport coefficients can be obtained within
  the collective response-function theory using the EGA. 
  The smooth 
ETF and fluctuating shell-structure components of the moments of inertia
are derived  
  in Section V 
for collective rotations of heavy nuclei.
The MI shell component is analytically  written 
in terms of the
periodic orbits and their bifurcations within the phase space approach of the
POT taking into account the ETF surface corrections as well as the 
temperature effects of a heated Fermi system. This component
is compared with the quantum results for the simplest case of the deformed
 spheroidal cavity.
Comments and conclusions are finally given in Sec.\ 6. 
Some details of the ETF and POT calculations are developed in the
Appendix A.


\section{THE EXTENDED GUTZWILLER APPROACH}

The periodic-orbit theory is a powerful semiclassical tool for the
  analytical description of the main static and dynamic properties of
  finite Fermi systems, such as nuclei, metallic clusters and quantum dots
  \cite{gutz,bablo,strumag,bt76,creagh} (for an introductory review see 
  also \cite{migdalrev,sclbook,MKApanSOL1}).
It provides us with the quantum-classical correspondence
where the quantum statistically averaged and  
fluctuative-shell properties of such systems   
can be described within one approach in terms of the classical 
objects, the short-time 
nearly local trajectories and the periodic  orbits of the classical 
Hamiltonian dynamics, respectively.
This theory answers,
sometimes even analytically, some fundamental questions
concerning the physical origins of the shell structure 
in any finite Fermi system, its pronounced strength depending on the
symmetries and symmetry breaking of the Hamiltonian, and the role of the
bifurcations of the POs. All these origins are of significant importance  
for a deeper understanding, based on classical pictures, the  
transport coefficients of the 
collective dynamics,
 and also the double-humped fission barrier, 
in particular, the existence of isomeric shapes at large deformations
\cite{smod,sclbook,brreisie,migdalrev,MKApanSOL1}. 
The chaos-order transitions and the chaotic nature of the nucleon 
  dynamics itself are at the center of the progress achieved by the POT.
 Some applications of the 
POT to the nuclear deformation energies were presented 
and discussed 
concerning the bifurcations of periodic orbits with pronounced shell effects
\cite{ellipseptp,spheroidpre,spheroidptp,MAFptp2006} (see also 
 \cite{ozoriobook,ullmo,creagh97,sie97,zphD,ssun,schomerus,hhun,brack2001,bmt,brackjorg,MAFptp2006,bridge,fedmagbr,kkmabPS2015,arita2012,sclbook,MKApanSOL1}
concerning the bifurcations and normal-form theories). 
Last but not least, the POT presents the
combined semiclassical macroscopic (ETF) and microscopic 
(PO shell-structure) dynamics, as the analytical version of the SCM
extended to the nuclear collective dynamics.

According to the SCM \cite{strut,fuhi}, 
the oscillating part of the total energy of a finite 
fermion system, the so-called shell-correction energy, is associated 
with an inhomogeneity of the s.p.\ energy levels near the 
Fermi surface. 
Depending on the level density at the Fermi energy -- and with 
it the shell-correction energy -- being a maximum or a minimum, 
the nucleus is particularly unstable or stable. This
situation varies with particle number and deformation, diffuseness
and other parameters of the nuclear mean field.
In consequence, the shapes of stable nuclei depend strongly 
on particle numbers and deformations. 
The SCM was successfully used to describe nuclear masses and deformation
energies and, in particular, fission barriers of heavy nuclei. 
One of the most
remarkable triumphs of the SCM is the description of the mass asymmetry
of fission fragments because of the shell effects beyond the LDM.
The miscroscopic 
foundations of the SCM are discussed in an 
early review by Strutinsky's group \cite{fuhi}.

On the way to a  more realistic semiclassical calculation, it is important
to account for a diffuseness of the nuclear surface. As found in 
\cite{aritapap,arita2012,MVApre2013}, 
the shell structure in the radial power-law potentials (RPLP)
and more general deformed power-law potentials (PLP) are 
good approximations 
to those of the corresponding familiar WS
potential for nuclei 
in the spatial domain where the particles are bound 
\cite{MKApanSOL1}. 

In this section, we present the POT within the EGA, focusing on the
nuclear collective dynamics and
 accounting for the symmetry-breaking and bifurcation phenomena
 by using the ISPM. The main  ingradients of the
POT concerning the semiclassical Green's functions based on the
Feynman path-integral representation of the mean-field dynamics
are presented in Section IIA. In Section IIB the ISPM trace formulas 
for the level densities are derived within the phase-space approach.
In Section IIC, we show the trace formulas for the averaged level
densities, and (free-) energy shell corrections as the PO sums being the
analytical versions of the SCM.
The specific expression of the oscillating level density for 
degenerate families in any integrable
potentials, in terms of the action-angle variables, is presented in Section 
IID. The POT will be applied in the next sections
for the calculations of the transport coefficients of the 
nuclear dynamics, such as the inertia, 
friction and stiffness of the collective vibrations, and the moments of
inertia of the collective rotations of nuclei.

\bigskip
\subsection{SEMICLASSICAL GREEN'S FUNCTIONS}
\label{semicl}
\medskip

The mean field approach can be founded on the one-body Green's function
formalism starting from the quantum Feynman path-integral propagator
\cite{gutz,sclbook}. 
This Feynman representation for the  time-dependent Green's function 
is conveniently used in order to develop the analytical 
semiclassical approximations
by applying the 
SPM to calculate the path 
integral. The stationary-phase conditions of this method 
reduce it to the sum over classical
trajectories (CT) giving the dominating contributions to the Green's 
functions. It is especially helpful for the calculations of their traces,
such as 
s.p. level and particle densities, 
partition functions, and free energies.

With the help of the SPM, Gutzwiller derived \cite{gutz} from
the Feynman  
path-integral propagator in the energy $\vareps$ representation
the semiclassical CT expansion of the Green's function
for a time independent Hamiltonian:
\bea\l{Gsem}
&G(\r^{}_1,\r^{}_2;\vareps) = \sum^{}_{\rm CT}G_{\rm CT}(\r^{}_1,\r^{}_2;\vareps)
\nonumber\\
&=G^{}_{\rm CT^{}_0}(\r^{}_1,\r^{}_2;\vareps)
+ \sum\limits_{CT\neq CT^{}_0} 
G^{}_{\rm CT}(\r^{}_1,\r^{}_2;\vareps), 
\eea
where
\bea\l{GsemGa}
\hspace{-0.5cm}G^{}_{\rm CT}(\r^{}_1,\r^{}_2;\vareps) &=& \mathcal{A}^{}_{\rm CT}\; 
e^{i \Phi^{}_{\rm CT}}, \nonumber\\
\Phi^{}_{\rm CT} &=&\frac{1}{\hbar}
S^{}_{\rm CT}\left(\r^{}_1,\r^{}_2;\vareps\right) - 
\frac{\pi}{2}\mu^{}_{\rm CT}\;. 
\eea
The summation index  in 
 (\ref{Gsem}) covers all the manifolds of the CTs  inside
the potential well which connect two
spatial points $\r^{}_1$ and $\r^{}_2$ of the nucleus 
for a given energy $\vareps$.  
These semiclassical derivations can be applied 
in the case
of $k^{}_FR \sim A^{1/3} \gg 1 $, where
$k^{}_F=p^{}_F/\hbar$ is the wave number at the Fermi energy 
$\vareps^{}_F$, $R$ 
the size, and $A$ 
the particle number of a finite Fermi system, as a heavy nucleus.

Among all these CTs, one can find
a short specific trajectory CT$_0$ without intermediate reflections 
from the 
nuclear boundary and
other trajectories like CT$_1$ with reflections, as shown in 
Fig.\ \ref{fig1} for the example of 
an infinitely deep spherical square-well potential.
The first term in (\ref{Gsem}) can be approximated by the Green's 
function of the locally free particle motion, 
\bea\l{GCT0}
G_{{\rm CT}_0} &\approx& G_0=
-\frac{m}{2 \pi \hbar^2 s}\; 
\exp\left[\frac{i p s}{\hbar}\right],\nonumber\\
s=|\s|,\quad \s&=&\r_2-\r_1,\quad p=\sqrt{2m\left[\vareps-V(\r)\right]},
\eea
 with the modulus of the particle momentum  $p$  
in the mean-field potential
$V(\r)$
($p=\sqrt{2m\vareps}$ inside of a billiard potential). 
In (\ref{GsemGa}),  $S_{\rm CT}$ 
is the action for the motion of the particle along such a CT, and 
$\mu^{}_{\rm CT}$ is the Maslov phase related to the
catastrophe (turning and caustic) points along the CT
\cite{strumag,migdalrev,MKApanSOL1,MAFptp2006,fed:jvmp,masl,fed:spm,masl:fed}. 
For the 
amplitude ${\cal A}_{\rm CT}$ in the semiclassical Green's function 
(\ref{GsemGa}) in the case of an unclosed {\it isolated}  
CT one has
\cite{strumag,sclbook,gutz}:
\begin{equation}
\!{\cal A}^{}_{\rm CT}(\r^{}_1,\r^{}_2;\vareps) \!=\!  
- \frac{1}{2\pi \hbar^2}\; \vert {\cal J}^{}_{\rm CT} 
(\p^{}_1, t^{}_{\rm CT} ;\r^{}_2, \vareps)\vert ^{1 / 2},
\label{ampiso}
\end{equation}
where ${\cal J}^{}_{\rm CT} (\p^{}_1,t^{}_{\rm CT};\r^{}_2, \vareps)$ is the Jacobian
for the transformation 
from the initial momentum $\p_1$ and time $t^{}_{\rm CT}$ of the particle 
motion along the CT 
to its final coordinate $\r^{}_2$ and energy $\vareps$. 
The specific expressions of the amplitudes ${\cal A}_{\rm CT}$
for the unclosed isolated trajectories (\ref{ampiso}) and
one-parametric
families of the degenerate closed periodic orbits in the 
infinitely deep
spherical square-well potential were derived in 
\cite{strumag}.  
For contributions of the one- and 
two-parametric degenerated families in the EGA amplitudes 
${\cal A}_{\rm CT}$ for the case of the harmonic oscillator 
potential, 
one can refer to \cite{magosc}.

The trace $\r_2 \to \r_1$ of the first term $G_{{\rm CT}_0}(\r_1,\r_2;\vareps)$
 in (\ref{Gsem}) corresponds to the smooth level density of the ETF model, 
$g^{}_{\rm ETF}(\vareps)$,  
\cite{sclbook,bablo,gutz,strumag}.

As well known \cite{strumag,sclbook}, other terms of the
 Green's function in Eq.\ (\ref{Gsem}) are strongly oscillating components,
due to $\hbar$ in the denominators of
the
exponents of (\ref{GsemGa}).
These oscillations become the stronger the smaller their period
with increasing $S/\hbar$ in the imaginary exponent 
for the semiclassical asymptotic limit $\hbar \to 0$.
The dimensionless parameter
related to $\hbar$
is $\hbar/S_{\rm CT}$, which,
    for potentials with sharp walls, like billiards,
is of the order of $ \sim 1/k^{}_{F}R \ll 1$ near the Fermi surface.
Therefore, the convergence of the second term
in Eq.\ (\ref{Gsem})
    with respect to
 this semiclassical parameter
 appears only after averaging of the Green's function traces
    (like level densities),
over energies
$\vareps$, or $k R$ for billiard-like potentials.   Near the Fermi energy,
this corresponds to
an averaging  over a large  enough interval of the
particle number $A$ through the radius $R$
[see Section IVA2, equation (\ref{akrTF})].
The corresponding Strutinsky averaging 
 \cite{strut,fuhi,sclbook,magNPAE2010,migdalrev,MKApanSOL1,MAFptp2006}
with a Gaussian width $\widetilde{\Gamma}$, which covers at least a 
few major shells in the energy spectrum (see Appendix B3)  
leads to a 
local ($\r_2 \to \r_1$) smooth quantity, e.g., the level and 
particle density and the free energy of the statistical Thomas-Fermi model. 
The nonlocal ($\r_2 \neq \r_1$) contributions to the ETF transport 
coefficients become also important (Section IV).
Therefore, we need a more extended statistical averaging in the phase 
  space (energy and spatial coordinate) variables. This is 
similar to the averaging used in  the  
derivation 
of the local hydrodynamical
equations from the semiclassical kinetic equation within the many-body 
  particle density or Green's function formalism
\cite{kadbaym,kolmagpl}.

\subsection{PHASE-SPACE TRACE FORMULA} 

 The level density, $g(\vareps)=\sum_i \delta(\vareps-\vareps_i)$, 
determined by the energy spectrum $\vareps_i$
for the Hamiltonian 
$H({\bf r},{\bf p})$, can be obtained as a semiclassical approximation by 
using the phase-space trace formula in $\mathcal{D}$ dimensions 
 \cite{ellipseptp,spheroidpre,spheroidptp,MAFptp2006}:
\bea\l{pstrace}
g_{\rm scl}(\vareps) &=& \frac{1}{(2\pi\hbar)^{\mathcal{D}}}\Re\sum_{\rm CT}
\int \d {\bf r}^{\prime\prime} \int \d {\bf p}^{\prime}\nonumber\\ 
&\times&
\delta \left(\vareps - H({\bf r}^{\prime\prime},{\bf p}^{\prime\prime})\right)
\left|\mathcal{J}_{\rm CT}({\bf p}^\prime_\perp,
{\bf p}^{\prime\prime}_\perp)\right|^{1/2} \nonumber\\
&\times&\exp\left\{\frac{i}{\hbar}\;
\Phi_{\rm CT}- i \frac{\pi}{2} \mu^{}_{\rm CT}\right\}\;, 
\eea
 where $\Phi_{\rm CT}$ is the phase integral related to the classical actions in
suitable variables by
\bea\l{legendtrans}
\hspace{-0.5cm}\Phi_{\rm CT} &\equiv& S_{\rm CT}({\bf p}^\prime,{\bf p}^{\prime\prime};t^{}_{\rm CT}) -
\left({\bf p}^{\prime}- {\bf p}^{\prime\prime}\right) 
 \cdot {\bf r}^{\prime\prime}\! \nonumber\\
&=& S_{\rm CT}({\bf r}^\prime,{\bf r}^{\prime\prime};\vareps) +
{\bf p}^{\prime} \cdot \left({\bf r}^{\prime} - {\bf r}^{\prime\prime}\right)\,  
\eea
(see the derivations in \cite{MKApanSOL1}). 
In (\ref{pstrace}), the sum is taken over all discrete CT
manifolds for particle motion from the initial
 ${\bf r}^\prime,{\bf p}^\prime$ to the final 
${\bf r}^{\prime\prime},{\bf p}^{\prime\prime}$ point with a given energy
$\vareps$ (see \cite{MAFptp2006}).
A CT can  uniquely be specified by fixing, for instance, 
the initial condition ${\bf r}^{\prime\prime}$ and the
final momentum ${\bf p}^{\prime}$ for a given time
$t^{}_{\rm CT}$ of the motion along the CT. 
$S_{\rm CT}({\bf p}^\prime,{\bf p}^{\prime\prime};t^{}_{\rm CT})$ 
is the action in the
momentum representation, 
\be\l{actionp}
S_{\rm CT}({\bf p}^\prime,{\bf p}^{\prime\prime};t^{}_{\rm CT}) = 
-\int_{{\bf p}^\prime}^{{\bf p}^{\prime\prime}}
\d {\bf p} \cdot {\bf r}({\bf p})\;. 
\ee
The integration by
parts relates (\ref{actionp}) to the action in the spatial coordinate space,
\be\l{actionr}
S_{\rm CT}(\r^\prime,\r^{\prime\prime};\vareps) = 
\int_{\r^\prime}^{\r^{\prime\prime}}
\d \r \cdot \p(\r)\; 
\ee
(or other generating functions)
by the Legendre transformation (\ref{legendtrans}).
The Maslov phase  $\mu^{}_{\rm CT}$  corresponds to the number 
of  conjugate ( turning and caustics) points
along the CT \cite{MAFptp2006,fed:jvmp,masl,fed:spm}).
We introduced here a local phase-space 3D coordinate system,
$\r=\{x,y,z\}$, $\,\p=\{p_x,p_y,p_z\}$, related to a
PO which gives the main contribution into the trace integral 
among the CTs. The variables $x, p_x$ are locally the parallel and 
$\{\r_\perp,\p_\perp\}$ the 
perpendicular ( with respect to a reference CT) 
phase-space coordinates  
($\r_\perp=\{y,z\} $, 
$\p_\perp=\{p_y,p_z\}$ ) \cite{gutz,smod,sclbook}. In (\ref{pstrace}),
$\mathcal{J}_{\rm CT}(\p^\prime_\perp,\p^{\prime\prime}_\perp)$ 
is the Jacobian
for the transformation from the initial 
$\p_\perp^\prime$ to the final $\p_\perp^{\prime\prime}$ momentum,
perpendicular to the CT.

For calculations of the trace integral by the 
SPM, one may write
the stationary phase conditions in both ${\bf p}^{\prime}$ 
and ${\bf r}^{\prime\prime}$ 
variables. According to the definitions (\ref{legendtrans}) 
and (\ref{actionp}), the stationary phase condition for the
${\bf p}^{\prime}$ variable
is the closing one in spatial coordinates:
\be\l{statcondp}
\left(\frac{\partial \Phi_{\rm PO}}{
\partial {\bf p}^{\prime}}\right)^* \equiv 
\left({\bf r}^\prime - {\bf r}^{\prime\prime}\right)^*=0\;.
\ee
Here and below a star on a quantity indicates that we evaluate that 
  quantity at the stationary point, 
${\bf p}^{\prime}={\bf p}^{\prime\;*}$.  Thus, one has the
closing condition, according to (\ref{statcondp}), 
$\Phi^\ast_{\rm PO}=S_{\rm PO}({\bf r}^\prime,{\bf r}^{\prime\prime};\vareps)$.
In the next integration over ${\bf r}^{\prime\prime}$  by the SPM 
we use the Legendre transformation (\ref{legendtrans}).
The stationary phase condition
for this integration over the Cartesian spacial coordinates 
${\bf r}={\bf r}^\prime={\bf r}^{\prime\prime}$ writes
\bea\l{statcondr}
\hspace{-0.5cm}\left(\frac{\partial \Phi_{\rm CT}}{
\partial {\bf r}^\prime}\!+\!\frac{\partial \Phi_{\rm CT}}{
\partial {\bf r}^{\prime\prime}}\right)^* 
\!&\equiv &\! \left(\frac{\partial S_{\rm CT}}{
\partial {\bf r}^\prime}\!+\!\frac{\partial S_{\rm CT}}{
\partial {\bf r}^{\prime\prime}}\right)^* \!\nonumber\\
&\equiv &
-\left({\bf p}^\prime - {\bf p}^{\prime\prime}\right)^*=0\;, 
\eea
where the star means ${\bf r}^\prime={\bf r}^{\prime\prime}={\bf r}^\ast$ 
along with 
${\bf p}^{\prime\;\ast}={\bf p}^{\prime\prime\ast}$,
and one has the closing condition for a CT 
in the momentum space, too. 
Therefore, the {\it stationary phase}
conditions are equivalent to 
the {\it periodic-orbit} equations (\ref{statcondp})
and (\ref{statcondr}). After applying these two conditions
we arrive at the trace formula in terms of a sum over POs 
\cite{migdalrev,sclbook}.

\subsection{THE TOTAL ISPM TRACE FORMULAS}

 The total ISPM trace formula is the sum of contributions of
all POs (the isolated families with the classical degeneracy
degree\footnote{The classical degeneracy degree $\mathcal{K}$ 
of a CT 
family is the number of independent parameters which determine a CT of the
manifold with a fixed action $S_{\rm CT}$ at a given particle energy.} 
$\mathcal{K}\geq 1$ and the isolated orbits with ${\cal K}=0$),
\begin{equation}\label{deltadenstot}
\delta g_{\rm scl}(\vareps)=\sum_{\rm PO}
 \delta g^{}_{\rm PO}(\vareps)
\end{equation}
where
\begin{equation}\label{dgPO}
 \delta g^{}_{\rm PO}(\vareps)= \Re 
\left\{{\cal B}_{\rm PO} \;
\exp\left[\frac{i}{\hbar} S_{\rm PO}(\vareps) -
\frac{i\pi}{2}\mu^{}_{\rm PO} \right]\right\},
\end{equation}
with ${\cal B}_{\rm PO}$ the amplitude of the density oscillations
depending on the PO classical degeneracy ${\cal K}$  and stability
factors through the Green's function amplitudes $\mathcal{A}_{\rm PO}$
\cite{gutz,strumag}.
In  (\ref{dgPO}), $S_{\rm PO}(\vareps)=\oint \d \r \cdot \p$ 
is the action and $\mu^{}_{\rm PO}$ the Maslov phase 
along the PO
\cite{gutz,strumag,sclbook,belyaev,migdalrev,MAFptp2006}. The Maslov phase 
$\mu_{\rm PO}$ is determined by the phase shifts through the turning and caustic
points, according to the catastrophe Maslov\&Fedoriuk theory 
\cite{fed:jvmp,masl,fed:spm,masl:fed}.

\subsubsection{The averaged level density}.

For comparison with quantum densities we need also to average the trace
formula  (\ref{deltadenstot}) over the spectrum in a given mean-field
potential. 
Since this trace formula has the simple form
of a sum of separate PO terms everywhere including
the bifurcations, one can take approximately analytically 
 the integral over energies 
with the Gaussian weight factor (folding
integral) 
 \cite{strumag,migdalrev,MVApre2013,sclbook,MAFptp2006,MKApanSOL1}. 
As a result for the
averaged density $\delta g^{\rm scl}_\Gamma(\vareps)$, one obtains
with this weight 
function for the averaging
parameter $\Gamma$, which is much smaller than
the Fermi energy $\vareps^{}_F$,  
\be\l{avdeltadentot}
\hspace{-0.5cm}\delta g^{\rm scl}_{\Gamma}(\vareps)
\!=\!\sum_{\rm PO} \delta g^{}_{\rm PO}(\vareps)\; 
\exp\left[-\left(t_{\rm PO} \Gamma/\hbar\right)^2\right],
\ee
where 
$t^{}_{\rm PO}=\partial S_{\rm PO}(\vareps)/\partial \vareps=Mt^{M=1}_{\rm PO}$ 
is the time period for a particle motion along the PO, 
  accounting for the number of periods $M$, {\bf where} $t^{(M=1)}_{\rm PO}$
  is the period for a primitive (one cycle, $M=1$) PO.

The total ISPM level density  as function of the 
energy $\vareps$ is given by 
\be\l{scldenstyra}
g^{\rm scl}_{\Gamma}(\vareps)= g^{}_{ETF}(\vareps) + 
\delta g^{\rm scl}_{\Gamma}(\vareps)\;,
\ee
where $g^{}_{\rm ETF}(\vareps)$ is the 
average part obtained within the ETF approximation \cite{sclbook}.
The convergence of the PO sum in (\ref{avdeltadentot}) is provided mainly
by the exponential Gaussian factor of this summand. Only the short-time
POs (small-length POs for billiard potentials) give the main contributions
to the PO sum  (\ref{avdeltadentot}) at a given finite  
averaging parameter $\Gamma$.
According to (\ref{avdeltadentot}),  with increasing the PO period, 
$t^{}_{\rm PO}$, and the averaging parameter, $\Gamma$, one  finds a similar  
smearing out of the long-time PO contributions. For
large $\Gamma$ (much larger than the distance $D$ 
between the neighbor s.p. 
levels but smaller or
of order of the distance between the neighbor major shells,
$D \ll \Gamma \siml \hbar \Omega \ll \vareps^{}_F$, where 
$\hbar \Omega \sim 2\pi/t_{\rm PO} \sim \vareps^{}_F/A^{1/3}$), 
one finds the coarse-graining
(major) shell-structure effects of short-time POs. For smaller 
$\Gamma $ ($D \siml \Gamma \ll  \hbar \Omega$), one observes a fine-reserved
shell structure involving the long-time POs. In this case, the amplitudes 
$\mathcal{B}_{\rm PO}$ can be enhanced by the bifurcation phenomena 
\cite{smod,spheroidptp,MAFptp2006,MVApre2013,MKApanSOL1}.

\subsubsection{Energy shell corrections}

 The PO expansion  for the energy shell corrections
$\delta E_{\rm scl}$ writes 
\cite{strumag,ellipseptp,spheroidptp,MVApre2013,sclbook,migdalrev,belyaev,MKApanSOL1}
\be\l{escscl}
\delta E = d_s\sum_{\rm PO} \frac{\hbar^2}{t_{\rm PO}^2}\,
\delta g^{}_{\rm PO}(\vareps^{}_F)\;, 
\ee
where $t^{}_{\rm PO}=M t^{M=1}_{\rm PO}(\vareps^{}_F)$
is the time of particle motion
along the PO (taking into
account its repetition number $M$) at the Fermi energy $\vareps=\vareps^{}_F$,
where $t^{M=1}_{\rm PO}$ is the time of particle motion along the primitive 
($M=1$) PO (at $\vareps=\vareps^{}_F$).
The factor $d_s$ takes into account the spin  
(spin-isospin) degeneracy for neutron and/or proton
Fermi systems.
 The Fermi energy $\vareps^{}_F$ is related to the conservation of the particle
number $A$ through the equation: 
\be\l{partnum}
A=d_s\int_0^{\vareps^{}_F} \d \vareps\,g(\vareps)\;. 
\ee

Note that the energy shell corrections $\delta E$ which are
the observed physical quantities do not
contain arbitrary averaging parameter $\Gamma$, in contrast to the
level density $g^{\rm scl}_{\Gamma}(\vareps)$ (\ref{scldenstyra}). 
The convergence of the PO
sum (\ref{escscl}) is ensured by the additional factor in front
of the density component $\delta g^{}_{\rm PO}$ which is inversely
proportional to the time $t^{}_{\rm PO}$ squared along the PO. 
Therefore, we need short-time POs if they occupy enough large
phase-space volumes.

Within the POT, at a given temperature $T$, after the statistical averaging
over the canonical ensemble, one obtains the PO sum for the semiclassical 
free-energy shell correction $\delta {\cal F}_{\rm scl}$ 
\cite{strumag,kolmagstr,fraukolmagsan,mskbPRC2010,belyaev}: 
\be\l{dfscl}
 \delta \mathcal{F} = \sum_{\rm PO} 
\frac{\pi t^{}_{\rm PO} T / \hbar}
{\sinh (\pi t^{}_{\rm PO} T / \hbar )}\;\delta E_{\rm PO}\;,  
\ee
where $\delta E_{\rm PO}$ is the PO component of the 
energy shell correction,
\be\l{descl}
 \delta E_{\rm PO} = d_s \, \frac{\hbar ^2}{t_{\rm PO}^2} \; 
                                          \delta g^{}_{\rm PO}(\lambda) \;,
\ee
and $\delta g^{}_{\rm PO} (\lambda)$ is the PO component 
(\ref{dgPO}) in the
oscillating level density (\ref{deltadenstot})  
at the chemical potential 
$\vareps=\lambda$. 
The oscillating (free) energy  
shell correction,
$\delta \mathcal{F}$ (\ref{dfscl}), is function 
of the particle number, $A$, 
through 
the chemical potential $\lambda$, 
which, at 
small temperatures $T$, equals approximately
the Fermi energy $\vareps^{}_F$, [$\lambda\approx \vareps^{}_F$, 
see (\ref{partnum})]. 
Notice that, in addition to the $1/t^{2}_{\rm PO}$ factor of the PO energy 
shell-correction component (\ref{escscl}), there is the temperature-dependent
factor, which leads to the exponential decrease of the contributions
of long-time POs, and ensures the convergence of the PO sum in the
free-energy shell correction (\ref{dfscl}). The temperature $T$ in 
(\ref{dfscl}) for $\delta \mathcal{F}$ takes a similar
role concerning the convergence of the PO sum as the averaging 
parameter $\Gamma$ in the averaged level-density
shell correction (\ref{avdeltadentot}). With increasing temperature $T$, one
finds the exponential decrease of the oscillating free-energy shell correction,
i.e., an exponential disappearance of the shell effects in the free energy.
For finite temperature, one obtains such a disappearance of the long-time
POs, such that only the short-time POs give the main contributions to the PO
sum (\ref{dfscl}). The critical temperature 
$T_{\rm cr} \approx \pi/\hbar\Omega
\approx 2-3$ MeV for the disappearance of the shell effects 
in heavy nuclei ($A=100-200$) (see, e.g., 
\cite{strumag,mskbPRC2010,belyaev,kolmagstr}) is
in good agreement with the
 quantum SCM
calculations \cite{fuhi}.
See more specific expressions of $\delta g^{}_{\rm PO}$ 
(\ref{dgPO}) in terms of the PO classical degeneracy,
the stability factors, and the action along the PO in 
\cite{gutz,strumag,sclbook,belyaev,migdalrev,MKApanSOL1}.
The POs appear through the stationary phase conditions 
(\ref{statcondp}) and
(\ref{statcondr})
(which, in the present context, is equivalent to the PO condition  
\cite{MAFptp2006})
for the calculation of integrals over $\r^{\prime\prime}$ and 
$\p^{\prime}$ in (\ref{pstrace})
by the ISPM
\cite{spheroidptp,MAFptp2006,migdalrev,MKApanSOL1}.

\subsection{SPHERICAL ACTION-ANGLE VARIABLES}

 We now transform the phase space trace formula (\ref{pstrace})
from the Cartesian phase space variables $\{\r,\p\}$ to the
canonical angle-action ones $\{{\bf \Theta},{\bf I}\}$. 
The latter variables have a clearer physical meaning, and are simpler 
  to use for integrable Hamiltonians.
For integrable systems, the action-angle variables are
particularly useful because 
the Hamiltonian $H$ does
not depend on the angle variables ${\bf \Theta}$, i.e., 
$H=H({\bf I})$. Since the angles ${\bf \Theta}$ are the cyclic
variables in this integrable case, the corresponding action variables
${\bf I}$ are the integrals of the particle motion. 
Therefore, from (\ref{pstrace})
one has
\bea\l{pstraceactang}
g_{\rm scl}(\vareps)
&=&\frac{1}{(2\pi\hbar)^3}  
\Re\sum_{\rm CT}
\int \d {\bf \Theta}^{\prime\prime}\, \int \d {\bf I}\,
\nonumber\\
&\times&\delta(\vareps-H({\bf I}))\left|
\mathcal{J}_{\rm CT}(\p''_\perp,\p'_\perp)\right|^{1/2} \times \nonumber\\
&\times& \exp\left[\frac{i}{\hbar}
\Phi_{\rm CT} - \frac{i \pi}{2}
\mu^{}_{\rm CT}\right]\;. 
\eea
The phase integral $\Phi_{\rm CT}$ (\ref{legendtrans}), 
as expressed 
in terms of the
action-angle variables through the actions 
(\ref{actionp}) or (\ref{actionr}) (standard generating functions)
are considered, in the mixed representation. 
The Jacobian ${\cal J}(\p''_\perp,\p'_\perp)$ is 
also transformed to the new variables. 
We took also into account 
explicitly that the actions ${\bf I}$ 
are constants
of motion for a spherical integrable Hamiltonian, omitting the upper 
subscripts
in ${\bf I}$ as related to their initial (prime) and final (double
prime) values. We used also usual Jacobian determinant transformations
from a set of variables to another set,
taking into account that there is no variations in a parallel
x direction along the PO, and the Jacobian of the canonical 
transformations equals one.
Note that in spite of the non-orthogonality of the angle-action
coordinate system there are still
the definite relations 
between the parallel (or perpendicular) components of the quantities
expressed in terms of the action $S_{\rm CT}(\r',\r'';\vareps)$
in the Cartesian and in the angle-action coordinate system, because of
the conservation of the actions $I_i$ for integrable
Hamiltonians along a trajectory CT \cite{MAFptp2006}. Therefore,
it makes sense to relate the $x$  components $I_x$ and $\Theta_x$ 
and the corresponding $y,z$ components 
of 
actions and the corresponding angles 
to the ``parallel'' and ``perpendicular'' components with
respect to the reference PO in the final trace formula 
(\ref{pstraceactang}), respectively.

PO solutions to the
stationary phase equations (\ref{statcondp}) and (\ref{statcondr})
are also invariants
with respect to the considered canonical transformation as Hamiltonian,
altogether that always can be expressed through both the 
Cartesian,  and the angle-action 
coordinate system 
by using the suitable transformation equations.


\section{THE ETF EFFECTIVE SURFACE APPROACH}

The explicit and accurate analytical expressions for the particle density
distributions were obtained within the nuclear 
ES approximation \cite{strtyap,strmagbr,strmagden}.
They take advantage of the saturation properties of nuclear matter
in the narrow diffuse-edge region in finite heavy nuclei.
The ES is defined as the location of points with a maximal density gradient.
An orthogonal coordinate system related locally to the ES is specified by
the distance $\xi$ of a given point from the ES and a tangent coordinate
parallel to the ES. Using the nuclear energy-density functional
theory \cite{rungegrossPRL1984,marquesgrossARPC2004}, 
one can simplify the variational condition derived from minimization of
the nuclear energy at some fixed integrals of motion in these coordinates
within the leptodermous approximation. In particular, in the 
ETF approach \cite{brguehak}, this can be done in sufficiently heavy nuclei
for any fixed deformation using the expansion in a small parameter
$a/R \sim A^{-1/3} \ll 1$,
where $a$ is of the order of the diffuse edge thickness of the nucleus,
$R$ is the mean-curvature radius of the ES, and
$A$ the number of nucleons. The accuracy of the ES approximation
in the ETF approach was checked \cite{strmagden}
without the spin-orbit (SO) and asymmetry terms by comparing the results
with those of Hartree-Fock (HF) and other ETF
models for some Skyrme forces. The ES approach \cite{strmagden} was also
extended by taking into account the SO and asymmetry effects
\cite{magsangzh,BMVnpae2012,BMRV,BMRPhysScr2014}.

Solutions for the isoscalar and isovector
particle densities and energies in the ES 
approximation of the ETF approach were applied to analytical
calculations of the surface symmetry energy, the neutron skin and isovector
stiffness coefficient in the leading order of the parameter $a/R$
\cite{BMRV}. Our results are compared with older investigations
\cite{myswann69,myswnp80pr96,myswprc77,myswiat85} within the 
LDM and with more recent works
\cite{vretenar1,vretenar2,ponomarev,danielewicz1,pearson,danielewicz2,vinas1,vinas2,vinas4,vretenar3,kievPygmy,nester1,nester2,nester3,vinas5}.

The splitting of the IVDR  into the main and satellite  peaks
  \cite{kievPygmy,nester1,nester2,BMRPhysScr2014,BMRps2015,BMRprc2015,endres} 
was obtained as function of the
isovector surface-energy constant within the FLDM
 \cite{kolmag,kolmagsh} in the ES approach.
The analytical expressions for the surface symmetry-energy 
constants have been tested by the IVDR energies and sum rules
within the FLDM  
\cite{BMRPhysScr2014,BMRps2015,BMRprc2015}
for some Skyrme forces neglecting derivatives of the nongradient
terms in the symmetry energy density per particle with respect to
the mean particle
density. In the present review, following \cite{BMRprc2015},
we shall extend the variational-ES 
method accounting for these derivatives introduced originally 
by Swiatecki and Myers
within the LDM \cite{myswann69}.

In Section IIIA, we give an outlook of the basic points of the ES approximation
within the density-functional theory. The
main results for the isoscalar and isovector
particle densities are presented in Section IIIB with
emphasizing the derivatives of the symmetry energy density per particle.
Section IIIC is devoted to analytical derivations of the
symmetry energy in terms of the surface energy coefficient, the neutron skin
thickness
and the isovector stiffness including these derivatives.
Sections IIID and IIIV are devoted to the collective dynamical description
of the IVDR structure in terms of the response functions and transition
densities. Discussions
of the results are given in
Section IIIVI and summarized at the end of this section.
Some details of calculations are presented in Appendix A.

\subsection{SYMMETRY ENERGY AND PARTICLE DENSITIES}

We start with the nuclear energy $E$ as a functional of
the isoscalar ($\rho_{+}$) and isovector  ($\rho_{-}$) densities
$\rho_{\pm}=\rho_n \pm \rho_p$ in the local density approach
\cite{brguehak,chaban,reinhard,bender,revstonerein,reinhardSV,ehnazarrrein,pastore,jmeyer}:
\begin{equation}\label{energy}
E=\int \d \r\; \rho_{+}\epsi\left(\rho_{+},\rho_{-}\right),
\end{equation}
where
$\epsi\left(\rho_{+},\rho_{-}\right)$
is the energy density per particle,
\bea\l{enerdenin}
&&\hspace{-1.0cm}\epsi\!\left(\rho_{+},\rho_{-}\!\right) \!=\!
- \!b^{}_{\rm V} \!+\! J I^2
\!+\!\vareps_{+}(\rho_{+}) \!+\! \vareps_{-}(\rho_{+},\rho_{-}\!)
\!\nonumber\\
&&+
\left(\frac{\mathcal{C}_{+}}{\rho_{+}} + {\cal D}_{+} +
\frac{\Gamma}{4 \rho_{+}^2}\right)
\left(\nabla \rho_{+}\right)^2 \nonumber\\
&+& \left(\frac{\mathcal{C}_{-}}{\rho_{+}} + {\cal D}_{-} \right)
\left(\nabla \rho_{-}\right)^2.
\eea
Here, $b^{}_{\rm V} \approx$ 16 MeV is the separation energy of a particle,
$J \approx $ 30 MeV is the main volume symmetry-energy constant of
infinite nuclear matter, and $~I=(N-Z)/A$ 
the asymmetry parameter;
$~N=\int \d \r \rho_n(\r)$ and $~Z=\int \d \r \rho_p(\r)$ are
the neutron and proton numbers, and $~A=N+Z$. These constants determine
the first two terms of the volume energy. The last four terms
are surface terms: The first two terms are independent of
the gradients of the particle densities, and the last two ones
depend on these gradients. For the first surface
term independent of the gradients,
$\vareps_{+}$, one  can simply use 
\be\l{epsilonplus0}
\vareps_{+}(\rho_{+})=\frac{K_{+}}{18}\;
e_{+}\left[\eps(w_{+})\right],
\ee
where $K_{+}\approx 220-245$ MeV (see Table 1) is the isoscalar
in-compressibility modulus of symmetric nuclear matter,
$w_{+}$ is the dimensionless isoscalar-particle density,
$w_{+}=\rho_{+}/\overline{\rho}$, and
\be\l{epsilonplus}
e_{+}\left[\eps\left(w_{+}\right)\right]=9 \eps^2 +
I^2\left[\mathcal{S}_{\rm sym}(\eps)-J\right]/K_{+}\;. 
\ee
The small parameter $\eps$,
\be\l{eps}
\eps=\frac{\overline{\rho}-\rho_{+}}{3\overline{\rho}}=\frac{1-w_{+}}{3}\;,
\ee
is used in the expansion,
\be\l{symen}
\mathcal{S}_{\rm sym}(\eps)= J - L \eps +\frac{K_{-}}{2} \eps^2 + \cdots\;,
\ee
around the particle density of infinite nuclear matter
$\overline{\rho}=3/4\pi r_0^3 \approx$ 0.16 fm$^{-3}$, and $r^{}_0$
is the commonly accepted constant in the $A^{1/3}$ dependence of a mean radius.
Several other constants, $L$ and $K_{-}$, 
which were
introduced by Myers and Swiatecki \cite{myswann69}, will be explained below.
The next isovector surface term $\vareps_{-}(\rho_{+},\rho_{-})$
in (\ref{enerdenin})
can be defined
through the same function $\mathcal{S}_{\rm sym}(\eps)$  
(\ref{symen}):
\be\l{surfsymen}
\vareps_{-}\left(\rho_{+},\rho_{-}\right)=
\mathcal{S}_{\rm sym}(\eps)\;
\left(\frac{\rho_{-}}{\rho_{+}}\right)^2 - J I^2.
\ee
For the first and second derivatives of $\mathcal{S}_{\rm sym}(\eps)$ with
respect to $\eps$,
one can take in (\ref{symen})  the derivative values
$L \approx 20 \div 120$ MeV and, even less known,
$K_{-}$ \cite{PR2008,vinas1,vinas5}.
The constants $\mathcal{C}_{\pm}$ and $\mathcal{D}_{\pm}$ in
(\ref{enerdenin}) are defined by the parameters of the Skyrme forces
\cite{brguehak,bender,chaban,reinhardSV,pastore},
\bea\l{Cpm}
\mathcal{C}_{+}&=&
\frac{1}{12} \left(t_1 - \frac{25}{12} t_2 - \frac{5}{3} t_2 x_2\right),
\nonumber\\
\mathcal{C}_{-}&=&-\frac{t_1}{48}\left(1 + \frac{5}{2}x_1\right) -
\frac{t_2}{36} \left(1 + \frac{19}{8} x_2\right).
\eea
The isoscalar SO gradient terms in (\ref{enerdenin}) are defined
with a constant:
$\mathcal{D}_{+} = -9m W_0^2/16 \hbar^2$, where
$W_0 \approx$ 100 - 130 MeV$\cdot$fm$^{5}$ and
$m$ is the nucleon mass.
The constant $\mathcal{D}_{-}$ is usually relatively
small and will be neglected below for simplicity.
Within the ETF, the terms proportional to $\Gamma$ of the gradient 
part in 
(\ref{enerdenin})  is coming from the $\hbar^2$ correction to the TF kinetic 
energy density \cite{sclbook}, $\Gamma=\hbar^2/18 m$ (Appendix D).
Equation (\ref{enerdenin})
can be applied in a semiclassical approximation for a realistic
Skyrme force \cite{chaban,reinhard,bender,revstonerein,ehnazarrrein},
in particular by neglecting higher $\hbar$ corrections
in the ETF kinetic energy \cite{brguehak,strmagbr,strmagden}
and also Coulomb terms.
All of them can be easily taken into account
\cite{strtyap,strmagden,magsangzh} neglecting relatively small 
Coulomb exchange terms.
Such exchange terms can be calculated numerically in extended Slater
approximations~\cite{GU-HQ2013_PRC87-041301}.

The energy density per particle in (\ref{enerdenin}) contains
the first two volume terms, and the surface components
including the new $L$ 
and 
$K_{-}$ derivative corrections of
$\vareps_{-}$ (\ref{surfsymen}), 
along with 
the isoscalar and isovector  density-gradient terms.
Both are important for finite nuclear systems.
These gradient terms, together with the other surface components
in the energy density (\ref{enerdenin}), within the ES approximation
are responsible for the surface tension in finite nuclei.

As usual, we minimize the energy $E$ under the constraints
of fixed particle number $A=\int \d \r\; \rho_{+}(\r)$ and
 neutron excess $N-Z= \int \d \r\; \rho_{-}(\r)$ using
the Lagrange multipliers $\lambda_{+}$ and $\lambda_{-}$ with the
isoscalar and isovector chemical-potential surface corrections
(see Appendix A).
Taking also into account additional deformation constraints
(like the quadrupole moment), our approach can be applied for any deformation
parameter of the nuclear surface, if its diffuseness $a$ is small
with respect to the curvature radius $R$.
Approximate analytical expressions of the binding energy will be obtained
at least up to order $A^{2/3}$. To satisfy the condition of particle
number conservation
with the required accuracy we account for relatively small
surface corrections ($\propto a/R \sim A^{-1/3}$ in first order)
to the leading terms in the Lagrange multipliers
 \cite{strmagbr,strmagden,magsangzh,BMRV}.
We take into account explicitly the diffuseness
of the particle density distributions. Solutions of the
variational Lagrange equations can be derived analytically
for the isoscalar and isovector surface-tension
coefficients (surface energy constants), instead of the
phenomenological constants of the standard LDM \cite{myswann69}
(the neutron and proton particle densities were considered
earlier to be distributions with a strictly sharp edge while,
in the ES approach, the ES is diffused).

\subsection{ISOSCALAR AND ISOVECTOR DENSITIES}

For the isoscalar particle density, $w =\rho_{+}/\overline{\rho}$, one has
up to 
leading terms in the leptodermous
parameter  $a/R$ the usual first-order
differential Lagrange equation  
\cite{strmagden,magsangzh,BMRV,BMRps2015,BMRprc2015}.
 Integrating this equation, one finds the solution:
\be\l{ysolplus}
x=-\int\limits_{w_r}^{w}\d y\; \sqrt{\frac{1 +\beta y}{y\;e_{+}[\eps(y)]}}\;,
\qquad
x=\frac{\xi}{a}\;,
\ee
for $x < x(w=0)$ and $w=0$ for $x \geq x(w=0)$, where $x(w=0)$
is the turning point.
$\beta=\mathcal{D}_{+} \overline{\rho}/\mathcal{C}_{+}$ is the dimensionless SO
parameter, see (\ref{epsilonplus}) for $e_{+}[\eps(y)]$
(for convenience, we often omit the lower index ``$+$'' in 
$w_{+}$ ).
For $w_r=w(x=0)$, one has the boundary condition, $\d^2 w(x)/\d x^2=0$
at the ES ($x=0$):
\be\l{boundcond}
\hspace{-0.2cm}e_{+}\![\eps(w_r)] \!+\! w_r(1 \!+\!\beta w_r)
\left[\frac{\d e_{+}[\eps(w)]}{\d w}\right]_{w=w_r}\!=\!0\;.
\ee

In (\ref{ysolplus}), $a \approx 0.5-0.6$ fm is the
diffuseness (mean-squared) parameter \cite{strmagden,magsangzh,BMRV},
\be\l{adif}
a=\sqrt{\frac{\mathcal{C}_{+} \overline{\rho} K_{+}}{30 b_V^2}},
\ee
found from the asymptotic behavior of the particle density,
$w \sim \exp(-\xi/a)$ for large $\xi$ ($\xi \gg a$).

As shown in \cite{strmagden,magsangzh},  the influence of the
 semiclassical $\hbar$ corrections (related to the ETF
kinetic energy) to $w(x)$ is negligibly small everywhere,
except for the quantum tail outside
the nucleus ($x \simg 1$). 
Therefore, all these corrections were neglected in
(\ref{enerdenin}).
With a good convergence of the expansion of 
$e_{+}[\eps(y)]$ in powers of $1-y$
 up to the quadratic term \cite{strmagden,magsangzh}
and small $I^2$ corrections in (\ref{epsilonplus}),
$e=(1-y)^2$,
one explicitly 
finds analytical solutions of (\ref{ysolplus}) in terms of the
algebraic, trigonometric and
logarithmic functions \cite{BMRV}.
For $\beta=0$ (i.e.\ , without SO terms), it simplifies to
the solution $w(x)=\tanh^2\left[(x-x_0)/2\right]$ for
$x < x_0=2{\rm arctanh}(1/\sqrt{3})$
and zero for $x$ outside the nucleus ($x \geq x_0$). 
As shown in Appendix A1,
for $\mathcal{C}_{\pm}=\mathcal{D}_{\pm}=0$, one obtains the well-know
solution $w(x)=1/[1+\exp(x)]$, symmetrical with respect to the ES, in contrast 
to the results mentioned above for a finite $\beta$.

After simple transformations of the isovector Lagrange equation
(\ref{lagrangeqminus}), one similarly finds up to the leading term in $a/R$  
in the ES approximation
for the isovector density, $w_{-}(x)=\rho_{-}/(\overline{\rho}I)$, the
equation and the boundary condition (\ref{yeq0minus}).
The analytical solution $w_{-}=w \cos[\psi(w)]$
can be obtained through the expansion (\ref{powser}) of $\psi$ in powers of
\be\l{csymwt}
\gamma(w)=\frac{3\eps}{c_{\rm sym}},\qquad
c_{\rm sym}=a \;\sqrt{\frac{J}{\overline{\rho}\;
\vert\mathcal{C}_{-}\vert}}\;.
\ee
Expanding up to the second order in $\gamma$, one obtains 
(Appendix A1)
\be\l{ysolminus}
\hspace{-0.3cm}w_{-}=  w\;\cos\left[\psi(w)\right]
\approx
w \left(1- \frac{\psi^2(w)}{2} \!+\! \cdots
\!\right),
\ee
with
\be\l{psi2}
\psi(w)=\frac{\gamma(w)}{\sqrt{1+\beta}}\;
\left[1 + \widetilde{c} \gamma(w) + \cdots\right],
\ee
\be\l{ctilde}
\widetilde{c}=\frac{\beta c_{\rm sym}^2 + 2 +
c_{\rm sym}^2 L (1+\beta)/(3J)}{2 c_{\rm sym}(1+\beta)};
\ee
see also the constant $c_3$ 
at higher (third) 
order corrections.
Notice 
that $w_{-}$ depends on $L$ in second order
in $\gamma$ but it is independent of $K_{-}$ at this order.

In Fig.\ \ref{fig2}, 
the $L$ dependence of the
function $w_{-}(x)$ is shown within approximately the total interval from
$L=0$ to $L=100$  MeV
\cite{vinas2}, and it is  compared to
that of the density $w(x)$ for the SLy5* force as a typical example.
As  shown in Fig.\ \ref{fig3} 
in a
larger (logarithmic) scale,
one observes notable differences in the isovector densities $w_{-}$ derived
from different Skyrme forces \cite{chaban,reinhardSV}
within the edge diffuseness. All these calculations
have been done with the finite
proper value of the slope parameter $L$. For
SLy forces this value is taken from 
\cite{jmeyer}, for $SGII$ from \cite{vinas2} and for others from
\cite{reinhardSV} (Table 1).  As shown below, this is in particular important
for calculations of the neutron skin of nuclei.
Notice that, with the precision of line thickness,
our results are almost the same
taking approximately $L=50$ MeV for SLy5* and $L=60$ MeV for SVsym32.
Note also that, up to second order in the small parameter $\gamma$,
the isovector particle density $w_{-}$ in (\ref{ysolminus})
does not depend on the symmetry energy in-compressibility $K_{-}$.
The $K_{-}$ dependence appears only at higher (third) order terms in the
expansion in $\gamma$ (Appendix A1).
Therefore, as a first step of the iteration procedure, it is possible to study
first the main slope effects of $L$ neglecting small $I^2$ corrections to the
isoscalar particle density $w_{+}$ (\ref{ysolplus}) through $e_{+}$
(\ref{epsilonplus}).
Then, we may study more precisely the effect of the second derivatives
$K_{-}$ taking into account higher order terms.

We emphasize that the dimensionless densities, $w(x)$
(see (\ref{ysolplus}) and \cite{BMRV,BMRps2015}) and $w_{-}(x)$
(\ref{ysolminus}), shown in Figs.\ 
\ref{fig2} and \ref{fig3}, 
were
obtained in the leading 
ES approximation ($a/R \ll 1$) as functions of
specific combinations
of Skyrme force parameters  as 
$\beta$ and $ c_{\rm sym}$
(\ref{csymwt}) accounting for the $L$-dependence
 (\ref{ctilde}).
These densities are at 
leading order in the leptodermous parameter
$a/R$ approximately universal functions, independent of the
properties of 
specific nucleus.
It yields largely the local density distributions in the
normal-to-ES direction $\xi$
with the correct asymptotic behavior outside of the deformed
ES layer  at $a/R \ll 1$,
as it is the case for semi-infinite nuclear matter.
Therefore, at the dominating order,
 the particle densities $w_{\pm}$
are {\it universal} distributions independent
of the specific properties of the nucleus while higher order 
corrections
to the densities $w_{\pm}$
depend, indeed, on its specific macroscopic 
properties. 

\subsection{ISOVECTOR ENERGY AND STIFFNESS}

The nuclear energy $E$ [equation 
(\ref{energy})]
in the improved ES approximation (Appendix A3) 
is split into the volume and
surface terms \cite{BMRV,BMRprc2015},
\be\l{EvEs}
E \approx -b^{}_{\rm V}\; A + J (N-Z)^2/A + E_{\rm S}\;.
\ee
For the surface energy $E_{\rm S}$ one obtains
\be\l{Es}
E_{\rm S} = E_{\rm S}^{(+)} + E_{\rm S}^{(-)}
\ee
with the
isoscalar (+) and isovector (-) surface components:
\be\l{Espm}
 E_{\rm S}^{(\pm)}= b_{\rm S}^{(\pm)} \frac{\mathcal{S}}{4\pi r_0^2}\;,
\ee
where $\mathcal{S}$ is the surface area of the ES, $b_{\rm S}^{(\pm)}$
are the
isoscalar $(+)$ and isovector $(-)$ surface-energy constants,
\be\l{sigma}
\hspace{-0.4cm}b_{\rm S}^{(\pm)} \approx 8 \pi r_0^2 \mathcal{C}_{\pm}
\!\int_{-\infty}^{\infty} \!\d \xi\!
\left(1 + \frac{\mathcal{D}_{\pm}}{\mathcal{C}_{\pm}} \rho_{+}\right)
\left(\frac{\partial \rho_{\pm}}{\partial \xi}\right)^2.
\ee
These constants are proportional to the corresponding surface tension
coefficients
$\sigma_{\pm}=b_{\rm S}^{(\pm)}/(4 \pi r_0^2)$ through the
solutions (\ref{ysolplus}) and (\ref{ysolminus})
for $\rho_{\pm}(\xi)$, which can be taken into account in leading
order of $a/R$ (Appendix A) . 
These
coefficients $\sigma_{\pm}$ are the same as found in the
expressions for  
capillary pressures of the macroscopic
boundary conditions; see Appendix A2, and 
\cite{strmagbr,strmagden,magsangzh,BMRV} 
with new values $\vareps_{\pm}$ modified
by $L$ and $K_{-}$ derivative corrections of 
(\ref{epsilonplus}) and (\ref{surfsymen}), 
also
 \cite{BMRps2015,BMRprc2015}).
Within the improved ES approximation where 
higher order corrections in the small parameter $a/R$
are taken into account, we derived in \cite{BMRV} equations for the
nuclear surface itself (see also \cite{strmagbr,strmagden,magsangzh}).
For more exact isoscalar and isovector particle densities we account for the
main terms at  
next order of the parameter $a/R$
in the Lagrange equations [see (\ref{lagrangeqminus}) for the isovector and
\cite{strmagbr,strmagden,magsangzh} for the isoscalar case].
Multiplying these equations by the corresponding 
$\partial \rho_{\pm}/\partial \xi$
and integrating them over the ES
in the normal-to-surface direction $\xi$ and using the
solutions for $w_{\pm}(x)$ up to the leading orders
[(\ref{ysolplus}) and
(\ref{ysolminus})],
one arrives at the ES equations in the form of
the macroscopic boundary conditions (Appendix A2 and
\cite{strmagbr,strmagden,magsangzh,kolmagsh,magstr,magboundcond,bormot,BMRV}.
They ensure equilibrium through the equivalence of the volume and
surface (capillary) pressure variations.
As shown  in Appendix A2, the
latter ones are proportional to the corresponding
surface tension coefficients $\sigma_{\pm}$.

For the energy surface coefficients $b_{\rm S}^{(\pm)}$ (\ref{sigma}), one obtains
\bea\l{bsplus}
b_{\rm S}^{(+)}&=& 6 \mathcal{C}_{+}
\overline{\rho} \mathcal{J}_{+}/( r_0 a),\nonumber\\
\mathcal{J}_{+}&=&\int\limits_0^1 \d w
\sqrt{w (1+\beta w)\; e_{+}[\eps(w)]},
\eea
\be\l{bsminus}
b_{\rm S}^{(-)}=k^{}_S\; I^2,\qquad
k^{}_{\rm S}= 6 \overline{\rho}\; \mathcal{C}_{-}\;\mathcal{J}_{-}/(r_0 a),
\ee
\bea\l{jminus}
&&\mathcal{J}_{-}=\int\limits_0^1 \d w\;
\sqrt{\frac{w\; e_{+}[\eps(w)]}{1+\beta w}} \left\{\cos(\psi)
\; \right. \nonumber\\
&+&\left. \frac{w\sin(\psi)}{c_{\rm sym}\sqrt{1+\beta}}
\left[1+2 \widetilde{c} \gamma(w)\right]\right\}^2
\approx\int_0^1 \d w (1-w) \nonumber\\
&\times&
\sqrt{\frac{w}{1+\beta w}}\;
\left\{1+\frac{2\gamma(w)}{c_{\rm sym}(1+\beta)} + 
\left(\frac{\gamma}{1+\beta}\right)^2  \right. \nonumber\\
&\times&\left.\left[\frac{1}{c_{\rm sym}^2} +
6 (1+\beta)
\left(\frac{\widetilde{c}}{c_{\rm sym}} -
\frac12\right)\right]\right\}, 
\eea
see (\ref{csymwt}) and (\ref{ctilde}) for $\gamma$ and $\widetilde{c}$,
respectively. Simple expressions for  the constants
$b_{\rm S}^{(\pm)}$ in (\ref{bsplus}) and
(\ref{bsminus}) can be easily derived explicitly  in terms of
algebraic and trigonometric functions by calculating analytically
integrals over $w$
for the quadratic form of $e_{+}[\eps(w)]$ [(\ref{Jp}) and (\ref{Jm})].
Note that in these derivations,  we neglected curvature terms
and, being of the same order,
shell corrections, which have been discarded from the very beginning
of Section III.
The isovector energy-density terms were obtained within the ES
approximation with high accuracy up to the product of two
small quantities, $I^2$ and $(a/R)^2$.

According to the macroscopic
theory \cite{myswann69,myswnp80pr96,myswprc77,BMRV,BMRprc2015},
one may define the isovector stiffness $Q$ with respect to the difference
$R_n-R_p$
between the neutron and proton radii as a dimensionless collective variable
$\tau$,
\bea\l{Esq}
E_{\rm S}^{(-)}&=&-\frac{\overline{\rho} r_0}{3} \oint
\d S\;Q \tau^2 \approx
-\frac{Q\tau^2 \mathcal{S}}{4 \pi r_0^2}\;,\nonumber\\
  \tau&=&\left(R_n-R_p\right)/r_0\;,
\eea
where $\tau$ is the relative neutron skin.
Comparing this expression to equation (\ref{Espm})
for the isovector surface energy written through
the isovector surface-energy constant
$b_{\rm S}^{(-)}$ (\ref{bsminus}), one obtains
\be\l{stifminus}
Q=-k^{}_{\rm S}\frac{I^2}{\tau^2}\;.
\ee
Defining the neutron and proton radii $R_{n,p}$ as positions of
maxima of the neutron and proton density gradients, respectively, one finds
 the neutron skin $\tau$ \cite{BMRV,BMRprc2015},
\be\l{skin}
\tau=\frac{8 a I}{r_0 c_{\rm sym}^2}\varsigma(w_r),
\ee
where
\bea\l{fw}
\hspace{-0.5cm}\varsigma(w)\!&=&\!\frac{w^{3/2}(1+\beta w)^{5/2}}{
(1\!+\!\beta)(3w\!+\!1\!+\!4 \beta w)}\left\{w
(1+2\widetilde{c} \gamma)^2 \! \right.\nonumber\\
&+&\left. 2\gamma  \left(1+\widetilde{c}
\gamma\right) \left[\widetilde{c}
w - c_{\rm sym}\left(1+
2 \widetilde{c} \gamma\right)\right]\right\} 
\eea
is taken at the ES value $w_{r}$  (\ref{boundcond}).
Finally, taking into account equations (\ref{stifminus}) and (\ref{bsminus}),
one arrives at
\be\l{stiffin}
Q=-\nu\; \frac{J^2}{k^{}_{\rm S}}, \qquad
\nu=\frac{k_{\rm S}^2 I^2}{\tau^2 J^2}=
\frac{9 \mathcal{J}_{-}^2}{16 \varsigma^2(w_r)}\;,
\ee
where $\mathcal{J}_{-}$ and $\varsigma(w)$ are given by (\ref{jminus})
and (\ref{fw}), respectively.
Note that $Q=-9J^2/4k^{}_{\rm S}$ has been predicted in
\cite{myswann69,myswnp80pr96}:  
For $\nu=9/4$, 
the first part of (\ref{stiffin}), which relates $Q$ with
the volume symmetry energy $J$, and the isovector surface-energy
constant $k^{}_{\rm S}$,
is identical to that used in
\cite{myswann69,myswnp80pr96,myswprc77,myswiat85,vinas1,vinas2}.
However, in our derivations $\nu$ deviates from $9/4$, and it is
proportional to the function
$\mathcal{J}_{-}^2/\varsigma^2(w_r)$. This function depends
significantly on the SO interaction
parameter while $\beta$ is approximately insensitive 
on the specific Skyrme force 
\cite{BMRV}.

The approximate universal functions $w(x)$,
((\ref{ysolplus})
and \cite{BMRV}), and $w_{-}(x)$
(\ref{ysolminus}) can be used in the leading order of the ES
approximation for calculations of the surface
energy coefficients
$b_{S}^{(\pm)}$ (\ref{sigma}), and 
the neutron skin $\tau \propto I$
(\ref{skin}). As shown
in \cite{BMRV} and in Appendix A3, here  only the particle
density distributions  $w(x)$  and $w_{-}(x)$ are needed
within the surface layer through their
derivatives.  The lower limit of the integration
over $\xi$ in (\ref{sigma}) can be then approximately extended to
$-\infty$ because there are no contributions from the internal volume region
in evaluations 
of the main surface terms of the pressure and energy.
Therefore,  the surface symmetry-energy coefficient
$k^{}_S$ in (\ref{bsminus}) and (\ref{Jm}) , the neutron skin
$\tau$ (\ref{skin}), and the
isovector stiffness
$Q$ (\ref{stiffin})
can be approximated analytically in terms of functions
of definite critical combinations of the Skyrme parameters  
$\beta$,
$c_{\rm sym}$, $a$, $\mathcal{C}_{-}$ and parameters of infinite nuclear matter
($b_{\rm S}$, $\overline{\rho}$, and $K_{+}$), 
also the symmetry energy constants
$J$, $L$ and $K_{-}$. Thus, in the considered ES approximation,
they do not depend on the specific properties
of the nucleus (for instance, the neutron and proton numbers),
the curvature and the deformation of the nuclear surface.

\subsection{THE FERMI-LIQUID DROPLET MODEL}
\l{fldm}

For IVDR calculations,  the
FLD model based on the linearized
Landau-Vlasov equations for the isoscalar [$ \delta f^{}_{+}(\r,\p,t)$]
and isovector [$ \delta f^{}_{-}(\r,\p,t)$] distribution functions
can be used  in phase space
\cite{kolmagsh,belyaev},
\bea\l{LVeq}
\frac{\partial \delta f^{}_{\pm}}{\partial t}
&+&
    \frac{\p}{m_\pm^*}
   {\bf \nabla}_r  \left[ \delta f^{}_{\pm} \;\right. \nonumber\\
&+& \left.\delta \left(\e-\e^{}_F\right)
    \left(\delta V_{\pm}
    + V_{\rm ext}^{\pm}\right)\right]
    =\delta St^{}_{\pm}.
\eea
Here $\e=p^2/(2m_\pm^*)$ is the equilibrium quasiparticle energy ($p=|\p|$)
and $\e_{{}_{\! F}}=(p_F^{\pm})^2/(2m_\pm^*)$ is the Fermi energy.
The isotopic dependence of the Fermi momenta
$p_F^{\pm}=p^{}_F\left(1 \mp \Delta\right)~$
is given by a small parameter
$\Delta=2\left(1+F_{0}'\right)\;I/3~$.
The reason of having $\Delta$
is the difference between
the neutron and proton potential depths because of
the Coulomb interaction.
 The isotropic isoscalar $F_0$  and isovector
 $F_0'$ Landau interaction constants are related
to the isoscalar in-compressibility $K=6\e_{{}_{\! F}}(1+F_0)$
and the volume symmetry energy $J=2\e_{{}_{\! F}}(1+F_0')/3 $
constants of nuclear matter, respectively. The effective masses
$m_{+}^*=m(1+F_1/3)$ and $m_{-}^*=m(1+F_1^\prime/3)$ are determined in terms of
the nucleon mass $m$ by anisotropic Landau constants $F_1$ and
$F_1^\prime$.  Equations (\ref{LVeq}) are coupled
by  the dynamical variation
of the quasiparticles' self-consistent interaction  $ \delta V_{\pm}$
with respect to the equilibrium
value $p^2/(2m_\pm^*)$. The   time-dependent external field
$V_{\rm ext}^{\pm} \propto \exp(-i \omega t)$
is periodic with a frequency $\omega$. For simplicity,
the collision term $\delta St_{\pm}$ is calculated
within the relaxation time $\mathcal{T}(\om)$ approximation
accounting for the retardation effects
due to the energy-dependent self-energy
beyond the mean-field approach,
$\mathcal{T}=4 \pi^2 \mathcal{T}_{0}/(\hbar \om)^2$
with the parameter $\mathcal{T}_{0} \propto A^{-1/3}$
(see (80) of \cite{belyaev} at zero temperature and 
\cite{kolmagsh}]).

The solutions of equations (\ref{LVeq}) are related to the dynamic multipole
particle-density variations, $\delta \rho^{}_{\pm}(\r,t) \propto
Y_{\lambda 0}(\hat{r})$, where $Y_{\lambda 0}(\hat{r})$ are
the spherical harmonics and
$\hat{r}=\r/r$.
These solutions can be found in terms of the
superposition of plane waves over the angle
of a wave  vector
${\bf q}~$,
\bea\l{plwave}
\hspace{-0.5cm}\delta f_{\pm}(\p,\r,t)
\!&=&\!
\int \d \Om_{{\bf q}} Y_{\lambda 0}(\hat{q})\;
\delta f_{\pm}(\p,\q,\om)\; \nonumber\\
&\times& \exp\left[-i\left(\om t - {\bf q} \r \right)\right]\;, 
\eea
where $\delta f_{\pm}(\p,\q,\om)$ is the Fourier transform of the
distribution function.
The time dependence (\ref{plwave}) is periodic as the external field
$V_{\rm ext}^{\pm}$ is also periodic with the same frequency
$\om=p_F^\pm s^\pm q/m_{\pm}^*$, where
$s^{+}=s$ and $s^{-}=s \left(NZ/A^2\right)^{1/2}$.
The factor $\left(NZ/A^2\right)^{1/2}$ accounts for conserving
the position of the mass center
for the isovector vibrations \cite{eisgrei}.
The sound velocity $s$ can be found from the dispersion equations
 \cite{kolmagsh}. The two solutions $s_n$ with $n=1,2$ are functions of
the Landau interaction constants and
$\omega \mathcal{T}$. 
The ``out-of-phase''
particle-density vibrations of the $s_1$ mode involve the ``in-phase'' mode
 $s_2$ inside of  
nucleus because of the symmetry interaction coupling.

For small isovector- and isoscalar-multipole
ES-radius vibrations of the finite neutron and proton
Fermi-liquid drops around the
spherical nuclear shape, one has
$\delta R_{\pm}(t) = R \alpha_S^{\pm}(t) Y_{\lambda 0} ({\hat r})\;$ with
 a small time-dependent amplitudes
$\alpha_{\rm S}^{\pm}(t) = \alpha_{\rm S}^\pm \exp(-i \omega t)$.
The macroscopic boundary conditions
(surface continuity and force-equilibrium equations) at the ES  
are given by
\cite{BMRV,kolmagsh,belyaev}:
\bea\l{bound2}
u_{r}^{\pm}\Big|_{r=R} &=&
R \dot{\alpha}_{\rm S}^{\pm} Y_{\lambda 0}({\widehat r}),\nonumber\\
\delta \Pi_{rr}^{\pm}\Big|_{r=R} &=&
\alpha_{\rm S}^\pm \overline{P}_{\rm S}^\pm\;Y_{\lambda 0}({\widehat r})\;.
\eea
The left hand sides (LHS) of these equations
are the radial components
of the mean-velocity field ${\bf u}={\bf j}/\rho$  (${\bf j}$ is
the current density)
and the momentum flux tensor $\delta \Pi_{\nu\mu}$ defined both
through the moments of $\delta f(\r,\p,t)$ in momentum space
\cite{kolmagsh,belyaev}. The RHS of (\ref{bound2})
are the ES velocities and capillary pressures.
These pressures are proportional to the isoscalar and isovector
surface-energy constants $b_{\rm S}^{\pm}$ in (\ref{sigma}),
\be\l{pressuresurf}
\overline{P}_{\rm S}^{\pm}=\frac23 \;
b_{\rm S}^{\pm} \;
\overline{\rho}\; \mathcal{P}_{\pm}\;  A^{\mp1/3},
\ee
where $\mathcal{P}_{+}= (\lambda -1)(\lambda + 2)/2~$, $\mathcal{P}_{-}= 1~$.
The coefficients $b_S^{\pm}$ are essentially determined by the constants
$\mathcal{C}_{\pm}$  (\ref{Cpm})
of the energy density (\ref{enerdenin}) in front of its gradient density terms.
The conservation of the center of mass is taken into
account in the derivations of the second boundary conditions
(\ref{bound2}) \cite{kolmagsh,belyaev}. Therefore,
one has a dynamical equilibrium of the forces acting at the ES.

\subsection{TRANSITION DENSITY AND NUCLEAR RESPONSE}

The response function,
$\chi_{\pm}(\om)$,
is defined as a linear reaction to the external
single-particle field $\hat{Q}(\r)$
with the frequency $\omega$.
For convenience, we may consider this field in terms of a similar
superposition of plane waves (\ref{plwave}) as
$\delta f_{\pm}$ \cite{kolmagsh,belyaev}.
 In the following, we will consider the
long wave-length limit with
\begin{equation}\label{vextpm}
\mathcal{V}_{\rm ext}^{\pm}(\r,t)=\alpha_{\rm ext}^{\pm,\om}(t) \hat{Q}(\r)\;,
\end{equation}
where
\begin{equation}\label{qextpm}
\alpha_{\rm ext}^{\pm,\om}(t)=\alpha_{\rm ext}^{\pm,\om}~e^{-i(\om+i0)t}\;,
\end{equation}
$\alpha_{\rm ext}^{\pm,\om}$  
the amplitude, and $\omega$ the frequency
of the external field. 
In this limit, the s.p.\  
operator  $\hat{Q}(\r)$ becomes the 
standard multipole
operator, $\hat{Q}(\r)=r^{\lambda}Y_{\lambda 0}(\hat{r})$ for $\lambda \geq 1$.
The response function $\chi_{\pm}(\om)$ is expressed through the
Fourier transform of the transition density
$\rho_{\pm}^{\om}(\r)$ as
\begin{equation}
\chi_{\pm}(\om)=
-\int \d\r\;\hat{Q}(\r)\;
\rho_{\pm}^{\om}(\r)/\alpha_{\rm ext}^{\pm,\om}.
\label{chicollrho}
\end{equation}
The transition density $\rho_{\pm}^{\om}(\r)$ is obtained
through the
dynamical part of the particle density  $\delta \rho_{\pm}(\r,t)$ in a
macroscopic model
in terms of solutions $\delta f_\pm(\r,\p,t)$
of the Landau-Vlasov equations
(\ref{LVeq}) with the boundary conditions (\ref{bound2})
as the same superpositions
of plane waves  (\ref{plwave})
\cite{kolmagsh}:
\begin{equation}
\delta \rho_{-}(\r,t) = \overline{\rho}\; \alpha^{-}_S
\rho_{-}^{\om}(x) \;Y_{10}(\hat{r})\;e^{-i \om t},
\label{drhomrt}
\end{equation}
where
\be\l{drhom}
\rho_{-}^{\om}(x)=\frac{qR}{j_{1}'(qR)}
\left[j^{}_{1}\left(\kappa\right)w(x)
+\frac{g_{{}_{\! V}}}{g_{{}_{\! S}}}\frac{\d w_{-}}{\d x}
\right],
\ee
\bea\l{gv}
g_{{}_{\! V}}&=&\int\limits_0^{w_0}
\d w 
\frac{\sqrt{w(1+\beta w)}}{1-w}\kappa^3 j_{{}_{\! 1}}(\kappa),\\
\l{gs}
g_{{}_{\! S}}&=&\int\limits_0^{w_0} \d w\;
\kappa^3\left[1 + \mathcal{O}(\gamma^2(w))
\right],\\
 \kappa&=&\kappa_o\left[1+\frac{a}{R} x(w)\right],\quad \kappa_o=qR\;.
\eea
The first term in (\ref{drhom}), proportional
to the dimensionless isoscalar density $w(x)$ 
[(\ref{ysolplus}), in units of $\overline{\rho}$]
accounts for volume density vibrations. 
The second term $\propto dw_{-}/dx$, where $w_{-}$
is a dimensionless isovector density 
[(\ref{ysolminus}), in units of $\overline{\rho} I$]
corresponds to the density variations due to
a shift of the ES. The particle number
and the center-of-mass position are conserved. 
In (\ref{gv}) and (\ref{gs}),
$j_\lambda(\kappa)$ and $j_\lambda'(\kappa)$ are the spherical Bessel functions
and their derivatives.
The upper integration limit $w_{{}_{\! 0}}$ in (\ref{gv})
and (\ref{gs}) is defined
as the root of a transcendent equation $x(w_{0})+R/a=0$.
As shown  
in Appendix A1 \cite{BMRprc2015}, 
the SO and $L$ dependent density $w_{-}(x)$
is of the same order as $w(x)$.
The dependencies of  $w_{-}(x)$ on
different Skyrme force parameters, mostly the isovector gradient-term constant
$\mathcal{C}_{-}$, the SO parameter $\beta$, and the
derivative of the volume symmetry energy $L$ are
the main reasons for the 
values of the neutron skin.

With the help of 
boundary conditions (\ref{bound2}), one can derive the response
function (\ref{chicollrho}) \cite{kolmagsh},
\bea\l{respfuni}
&&\chi^{}_{\lambda}(\om)=\sum_n \chi_{\lambda}^{(n)}(\omega)
\nonumber\\
&&=\sum_n \mathcal{A}_{\lambda}^{(n)}(\kappa_o)/ \mathcal{D}_{\lambda}^{(n)}
\left(\om - i\frac{\Gamma}{2}\right)\;, 
\eea
with $\om=p^{}_Fs_n \kappa_o \left(NZ/A^2\right)^{1/2}/(m^*R)~$
($m_{-}^*\approx m_{+}^*=m^*$).
This response function describes two modes, the main ($n=1$)
IVDR and its satellite  ($n=2$)
as related to the out-of-phase $s_1$ and in-phase $s_2$ sound velocities,
respectively. We assume here that the ``main''
peak exhausts mostly
the 
energy weighted sum rule (EWSR), and the ``satellite'' corresponds
to a 
significantly smaller part of the EWSR. This two-peak structure is due to
the coupling of the
isovector and isoscalar density-volume vibrations because of
 the neutron and proton
quasiparticle interaction $\delta V_{\pm}$ in (\ref{LVeq}).
Therefore, one takes into account
an admixture of the isoscalar mode to the isovector IVDR excitation.
The wave numbers $q=\kappa_o/R$ of the lowest poles ($n=1,2$)
 in the response function (\ref{respfuni})
are determined by the secular equation,
\bea\l{seculeq}
&&\mathcal{D}_{\lambda}^{(n)} \equiv  j_{\lambda}'(\kappa_o)
\nonumber\\
&&-\frac{3 \e_{{}_{\! F}}\kappa_oc_1^{(n)}}{2 b_S^{-} A^{1/3}}
\left[j_{\lambda}(\kappa_o) + c_2^{(n)}j_{\lambda}''(\kappa_o)\right]
=0\;.
\eea
The width of an IVDR peak $\Gamma$ in (\ref{respfuni})
corresponds to an imaginary part
of the pole having its origin in the  collision term
$\delta St_{\pm}$
of the Landau-Vlasov equations. At this pole, for the relaxation time
one has
\be\l{relaxtimen}
\mathcal{T}_n=4 \pi^2 \mathcal{T}_{0}/(\hbar \om_n)^2
\ee
with an $A$ dependent constant, $\mathcal{T}_0 \propto A^{-1/3}$.
For the amplitudes one finds  
$\mathcal{A}_{\lambda}^{(n)} \propto \Delta^{n-1}$.
The complete expressions for  
amplitudes $\mathcal{A}_{\lambda}^{(n)}$ and 
constants $c_i^{(n)}$ are given in \cite{kolmagsh,belyaev}.
Assuming a small value of $\Delta$, one may
call the $n=2$ mode as a ``satellite'' (or some kind 
of the pygmy resonance) in comparison
with the ``main'' $n=1$ peak.
On the other hand, other
factors such a  collisional relaxation  time,
 the surface symmetry energy constant $b_{S}^{-}$, and the particle
number $A$ lead sometimes to a re-distribution
of the EWSR values among these two IVDR peaks. The slope $L$ dependence
of the transition densities $\rho_{-}^{\om}(x)$ (\ref{drhom}), and
the strength of the response function (\ref{respfuni}),
\be\l{strength}
S(\omega)=\Im \chi^{}_{\lambda}(\om)/\pi
\ee
have its origin in the symmetry energy
coefficient $b_S^{(-)}$ (\ref{bsminus}) and (\ref{jminus}); see also
(\ref{epsilonplus}), (\ref{symen}), and (\ref{ctilde}).  Thus,
one may evaluate the EWSR 
contribution of the $n$th peak by
integration
over the region $\hbar \Delta \omega$ around the peak energy
$E_n=\hbar \om_n$,
\be\l{sumruleewsr}
S_n^{(1)}= 
\hbar^2\int  
\d \omega\; \omega\;S_n(\omega).
\ee

In accordance with the time-dependent HF approaches based on the 
Skyrme forces, see for instance \cite{nester1,nester2,kievPygmy},
we may expect that the energies of the satellite resonances in the IVDR and 
ISDR channels can be close. Therefore, we may calculate separately the
neutron, $\rho_n^{\omega}$, and proton, $\rho_p^{\omega}$, transition densities
for the satellite by calculating the isovector and isoscalar transition 
densities at the same energy $E_2~$,
\bea\l{rhonp}
\rho^{\om}_{n}(x)&=&
\left[\rho_{+}^{\om}(x) + \rho_{-}^{\om}(x)\right]/2\;,\nonumber\\
\rho^{\om}_{p}(x)&=&
\left[\rho_{+}^{\om}(x) - \rho_{-}^{\om}(x)\right]/2\;.
\eea

\subsection{DISCUSSIONS OF RESULTS}

In Table 2 we show the isovector surface-energy 
coefficient $k^{}_{\rm S}$  (\ref{bsminus}),
the stiffness parameter $Q$ (\ref{stiffin}), its constant $\nu$ and
the neutron skin $\tau$ (\ref{skin}) \cite{BMRprc2015}. 
They are obtained within the ES
approximation with the quadratic expansion for $e_{+}[\eps(w)]$ and neglecting
the $I^2$ slope corrections, for several Skyrme forces
\cite{chaban,reinhard} whose parameters are presented in Table 1.
Also shown are the quantities $k_{{\rm S}\;0}$, $\nu_0$, $Q_0$ and $\tau_0$
neglecting the slope corrections ($L=0, K_{-}=0$). This is in addition to
results of \cite{BMRV} where another important dependence
on the SO interaction measured by $\beta$ was presented.
In contrast to a fairly good agreement for the analytical isoscalar
surface-energy constant $b_{\rm S}^{(+)}$ (\ref{bsplus}), 
the isovector surface-energy coefficient $k^{}_{\rm S}$ is more sensitive to the choice
of the  Skyrme forces than the isoscalar one $b_{\rm S}^{(+)}$ 
\cite{magsangzh,BMRV}.
The modulus of  $k^{}_{\rm S}$  is significantly larger for most of
the Skyrme forces SLy...
\cite{chaban} and SV... \cite{reinhard} than for the other ones. However,
the $L$ dependence of $k^{}_{\rm S}$ is 
not strong
(cf. the first two rows of Table 2,  
where low subscript shows the quantities
obtained with $L=0$)
as it should be for  a small parameter $\eps$ of the symmetry energy density
expansion (\ref{symen}). For SLy and SV forces,
the skin stiffnesses $Q$ 
are correspondingly significantly smaller in absolute value
being closer to the well-known empirical
values $Q \approx 30-35$ MeV \cite{myswprc77,myswnp80pr96,myswiat85}
obtained by Swiatecki and collaborators.
Note that the isovector stiffness $Q$ is even much more sensitive
to the parametrization of the Skyrme force and
to the slope parameter $L$ than the
constants $k^{}_{\rm S}$. In \cite{BMRV}, we studied the
hydrodynamical results for $Q$
as compared to the FLDM for the averaged properties of  the giant
IVDR (IVGDR) at zero slope $L=0$.
The IVDR structure in terms of the two (main and satellite)
peaks was discussed in \cite{BMRPhysScr2014,BMRps2015,BMRprc2015}  
in some magic nuclei  with a large neutron excess within the semiclassical FLDM
based on the effective surface approach.
For the comparison with experimental data and other theoretical results
we present in Table 2 (rows 9 and 11) a small $L$ 
dependence of the
IVGDR energy parameter
\be\l{Dconst}
D=E_{\rm IVGDR}\;A^{1/3}\,
\ee
where $E_{\rm IVGDR}$ is the IVGDR energy 
averaged over the strength distribution $S_n$ for a given nucleus,
\be\l{Eivdgdr}
E_{\rm IVGDR}=\frac{E_1 S_1(\om^{}_1) +E_2 S_2(\om^{}_2)}{
S_1(\om^{}_1) + S_2(\om^{}_2)}
\ee
[see also
(\ref{strength}) for the definition of the strength $S(\om)$, 
$D_0$ is 
obtained with $L=0$].
A more precise reproduction of the $A$-dependence of the IVGDR energy
parameter $D$
for finite values of $L$ (see the last three rows for several isotopes)
might determine more consistent values of $Q$, but, at present, it seems
to be beyond the accuracy of both the hydrodynamical (HD) and the FLD models.
The IVGDR energies obtained by solving the semiclassical Landau-Vlasov 
equations
(\ref{LVeq}) with the macroscopic FLDM boundary conditions (\ref{bound2})
\cite{BMRV} are also basically insensitive
to the isovector surface-energy constant $k^{}_{\rm S}$ 
\cite{BMVnpae2012,BMRV,BMRPhysScr2014,BMRprc2015}.
They are in a good agreement with the experimental data, and do not depend much
on the Skyrme forces, even if we take into account the symmetry energy
slope $L$ (last three rows in Table 2).

More realistic self-consistent  HF calculations taking into account the Coulomb
interaction, the surface-curvature, and quantum-shell effects have led to
larger
values of $Q\approx 30-80$ MeV \cite{brguehak,vinas2}.
For larger $Q$ (Table 2)
the fundamental parameter $(9 J /4Q) A^{-1/3}$ of the LDM expansion
in \cite{myswann69} is really small for $A \simg 40$, and therefore,
the results obtained using  the leptodermous expansion are better
justified.

An investigation within the ES approach 
shows that the IVDR strength
is split into a main peak which
exhausts an essential part of the EWSR independent of the model
and a satellite peak
with a much smaller contribution into this quantity Figs.\ 
\ref{fig4}--\ref{fig6}.  Focusing on a 
more
sensitive  $k_S$ dependence of the 
IVDR satellite 
resonances,
one may take now into account the slope $L$ dependence of the
symmetry energy density (\ref{symen})
\cite{kievPygmy,nester1,nester2,BMRps2015}.
The total IVDR strength function, being respectively
the sum of the  ``out-of-phase'' $n=1$ and
``in-phase'' $n=2$ modes for the isovector- and
isoscalar-like 
particle density vibrations in the nuclear volume, respectively
(solid lines in Figs.\ \ref{fig4} and \ref{fig5} 
for the zero $L$, and dotted and dashed ones for the finite $L$),
has a rather remarkable shape asymmetry 
\cite{BMRPhysScr2014,BMRps2015,BMRprc2015}.
For SLy5$^*$ (Fig.\ \ref{fig4}) and  for SVsym32 (Fig.\ \ref{fig5}) 
one has the ``in-phase'' satellite to the right of the
main ``out-of-phase'' peak. An enhancement to the left of the main peak
for SLy5* is due to the increasing of the
"out-of-phase" strength (rare dotted curve in Fig.\ \ref{fig4})
at small energies
because of appearance of a peak at the energy about a few
MeV, in contrast to the SVsym32 case. The semiclassical FLDM
calculations at the lowest $\hbar$ order should be improved here, for
instance by taking into account the quantum effects  
as shell
corrections within a more general POT \cite{belyaev,BMps2013}.
In the nucleus $^{132}$Sn the IVDR energies of the two peaks
 do not change much with $L$ in both cases: $E_1=17$ MeV, $E_2=20$ MeV
for SLy5$^*$ (Fig.\ \ref{fig4}) and $E_1=15$ MeV, $E_2=18$ MeV for SVsym32
(Fig.\ \ref{fig5}). We find only an essential re-distribution of the EWSR 
contributions (normalized to 100\% for 
the EWSR sum of the main and satellite peaks) [(\ref{sumruleewsr}) 
for $S_n^{(1)}$] ,
This is due to a significant enhancement of
the main ``out-of-phase'' peak
with increasing $L$,  $S^{(1)}_1=89$\% and $S^{(1)}_2=11$\% for SLy5$^*$
(Fig.\ \ref{fig4}) and more pronounced EWSR
distribution $S^{(1)}_1=76$\% and $S^{(1)}_2=24$\% for SVsym32
(Fig.\ \ref{fig5}) [cf.\ with the corresponding $L=0$ results:
$S^{(1)}_1=88$\% and $S^{(1)}_2=12$\% for SLy5* and
$S^{(1)}_1=73$\% and $S^{(1)}_2=27$\% for SVsym32]. 
These more precise calculations
change essentially the IVDR strength distribution
for the SV forces  because of the smaller $c_{\rm sym}$ value as compared
to other Skyrme interactions (Table 1). The collision relaxation time,
$\mathcal{T}=4.3 \cdot 10^{-21}$ s, is taken in Figs. \ref{fig4}--
\ref{fig6} in agreement with the IVGDR  widths
\cite{belyaev}.
Decreasing the relaxation time $\mathcal{T}$ by
a factor of about 1.5 almost does not change the IVDR strength structure.
However,
we found a strong dependence on the relaxation time $\mathcal{T}$ in a wider
region of $\mathcal{T}$ values. The ``in-phase'' strength component with
a wide maximum does not depend much on the Skyrme force
\cite{chaban,reinhardSV,pastore}, the slope parameter
$L$, and the relaxation time $\mathcal{T}$.
We found also a regular change of the
IVDR strength for different double-magic
isotopes (Fig.\ \ref{fig6}). Besides of a big
change for the energy (mainly because of $E_1$) and the strength
[$S_1(\omega$)], one also obtains more asymmetry
for $^{68}$Ni than for the other isotopes.
Calculations for nuclei with different mass $A$
were performed with the
relaxation time  $\mathcal{T}$ (\ref{relaxtimen})
where
$\mathcal{T}_{0}=\mathcal{T}_{0\rm Pb} (208/A)^{1/3}$ with the parameter
$\mathcal{T}_{0\rm Pb}=300 $ MeV$^2\cdot$s derived from the IVGDR width
of $^{208}$Pb, in agreement with experimental data for the averaged
$A$ dependence of the IVGDR widths ($\propto A^{-2/3}$).
 In this way the IVDR width becomes
larger with decreasing
$A$ as $A^{1/3}$, and at the same time, the height of peaks  decreases.
The $L$ corrections are also changing much in the same scale of all three
nuclei.

The essential parameter of the Skyrme HF approach leading to the
significant differences in the $k_S$ and $Q$ values is the constant
$\mathcal{C}_{-}$
[(\ref{enerdenin}) and Table 1].
Indeed, $\mathcal{C}_{-}$ is the key quantity in the expression for $Q$
(\ref{stiffin})
and the isovector surface-energy constant $k^{}_{\rm S}$ [or $b_{\rm S}^{(-)}$
(\ref{bsminus})],
because $Q \propto 1/k^{}_{\rm S} \propto 1/\mathcal{C}_{-}$
and $k^{}_{\rm S}\propto \mathcal{C}_{-}$ \cite{BMRV}. 
Concerning $k^{}_{\rm S}$
and the IVDR strength structure,  this is even more
important than the $L$ dependence; though the latter changes significantly the
isovector stiffness $Q$, and the neutron skin $\tau$.
As seen in Table 1,  the constant $\mathcal{C}_{-}$
is very different in absolute value and in sign for different Skyrme
forces whereas $\mathcal{C}_{+}$ is almost constant.
The isoscalar energy-density constant $ b_{\rm S}^{(+)}$
is proportional to $C_{+}$ (\ref{bsplus}),
in contrast to the isovector one.
All of Skyrme parameters are fitted to the well-known experimental
value  $b_{\rm S}^{(+)} =17-19$ MeV  while
there are so far no clear experiments which would determine $k^{}_{\rm S}$
well enough because the mean energies of the IVGDR (main peaks) do not
depend very much on $k^{}_{\rm S}$ for  
different Skyrme forces (the last three rows of Table 2).
Perhaps, the low-lying isovector collective states are more sensitive 
but,
at the present time, there is no careful systematic study of their
 $k^{}_{\rm S}$ dependence.
Another reason for so different $k^{}_{\rm S}$ and $Q$ values might be due
to difficulties in deducing $k^{}_{\rm S}$ directly from the HF calculations 
because of the curvature and quantum effects.
In this respect, the semi-infinite Fermi system with a hard plane wall
might be more adequate for the comparison of the
HF theory and  the ETF effective surface approach.
We have also to go far away from the nuclear stability line to
subtract uniquely the coefficient $k^{}_{\rm S}$ in the dependence of
$b_{\rm S}^{(-)} \propto I^2=(N-Z)^2/A^2$, according to (\ref{bsminus}).
For exotic nuclei one has more problems to  derive $k^{}_{\rm S}$  from the
experimental data with enough precision.
Note that, for studying the IVDR structure, the quantity $k^{}_S$
is more fundamental
 than the isovector stiffness $Q$  because of  
the direct relation to the
tension coefficient $\sigma_{-}$ of the isovector capillary pressure.
Therefore, it is simpler to analyze
the experimental data for the IVGDR within the macroscopic HD or FLD models
in terms of the constant $k^{}_S$. The quantity $Q$ involves also the
ES approximation for
the description of the nuclear edge through the neutron skin $\tau$ in
 (\ref{stifminus}). The $L$ dependence of the neutron skin $\tau$
is essential but not so dramatic in the case of SLy and SV
forces (Table 2),
besides of the SVmas08 forces with the effective mass
0.8. The precision of such a description depends more on the specific
nuclear models \cite{vinas1,vinas2,vinas5}.
On the other hand, the neutron skin thickness $\tau$, 
as the stiffness $Q$,
is interesting in many aspects for an investigation of exotic nuclei,
in particular, in nuclear astrophysics.

We emphasize that for specific Skyrme forces there exists an
abnormal behavior of the
isovector surface constants $k^{}_{\rm S}$ and $Q$. It is related to the
fundamental constant $\mathcal{C}_{-}$ of
the energy density (\ref{enerdenin}) but not to
the derivative corrections to the symmetry energy density.
For the parameter
set T6  ($\mathcal{C}_{-}=0$) one finds $k^{}_{\rm S}=0$ \cite{BMRV}.
Therefore,
according to (\ref{stiffin}), the value of $Q$ diverges
($\nu$ is almost independent from $\mathcal{C}_{-}$ for SLy and SV forces;
 Table 2 and 
\cite{BMRV,BMRPhysScr2014,BMRps2015}).
The isovector gradient terms which are important for
the consistent derivations within the ES approach are also not included
($\mathcal{C}_{-}=0$) into the symmetry energy density in
\cite{danielewicz1,danielewicz2}.
 In relativistic investigations \cite{vretenar1,vretenar2,vretenar3} of
the pygmy modes and the structure
of the IVGR distributions, the dependence of these quantities
on the derivative terms has not
been investigated so far. It therefore remains an interesting
task for the future to
apply similar semiclassical methods such as the  ES approximation
 used in here also in
relativistic models.
Moreover, for RATP \cite{chaban} and SV \cite{reinhard}
(like for SkI) Skyrme forces, the isovector stiffness $Q$ is even negative
as $\mathcal{C}_{-}>0$ ($k^{}_{\rm S}>0$) in contrast to other Skyrme forces.
This would lead to an instability of the vibration of the neutron skin.

Table 2 shows also the coefficients $\nu$ of (\ref{stiffin})
for the isovector stiffness $Q$.
They are almost constant for all SLy and SV
Skyrme forces, unlike other forces \cite{BMRV}.
However, these constants $\nu$, being sensitive to the SO
($\beta$) dependence through (\ref{fw}),  (\ref{skin}) and
(\ref{jminus}), change also with $L$ (Table 2).
As compared to 9/4 suggested in \cite{myswann69}, they are
significantly smaller
in magnitude for the most of the Skyrme forces.

Figs.\ \ref{fig6} and \ref{fig7} 
show more systematic study for several isotopes and for the chain of the
Sn isotopes, respectively. In Fig.\ \ref{fig7}, we compare the results
of our calculations with the experimental data  
\cite{bermanSNexp,vanderwoudeSNexp,varlamovbook,varlamovprep,dietrichdata,adrich}. 
The latter
were obtained by the fitting of the
experimental strength curve for a given almost spherical Sn isotope 
by the two Lorenzian oscillator-strength functions as described in 
\cite{kolmagsh,belyaev}. It is always possible in the case of the asymmetric 
shapes of the strength curves with usual enhancement on right of the main 
peak, even in the case if the satellite cannot be distinguished well from
the main peak in almost spherical nuclei (unlike the clear shoulders
for the IVDRs in deformed ones). 
Each of these functions has three fitting parameters
such as the inertia, stiffness and width of the peak \cite{belyaev}. 
We found rather a good agreement
of our ETF ES results with these experimental data for the energies, ratio of
the strengths at the satellite to the main modes and the EWSR contributions.

More precise $L$-dependent calculations
change essentially the IVDR strength distribution
for the SV forces  because of the smaller $c_{\rm sym}$ value as compared
to other Skyrme interactions (Table 1). For $^{208}$Pb one obtains
$E_1=15$ MeV, $S^{(1)}_1=91$\% for the main peak 
and $E^{}_2=17$ MeV, $S^{(1)}_2=9$\% the satellite for SLy5$^*$; and 
$E_1=13$ MeV, $S^{(1)}_1=83$\% for the main peak 
and $E_2=16$ MeV, $S^{(1)}_2=17$\% the satellite for SVsym32 forces. These calculations
are qualitatively in agreement with the experimental
results: $E_1=13$ MeV, EWSR $_1=98$\% for the main peak 
and $E_2=17$ MeV, EWSR $_2=2$\% the satellite. Descrepances might be related
to the strong shell effects in this stable double magic nucleus which are neglected 
in the ETF ES approach.

In Fig.\ \ref{fig8} 
we show, in the case of the Skyrme forces SLy5*
and SVsym32, the transition densities $\rho_{\mp}^\omega(x)$ of
(\ref{drhom}) for the ``out''- of-phase (-) and the ``in''- phase (+)
modes of the volume vibrations at the excitation
energy $E_2$ of the satellite.
These are the key quantities for the calculation of the IVDR strengths,
according to (\ref{chicollrho}).
The $L$ dependence is rather small, slightly notable mostly near the ES
($|x|\siml 1$). From Fig.\ \ref{fig9}, 
one finds a remarkable neutron versus 
proton excess  near the nuclear edge for the same forces, which is however,
very slightly depending on the slope parameter $L$. A small dependence
of the transition densities on $L$ comes through
the symmetry-energy constant $k^{}_{\rm S}$ which is almost the same
in modulus for these forces. We did not find a dramatic change of
the transition densities with the sign of $k^{}_{\rm S}$.
Therefore, there is a weak sensitivity of the transition densities on $L$
through the energy $E_2$. We would have expected a stronger influence of
the sign of $k_{\rm S}^{}$ on the vibrations of the neutron 
skin rather than on the IVDR.
This different sign leads to the opposite, stable and unstable,
neutron skin vibrations.  One observes also other differences
between the upper (SLy5*) and the
lower (SVsym32) panels in both figures:
We find a redistribution of the surface-to-volume
contributions of the transition densities for these two modes.
In Figs.\ \ref{fig10} and \ref{fig11}, one finds
a considerable change of
the neutron-proton transition densities for the same 
different isotopes for SLy5*
and SVsym32 forces as in Fig.\ \ref{fig6}.

The last three figures show theoretical (Figs.\ 
\ref{fig12} and \ref{fig13}) and experimental 
(Fig.\ 
\ref{fig14}) evaluations of the neutron skin. Fig.\ \ref{fig12} presents
our calculations of the dimensionless skin $\tau/I$.
Being independent of the specific properties of the nucleus,
this quantity is universal.
Fig.\ \ref{fig13} 
shows the absolute values of
the skin obtained from $\tau/I$ multiplying the mean-square evaluations of
the nuclear radii by the factor $\sqrt{3/5}$ for an easy comparison with  experimental data in Fig.\ 
\ref{fig14}. For $^{208}$Pb, one finds that the experimental values $\Delta r_{np}^{exp}=0.12-0.14$ fm in Fig.\  
\ref{fig14}
(0.156$_{-0.021}^{+0.025}$ fm, see  \cite{rcnp}) are in good agreement with our
calculations $\Delta r_{np}^{\rm theor}\approx 0.10-0.13$ fm
within the ES approximation
(the limits show values from SLy5* to SVsym32). For the isotope
$^{124}$Sn one obtains $\Delta r_{np}^{\rm theor}\approx 0.09-0.12$ fm, 
also in good agreement with
experimental results (Fig.\ 
\ref{fig14}). For the isotope
$^{132}$Sn, we predict the value
$\Delta r_{np}^{\rm theor}\approx 0.11-0.15$.
Similarly, for $^{60}$Ni and $^{68}$Ni, one finds
 $\Delta r_{np}^{\rm theor}\approx 0.03-0.04$ (as in Fig.\ 
\ref{fig14}) and $0.08-0.11$, respectively.

Thus, in this section, 
the slope parameter $L$ was taken into account in the leading ES
approximation in order to derive simple analytical expressions  for
the isovector
particle densities and energies. These expressions were used for calculations of
the surface symmetry energy, the neutron skin thickness
and the isovector stiffness coefficients as functions of
$L$. For the derivation of the surface symmetry energy and its dependence
on the particle density we have to include main higher order terms
in the parameter $a/R$.
These terms depend on the well-known parameters of
the Skyrme forces. Results for
the isovector surface-energy constant $k^{}_{\rm S}$,  the neutron skin
thickness $\tau$, and the neutron skin stiffness $Q$
depend in a sensitive  way on the parameters of the
Skyrme functional (especially on the
parameter $\mathcal{C}_{-}$) in the gradient terms of the density in the
surface symmetry energy [see (\ref{enerdenin})].
The isovector constants $k_{\rm S}$, $\tau$ and $Q$ depend also
essentially on the slope parameter $L$,  
in addition to the SO 
interaction constant $\beta$.
For all Skyrme forces, the isovector stiffness constants
$Q$ are significantly larger than those obtained in earlier investigations.
However, taking into account their $L$ dependence they come closer
to the empirical data. It
influences 
more on the isovector stiffness $Q$ and on the neutron skin $\tau$,
than on the surface symmetry energy constant $k^{}_{\rm S}$.
The mean IVGDR energies and sum rules calculated in the
macroscopic models like the FLDM \cite{kolmagsh,BMVnpae2012}
in Table 2 are in a fairly good agreement with the
experimental data for most of the $k^{}_{\rm S} $ values. As compared with
the experimental data and other recent theoretical works, we found a
rather reasonable two-peak
structure of the IVDR strength within the FLDM.
According to our results for  the neutron and proton transition densities
[Figs.\ 
\ref{fig8}, \ref{fig9} and \ref{fig10}], we may interpret 
semiclassically the IVDR satellites as
some kind of pygmy resonances \cite{endres} on right
of the main IVDR peak, though they might be of different nature 
from the so called Pygmy Dipole Resonances (PDRs) found on left of this peak
\cite{vretenar1,vretenar2,ponomarev,vretenar3,nester1,nester2,kievPygmy,adrich,wieland,savran}.
The IVDR energies, sum rules and n-p transition densities
obtained analytically within the semiclassical FLD approximation
are sensitive to the surface symmetry energy constant
$k_{{}_{\! S}}$ and the slope parameter $L$. Therefore, their comparison with the
experimental
data can be used for the evaluation of $k_{{}_{\! S}}$ and $L$.
It seems helpful to describe them in terms of only few critical
parameters, like $k^{}_{\rm S}$ and $L$.

For further perspectives, it would be worthwhile to
apply our results to calculations of the satellite resonances in the IVDR
strength within the FLDM \cite{kolmagsh} in a more systematic way.
In this respect it is also interesting that the low-lying collective
isovector states are expected to be even
more sensitive to the values of $k^{}_{\rm S}$ within 
the POT  
\cite{GMFprc2007,BMYijmpe2012,BMps2013}.
More general problems of classical and quantum chaos
in terms of the level statistics and Poincar\'e and Lyapunov exponents
(see \cite{BMstatlevPRC2012} and references therein) might lead to a progress
in studying the fundamental
properties of collective dynamics like nuclear fission within the 
Swiatecki\&Strutinsky
macroscopic-microscopic model. Our approach is helpful also
for further study of the effects in the surface symmetry energy because it
gives analytical
universal expressions for the  constants $k^{}_{\rm S}$, $\tau$ and $Q$
as functions of the slope parameter $L$ which do not depend on
specific properties of nuclei as they are directly connected with
a few critical parameters of the Skyrme interaction without 
any fitting.


\section{COLLECTIVE EXCITATIONS\\ AS A SEMICLASSICAL RESPONSE}
\l{collexcresptheor}

The collective dynamics of complex nuclei at low excitation energies, such as the vibration modes, can be described within several theoretical approaches \cite{bormot,SOLbook1981,SOLbook1992,SOLQPMV-EPNP1985,migdalrev,berborbro}. One of the most powerful tools for its description is based on the response function theory \cite{bormot,hofbook}, 
in particular, within a semiclassical kinetic approach \cite{belyaev}. 
This theory basically equivalent to the Random Phase Approximation
(RPA) in the QPM  \cite{SOLbook1981,SOLbook1992,SOLQPMV-EPNP1985}.  
The collective variables are introduced  
\cite{bormot,hofbook} explicitly as deformation parameters of a mean single-particle field. The nuclear collective excitations are parametrized in terms of the transport coefficients as the stiffness, the inertia, and the friction parameters through the adequate collective-response functions 
\cite{hofbook}. The quantum formulation of this problem can be 
significantly simplified by using the SCM
\cite{strut,fuhi,sclbook,magNPAE2010} 
within the semiclassical approximation of the 
POT (Section II, and 
\cite{sclbook,gutz,bablo,strumag,magvvhof}). It would be worth to apply 
first the ideas of the 
SCM averaging and POT at a few leading orders in $\hbar$, as 
the ETF approach \cite{sclbook}, for calculations of the smooth transport coefficients at low excitation energies.

The semiclassical derivations of the famous wall formula for the average friction, owing to collisions of particles of the perfect Fermi-gas with a slowly moving surface of the mean-field edge-like potential, were suggested in 
\cite{wall,hatkoonran,koonran}, see also its derivations in 
\cite{abrdelmat,yanbro,hofivyam}. The explicit analytical expressions of a smooth friction and inertia for the low-lying nuclear collective excitations within the semiclassical Gutzwiller path-integral approach to the POT \cite{gutz,strumag} at leading orders in $\hbar$, with the main focus on the consistency condition 
\cite{bormot,hofbook,ivhofpashyam} between the variations of potential and particle number density, were considered in 
\cite{MGFyaf2007,belyaev,MGFnpae2005,GMFprc2007,MGFnpae2008}.
 In the first Section IVA we derive the 
response function at small frequencies in terms of the averaged transport coefficients  
for studying the low-lying collective-vibration states.

\subsection{LOW-LYING COLLECTIVE EXCITATIONS OF NUCLEI}

 Following 
\cite{hofbook}, we begin with a general response function formalism 
for transport coefficients
in Section IVA1. 
They are expressed in terms of the  semiclassical Green's functions 
(Section II) and
averaged in the phase 
space variables 
over many s.p.\ states near the Fermi surface  
in the simple case of a spherical cavity-like potential for 
a mean field at equilibrium in Section IVA2.
The self-consistent relations between the transport coefficients and
the coupling constant are presented in Section IVA3. 
The mean statistically vibration energies of the low-lying collective states and their EWSR contributions are derived in terms of analytical functions of nuclear particle number in Section IVA4. 
The reduced probabilities for the direct radiation decay of gamma quanta and the corresponding lifetimes of nuclei are discussed in Section IVA5.  
These analytical results for the low-lying quadrupole and octupole collective modes are compared with experimental data \cite{raman,kibedi,table} in 
Section IVA6  
and are summarized at the end of this section. 
Some details of the derivations are presented in Appendix B.

\bigskip
\subsubsection{Response theory and transport coefficients}
\label{theory}
\medskip

Many-body collective excitations can be described  
in terms of the nuclear response to an external perturbation
(\ref{vextpm}). For the symmetric nuclei, one has
\begin{equation} \label{vext}
V_{{\rm ext}} = \hat{\rm Q}\; q_{\rm ext}^\omega e^{- i\omega t}\;, 
\end{equation}
with a vibration amplitude, $q_{\rm ext}^\omega$, and the multipole 
s.p.\  
operator, $\hat{\rm Q}= r^\lambda {\rm Y}_{\lambda 0}(\theta)$, 
 ($\lambda \geq 2$) 
\cite{bormot}. 
Its quantum average perturbation,  
$\delta \langle \hat{\rm{Q}}\rangle _t$, at time $t$ is calculated 
through the Fourier transform $\delta \langle \hat{\rm{Q}}\rangle _\omega$ 
obtained within the linear response theory \cite{bormot,hofbook,belyaev},
\begin{equation}\label{colresp}
\delta \langle \hat{\rm Q}\rangle _\omega = 
- \chi_{\rm QQ}^{\rm coll} (\omega)q_{\rm ext}^\omega\;,
\end{equation}
where $\chi_{\rm QQ}^{\rm coll} (\omega)$ is the collective response function
in the $Q$ mode. 
[In Section IV, $Q$ should not be confused with 
the neutron skin stiffness of the 
previous Section III.]
The total Hamiltonian,
$H_{\rm tot}=H + V_{\rm ext}$ at $q_{\rm ext}^{\omega} = 0$, i.e.\ , $H$, 
depends on a collective variable $q$ defined as the time-dependent 
amplitude $q(t)$ of the potential $V(q)$.
The vibrations of the axially-symmetric nuclear surface with a multipolarity 
$\lambda$ near the spherical shape can be described by 
\bea\label{surf}
&R(\theta,q) = R[1 + q(t)Y_{\lambda 0} (\hat{r})]\;,\nonumber\\
&q(t) = q_\omega e^{- i\omega t}\;, \quad  \hat{r}=\r/r=\cos\theta\;,
\eea
in the spherical coordinates, 
$Y_{\lambda 0} (\hat{r})$ is the spherical function of
$\hat{r}$. 
The unperturbed quantities in dynamical variations 
are zero in this case. The consistency condition writes
\begin{equation}\label{consisteq}
\delta \langle \hat{\rm Q}\rangle _\omega = \kappa^{}_{\rm QQ} 
\,\delta q_\omega\;,
\end{equation}
where $\kappa^{}_{\rm QQ}$ is the coupling constant, see Appendix B1. 
With help of the condition (\ref{consisteq}), 
the collective response \cite{bormot},
\begin{equation}\label{colrespintrans}
\chi _{\rm{QQ}}^{{\rm coll}} (\omega) = \kappa^{}_{\rm{QQ}} 
\frac{\chi^{}_{\rm{QQ}} (\omega)}{\chi^{}_{\rm QQ}(\omega) + \kappa_{\rm QQ}}\;,
\end{equation}
is expressed in terms of the so called intrinsic response function, 
$\chi^{}_{\rm QQ}(\omega)$, defined by
\begin{equation}\label{colrespin}
\delta \langle \hat {\rm Q}\rangle_{\omega} 
= -\chi_{\rm QQ}(\omega) \left(\delta q_{\omega} + q_{\rm ext}^\omega\right)\;. 
\end{equation}
One dominating peak in the collective strength function,
\begin{equation}\label{stren}
S(\omega)= \frac{1}{\pi}\; \Im \chi _{\rm QQ}^{\rm coll} (\omega)\;,
\ee
based on (\ref{colrespintrans}), at low excitation energies, 
$\hbar \omega_{\lambda}$, is assumed to be well 
separated from all other solutions of the secular 
equation $\chi^{}_{\rm QQ}(\omega) + \kappa^{}_{\rm QQ}=0$ 
for $\omega=\omega_{\lambda}$. 
See more detailed explanations of this approach for the case of 
another s.p.\ operator 
\begin{equation}\label{foperdV}
\hat{{\rm F}}=(\partial V/\partial q)_{q=0}
\end{equation}
with a mean field $V(q)$ in (\ref{consisteqF}) and for 
its applications to the 
collective nuclear dynamics in 
\cite{hofbook,ivhofpashyam,MGFyaf2007,belyaev}. 
The corresponding oscillator 
response function in the q-mode, $\chi _{\rm qq}^{\rm coll} (\omega)$, can be conveniently written in an inverted approximate form 
\cite{hofbook,ivhofpashyam}:
\begin{eqnarray}\label{invcolresin}
\frac{1}{\chi _{\rm qq}^{\rm coll} (\omega)} &=& \frac{1}{\chi _{\rm qq} (\omega)} 
+ \kappa^{}_{\rm FF}\;\nonumber\\
&=& - M_{\rm FF}\omega ^2 - i\gamma^{}_{\rm FF} \omega + C_{\rm FF}\;,
\end{eqnarray}
where $\kappa^{}_{\rm FF}$ is the coupling constant in the ${\rm F}$ mode, as 
shown in (\ref{consisteqF}),
\begin{equation}
\chi^{}_{\rm qq}(\omega) = \frac{\chi^{}_{\rm FF}(\omega)}{\kappa_{\rm FF}
^2}= \frac{\chi^{}_{\rm QQ}(\omega)}{\kappa_{\rm QQ}^2}\;.
\label{chiqQF}
\end{equation} 
According to the consistency conditions (\ref{consisteq}) 
and (\ref{consisteqF}), we used in (\ref{chiqQF}) the approximate transformations between the quantities defined in different variables F and Q, corresponding to the 
 s.p.\  operators $\hat{F}$ and $\hat{Q}$. 
These transformations will be used for presentation of the results satisfying the consistency condition for variations of the nuclear 
potential, and the particle density in suitable units. Thus, according to (\ref{chiqQF}), the inverse collective-response function for low frequencies $\omega$ is approximated by (\ref{invcolresin}) through the response function of a damped harmonic oscillator, $\chi_{\rm qq}^{\rm coll} (\omega)$, with the stiffness of the nuclear free energy ${\cal F}$, $C_{\rm FF} \approx C_{\rm FF}(0) = \left({\partial ^2{\cal F} / \partial q^2} \right)_{q = 0}$, see
\cite{hofbook}, the friction $\gamma^{}_{\rm FF}$ and the inertia 
$M_{\rm FF}$ parameters,
\begin{eqnarray}\label{trcollQF}
C_{\rm FF}&=&C_{\rm QQ}\;\frac{\kappa_{\rm FF}^2}{\kappa_{\rm QQ}^2}\;,\nonumber\\
\gamma^{}_{\rm FF}&=&\gamma^{}_{\rm QQ}\;\frac{\kappa_{\rm FF}^2}{k_{\rm QQ}^2}\;,
\nonumber\\
M_{\rm FF}&=&M_{\rm QQ}\;\frac{\kappa_{\rm FF}^2}{\kappa_{\rm QQ}^2}\;.
\end{eqnarray}
The consistent transport coefficients $C_{\rm QQ}$, $\gamma^{}_{\rm QQ}$ and 
$M_{\rm QQ}$ in a variable Q
are related to the auxiliary intrinsic parameters $C_{\rm QQ}(0)$, 
$\gamma^{}_{\rm QQ}(0)$ and $M_{\rm QQ}(0)$, as those of expansion of the intrinsic response function, $\chi^{}_{\rm QQ}(\omega)$, in $\omega$ in the 
``zero-frequency limit'', $\omega \to 0$, for a 
slow 
collective motion \cite{hofbook},
\bea\l{chiexp}
\chi^{}_{\rm QQ}\left(\omega\right)&=&\chi^{}_{\rm QQ}(0) - 
i \gamma^{}_{\rm QQ}(0)\,\omega \nonumber\\
&-& M_{\rm QQ}(0)\;\omega^2 + \cdots\;,\nonumber\\
\chi^{}_{\rm QQ}(0)&=&-\kappa^{}_{\rm QQ}-C_{\rm QQ}(0)\;. 
\eea
Thus, for the transport parameters of the oscillator $\omega$-dependence 
in (\ref{invcolresin}) and (\ref{trcollQF}), one has 
\begin{eqnarray}\label{trcoll}
C_{\rm QQ} &=& \left[1 + C_{\rm QQ}(0)/\chi^{}_{\rm QQ} (0) 
 \right]C_{\rm QQ}(0)\;, \nonumber\\
\gamma^{}_{\rm QQ} &=& \left[ 1
+ C_{\rm QQ}(0)/\chi_{\rm QQ}(0) \right]^2\gamma_{\rm QQ}
(0)\;, 
\nonumber\\
 M_{\rm QQ} &=& \left[ 1 +
C_{\rm QQ}(0)/\chi_{\rm QQ} (0) \right]^2\;\nonumber\\
&\times&\left[M_{\rm QQ}(0) + 
 \gamma_{\rm QQ}^{2}(0)/\chi^{}_{\rm QQ}(0) \right]. 
\end{eqnarray}
With (\ref{chiqQF}) and (\ref{trcollQF}), the poles of the 
oscillator response function of (\ref{invcolresin}) are determined by the Newtonian equation of motion with a friction,
\begin{equation}\label{langeq}
M_{\rm FF}\ddot{Q} + \gamma^{}_{\rm FF}\dot{Q} - C_{\rm FF} =0,
\end{equation}
which is helpful to clarify the physical meaning of the inertia $M_{{\rm FF}}$, the friction 
$\gamma^{}_{{\rm FF}}$, 
and the stiffness $C_{{\rm FF}}$ of the collective motion.

The intrinsic response $\chi^{}_{\rm{QQ}}(\omega)$ 
in equations (\ref{colrespintrans}) and (\ref{colrespin})
can be expressed in terms of the s.p.\  Green's function 
$G(\r_1,\r_2;\vareps)$ (see (\ref{chiGGdef}) after the replace 
$\hat{Q}=\hat{F}$ \cite{magvvhof,hofbook}).
\begin{eqnarray}
\label{chiGGdef}
&&\hspace{-0.5cm}\chi^{}_{\rm QQ}(\omega) = \frac{d_{s}}{\pi} \int \limits_0^{\infty} 
{\d \vareps \,n(\vareps) \int {\d\r_1}} \int {\d \r_2} \;\nonumber \\ 
&\times&\hat{\rm Q}(\r_1)\,\hat{\rm Q}(\r_2) \,\Im G(\r_1,\r_2;\vareps)
\nonumber\\
& \times& \left[\;\overline{G} \left(\r_1,\r_2;\vareps - \hbar \omega\right)
+ G\left(\r_1,\r_2;\vareps + \hbar \omega\right) \right],
\end{eqnarray}
where $n(\vareps)$ is the Fermi occupation numbers at the energy $\vareps$ for temperature $T$, $n(\vareps)=\{1+\exp[(\vareps-\lambda)/T]\}^{-1}$, 
with the chemical potential, $\lambda\approx\vareps_{\rm F}$, $\vareps_{\rm F}$ is the Fermi energy. The factor of $d_{s}$ accounts again for the 
spin (spin-isospin) degeneracy by neglecting differences between the neutron and the proton potential wells (Section II). 
For the Green's function $G(\r_1,\r_2;\vareps)$ 
(bar above $G$ in (\ref{chiGGdef}) means
the complex conjugation) we use the energy spectral representation, 
\begin{equation}
\label{GFdef} 
G\left(\r_1,\r_2;\vareps\right) = 
\sum_i \frac{\overline{\psi}_i(\r_1)\psi_i(\r_2)}{\vareps - \vareps_i + i \eps_o}\;,
\end{equation}
where $\vareps_i$ is eigenvalues, $\psi_i$  
eigenfunctions, and 
$\epsilon_o \rightarrow +0$ in the quantum mean-field approximation.

With the help of (\ref{chiGGdef}), the intrinsic response function, 
$\chi^{}_{\rm QQ}(\omega) $, in 
the ``zero-frequency limit'' (\ref{chiexp}) ($\omega \rightarrow 0$) can 
be expressed in terms of the Green's function $G$ 
through the intrinsic parameters 
\cite{hofbook,ivhofpashyam},
\bea\l{gtfdef}
&&\hspace{-0.5cm}\gamma^{}_{\rm QQ}(0) = 
-i\left(\frac{\partial \chi_{\rm QQ}(\omega)}{\partial\omega} \right)_{\omega=0} 
\;\nonumber\\
&=& \frac{d_s\hbar}{\pi} \int \limits_0^{\infty} \d\vareps\; 
n(\vareps)\int\d\r_1\int\d\r_2
\;\hat{\rm Q}\left(\r_1\right)\,\hat{\rm Q} \left(\r_2\right)\, \nonumber\\
&\times& \frac{\partial}{\partial \vareps}\left[\Im \,G
\left(\r_1,\r_2;\vareps\right)\right]^2, 
\eea
\bea\l{btfdef}
&&\hspace{-0.5cm}M_{\rm QQ}(0)= \frac{1}{2}\,\,\left(\frac{\partial^2
\chi_{\rm QQ}(\omega)}{\partial\omega^2}
 \right)_{\omega = 0} \;\nonumber\\
&=& \frac{d_s\hbar^2}{\pi}\,\int\limits_0^{\infty} \d \vareps\, n(\vareps) \,
\int \d \r_1\,\int \d \r_2\,\hat{\rm Q}(\r_1)\,\hat{\rm Q}(\r_2)
\;\nonumber\\
&\times&\rm{Im} \, G\left(\r_1,\r_2;\vareps\right) \,
\frac{\partial^2}{\partial \vareps^2} \Re\,G\left(\r_1,\r_2;\vareps\right).
\end{eqnarray}
Using the spectral representation (\ref{GFdef}) for the Green's function $G$ 
one reduces equivalently equation (\ref{btfdef}) to the well-known 
cranking model inertia in the mean-field limit in the ${\rm F}$
mode ($\epsilon^{}_0 \to +0$) 
\cite{hofbook},
\begin{equation}\label{mcran} M(0) = 
d_s\hbar^2{\sum_{ij}}^{\prime}\, 
\frac{ n_i - n_j}{(\vareps_j - \vareps_i)^3}\, |<i|\hat{{\rm F}}|j>|^2,
\end{equation}
where $~<i|\hat{{\rm F}}|j>~$ is the matrix element of the operator 
$\hat{{\rm F}}$ (\ref{foperdV}); 
see, e.g.\ , \cite{hofbook}. The prime means that the
diagonal terms, $~\vareps_i=\vareps_j$, are excluded in these summations. 

For the collective response function at the low-lying excitation energies, 
(\ref{colresp}), with the help of
(\ref{invcolresin}) and (\ref{chiqQF}), one has 
\begin{equation}\label{chicollQ}
\chi_{\rm QQ}^{\rm coll}\left(\omega\right) = 
\frac{\kappa_{\rm QQ}^2}{-M_{\rm FF}\omega^2 
-i\gamma^{}_{\rm FF}\omega + C_{\rm FF}}\;,
\end{equation}
where the inertia $M_{\rm FF}$, the friction $\gamma^{}_{\rm FF}$,
 and the 
stiffness $C_{\rm FF}$ are given by (\ref{trcollQF}), (\ref{trcoll}), (\ref{gtfdef}) and (\ref{btfdef}). The coupling constant $\kappa^{}_{\rm QQ}$ is defined by the consistency condition (\ref{consisteq}), see Appendix B1. 
According to (\ref{chicollQ}) for the response function 
$\chi_{\rm QQ}^{\rm coll}(\omega)$ ($\lambda=2,3,\cdots$) , the strength function $S_{\lambda}(\omega)$ (\ref{stren}) 
for the first lowest peak is given by 
\begin{equation}\label{strength1}
\hspace{-0.5cm}S_{\lambda}(\omega) \!=\! \frac{\kappa_{\rm QQ}^2}{\pi} 
\frac{\gamma^{}_{\rm FF}\,\omega}{\left(
-M_{\rm FF}\omega^2 +C_{\rm FF} \right)^2 + \gamma_{\rm FF}^2\omega^2}\;.
\end{equation}
Substituting this strength function into its moments,
\begin{equation}\label{momstren}
S_{\lambda}^{(l)}
=\hbar^{{\it l}+1} \int\limits_0^{\infty} \d \omega \;\omega^{{\it l}}\;
S_{\lambda}(\omega)\;,
\end{equation}
${\it l} = 0, 1, ...$, one can evaluate the probability distributions for excitations of the low-lying collective states.

\subsubsection{Semiclassical EGA 
for transport coefficients}

The trace of the first term $G_{{\rm CT}_0}(\r_1,\r_2;\vareps)$, 
$\r_2 \to \r_1$, in (\ref{Gsem}) corresponds to a smooth level density, 
$g^{}_{\rm ETF}(\vareps)$, and the ETF particle number conservation writes 
\cite{sclbook,bablo}
\begin{eqnarray}\label{akrTF}
A&=& d_s \int \limits_0^{\infty} \d \vareps\, \widetilde{n}(\vareps)\, 
g^{}_{\rm ETF}(\vareps)\nonumber \\
&\approx& 
d_s\left[\frac{2 (k^{}_FR)^3}{9 \pi} - \frac{(k^{}_FR)^2}{4} + 
\frac{2 k^{}_FR}{3 \pi}\right]\;, 
\end{eqnarray}
where $\widetilde{n}$ is the occupation number averaged by using the Strutinsky
smoothing \cite{strut,fuhi}, 
$k^{}_{F}$ is the Fermi momentum in units of $\hbar$, 
$k^{}_{F}=\sqrt{2 m \vareps_{F}/\hbar^2}$ for billiards. 
Equation (\ref{akrTF}) 
determines the semiclassical parameter $k^{}_{F}R$ as function of the 
particle number $A$. The second and third terms in the very right approximation 
in (\ref{akrTF}) for 
spherical 
cavity-like mean fields, which becomes exact for 
the infinitely deep square-well potential, account for 
important surface and curvature corrections to the first main 
volume component, respectively. 
The temperature corrections, $\sim (T/\vareps^{}_{F})^2$, might be taken into account through the usual Sommerfeld expansion, too, see below and \cite{hofbook,MGFyaf2007}. We shall omit such small corrections because the applications will 
be applied
to the low-lying collective excitations at zero temperature.

As well known \cite{strumag,sclbook}, due to $\hbar$ in the denominators of exponents of (\ref{GsemGa}) in the oscillating terms of the Green's function traces, their semiclassical expansions in $\hbar$, or in dimensionless parameter, $\hbar/S_{\rm CT} \sim 1/k^{}_{F}R$, converge after averaging in $k^{}_{F}R$, for instance, over a large enough interval of the particle number $A$ through the radius $R$ in accordance with (\ref{akrTF}). The Strutinsky averaging 
 \cite{strut,fuhi,magNPAE2010,migdalrev}
with a Gaussian width $\widetilde{\Gamma}$, which covers at least a few major shells in energy spectrum, see Appendix B1, 
leads to the local ($\r_2 \to \r_1$) averaged 
quantities; in particular, the smooth level and particle 
density, and 
 free energy. 
According to (\ref{gtfdef}), (\ref{btfdef}), 
the unlocal ($\r_2 \neq \r_1$) contributions into the ETF transport coefficients become also important. Therefore, we need more extended statistical averaging in the phase space (energy and spatial coordinate) variables, as in the semiclassical (moreover, local hydrodynamical) 
derivations within the many-body particle density or Green's function 
formalism 
\cite{kadbaym,kolmagpl}.

The averaged semiclassical inertia, ${\tilde M}_{\rm QQ}(0)$, and friction, 
$\widetilde{\gamma}_{\rm QQ}(0)$, parameters can be found by substitution of the trajectory expansion of Green's function (\ref{Gsem}) into (\ref{gtfdef}) and (\ref{btfdef}),
\begin{eqnarray}
&&\hspace{-0.2cm}\widetilde{M}_{\rm QQ}(0) = \frac{d_s \hbar^2}{\pi}\,
\sum_{\rm CT, CT'}\langle \int \limits_0^{\infty} \d \vareps n(\vareps) \nonumber\\
&&\times \int \d \r_1\!\int \d \r_2\,\hat{\rm Q}\left(\r_1\right)
\hat{\rm Q}\left(\r_2\right)\nonumber\\
 &\times& \Im \, 
G_{\rm CT}\left(\r_1,\r_2;\vareps\right)\,\, 
\frac{\partial^2} {\partial \vareps^2}
\Re \,G_{\rm CT'}\left(\r_1,\r_2;\vareps\right) \rangle_{\rm av}\;,
\label{m0}
\end{eqnarray}
\begin{eqnarray}
&&\widetilde{\gamma}_{\rm QQ}(0) = 
 \frac{d_s \hbar}{\pi} 
\sum_{\rm CT,CT'} \langle \int\limits_0^{\infty} \d \vareps 
\;n(\vareps) \nonumber\\
 &&\times \int \d\r_1 \,
\int \d\r_2\; 
\hat{\rm Q}\left(\r_1\right)\,
\hat{\rm Q} \left(\r_2\right)
\nonumber\\
&\times& 
\frac{\partial}{\partial \vareps}
\left[\Im\,G_{\rm CT}\left(\r_1,\r_2;\vareps\right)\;
\Im\,G_{\rm CT' }\left(\r_1,\r_2;\vareps\right)\right]\rangle_{\rm av}\;,
\label{gamma0}
\end{eqnarray}
where the angle brackets $<\cdots>_{\rm av}$ mean an averaging over the phase space coordinates, including the SCM averaging in $k^{}_{F}R$ variable
with a width related to $\widetilde{\Gamma}$, as mentioned above. In order to calculate analytically these quantities, we need to distinguish the two limit cases \cite{GMFprc2007,MGFyaf2007}: 

(i) the nearly local part, $S_{\rm CT}(\r_1,\r_2;\vareps_F)/\hbar 
\approx k^{}_F {\cal L}_{\rm CT} \siml 1\;$,

and 

(ii) nonlocal contributions, $k^{}_{F} {\cal L}_{\rm CT} >> 1\;$,

\noindent where ${\cal L}_{\rm CT}$ is the length of the CT in the edge-like potential wells. We emphasize that the averaging over phase-space variables (the spatial coordinates, and the energy spectrum) leads to the nearly local approximation (NLA) (i) for the inertia (\ref{m0}) and friction (\ref{gamma0}) coefficients, in contrast to the case (ii). For the case (i), the partial SCM averaging in $k^{}_{F} R$ (for instance, in nuclear sizes $R$ or particle numbers $A$ for 
the constant 
$k^{}_{F}$ fixed by the particle density of  
infinite matter) ensures a convergence of the semiclassical expansions of smooth quantities in $1/k^{}_{F}R$ within the ETF model \cite{sclbook}. The strong energy-dependent exponential factor of (\ref{GsemGa}), $\exp(i S_{\rm CT}(\vareps)/\hbar)$, for 
$\hbar \to 0$ serves appearance of the damping factor, 
$\propto \exp[-({\cal L}_{\rm CT} \Gamma/4R)^2]$ after such averaging 
with a width of the Gaussian weight function, 
$\Gamma \simg (2 \div 4) k^{}_{F}R\, \hbar \Omega/2$, 
see Appendix B3. 
This averaging, corresponding to the $2 \div 4$ distances between the major shells in energy spectrum, 
$\hbar \Omega \approx \vareps^{}_{F}/A^{1/3}= 7 \div 10 ~$ MeV for heavy nuclei,
$A= 200 \div 50$, respectively, removes shell effects, like in the ETF both level and particle densities 
\cite{strut,fuhi,sclbook,MGFyaf2007,MGFnpae2008}. The most important contribution is coming then from the trajectory, CT$=$CT$^\prime$=CT$_0$, see 
Fig.
\ref{fig1}, with a short length smaller than a few wave lengths, 
$1/k^{}_{F}$, ${\cal L}_{{\rm CT} 0}=s=|\r_2-\r_1| \siml 1/k^{}_{F} << R$, at large semiclassical parameter, $k^{}_{F}R >> 1$.

As in \cite{kadbaym,kolmagpl,MGFyaf2007}, it is convenient now to transform the variables $\{\r_1,\r_2\}$ to the Wigner coordinates $\{\r, \s \}$,
\begin{equation}
\r = (\r_1 + \r_2)/2, \qquad\qquad \s=\r_2-\r_1, 
\label{transform}
\end{equation}
to simplify calculations of the inertia $\widetilde{M}_{\rm QQ}(0)$ (\ref{m0}) 
and the friction $\widetilde{\gamma}_{\rm QQ}(0)$ (\ref{gamma0}) 
by separating a  
slow motion of particles in 
variable $\r$ and their fast motion in $\s$. With the transformation (\ref{transform}) and exchange of the energy and spatial integrations, in the case 
(i),
 one has 
\bea\l{mQQ0}
&&\widetilde{M}_{\rm QQ}(0) = \frac{d_s \hbar^2}{\pi}\nonumber\\
&\times& \langle 
\int \d \r\,\int \d
\s\,\hat{Q}\left( \r + \frac{\s}{2}\right)\, 
\hat{Q}\left( \r - \frac{\s}{2}\right)\;
\nonumber\\
 &\times& \int\limits_0^{\infty} \d \vareps\, n(\vareps)\;
\Im \, G_{{\rm CT}_0 } \left(\r+\frac{\s}{2},\r-\frac{\s}{2};\vareps\right)
\nonumber\\ 
&\times& \frac{\partial^2} {\partial \vareps^2}
\Re \,G_{{\rm CT}_0 }\left(\r+\frac{\s}{2},\r - \frac{\s}{2};\vareps\right)
\rangle_{\rm av}, 
\eea
\bea\l{gammaQQ0}
&&\widetilde{\gamma}_{\rm QQ}(0) = \frac{d_s \hbar}{\pi}\nonumber\\
&\times& \langle \int \d\r 
\int \d\s \,\,\hat{Q}\left(\r + \frac{\s}{2}\right)\,
\hat{Q}\left( \r - \frac{\s}{2}\right)\; 
\nonumber\\
&\times& \!\!\!\int\limits_0^{\infty} \d \vareps \;n(\vareps)
\frac{\partial}{\partial \vareps} \left[\Im\,G_{{\rm CT}_0}
\left(\r+\frac{\s}{2},\r-\frac{\s}{2};\vareps\right)
\right]^2 \rangle_{\rm av}\;. 
\eea

As shown in Appendix B3,  
for small enough 
length $s$ of the trajectory CT$_0$, $s/R << 1$ in the case (i), the corresponding component $G_{{\rm CT}_0}$ of Green's function 
(\ref{Gsem}) in (\ref{mQQ0}), (\ref{gammaQQ0}) 
in terms of the new integration variables $\{\r, \s\}$ is reduced approximately to its simple analytical form (\ref{GCT0}) \cite{belyaev,MGFyaf2007}.
Formally, $G_0$ coincides with the well-known Green's function 
for a free particle motion 
\cite{bablo,hatkoonran,koonran,strumag}.

The internal integral over $\vareps$ in (\ref{mQQ0}) and (\ref{gammaQQ0}) can be taken analytically within the nearly local approximation (\ref{GCT0}). 
For the analytical integrations 
over the Wigner coordinates $\r$ and $\s$, the integrands depending on $k^{}_{F}s$ in (\ref{mQQ0}), (\ref{gammaQQ0}) are simplified by means of averaging in the phase space variables. As shown in Appendix B3, 
using the approximation 
(\ref{GCT0}) one may identically transform the expressions for the
inertia $M_{\rm QQ}(0)$, the friction $\gamma^{}_{{\rm QQ}}(0)$ and the isolated susceptibility $\chi^{}_{\rm QQ}(0)$ to sums of local 
(volume) terms and their small nonlocal (surface) corrections. The integrands, proportional linearly to the nonlocal (correlation-like) components depending on $k^{}_{F}R$, are zeros within the considered approach (i). The inertia terms expressed linearly through the correlation function, $<Q(\r+\s/2)Q(\r-\s/2) - Q^{2}(\r)>_{\rm av}$, 
averaged in phase-space variables, are neglected like in the derivations of the hydrodynamic model (HDM) starting from a many-body system of strong interacting particles.  
A more general statistical principle of the weakness of correlations is used usually in the semiclassical derivations of the kinetic equations with 
integral collision terms, by separating a slow motion along the mean coordinate $\r$ within nearly local condition (i) from a fast dynamics in the relative coordinate $\s$ in terms of its collisional correlations \cite{kadbaym,kolmagpl,belyaev}. Such correlation-like functions are concentrated at small $s$ of the order of a few wave lengths, $1/k^{}_{F}$, as explained in Appendix B3. 
The transport coefficients are simplified by 
averaging in fast oscillations of functions of the relative coordinate 
$\s$ at a given mean coordinate $\r$. The integrations over angles of vectors $\s$ and $\r$ can be approximately performed analytically in the NLA (i), see Appendix B3. 

Thus, the main contribution into remaining integrals over $\s$ and $\r$ 
in
 (\ref{mQQ0}) and (\ref{gammaQQ0}) within the NLA (i) is coming from $s \siml 1/k^{}_{F}$ for $1/k^{}_{F}R <<1$. The major terms can be found in the perfect local case when a smooth product of the multipole operators in these equations can be approximately taken off the integral over $\s$ at $\s=0$, according to
the property of the phase-space averaging of the correlation functions mentioned above. For the main local approximation in the case (i), after the integration over $\vareps$ (or corresponding $kR$), one may take also 
analytically the integrals over $\s$ and $\r$ in (\ref{mQQ0}) and (\ref{gammaQQ0}), 
see Appendix 
B3. 
As the final result, for $\lambda \geq 2$ one finally arrives at the inertia,
\begin{equation}
\widetilde{M}_{\rm QQ}(0) = \frac{d_s m^3 R^{2 \lambda + 6}}{12\pi \hbar^4}\; 
f_\lambda^{(3)},
\label{mQQ}
\end{equation} 
the friction,
\begin{equation}
\widetilde{\gamma}_{\rm QQ}(0) = \frac{d_s m^2 R^{2 \lambda +4}}{2 \pi^2 \hbar^3}\, 
f_{\lambda}^{(1)},
\label{gammaQQ}
\end{equation}
and the isolated susceptibility, 
\begin{equation}
\widetilde{\chi}^{}_{\rm QQ}(0)=\frac{d_s m R^{2 \lambda + 2}\; k^{}_{F}R}{2\pi^2\; \hbar^2}\,f_{\lambda}^{(0)},
\label{chiQQ}
\end{equation}
see Appendix B,  
respectively.
Here, $f_{\lambda}^{(n)}$ are the integrals over the 
dimensionless radial variable,
\begin{equation}
f_{\lambda}^{(n)}=\int \limits_{0}^{1}\d \wp\;\wp^{2 \lambda + 2}\;\left(\wp +
1\right)^{n},
\label{flkdef}
\end{equation}
\bea\l{rint}
&&\hspace{-0.5cm}f_{\lambda}^{(0)}=\frac{1}{2\lambda+3},\quad 
f_{\lambda}^{(1)}=\frac{4 \lambda +7}{2(\lambda+2)(2 \lambda +3)}\;,\nonumber\\ 
&&f_{\lambda}^{(3)} =
\frac{(4 \lambda + 9)
[(4 \lambda + 9)^2-7]}{4(\lambda+2)(\lambda+3)(2 \lambda + 3)(2 \lambda + 5)}\;,
\eea
where $n=0,1,2,...,\quad \wp=r/R$. The next high-order terms of expansion of the product of multipole operators in powers of the dimensionless variable $s/R$ 
lead to some small surface corrections, relatively $\sim 1/k^{}_{F}R$, 
at large particle number A. In particular, it is shown that the next order curvature corrections, $\sim 1/(k^{}_{F}R)^2$, for a given large $k^{}_{F}R$ 
can be neglected within the considered almost local (TF) approximation in the case (i). 
Notice that more important surface ($\sim 1/k^{}_{F}R$) and curvature 
($\sim 1/(k^{}_{F}R)^2$) corrections are originated from those of the ETF relationship between 
$k^{}_{F}R$ and particle number A (\ref{akrTF}). The derivation of the surface and curvature corrections are considered in Appendix B 
shown for shortness only at the very end
of Section \ref{transport}. All of them will be discussed in last Sections \ref{enewsr}--\ref{disc}.

For evaluation of the contributions (ii) of longer trajectories, the Gutzwiller expansion (\ref{Gsem}) with (\ref{GsemGa}) valid for the {\it isolated} classical paths, fails because we have to account for {\it a continuous symmetry} of the spherical Hamiltonian, e.g.\ , appearance of the axially-symmetric degenerated families of planar POs with their points fixed inside of the spherical reflection boundary \cite{strumag}. For such a family, due to the integration over its continuous parameter, the amplitude of the Green's function term, $G_{\rm CT}$, in expansion (\ref{Gsem}) over trajectories CTs is enhanced in order of $(k^{}_{F} {\cal L}_{\rm CT})^{1/2}$ (or $\hbar^{-1/2}, ~k^{}_{F}{\cal L}_{\rm CT} >> 1$), as compared to (\ref{Gsem}) and (\ref{GsemGa})  
\cite{strumag,magosc}. 
For the case of higher classical degeneracy ${\cal K}$ 
(Section II) the CTs closed in the phase space, i.e.\ , PO families, 
yield the contributions into the Green's function 
amplitude enhanced by the factor $(k^{}_F{\cal L}_{\rm CT})^{{\cal K}/2}$ for 
billiards, or $[S_{\rm CT}(\vareps^{}_F)/\hbar]^{{\cal K}/2}$ for potentials
with the edge diffuseness as RPLP in \cite{MKApanSOL1}. 
For the nondiagonal (CT$ \neq $CT$^\prime$) contributions (ii) 
into the integrals over $\r_2$ (or $\s$ in the $\r$, $\s$ coordinates) in 
(\ref{m0}) and (\ref{gamma0}) with parameter $\Gamma$ of the SCM averaging, much smaller than that related to the distance between major shells $\hbar \Omega$, the leading terms in semiclassical parameter $k^{}_{F}R$ are the POs, according to the stationary phase conditions \cite{magvvhof}. These nondiagonal  
terms 
provide,
through the stationary phase (PO) conditions provide mainly the shell (nonlocal) corrections to the inertia $M_{\rm QQ}(0)$, the 
friction $\gamma^{}_{\rm QQ}(0)$, and the isolated susceptibility 
$\chi^{}_{\rm QQ}(0)$ \cite{MGFnpae2008}. They will be discussed in details in further publications. In following,
within this Section IVA, we shall 
consider only the smooth transport coefficients, and therefore, 
for simplicity,
the tilde above them will be omitted everywhere.

\bigskip
\subsubsection{Coupling constants and transport coefficients}
\label{transport}
\medskip

As shown in \cite{hofbook,GMFprc2007,MGFyaf2007}, the 
consistent collective-transport coefficients $\gamma^{}_{\rm QQ}$, $M_{\rm QQ}$ and $C_{\rm QQ}$ of (\ref{trcoll}) differ from their auxiliary ``intrinsic'' 
parameters $\gamma^{}_{\rm QQ}(0)$ (\ref{gammaQQ}), $M_{\rm QQ}(0)$ (\ref{mQQ}) and 
$C_{\rm QQ}(0)$, see also $\chi^{}_{\rm QQ}(0)$ (\ref{chiQQ}), 
in the low frequency expansion (\ref{chiexp})
 by small semiclassically corrections, 
\be\l{cchi}
\frac{C_{\rm QQ}(0)}{\chi^{}_{\rm QQ}(0)} = 
\frac{C_{\rm FF}(0)}{\chi^{}_{\rm FF}(0)} \sim\,
\frac{1}{(k^{}_{F} R)^2} << 1\;,
\ee
as shown in \cite{hofbook,MGFyaf2007}, and 
\bea\l{g2chm}
&&\hspace{-0.5cm}\frac{\gamma^2_{\rm QQ}(0)}{\chi_{\rm QQ}(0) M_{\rm QQ}(0)} 
= \frac{6}{\pi}\, 
\frac{[f_{\lambda}^{(1)}]^2}{f_{\lambda}^{(0)}\,f_{\lambda}^{(3)}}\,\frac{1}{k^{}_{F} R}
\nonumber\\ 
&&\approx \frac{1}{k^{}_{F} R} << 1\;,
\eea
in contrast to the corresponding result within the approach of 
\cite{MGFyaf2007}. 
Therefore, in the following derivations, one may neglect the 
curvature corrections  
(\ref{cchi}) in (\ref{trcoll}) but keep (\ref{g2chm}) 
for the surface terms ($\sim 1/k^{}_{F}R$),
\bea\l{coltrin}
C_{\rm QQ} &\approx& C_{\rm QQ}(0), \quad 
\gamma^{}_{\rm QQ} \approx \gamma^{}_{\rm QQ}(0)\;,\nonumber\\ 
M_{\rm QQ} &\approx& M_{\rm QQ}(0)+\gamma_{\rm QQ}^{2}(0)/\chi^{}_{\rm QQ}(0)\;.
\eea

For calculations of the response function $\chi_{\rm QQ}^{\rm coll}(\omega)$ 
(\ref{chicollQ}),
one has to transform the transport coefficients 
$\gamma^{}_{\rm QQ}$, $M_{\rm QQ}$ and $C_{\rm QQ}$ 
from the variable $Q$ related to the s.p.\ 
operator 
$\hat{Q}$ to another variable associated with the operator $\hat{{\rm F}}$ 
by using (\ref{trcollQF}) 
\cite{bormot,hofbook,MGFyaf2007}.  
The coupling constants 
$\kappa^{}_{\rm FF}$ and $\kappa^{}_{\rm QQ}$ appearing both 
in the transformation equations (\ref{trcollQF}) are given by
\begin{eqnarray}\label{kappaF}
\kappa^{}_{\rm  FF}&=&-\frac{32 \overline{\rho} 
b^{}_{\rm V} K R^4}{675 b^{}_{\rm S} r_0}
\nonumber\\
&\approx& -\frac{8\;KR}{225 \pi\;b^{}_{\rm S}r_0}\; b^{}_{\rm V} A
\end{eqnarray}
and 
\begin{eqnarray}\label{kappaQres}
\kappa_{\rm QQ}&=&\rho R^{L+3}\;, \nonumber\\ 
\rho&=&
\overline{\rho}\left(1+\frac{6 b^{}_{\rm S}r_0}{K R}\right) \approx 
\overline{\rho}\;,
\end{eqnarray}
with the parameters $b^{}_{\rm V}$, $K$, 
 $r_0~~$  
($\overline{\rho}$) 
of infinite 
nuclear matter  
defined in Section IIA.
The energy surface constant, 
$b^{}_{\rm S}=b^{(+)}_{\rm S}$, is given by 
(\ref{bsplus}). 
For the typical nuclear parameters, the surface-tension particle-density correction in (\ref{kappaQres}) inside of nucleus is small relatively for  
heavy  nuclei, 
$6 b^{}_{\rm S}r_0/KR \approx 6 b^{}_{\rm S}/KA^{1/3} <<1$, in the ES approximation \cite{strmagbr,strmagden}.

Substituting (\ref{trcoll}), (\ref{kappaF}), (\ref{kappaQres}), (\ref{mQQ}) and (\ref{gammaQQ}) into (\ref{trcollQF}) for the consistent transport coefficients in the Thomas-Fermi approach,
one approximately finds
\begin{eqnarray}\label{mFmirr}
&&\hspace{-0.5cm}M_{\rm FF}= 6 \pi^2 L f_{\lambda}^{(3)}\; \rho\;\left( \frac{16 b_{v}\;KR}{2025 
\vareps_{F} \; b^{}_{\rm S} r_{0}}\right)^2\nonumber\\
&&\times\frac{ (k^{}_{F} R)^4}{A}\; M_{\rm irr}\;, \qquad
M_{\rm irr}=\frac{3 A m R^2}{4 \lambda \pi}\;, 
\end{eqnarray}
\begin{eqnarray}\label{gammaFwf}
\gamma^{}_{\rm FF}&=& \frac{1}{2}\;
f_{\lambda}^{(1)}\;\left(\frac{32 b^{}_{\rm V} \;KR}{675 \vareps^{}_{F}\;b^{}_{\rm S}r_{0}}
\right)^2\gamma_{\rm wf}\;,\nonumber\\ 
\gamma_{\rm wf}&=&\frac{3}{4} \hbar \rho k^{}_{F} R^4\;,
\end{eqnarray}
where $M_{\rm irr}$ is the irrotational flow inertia of the  
hydrodynamical
LDM (HDM) \cite{bormot}, and
$\gamma_{\rm wf}$ is the wall formula 
\cite{wall,koonran} re-derived in \cite{MGFyaf2007} for the operator $\hat{\rm F}$ of (\ref{consisteqF}) in the NLA (i). 
They both are used below as convenient units. Notice, the convergence of the expansions in semiclassical parameter, $1/k^{}_{F}R$, and  
leading terms (\ref{mFmirr}) for $M_{\rm FF}$ and (\ref{gammaFwf}) for $\gamma^{}_{\rm FF}$ survive owing to the Strutinsky averaging in $k^{}_{F}R$ [see (\ref{Jwfavdef})] with a  
Gaussian width $\widetilde{\Gamma}$ corresponding at least to a few major shells $\hbar \Omega$ of the energy spectrum 
 \cite{strut,fuhi,sclbook}.
The in-compressibility $K$ and the surface energy constant $b^{}_{\rm S}$ appear in these equations in 
terms of the same semiclassical and leptodermous parameter, 
$1/k^{}_{F}R \sim a/R \propto b^{}_{\rm S}/K A^{1/3}$ ($k^{}_{F}a \sim k^{}_{F}r_0 \sim 1$, 
$R=r_0 A^{1/3})$ up to a number constant, 
through (\ref{kappaF}) for the coupling constant $\kappa^{}_{\rm FF}$, 
see Appendix B1  
and Section III. 
Such general common 
properties of the  
ETF and liquid-droplet models based both on expansion in  
small parameter $a/R$ were found, for instance, within the ES approach \cite{strmagbr,strmagden}. 

However, the ETF inertia (\ref{mFmirr}) is much larger than that of irrotational flow, $M_{\rm irr}$, in terms of the parameter $k^{}_{F}R \sim A^{1/3} >> 1$,
\begin{eqnarray}\label{mfmirrA}
\frac{M_{\rm FF}}{M_{\rm irr}} &=& \overline{M}_{\rm FF} A\;,\\
\overline{M}_{\rm FF} &\approx&
0.036, \quad \mbox{for}\quad \lambda=2\;, \nonumber\\
\overline{M}_{\rm FF} &\approx&
0.043, \quad \mbox{for}\quad \lambda=3\;. 
\end{eqnarray}
In these evaluations of the relative inertia for the symmetric 
Fermi-system of $A$ nucleons we used the nuclear data mentioned above
($b^{}_{\rm V}=16$ MeV, $K=220$ MeV, $\overline{\rho}=0.16$ fm$^{-3}$,
$b^{}_{\rm S}=18$ MeV). The Fermi energy
$\vareps^{}_{F}$ is determined through the Fermi wave number $k^{}_{F}$ by particle density of 
infinite nuclear matter, $\overline{\rho}= 2 k_{F}^3/3 \pi^2$, 
as usually in the Thomas-Fermi model 
\cite{sclbook}. For particle number $A=100 \div 200$, the inertia values 
$M_{\rm FF}$ are larger than the irrotational flow one by factor of about $4\div 7$ for the quadrupole (L=2) and almost $4 \div 9$ for the octupole (L=3) vibrations, respectively. Note that an inertia enhancement with respect to the quantity $M_{\rm irr}$ was found in \cite{kolkon} within the stochastic cranking model.

Taking into account the unlocal surface corrections (\ref{mQQ1cor12}) of (\ref{mQQ0}) and (\ref{g2chm}) of (\ref{trcoll}) ($\propto \gamma_{\rm QQ}^2(0)$), 
which are both small relatively as $1/k^{}_{F}R \sim A^{-1/3}$, for the inertia 
$M_{\rm FF}$, see also (\ref{trcollQF}), up to small negligibly curvature corrections of (\ref{cchi}), one obtains
\begin{eqnarray}\label{mFmirrsurfcor}
&&M_{\rm FF}=\frac{\pi \rho m^3R^{2L+6}}{2 \hbar^4 k_F^3}\,
\frac{\kappa^2_{\rm FF}}{\kappa^2_{\rm QQ}}\;\nonumber\\
&\times&\left(f_{\lambda}^{(3)}+\frac{\zeta_{\lambda}}{\pi\; k^{}_{F}R}\right) \left(1+
\frac{6 [f_{\lambda}^{(1)}]^2}{\pi k^{}_{F} R f_{\lambda}^{(0)}\;f_{\lambda}^{(3)}}\right)\;. 
\end{eqnarray}
Here, $\kappa^{}_{\rm FF}$ and $\kappa^{}_{\rm QQ}$ are the coupling constants
(\ref{kappaF}) and (\ref{kappaQres}) [(\ref{kld}) and (\ref{coupconsurfcor})] 
completed by small surface and curvature corrections, respectively; $f_{\lambda}^{(n)}$ are given by (\ref{flkdef}) and (\ref{rint}); 
$\zeta_{\lambda}=\zeta_{\lambda}^{(1)} + \zeta_{\lambda}^{(2)}>0$, where
$~\zeta_{2}^{(1)}=- 3 f_{2}^{(2)}=-127/84$, $~\zeta_{2}^{(2)}=1279/576~$ for $~\lambda=2~$,
$~\zeta_{3}^{(1)}=- 3 f_{3}^{(2)}=-199/165$, and 
$~\zeta_{3}^{(2)}=67031/42240~$ 
at $~\lambda=3$, $f_{\lambda}^{(2)} = 
(8\lambda^2+32\lambda+31)/[(\lambda+2)(2\lambda+3)(2\lambda+5)]$, 
see Appendix B,
(\ref{mQQVSsplit}) and (\ref{mQQ1cor12}).

For the stiffness, $C_{\rm FF}$, at leading order of expansion in $A^{-1/3}$ within the ES approximation \cite{strmagbr,strmagden,GMFprc2007,MGFyaf2007}, 
as for the coupling constant, $\kappa^{}_{\rm FF}$, one has the values of the HDM for the vibration multipolarity $\lambda$ \cite{bormot}, 
\begin{equation}\label{stifLDcoul}
C_{\rm FF} \equiv C_{\lambda}^{(S)} + C_{\lambda}^{\rm (Coul)}\;.
\end{equation} 
The surface component of the HD stiffness,
\begin{equation}\label{stifLD}
C_{\lambda}^{({\rm S})} = \frac{b^{}_{\rm S}}{4\pi r_0^2}\,
(\lambda - 1)(\lambda + 2)\,R^{2}\;,
\end{equation}
is complemented by the Coulomb term along the $\beta$-stability line 
\cite{bormot,bormot1}, 
\begin{eqnarray}\label{Ccoul} 
C_{\lambda}^{({\rm Coul})}&=&-\frac{3\,(\lambda-1)}{2 \pi\,(2\lambda+1)}\,
\frac{Z^2e^2}{R}\;,\nonumber\\
 Z &=& \left[
\frac{A}{2+3e^2\,A^{2/3}/10 r_0 J}\right]\;,
\end{eqnarray}
where $Ze$ is the charge of nucleus. The square brackets in (\ref{Ccoul}) mean the integer part of number, and $J$ is the coefficient of symmetry term in the nuclear binding energy (Section III). The approximation 
$C_{\rm FF} \approx C_{\rm FF}(0)$ up to small semiclassically curvature corrections, $\sim A^{-2/3}$, see (\ref{cchi}), was used here as in (\ref{mFmirrsurfcor}) \cite{hofbook,GMFprc2007,MGFyaf2007}.

\bigskip
\subsubsection{Vibration energies and sum rules}
\label{enewsr}
\medskip

The energies of the collective vibration modes are determined by poles of the response function (\ref{chicollQ}) with the inertia $M_{\rm FF}$ 
(\ref{mFmirr}),
 the friction $\gamma^{}_{\rm FF}$  (\ref{gammaFwf}), the stiffness 
$C_{\rm FF}$ (\ref{stifLDcoul}), and the coupling constant
$\kappa^{}_{\rm QQ}$ (\ref{kappaQres}). These poles are given by
\begin{equation}
\omega_{\pm}=\varpi \left(\pm \sqrt{1-\eta_{\rm FF}^2} - i \eta_{\rm FF}\right),
\label{roots}
\end{equation}
where
\bea\l{freqeta}
\varpi&=&\sqrt{\frac{C_{\rm FF}}{M_{\rm FF}}}\;, \nonumber\\
\eta_{\rm FF} &=& 
\frac{\gamma^{}_{\rm FF}}{2 \sqrt{M_{\rm FF} \;C_{\rm FF}}}\;.
\eea
Subscript ``$FF$'' in $\eta^{}_{\rm FF}$ will be omitted within this
section for simplicity. According to (\ref{gammaFwf}), (\ref{mFmirr}), (\ref{stifLD}) and (\ref{Ccoul}), for the effective damping parameter 
$\eta$, 
see (\ref{freqeta}), one finds $\eta \siml$ 0.4 
and 0.2 for $\lambda=2$ and $3$, 
respectively. The last estimates were obtained for the same nuclear parameters shown above at $A \siml 200$ with accounting for surface and curvature corrections. As seen from these estimates, the collective motion under consideration is underdamped, $\eta < 1$, 
for any realistic
particle numbers, $A \siml 200$. Note that the residue interaction was zero from the very beginning in (\ref{GFdef}) for the Green's function $G$, 
$\epsilon^{}_0=+0$. The averaging over $k^{}_{F}R$ which guarantees a convergence of smooth transport coefficients in the semiclassical expansion over $\hbar$ (or $1/k^{}_{F}R$ in dimensionless units), leads to a finite friction coefficient, $\gamma^{}_{\rm FF}$, or an effective damping constant $\eta$, as formally
with $\eps_o\neq 0$ in (\ref{GFdef}). More precisely, it takes place for the formal averaging with Lorentzian weight function, see \cite{bablo,koonran}. However, as shown in this and two next sections, the influence of the effective damping parameter $\eta$ on calculations of the excitation energies, transitions probabilities, and EWSR contributions 
can be neglected in the following derivations.

Neglecting now by small 
$\eta^2$ term in the real part of (\ref{roots}) for calculations of the smooth low-lying collective vibration energy, $\hbar \omega = \hbar\Re\; \omega_{+} \approx
\hbar \varpi$, from (\ref{roots}), (\ref{freqeta}), (\ref{mFmirr}) 
and 
(\ref{stifLDcoul}), at $\lambda \geq 2$ one approximately obtains 
\bea\l{hw} 
\hbar \omega_{\lambda}&=&\frac{D_{\lambda}}{A}\;,\nonumber\\ 
{\cal D}_{\lambda}&=& \overline{D}_{\lambda} \left(1 + \frac{
C_{\lambda}^{({\rm Coul})}}{C_{\lambda}^{({\rm S})}}\right)^{1/2}, 
\eea
\bea\l{hw1} 
\overline{D}_{\lambda}&=&\frac{75\,b^{}_{\rm S} 
\vareps^{}_{F}}{4 \pi b^{}_{\rm V} K}
\nonumber\\ 
&\times&
\sqrt{\frac{3 \vareps^{}_{F} 
b^{}_{\rm S} (\lambda-1)(\lambda+2)} {f_{\lambda}^{(3)}}}\;. 
\eea 
For nuclear parameters mentioned above, the constant $\overline{\cal
D}_{\lambda}\;$, independent of the particle number A, 
is given approximately 
by $\overline{D}_{2} = 100$ MeV and $\overline{D}_{3}= 180$ MeV. 
With the Coulomb corrections of (\ref{hw}), these constants 
become slightly almost linearly decreasing functions of A within the 
interval about $A=100 \div 200$. For this A interval they are 
modified approximately to the values $D_{2}=90 \div 70$ MeV, and 
$D_{3}=170 \div 150$ MeV, respectively, 
see (\ref{stifLD}) and (\ref{Ccoul}).

We may now evaluate the EWSR $S_{\lambda}^{(1)}$ for contribution of the first low-lying excitation, see the integral (\ref{momstren}) 
for ${\it l}=1$ with the strength function 
$S_{\lambda}(\omega)$ (\ref{strength1}). Using now the even parity of its integrand, one can extend the low integration limit to $-\infty$ and integrate over $\omega$ by the residue method in complex plane of $\omega$. Closing the integration contour in lower plane of $\omega$ we calculate the contributions of the two poles $\omega=\omega_{\pm}$ inside of this closed contour. Finally, for the EWSR of the low-lying collective excitation, one finds [see also 
(\ref{momstren}), (\ref{kappaQres}) and (\ref{mFmirr})]
\begin{eqnarray}\label{sumrules}
S_{\lambda}^{(1)}&=&\frac{\hbar^2\;\kappa_{\rm QQ}^2}{2 M_{\rm FF}} \nonumber\\
&=&
\frac{M_{\rm irr}}{M_{\rm FF}}\, S_{\lambda,{\rm cl}}\;, 
\end{eqnarray}
where 
\begin{equation}\label{EWSR1}
S_{\lambda,{\rm cl}}= \frac{\hbar^2\;\kappa_{\rm QQ}^2}{2 M_{\rm irr}}= \frac{3
\lambda;\vareps^{}_{F}}{4 \pi\,(k_{F}R)^2}\;A R^{2\lambda}.
\end{equation}
This $S_{\lambda,{\rm cl}}$ appears to be exactly the same as the contribution of the low-lying peak in the HDM of the irrotational flow in a classical liquid droplet. 
It is equivalent to the EWSR estimation independent of the model, see 
(6.179a) in \cite{bormot}. The ratio of the inertias, $M_{\rm irr}/M_{\rm FF}$, in 
(\ref{sumrules}) is given by (\ref{mFmirr}). Note that the last equation in 
(\ref{sumrules}) recalls the EWSR relation (6.183) of \cite{bormot}. Thus, we may evaluate the relative contribution of the low-lying collective state into this total EWSR estimation $S_{\lambda,{\rm cl}}$, see (\ref{sumrules}) 
with (\ref{mFmirr}) and 
(\ref{mfmirrA}),
\begin{eqnarray}\label{EWSR}
\frac{S_{\lambda}^{(1)}}{S_{\lambda,{\rm cl}}}&=
&\frac{\overline{S}_{\lambda}^{(1)}}{A}\;,\nonumber\\
 \overline{S}_{\lambda}^{(1)}&=& \frac{2}{\lambda f_{\lambda}^{(3)}}\;
\left(\frac{225\, 
\vareps^{}_{F} b^{}_{\rm S} k^{}_{F} r_0}{8\pi b^{}_{\rm V} K}\right)^2,
\end{eqnarray}
with constants $\overline{S}_{2}^{(1)}\approx 7$ and $\overline{S}_3^{(1)} 
\approx 6$ for the same nuclear parameters. For $A\sim 100 \div 200$ one has a small relatively EWSR contribution of the low-lying collective excitations in the framework of the ETF model. According to (\ref{mfmirrA}), this is obviously because of small  
values of the ratio of inertia parameters, $M_{\rm irr}/M_{\rm FF}$, for large particle numbers A. It is in contrast to the HDM where the first low-lying peak exhausts erroneously 100\% of the EWSR $S_{\lambda,{\rm cl}}$ 
independent of the model \cite{bormot}. By this reason, the ETF approach to the collective nuclear vibrations is much improved with respect to the HDM results: In addition to the low-lying collective states, one has a possibility for the giant multipole-resonance contributions
which mainly exhaust the EWSR. 

The small relatively surface and curvature corrections can be taken into account in the vibration energies (\ref{freqeta})  and sum rules (\ref{sumrules}) through (\ref{mFmirrsurfcor}) for the inertia and (\ref{akrTF}) for the ETF relationship of the parameter $k^{}_{F}R$ to the particle number $A$, see 
Appendices B1 
[(\ref{kld}) and (\ref{coupconsurfcor})] and B3.  
The last kind of the relative surface ($\sim A^{-1/3}$) and curvature ($\sim A^{-2/3}$) corrections from (\ref{akrTF}) yield the major contribution into the $A$-systematics of the vibration energies (\ref{hw}) for large $A$,
\begin{equation}\label{hwsurfcoras}
\hbar \omega_{\lambda}= \frac{D_{\lambda}}{A} 
\left(1 - \frac{w^{(s)}_{\lambda}}{A^{1/3}} + \frac{w^{(c)}_{\lambda}}{A^{2/3}}\right),
\end{equation}
where $D_{\lambda}$ is defined in (\ref{hw}) and (\ref{hw1}),
\bea\l{DLS}
w^{(s)}_{\lambda}&=& \left(\frac{9 \pi}{8}\right)^{2/3} \approx 2.3\;,\nonumber\\
w^{(c)}_{\lambda}&=&3\left(\frac{8}{9 \pi}\right)^{2/3} \approx 1.3\;.
\eea
Similarly, from (\ref{sumrules}) one obtains the EWSR ratio: 
\begin{eqnarray}\label{sumrulsurfcor}
&\frac{S_{\lambda}^{(1)}}{S_{\lambda,{\rm cl}}}=
\frac{\overline{S}_{\lambda}^{(1)}}{A} 
\nonumber\\
&\times\left(1 - \frac{2 w^{(s)}_{\lambda}}{A^{1/3}}
+\frac{[w_{\lambda}^{(s)}]^2+2w_{\lambda}^{(c)}}{A^{2/3}}\right),
\end{eqnarray}
where $\overline{S}_{\lambda}$ is the constant independent of the particle numbers A in (\ref{EWSR}). Other surface and curvature components do not contribute almost due to their smallness or mutual compensations. Note that the surface corrections decrease essentially
both vibration energies $\hbar \omega_{\lambda}$ (\ref{hwsurfcoras}), and EWSR contributions $S_{\lambda}^{(1)}$ (\ref{sumrulsurfcor}) for large sufficiently values of particle numbers, $A=100 \div 200$, because of positive sign of the dominating surface correction constant $w_{\lambda}^{(s)}$, Eq.\  (\ref{DLS}), that is
important especially for the EWSR.

\bigskip
\subsubsection{Transport probabilities and direct radiation decay}
\label{radiation}
\medskip
The radiation decay of the low-lying collective states can be considered as a direct emission of the gamma-quantum from nucleus with the semiclassical description of a charged system within our ETF
approach. A similar process for the case of the direct radiation decay of isoscalar giant multipole resonances by using the Landau-Vlasov approach with the 
macroscopic boundary conditions (Appendix A2) 
within the framework of the semiclassical description of a nuclear system was studied in \cite{magpl}. For the probability of the transitions per unit of time with the electric radiation of the gamma-quantum energy $\vareps^{}_{\gamma}$ at 
multipolarity $\lambda$, one has \cite{bormot1}
\begin{eqnarray}\label{probabEL}
&&\hspace{-0.7cm}W(E\lambda,I_1 \rightarrow I_2 )= 
\frac{8\pi(\lambda+1)}{\lambda[(2\lambda+1)!!]^2 \hbar} \nonumber\\
&\times&\left(\frac{\vareps_{\gamma}}{\hbar c}\right)^{2\lambda+1}B(E\lambda,I_1
\rightarrow I_2)\;. 
\end{eqnarray}
Here, $I_1$ and $I_2$ are the spins of the initial and the final states, 
$B(E\lambda,I_1 \rightarrow I_2)$ is the reduced probability,
\bea\label{redprobEL}
&&\hspace{-0.5cm}B(E\lambda,I_1 \rightarrow I_2)\nonumber\\
&&= \sum_{\mu M_2}\vert \langle I_2 M_2 \vert {\cal M}(E\lambda,\mu) 
\vert I_1 M_1 \rangle \vert^2\;,
\eea 
$\mu$ and $M_{2}$ are the projections of the gamma-quantum spin 
$\lambda$ and the angular momentum of a final nuclear state $I_{2}$, 
respectively, and
\begin{equation}\label{operEL}
\mathcal{M}(E\lambda,\mu)= \frac{e Z}{A}\;\int \d \r \rho(\r)\;
r_{}^{\lambda} Y_{\lambda
\mu}(\theta, \varphi)\;.
\end{equation}
The effective charge factor can be approximately put one for the isoscalar collective excitations with $\lambda \geq 2$.

The quantum reduced probability $B(E\lambda, 0 \rightarrow \lambda)$ 
(\ref{redprobEL}) 
can be evaluated through the zero moment $S_{\lambda}^{(0)}$ 
(\ref{momstren}) of the strength function $S_{\lambda}(\omega)$  
(\ref{strength1}). Taking into account the conservation 
equations for the energy, 
$\vareps_{\gamma}=\hbar \omega_{\lambda}$, and the angular momentum, 
$I_2=\lambda$ ($I_1=0$) in the direct nuclear-gamma decay 
\cite{bormot1,magpl} for the zero moment of (\ref{momstren}) 
by using a handbook one obtains 
\bea\l{semiBLint}
S_{\lambda}^{(0)}&=& \frac{\hbar\;\kappa_{\rm QQ}^2}{2\pi
M_{\rm FF}\omega_{\lambda}\;\sqrt{1-\eta^2}} \nonumber\\
&\times&\mbox{arccot}
\left[\frac{2\eta^2-1}{2\eta\sqrt{1-\eta^2}}\right] \nonumber\\
&=&
\frac{\hbar\;\kappa_{\rm QQ}^2}{2 M_{\rm FF}\,\omega^{}_{\lambda}} \left[1 -
\frac{2 \eta}{\pi} + {\cal O}\left(\eta^2\right)\right],
\end{eqnarray}
where $M_{\rm FF}$, $\kappa^{}_{\rm QQ}$ and $\eta=\eta_{\rm FF}$ are given by 
(\ref{mFmirrsurfcor}), (\ref{coupconsurfcor}) and (\ref{freqeta}), respectively, see also (\ref{gammaFwf}) and 
(\ref{stifLDcoul}), $\eta < 1/\sqrt{2}$. As shown in the previous section, the contribution of $\eta$ terms is really small enough for particle numbers
 $A \siml 200$ and all multipolarities $\lambda \geq 2$ under consideration such that one may neglect all $\eta$ corrections in (\ref{semiBLint}). The averaged probability $B(E\lambda,0 \rightarrow \lambda)$ (\ref{redprobEL}) can be then approximated semiclassically within the ETF model,
\bea\l{semiBLappr}
&&\hspace{-0.5cm}B(E\lambda, 0\rightarrow \lambda) 
\approx B_{\rm scl}(E\lambda, 0 \rightarrow \lambda)\nonumber\\
&&\!=\!(2
\lambda+1)\left(\frac{e Z}{A}\right)^2 S_{\lambda}^{(0)} \!\approx\! 
(2 \lambda+1)
\left(\frac{e Z}{A}\right)^{2}\!\nonumber\\
&&\times \frac{\hbar\;\kappa_{\rm QQ}^2}{2
M_{\rm FF}\,\omega^{}_{\lambda}}\;.
\eea
The degeneracy factor
$2 \lambda +1$ was accounted because of the additional summation over the
projections $M_{2}$ of the final angular momentum $I_{2}=\lambda$ in
(\ref{redprobEL}) with (\ref{operEL}), as compared to the simplest
multipole operator for the isoscalar excitations of $A$ nucleons,
$\int \d \r \rho(\r) \;r^{\lambda} Y_{\lambda 0}(\theta)$, considered in the
previous sections. The factor $(e Z/A)^2$ must be also taken into
account in the last two equations in (\ref{semiBLappr}), see
(\ref{redprobEL}), (\ref{operEL}), and (6.61), (6.182) of
\cite{bormot}. The semiclassical energy of the low-lying
collective state, $\hbar \omega_{\lambda}$, was derived in Section
\ref{enewsr}, see the first equation in (\ref{freqeta}) and its
approximations (\ref{hw}) and (\ref{hwsurfcoras}), 
as applied for the
considered direct radiation decay like in \cite{magpl}. Other
denotations are the same as in the sum rule
(\ref{sumrules}). Thus, from comparison of the transition probability
(\ref{semiBLappr}) and the EWSR (\ref{sumrules}) modified with the
operator (\ref{operEL}), 
one has the following approximate relationship
between the probability (\ref{semiBLappr}) and the corresponding sum
rule \cite{bormot,magpl}:
\bea\l{SBwrel}
{S}_{\lambda,{\rm scl}}&\equiv& (2\lambda+1)\; \left(\frac{e
Z}{A}\right)^{2}\; S_{\lambda}^{(1)} \nonumber\\
&=& \hbar \omega^{}_{\lambda} B_{\rm scl}(E\lambda,
0 \rightarrow \lambda). 
\eea

According to (\ref{SBwrel}), (\ref{sumrules}), (\ref{mfmirrA}) and 
(\ref{hw}), the reduced probability $B(E\lambda)$, for example, for the radiation process $\lambda \to 0$ in units of the s.p.\
estimation \cite{bormot1} mainly writes
\begin{equation}\label{be2colweiskopf}
\hspace{-0.5cm}\frac{B_{\rm scl}(E\lambda)}{B_{\rm s.p.}(E\lambda)}
\approx \frac{\hbar^2
\lambda(3+\lambda)^2}{6 m r_0^2 \;D_{\lambda}
\overline{M}_{\rm FF}}\left(\frac{Z}{A}\right)^{2}A^{1/3}\;.
\end{equation}
As seen from this simple evaluation, the particle dependence of the
reduced probability $B_{\rm scl}(E\lambda)$ divided by the factor $(Z/A)^2$
in the s.p.\ units is roughly proportional to a large
semiclassically parameter $A^{1/3}$. In these derivations we used the
approximate relation (\ref{Ccoul}) of the proton number $Z$ to the
total particle number $A$. Taking into account also the coefficient
in front of the $A^{1/3}$ dependence (\ref{be2colweiskopf}), one
obtains even larger magnitude for this relative probability, $~\approx
80 \div 130~$ for the quadrupole and $~\approx 80 \div 90~$ for
the octupole low-lying collective states at large particle numbers, $A
= 100 \div 200$, respectively. With the main surface and curvature
corrections of (\ref{sumrulsurfcor}) and (\ref{hwsurfcoras}), owing
to the ones of the ETF relationship (\ref{akrTF}) 
between $k^{}_{F}R$ and $A$, these
quantities are decreased with respect to their local (volume)
approximation, $~\approx 50 \div 70~$ for $\lambda=2$ and $~\approx 50 \div
60$ for $\lambda=3$. Thus, in any case the quadrupole and
octupole electric transitions within our semiclassical model are the
well pronounced sufficiently 
collective excitations.

For the mean semiclassical lifetime with respect to the direct gamma decay, one has
\begin{eqnarray}\label{lifetime}
&&t_{\lambda,\rm scl}=W_{\rm scl}^{-1}(E\lambda,\lambda \to 0) \;\nonumber\\
&\propto& \omega_{\lambda}^{-(2 \lambda + 1)} 
B_{\rm scl}^{-1}(E\lambda, \lambda \rightarrow 0) \nonumber\\
&\propto& (A/eZ)^2
A^{2(2\lambda+1)/3}. 
\end{eqnarray}
With the semiclassical ETF probability per unit of time 
$W_{\rm scl}(E\lambda, \lambda \to 0)$ (\ref{probabEL}) corresponding to the reduced probability $B_{\rm scl}(E\lambda, \lambda \rightarrow 0)$ 
(\ref{semiBLappr}), 
one obtains $t_{2,\rm scl} \approx 80 \div 1100$ and $t_{3,\rm scl} \approx (17 \div 1100)\cdot
10^6$ ps for the quadrupole and octupole low-lying collective vibration states within the
same particle number interval. We accounted for the main surface and curvature corrections which increase significantly lifetimes $t_{\lambda,\rm scl}$ 
(\ref{lifetime}). In these evaluations we neglected
the corrections owing to the conversion processes which become important for much smaller excitation energies.

\bigskip
\subsubsection{Comparison with experimental data 
}
\label{disc}
\medskip

Figures 
\ref{fig15} and \ref{fig16} 
show the local Thomas-Fermi approach to the low-lying collective quadrupole, $\hbar \omega_{2}$, and octupole, $\hbar \omega_{3}$, excitation 
energies; see (\ref{hw}) with 
(\ref{stifLDcoul}) for the stiffness, $C_{\rm FF}$, and
(\ref{mFmirr}) for the inertia, $M_{\rm FF}$, without surface and curvature corrections, as compared to the experimental data \cite{raman} and \cite{kibedi}, respectively, see also
\cite{table}. 
The calculations are performed for the following nuclear 
parameters: $\overline{\rho} = 0.16$ fm$^{-3}$ ($r_0=1.14$ fm),
    $b^{}_{\rm V}=16$ MeV, $b^{}_{\rm S}=18$ MeV, $K=220$ MeV,
    $J_{\rm sym}=30$ MeV (Section III). 
The almost spherical nuclei with quadrupole deformations, $q_2 \siml 0.05$, are selected from these experimental data \cite{table,plugor,plugorkad,MGFyaf2007}. The Thomas-Fermi
results for smooth
vibration energies are significantly improved with respect to the
hydrodynamic (HD) 
behavior with the same stiffness $C_{\rm FF}$
(\ref{stifLDcoul}) but the irrotational flow inertia, $M_{\rm irr}$ of
(\ref{mFmirr}), especially for the quadrupole case.

More complete ETF approach (\ref{hw}), (\ref{stifLDcoul}) and
(\ref{mFmirrsurfcor}) with accounting for the surface and curvature
corrections of the function, $k^{}_{F}R(A)$, found from 
(\ref{akrTF}) and those of the inertia $M_{\rm FF}$,
(\ref{mFmirrsurfcor}), (\ref{kld}) and (\ref{coupconsurfcor}), see
Appendices B1 and B3,  
are shown as ETF curves in Figs.\ 
\ref{fig15} and \ref{fig16}. As expected, the comparison 
with experimental data, except
for several doubly-closed-shell (magic) nuclei which appear above the
regular A-systematics, is essentially improved by these corrections, 
mainly for smaller particle numbers A. The Coulomb stiffness component
becomes important for larger A. The reason of better agreement of the ETF 
approach, as compared to the
HD model, with experimental data for nonmagic nuclei can 
be explained
by larger ETF inertias $M_{\rm FF}$, see (\ref{mfmirrA}), than
that of the irrotational flow, $M_{\rm irr}$, for 
heavy
nuclei. As seen from these Figures, the explicit analytical formula
(\ref{hwsurfcoras}) (ETFA, dashed), where the only surface and
curvature corrections in the $k^{}_{F}R(A)$ ETF dependence (\ref{akrTF}) were
taken into account, is a good asymptotics for large particle numbers,
$A \simg 40$. We should emphasize that this asymptotics versus the TF
and ETF curves shows importance of namely these corrections, as
compared to all other ones.

Figs.\ 
\ref{fig17} and \ref{fig18} 
show the semiclassical reduced
probability, $$~~~B_{\rm scl}(E\lambda,0 \rightarrow \lambda)=
(2 L+1) B_{\rm scl}(E\lambda,\lambda \rightarrow 0)\nonumber$$
[see (\ref{semiBLappr})], related to the zero moment of the strength function
(\ref{stren}), see also (\ref{momstren}) at ${\it l}=0$, versus
experimental data \cite{table,raman,kibedi} for the quadrupole, $0^{+}
\to 2^{+}$, and the octupole, $0^{+} \to 3^{-}$, electric collective
transitions in the low-lying energy region for the same nearly
spherical nuclei. The logarithmic scale is used in order to show this
comparison in a wide region of particle numbers. As seen from these
Figures, one has a 
good agreement between the averaged
semiclassical reduced transition probabilities and the global behavior
of experimental data \cite{raman,kibedi} (besides of magic nuclei).
The surface and curvature correction effects measured by differences
between TF and ETF curves are in fact smaller than both of them in
comparison with the HD values. Our semiclassical smooth A-systematic
results look better versus the experimental data than that of the 
HDM.
The agreement between the full ETF (thin solid) and the
analytical asymptotics ETFA (thick dashed) with the main
surface and curvature corrections coming from (\ref{akrTF}) is really
perfect for almost all particle numbers, except for very small
particle numbers, $A \siml 40$. As shown in Section \ref{radiation},
the comparison in Figs.\ 
\ref{fig17} and \ref{fig18} 
is a basis in
experimental and theoretical analysis of the direct radiation
decay of the low-lying collective states. For instance, 
the quadrupole lifetime $t^{}_{\lambda,\rm{scl}}$
(\ref{lifetime}) (Fig.\ \ref{fig19})
associated with the direct processes of the
gamma-quantum emission \cite{bormot1,magpl} are reasonable in order
of the magnitude as compared to the experimental data \cite{raman}.
As seen from comparison of the 
frequent dashed (ETF$^*$) and solid lines (ETF) in
Figs.\ 
\ref{fig17} and \ref{fig18}, one may really neglect the $\eta$
corrections of (\ref{semiBLappr}) which arise from the 
averaging procedure. The dynamic surface effects improve our
semiclassical lifetime $t^{}_{\lambda,\rm scl}$ (\ref{lifetime}) towards
the experimental data \cite{table,raman}.

One of the most important characteristics of the low-lying collective states is their energy weighted sum rule contribution (\ref{sumrules}) into the total value $S_{\lambda,{\rm cl}}$ (\ref{EWSR1}), 
independent of the model \cite{bormot}, see 
Figs.\ 
\ref{fig20} and \ref{fig21}. The experimental EWSRs were evaluated as the products of the measured transition probabilities $B(E\lambda)$, plotted in 
Figs.\ 
\ref{fig17} and \ref{fig18}, 
and the corresponding vibration energies $\hbar \omega_{\lambda}$ 
(Figs.\ 
\ref{fig15} and \ref{fig16}) taken both 
from \cite{raman} and \cite{kibedi} for the quadrupole (Fig.\ \ref{fig17}
) and octupole (Fig.\ 
\ref{fig18}) vibrations, respectively. 
According to (\ref{mfmirrA}), by the same reason of enhancement of the inertia $M_{\rm FF}$ with respect to the irrotational flow $M_{\rm irr}$, the relative contribution of the low-lying collective states into the EWSR within the 
Thomas-Fermi approach (\ref{sumrules}), (\ref{mFmirr}) and more complete ETF model (\ref{sumrules}), (\ref{mFmirrsurfcor}), (\ref{kld}) and (\ref{coupconsurfcor}), 
become basically correct for larger particle numbers A, in contrast
 to the HDM. As displaced in Figs.\ 
\ref{fig17} and \ref{fig18}, within the ETF approximation we
 obtained much smaller relatively EWSR contributions of these states
 [(\ref{sumrules}), (\ref{sumrulsurfcor}) and (\ref{DLS})] into
 the total EWSR (\ref{EWSR1}) for large  
particle
 numbers, $A \simg 70$. This is mainly in agreement with experimental
 data \cite{raman,kibedi} for the EWSR at such particle numbers
 $A$, especially better with accounting for surface and curvature corrections. 
 Again, a good EWSR asymptotics (\ref{sumrulsurfcor})
 (ETFA) for large $A$ takes only into account the surface and curvature
 corrections of the $k^{}_{F}R(A)$ function determined by the ETF particle
 number conservation equation (\ref{akrTF}).

However, Figs.\ 
\ref{fig15}-\ref{fig18} 
show also obvious importance of
other contributions, first of all arising from the shell effects,
pronounced especially for the quadrupole case, see Figs.\ 
\ref{fig15}, \ref{fig17} and \ref{fig20}.
We should not expect that the smooth ETF model 
approximates the statistically
averaged experimental data \cite{raman,kibedi} (without
shell fluctuations) depending on particle
number, 
in the characteristics of the low-lying collective states.
In contrast to the stiffness
which is almost linear in 
oscillating shell components as the 
free energy, 
$\mathcal{F}=\tilde{{\cal F}}
+ \delta {\cal F}$, see next section and \cite{MGFnpae2008}, 
the energies $\hbar \omega^{}_{\lambda}$ (\ref{freqeta}),
probabilities $B(E\lambda)$ (\ref{semiBLappr}) and EWSRs  
(\ref{strength1}) are
positive and depend on these components, through the inertia and  
coupling constants, in a more complicate (nonlinear)
way. They are certainly beyond the smooth ETF
approximation \cite{MGFnpae2008}. As expected, the magic nuclei
like $^{208}$Pb (the full point much above others on right of
Figs.\ 
\ref{fig15}, \ref{fig17} 
and near the very right of minimum of
the $B(E\lambda)$ in Fig.\ 
\ref{fig17}), for example, should be excluded from
comparison with the ETF approach for vibration energies, the reduced
transition probabilities and the EWSR contributions. They are
certainly out of the smooth A systematics. The deflection of the
experimental data for $B(E\lambda)$ from the averaged semiclassical
particle-number dependencies can be assumed to be referred to those of
the matrix elements in (\ref{redprobEL}) within more exact RPA and
POT approaches taking both into account the shell effects. As noted
in \cite{MGFyaf2007,MGFnpae2008}, the pairing effects in calculations of the
inertia within the cranking model 
\cite{broglia,fuhi} for
even--even nuclei (all full heavy experimental dots), lead basically
to the $A^{-2/3}$ behavior for not too both heavy and light nuclei
\cite{plugor,plugorkad}. Notice, the mean vibration energy, $\hbar
\omega^{}_{\lambda}$, as function of its multipolarity $\lambda$ 
[(\ref{hw}),
(\ref{hw1}) and (\ref{hwsurfcoras})] differs essentially from that 
predicted by the
hydrodynamic approach \cite{bormot}, the pairing cranking model
\cite{broglia} and that found in \cite{MGFyaf2007} with the same surface HD
stiffness because of different evaluations of the inertia. As seen
from all Figures 
\ref{fig15}--\ref{fig21}, the ETF approximation 
accounting for the Coulomb, surface and curvature corrections are
largely in good agreement with the experimental data for almost
spherical heavy  
nuclei, except for enhancement due to the
pronounced obviously shell effects in a few doubly closed shell
nuclei.

As conclusions from this section, 
for low-lying nuclear collective excitations related to the standard
multipole transitions, within a few lowest orders of the POT in $\hbar$
corresponding to the ETF approximation, we derived
smooth inertia for the vibrations near a spherical shape of the
mean edge-like field. The consistent collective ETF inertia is 
significantly larger than
that of irrotational flow. The smooth low-lying collective vibration
energies in spherical nuclei might roughly satisfy the $A^{-1}$
particle-number dependence with the relative $A^{-1/3}$ surface and
$A^{-2/3}$ curvature corrections for heavy enough nuclei, in contrast
to the mainly $A^{- 1 / 2}$ behavior predicted by the HDM and
$A^{- 1 / 3}$ dependence obtained in \cite{MGFyaf2007}.
We emphasize importance of the surface corrections, coming mainly from
the ETF dependence of the semiclassical parameter $k^{}_{F}R$ on particle
number A, in comparison with experimental data for the quadrupole and
octupole vibration energies, the lifetime of the low-lying collective
states and their EWSR contributions, besides of doubly magic
nuclei. The major behavior of the electric reduced transition
probabilities in nonmagic almost spherical nuclei are in rather good
agreement with averaged semiclassical estimations. In particular, the
semiclassical lifetimes with respect to the direct radiation decay are
reasonable as compared to their experimental data in order of the
magnitude. We found simple analytical asymptotics for the vibration
energies, the reduced probabilities, the corresponding lifetimes, and
EWSRs with explicit analytical A-dependence (ETFA)
for larger particle numbers A in good
agreement with more complete ETF approach. As the ETF inertia
$M_{\rm FF}$ is significantly larger than $M_{\rm irr}$ for the
irrotational flow, our vibration energies and
contributions to the EWSR are basically in agreement with their
experimental data (besides of magic nuclei) than those found in the HD
approach for large  
particle numbers. We proved semiclassically
that the reduced transition probabilities in the Weisskopf units for the
low-lying vibration excitations are very large 
that allows us to refer them to the
collective states. The effect of surface corrections on the smooth
vibration energies, sum rules and lifetimes of these states within the
ETF approach is emphasized much as compared to the TF approximation
leading in semiclassical expansion over $1/k^{}_{F}R$. We point out also
importance of the shell effects in all such characteristics of
low-lying collective states in magic nuclei, especially for the
collective quadrupole-vibration modes which are certainly out the
smooth A-systematics predicted by the ETF model.

For further perspectives, the Gutzwiller trajectory expansion
(\ref{Gsem}) with account for symmetries of the Hamiltonian
\cite{strumag} can be used in the semiclassical derivations of the
inertia and friction by applying the stationary-phase method for
calculations of the transport coefficients in order to study their
shell corrections related to the periodic orbits \cite{magvvhof}. We
may hope to overcome semiclassically some difficulties with the inertia
and the friction calculations which take into account the dissipative
width of the multipole strength functions.

\subsection{CORRECTIONS TO
ETF TRANSPORT COEFFICIENTS}

For calculations of the static nuclear properties of 
the total binding and
deformation energy, the famous SCM 
was successfully applied in many 
successful works, see
for instance \cite{fuhi,sclbook}. The nuclear energy was defined 
\cite{strut} as a sum of the phenomenological macroscopic part given by the
liquid-drop energy and the (semi)microscopic shell correction. 
The SCM is based on the
concept of existence of the quasiparticle spectrum near the Fermi surface
by the Migdal theory of finite fermion systems with  
the strong interaction of  
particles \cite{migdal}. Within this concept,  
the nuclear shell-correction free
energy can be considered perturbatively as a quasiparticle correction to the
total statistically smoothed
(macroscopic) 
free energy 
described phenomenologically through
the LDM or the ETF approach
\cite{sclbook}.

A first attempt to generalize these ideas to the collective dynamics  
were suggested 
as the so called liquid-particle (liquid-gas) model
\cite{strmagbr} for time evolution of the one-body density matrix
by extracting the quasiparticle effects of the quantum Fermi gas  
of almost independent particles moving in a mean field 
from the macroscopic time-dependent liquid-drop state.  
For simplicity, the latter was assumed to be approximately as the 
in-compressible condensed matter. In this model of the combined
microscopic-macroscopic dynamics, the macroscopic quantities,
 determined by means of their averaging
over the particle phase space, describe the short-length correlation 
(liquid-like)
properties, as compared to the 
mean radius of a heavy nucleus. 
Small nuclear quasiparticle excitations near the Fermi surface
\cite{migdal} are responsible for the long-length 
(quasiparticles' gas) correlations. Thus,
they arise as the self-consistent shell corrections to 
a phenomenological LDM  
 in line of the quasiparticle Fermi-liquid theory by Landau \cite{landau}
and the self-consistent finite Fermi systems by Migdal \cite{migdal}. 
In the semiclassical
approximation to the  
liquid-gas model 
\cite{magstr}, such a splitting
into the two components, described by the suitable liquid and gas
properties, is realized in the nuclear volume. The collective dynamics within a
relatively small nuclear-surface layer is reduced self-consistently to the
macroscopic boundary conditions for the Landau-Vlasov equation of
the quasiparticle motion inside the nucleus (Section III and Appendix 
A2).

For the description of the low-lying nuclear-energy 
collective excitations
\cite{migdal,bormot,khodsap,berborbro,hofbook}, 
more simple proposals were suggested
in  
 \cite{magvvhof,MGFnpae2005,MGFyaf2007} by
employing the response theory (Section IVA).
The collective variables were introduced there explicitly as deformation
parameters of a mean s.p.\ field. The nuclear excitations were
parametrized in terms of the transport coefficients, such as the stiffness,
the inertia, and the friction parameters defined 
through the adequate
collective-response functions. In analogy with the SCM, the response function
was split into the smooth macroscopic and the fluctuating shell
(quasiparticle) components. Its fluctuating part was calculated
semiclassically within the POT (Section III and
\cite{sclbook,bablo,strumag,smod,spheroidptp}), which is a powerful
analytical tool for study of the shell effects in the level density and the 
energy
shell correction.

The main scope of this Section is a suggestion of a simple version of the SCM
splitting with applying it immediately to the transport coefficients
for slow collective motion, as the sum of their macroscopic 
(statistically averaged) ETF part and
shell corrections. 
As shown in \cite{GMFprc2007,MGFyaf2007} (Section IVA), the averaged transport
coefficients can be simplified analytically with the help of the POT at 
leading orders in $\hbar$ in the nearly local approximation corresponding to the
ETF approach. Following the response function theory
of the transport coefficients (Section IVA with replacing 
$\hat{Q}=\hat{F}$, see also \cite{hofbook}) 
we are going now to formulate 
the modified SCM for calculations of
transport coefficients, such as the nuclear inertia and friction
 in a slow collective dynamics. Section IVB1 shows the well-known 
SCM as applied for the semiclassical POT free energy and stiffness
taking the example of small multipole vibrations near  the spherical surface
of a sharp-edged potential. 
In Section IVB2, we suggest to extend the SCM to the
calculations of transport coefficients for a slow collective motion by making
use of the consistent ETF approximation \cite{GMFprc2007,MGFyaf2007} 
for the macroscopic part of the combined SCM dynamics. 
For calculations of their shell corrections, the infinitely
deep spherical square-well potential is used as a mean equilibrium
field. Our SCM results
for the temperature dependence of the transport coefficients as well as
the quadrupole vibration energies, the reduced and effective friction
parameters are compared in Section IVB3 with their contrpartners
of the quantum independent-particle cranking model \cite{hofbook,hofivyam} 
and experimental data discussed in \cite{MGFyaf2007,gonfro,hilgonros,frgon}.
The smooth trajectory corrections to the ETF components of the transport 
coefficients, in addition to the shell effects, are derived
 in Section IVB4. 
The conclusion remarks are given at the end of this Section.

\bigskip
\subsubsection{Shell corrections to the free \\ energy and stiffness}

 For calculations of the quasi-static quantities as the free energy
$\mathcal{F}$ and the stiffness $C$, the SCM 
\cite{strut,fuhi,sclbook} was successfully applied in
 many works, see for instance 
 \cite{sclbook,hofbook}.
The basic point of
these calculations is the Strutinsky renormalization for heated Fermi
systems, which is similar to that for the
SCM binding and deformation energies of nuclei,
\bea\label{fSQM}
\mathcal{F}
&=& \mathcal{F}_{\rm LD} +\delta \mathcal{F}, \nonumber\\
\delta \mathcal{F}&=&
\mathcal{F}_{\rm IP}(T,A)-\widetilde{\mathcal{F}}_{\rm IP}(T,A)
\nonumber\\
 &\equiv& \delta 
\Omega(T,\lambda) = \Omega_{\rm IP}(T,\lambda) - 
{\tilde \Omega}_{\rm IP}(T,\lambda), 
\eea
where $\mathcal{F}_{\rm LD}$, $T$, $A$, and $\lambda$ are
the LDM free energy, temperature, particle number and the chemical 
potential, respectively. For
the free energy $\mathcal{F}_{\rm IP}(T,A)$, and the grand 
canonical thermodynamical 
potential $\Omega_{\rm IP}(T,\lambda)$ of the quantum
independent particle (IP) model (IPM), one has
\begin{eqnarray}\label{fomsp} 
\mathcal{F}_{\rm IP}(T,A) &\equiv& 
\Omega_{\rm IP}(T,\lambda) + \lambda A, \nonumber\\
\Omega_{\rm IP}(T,\lambda) &=& -2T \sum_i \ln\left(1+e^{(\lambda -
 \vareps_i)/T}\right),\nonumber\\
A&=& 2 \sum_i n_i=2 \sum_i {\tilde n}_i\;,
\end{eqnarray}
where $\vareps_i$ is the energy spectrum in a given potential well.
(The factor 2 accounts for the spin degeneracy of the symmetrical system
of nucleons.)
Their SCM averaged quantities $\widetilde{\mathcal{F}}_{\rm IP}(T,A)$, 
$\widetilde{\Omega}_{\rm IP}(T,\lambda) $ and smooth 
occupation numbers of quantum $i$ states $\widetilde{n}_i$ 
are determined with the help of the Strutinsky
local-energy averaging procedure (Appendix B3), 
see the analytical expressions
for $\widetilde{n}_i $ at a given temperature $T$ 
for 
the infinitely deep spherical square-well (spherical box) 
potential in \cite{MGFyaf2007,MGFnpae2008}. 
They are valid under the conditions
$T/\vareps^{}_F A^{1/3} << \Delta << k^{}_FR \approx 2 A^{1/3}$, where $\Delta$
is the width of Gaussian weight function, 
$~f_{\rm av}^{(2\mathcal{M})}\left[\left(k^{}_FR-k_F'R\right)/\Delta\right]$, 
with the correction polynomials of the order of $2\mathcal{M}$ for the
averaging over a suitable current variable, $k_F'R$, for a  billiard 
(cavity potential) system. 
(The tilde means traditionally the SCM averaging \cite{strut,fuhi}).
The shell structure components of the SCM, $\delta \mathcal{F}$ and
$\delta \Omega$, are determined in (\ref{fSQM}) in terms of quantities of
the IPM.
The stiffness $C$ can be then calculated by
\bea\label{stSQM}
C &=& \left(\frac{\partial^2 \mathcal{F}}{\partial q^2}\right)_{q=0}
=C_{\rm{LD}} +\delta C, \nonumber\\
\delta C&=&C_{\rm{IP}}(0)-{\widetilde C}_{\rm{IP}}(0),\nonumber\\
C_{\rm LD}&=&\frac{b^{}_{\rm S}}{4\pi r_0^2}\,(\lambda - 1)(\lambda + 2)\,R^{2}, 
\eea
where $C_{\rm LD}$ is the stiffness of the LDM ($\lambda \geq 2 $) \cite{bormot}.
The surface energy constant $b^{}_{\rm S}$ is
derived in Section III with
the help of Appendix A3 ($b^{}_{\rm S}\approx 18$ MeV 
in this section),
see also the usual definition for $ r_0$ 
through the particle density $\overline{\rho}$  of 
infinite nuclear matter in Section III, $\overline{\rho} =k_F^3/3 \pi^2 $
in this Section.
The IPM stiffness,
\begin{equation}\label{stiffnessC0}
C_{\rm IP}(0)= (\partial^2 \mathcal{F}_{\rm IP}/
\partial q^2)_{q=0}\;,
\end{equation}
 can be also decomposed in terms of
the two components $\widetilde{C}_{\rm IP}(0)$ and $\delta C$ in line of 
that for the SCM free energy $\mathcal{F}$ (\ref{fSQM}), i.e.\ , through
 the average $\widetilde{\mathcal{F}}$ (in $k^{}_FR$),
and the oscillating shell component,  $\delta \mathcal{F}$, respectively. 
The averaged quantum 
stiffness $\widetilde{C}_{\rm IP}(0)$ (\ref{stSQM})
can be found with the averaging parameters, $\Delta=1.5$ and $\mathcal{M}=3$,
for which we found a 
good plateau condition at temperatures
$T \approx 1 \div 4$ MeV. Notice that the smooth stiffness
of the perfect Fermi gas, $\widetilde{C}_{\rm IP}(0)$, differs essentially from
the liquid-drop quantity, $C_{\rm LD}$, which takes phenomenologically 
into account a strong interaction of the nucleons, in contrast to 
$\widetilde{C}_{\rm IP}(0)$.
The total stiffnesses $C$ and $C_{\rm IP}(0)$ (\ref{stSQM})
as $\mathcal{F}$ and $\mathcal{F}_{\rm IP}$ in (\ref{fSQM})
describe different (liquid and gas) physical systems through the 
corresponding SCM and IPM.
Note also that in the IPM calculation of $\delta C$ (\ref{stSQM}) 
we shall neglect approximately the curvature of the s.p.\ levels 
with respect to the dominating contributions of squares of their slopes
\cite{fedjaslopes}.  

Fig.\ \ref{fig22} 
shows good agreement of the IPM 
shell corrections $\delta {\cal F}$ (\ref{fSQM}) to the free
energy ${\cal F}$ (see bottom) and $\delta C$ (\ref{stSQM}) to the
stiffness $C$ (top) for the spherical box potential (at equilibrium)
 with the corresponding
semiclassical POT results, $\delta \mathcal{F}_{\rm scl}$ 
and $\delta C_{\rm scl}$, as functions of $k^{}_FR$ at a small 
temperature
 $T=1$ MeV as example \cite{MGFnpae2005,MGFyaf2007}. 
The calculations
were performed for the averaging parameters $\Delta =1.5$ and $\mathcal{M}=3$, 
and the arrow shows a magic value $k_FR=13.36$ for this billiard potential
[$\overline{\rho} = 0.16$ fm$^{-3}$
 ($r_0=1.14$ fm), $b^{}_{\rm S}=18$ MeV]. For such a potential,
one obtains
\bea\l{dfdcpot}
\delta {\cal F}_{\rm scl} &=& \sum_{\rm PO} \delta {\cal F}_{\rm PO}
\left(\lambda,q\right),\nonumber\\
\delta C_{\rm scl} &=& 
\left(\frac{\partial^2 {\delta {\cal F}_{\rm scl}}}{\partial
 q^2}\right)_{Q=0}\nonumber\\
&=& -\frac{5}{84 \pi}\,\sum_{\rm PO} (k_F\mathcal{L}_{\rm PO})^2 
\delta {\cal F}_{\rm PO}\left(\lambda,0\right),
\eea
where
\bea\l{fbeta}
\hspace{-0.5cm}\delta {\cal F}_{\rm PO}\left(\lambda,q\right) &=&
 \left(\frac{2 \vareps^{}_F}{k^{}_F \mathcal{L}_{\rm PO}}\right)^2\nonumber\\
& \times& \Phi\left(\frac{\pi k^{}_F\mathcal{L}_{\rm PO} T}{2\vareps^{}_F}\right)
 \;\delta g_{\rm scl}^{({\rm PO})}(\lambda,q),\nonumber\\
 &&\Phi(z)=z/\sinh(z)\;,
 \eea
where $~\mathcal{L}_{\rm PO}(q)~$ is the PO length,  
$k^{}_F=p^{}_F/\hbar$ the wave number at the Fermi energy $\vareps^{}_F$.
Here, the sums are taken over POs in the spherical
cavity, and $\delta g_{\rm scl}^{({\rm PO})}(\vareps,q)$ is the
PO-component of the oscillating part of the POT level density 
(\ref{deltadenstot}), 
\bea\label{dgPO1}
&\delta g_{\rm scl}(\vareps,q)=
\sum_{\rm PO}\delta g_{\rm scl}^{({\rm PO})}(\vareps,q)\;,\nonumber\\
\delta g_{\rm scl}^{({\rm PO})}(\vareps,q) 
&\!= \!\!\mathcal{B}_{\rm PO}\;\cos\left[S_{\rm PO}(\vareps,q)/\hbar -
\pi \mu^{}_{\rm PO}/2\right],
\eea
with the explicitly written deformation argument $q$ and 
the amplitude $\mathcal{B}_{\rm PO}$ from Section II, see 
\cite{bablo,strumag,smod,spheroidptp}.
$~S_{\rm PO}(\vareps,q)~$ and $~\mu^{}_{\rm PO}~$ 
are the classical action and the Maslov
phase for the PO, respectively,
\bea\l{SPO}
S_{\rm PO}\left(\lambda,q\right)&=&p^{}_F \mathcal{L}_{\rm PO}(q), \nonumber\\
p^{}_F&=&\sqrt{2 m \lambda} \approx \sqrt{2m \vareps^{}_F}\;. 
\eea
In order to obtain the stiffness shell correction $\delta
C_{\rm scl}$ in (\ref{dfdcpot}) we used the POT sum
for $\delta \mathcal{F}_{\rm scl}$ (\ref{dfdcpot})
over POs in the slightly deformed spheroidal-box
potential which is a good approximation for the quadrupole
shapes at small static deformations $q$. As shown in \cite{magvvhof}, 
the $q$ derivatives of the strong oscillating
cosine in the level density component of (\ref{fbeta}), 
$\delta g_{\rm scl}^{({\rm PO})}(\lambda,q)$, yield the semiclassically leading
(in order of $~\hbar$) contribution of the main POs (triangles,
quadrangles and so on) in the meridian plane of the spheroidal cavity at these
deformations \cite{smod,spheroidptp,migdalrev}. 
All quantities in (\ref{dfdcpot}),
(\ref{fbeta}) shown in Fig.\ \ref{fig22}, 
are taken finally at the spherical
equilibrium, $q=0$. 
Fig.\ \ref{fig22} 
shows
the strong major-shell oscillations of $\delta \mathcal{F}$ and $\delta C$ as
functions of $k^{}_FR$  
in the s.p.\
spectrum. The convergence of the POT sum for the stiffness,
$\delta C_{\rm scl}$, is more slow than that for the free energy,
$\delta \mathcal{F}_{\rm scl}$, shell corrections because of the additional
factor of $(k^{}_F\mathcal{L}_{\rm PO})^2$ in $\delta C_{\rm scl}$, 
see (\ref{dfdcpot}).
The exponential convergence of the both PO sums in (\ref{fbeta}) 
is provided by the temperature-dependent factor, 
$\Phi(\pi k^{}_F \mathcal{L}_{\rm PO} T/2 \vareps^{}_F)$, written explicitly in 
(\ref{fbeta})
for $\delta \mathcal{F}_{\rm PO}$. The shell corrections to the free energy 
$\delta \mathcal{F}_{\rm PO}$ (\ref{dfdcpot}) decrease exponentially
 due to this factor 
with growing both temperature $T$ and semiclassical parameter, 
$k^{}_F \mathcal{L}_{\rm PO} \propto k^{}_FR\gg 1$. 
Therefore, for large 
temperature and particle numbers, $k^{}_FR \sim 2 A^{1/3}$,
the shortest POs give the major contribution into the POT sums 
(\ref{dfdcpot}) for $\delta \mathcal{F}_{\rm scl}$ and $\delta C_{\rm scl}$,
and, as seen from Fig.\ \ref{fig22}, approximately
one finds
$\delta C \propto -\delta {\cal F}$, as seen from
Fig.\ \ref{fig22}.  
The critical temperature, $T_{\rm{cr}}$, for disappearance
of the shell effects in the free energy $\mathcal{F}$  
(\ref{fSQM})
and the stiffness $C$ (\ref{stSQM}) can be related to the distance
between the major shells $~\hbar \Om$ through the period $t^{}_{\rm PO}$
of the shortest POs, $~t^{}_{\rm PO}=m \mathcal{L}_{\rm PO}/p^{}_F$, see
(\ref{dfdcpot}), (\ref{fbeta}) and 
\cite{strumag}, 
\bea\label{Tcr}
\hspace{-1.0cm}T_{\rm cr} &\approx& \hbar\Om/\pi,
\qquad \hbar\Om \approx 2 \pi \hbar/\langle \tau_{\rm PO} \rangle\nonumber\\
&=&\frac{4 \pi \vareps^{}_F}{k^{}_F \langle {\cal L}_{\rm PO} \rangle} 
\approx \frac{\vareps^{}_F}{\,A^{1/3}}\;,
\eea
where $\langle t^{}_{\rm PO} \rangle$ and 
$~\langle \mathcal{L}_{\rm PO} \rangle~$ are the
mean period and length of the main short orbits, respectively.

\subsubsection{The inertia and friction}

For the example of the infinitely deep spherical
square-well potential, $V(q)$, the s.p.\ operator,
$\hat{F}$, in (\ref{chiGGdef}) is given analytically by (\ref{foperdV}), 
\begin{equation}\label{Finf} 
\hat{{\rm F}}(\r) =
- V_0
R\,\delta (r - R)\,Y_{\lambda 0} (\hat{r}), 
\end{equation}
with the boundary condition:
\begin{equation}\label{boundinf}
\frac{2m}{\hbar^2}\, V_0 G \rightarrow
\left(\frac{\partial^2G}{\partial
r_1\partial r_2}\right)_{r_j=R}, \,\,\, j=1,2\;.
\end{equation} 
Here $V_0$ is the depth of the potential well, and 
$V_0 \rightarrow \infty$ at its edge.
With the spectral representation of (\ref{chiGGdef}) for the Green's
function, $G(\r_1,\r_2;\vareps)$, in the mean-field limit 
$\epsilon^{}_0 \rightarrow +0$, the
equation (\ref{trcoll}) for the inertia $M(0)$ 
is equivalent to the well-known 
cranking model inertia (\ref{mcran}), which we use now with the replacement
$\hat{Q}=\hat{{\rm F}}$.

 In close analogy with the nuclear SCM relationships (\ref{fSQM}) and
(\ref{stSQM}) for the free energy ${\cal F}$ and stiffness $C$, 
one can obtain
the renormalized SCM inertia,
\begin{eqnarray}\label{SQM}
&& M=M_{\rm ETF} + \delta M,\nonumber\\
&& \delta M = M_{\rm IP}(0) - \widetilde{M}_{\rm IP}(0)\nonumber\\
&=&
2\hbar^2{\sum_{ij}}^{\prime}\, 
\frac{\delta n_i-\delta n_j}{(\vareps_j - \vareps_i)^3}\, 
\left|<i|\hat{F}|j>\right|^2,
\end{eqnarray}
where $\delta n_i=n_i-\widetilde{n}_i$, see \cite{strmagbr} for 
the inertia shell
correction $\delta M$, $M_{\rm IP}(0)$ is given by (\ref{mcran}). For the
smooth part of inertia $M_{\rm ETF}$ in
(\ref{SQM}), one may apply the macroscopic approach 
(\ref{coltrin}) \cite{MGFyaf2007}
\begin{eqnarray}\label{bcol} 
&M_{\rm ETF} = \left(1 + 
C_{\rm LD}/\chi^{}_{\rm ETF}(0)\right)^2\;\nonumber\\ 
 & \times \left(M_{\rm ETF}(0) + M_{\rm corr}^{(1)}\right) 
 \approx M_{\rm ETF}(0)
 + M_{\rm corr}^{(1)}\;,\nonumber\\
& M_{\rm corr}^{(1)}
= \gamma_{\rm TF}(0)^2/\chi_{\rm ETF}(0) \approx
 -\gamma_{\rm wf}^2/\kappa^{}_{\rm ETF}\;. 
\end{eqnarray}
Here, $C_{\rm LD}$ is the LDM stiffness given in (\ref{stSQM}),
$\gamma_{\rm ETF}(0)$ and $M_{\rm ETF}(0)$, (\ref{trcoll}), 
are the auxiliary ``intrinsic'' 
friction and inertia parameters in the ETF approximation
\cite{MGFyaf2007} (Appendices B1, and B3). 
We assume here that the shell fluctuations of the inertia $M$ are dominating 
from the occupation number variations $\delta n_i$ as compared to 
the squares of matrix elements in the cranking formula (\ref{mcran}).
For the ETF isolated susceptibility $\chi_{\rm ETF}(0)$ in (\ref{bcol}),
one finds \cite{MGFyaf2007}
\bea\l{parck} 
&\chi^{}_{\rm ETF}(0) = -\kappa^{}_{\rm ETF}-C_{\rm LD}\; \nonumber\\
&= 
\kappa^{}_{\rm ETF} \,\left[1 + {\cal O}\left(A^{-2/3}\right)\right], 
\eea
where $\kappa^{}_{\rm ETF}$ is the ETF coupling constant,
\be\l{parckcoup}
\kappa^{}_{\rm ETF}
= - 8 b^{}_{\rm V}\, K\, R^4/225 \pi\, b^{}_{\rm S}\, r_0^4\;,
\ee 
with $b^{}_{\rm V}$ being the binding energy per nucleon, $b_{\rm S}$
the surface energy constant,
and $K$ the in-compressibility
modulus of infinite symmetric nuclear matter (see Section III).
Finally, with (\ref{parckcoup}) one obtains
\bea\l{gcol}
&\gamma^{}_{\rm ETF} \approx \gamma^{}_{\rm ETF}(0) \nonumber\\
&= 
\gamma_{\rm wf}\left(1+\gamma_{\rm corr}^{(1)}\right)
\,\left(1 + \frac{\pi^2\,
 \bar{T}^2}{3}\right), 
\eea
with
\begin{equation}\label{gtf}
\gamma_{\rm wf}=\frac{\hbar(k^{}_FR)^4}{4
 \pi^2},\quad \gamma_{\rm corr}^{(1)}=-\frac{1}{42}\;,
\end{equation}
\be\l{bcolmcorr} 
M_{\rm corr}^{(1)} 
= \frac{225 mR^2\,(k^{}_F r_0)^4\,b^{}_{\rm S}\,
\vareps^{}_F\,(k^{}_F R)^2}{(64 \pi^3\,
 b^{}_{\rm V}\,K)}\;,
\ee
and 
\bea\l{btf}
 &\hspace{-0.5cm}M_{\rm ETF}(0) = \frac{mR^2}{8 \pi}\left[\frac{16(k^{}_F R)^3}{385 \pi} 
\left(1 +
\frac{\pi^2\, \bar{T}^2}{8}\right) -(k_FR)^2 \right. \nonumber\\
 &\left. +
\frac{87368\,k^{}_FR}{9009 \pi}\left(1 -
\frac{\pi^2 \,\bar{T}^2}{24}\right)\right],\quad \overline{T}=
\frac{T}{\vareps^{}_F}\;. 
\end{eqnarray}
Notice that the nuclear dynamical ES approximation 
(Section III)  used in the
derivation of $\kappa^{}_{\rm ETF}$ in \cite{MGFyaf2007} relates the
statistical ETF approach to the LDM. They are described both in
terms of the local quantities, such as the particle, current and energy
densities, the volume (in-compressibility modulus $K$), and 
surface capillary 
pressures (surface energy constant $b_{\rm S}$), 
in contrast to the IPM.
As shown in Section III, the ES is defined as the positions of maxima of
the particle density gradient \cite{strmagbr,strmagden}. 
The ES approximation is based on the leptodermous expansion in small
parameter, $~a/R \sim A^{-1/3}$, where $~a~$ is the diffuseness of
the nuclear edge and $R$ the curvature radius of the ES.
 In particular,
the particle density inside the nucleus is almost constant (with small 
$~A^{-1/3}~$ surface corrections) for large particle numbers.
Therefore, as shown in \cite{MGFyaf2007}, one can 
neglect a small ratio, $C_{\rm LD}/\kappa^{}_{\rm LD} \sim A^{-2/3}$,
in (\ref{parck}) and (\ref{bcol}).  
The small nonlocal friction correction 
$\widetilde{\gamma}_{\rm cor}$ (\ref{gtf}) 
was omitted in (\ref{bcol}), too. The ``intrinsic'' inertia
$M_{\rm ETF}(0)$ (\ref{btf}) in the limit $k_FR \rightarrow \infty$ is
the sum of the ``volume'', $\propto (k^{}_F R)^3 $, the ``surface'',
$\propto (k^{}_F R)^2$, and the ``curvature'', $\propto k^{}_F R$, terms. It is
similar to the ETF expansion (\ref{akrTF}) 
in $1/k^{}_FR$ for the relationship of
$k^{}_FR$ to the particle number $A$. However, it is modified now by a 
temperature dependence through the Fermi distribution (occupation number)
$n(\vareps)$ as
\bea\l{akrTFT}
A&=&2 \int_0^{\infty} \d \vareps\,n(\vareps)\, g^{}_{\rm ETF}(\vareps)\nonumber\\
&=& 2\left[\frac{2 (k^{}_FR)^3}{9 \pi}\,\left(1+\frac{\pi^2 {\bar
	T}^2}{8}\right)\right.\nonumber\\
&-&\left.\frac{(k^{}_FR)^2}{4} +
\frac{2 k^{}_FR}{3 \pi}\,\left(1-\frac{\pi^2 \overline{T}^2}{24}\right)\right],
\eea
with the ETF level density $g^{}_{\rm ETF}(\vareps)$ \cite{sclbook,bablo}. 
The factor of 2 accounts for the spin degeneracy and
the weak ${\bar T}^2$ temperature dependence was found by
the Sommerfeld expansion as in the derivation of (\ref{gtf}) and
(\ref{btf}) 
\cite{MGFyaf2007}. Therefore,
we may call (\ref{btf}) for the inertia, and similarly, (\ref{gtf})
for the friction, as the ETF approximation for a heated Fermi system
\cite{sclbook}. It should
be noted that from (\ref{akrTFT}) one finds $k^{}_FR$ as function of the
particle number $A$ and temperature parameter $\overline{T}$. This
temperature-dependent function can be substituted into 
 (\ref{gtf}) and (\ref{btf}) to obtain 
the $A$ dependence of 
$\gamma^{}_{\rm TF}(0)$ and $M_{\rm TF}(0)$. The other temperature corrections
proportional to $\overline{T}^2$ are written explicitly in these equations.

In order to compute the inertia shell corrections $\delta M$ by
(\ref{SQM}) we note that the averaged IPM ``intrinsic'' inertia
$\widetilde{M}_{\rm IP}(0)$ (AIM) as function of $k^{}_FR$ [or particle numbers
$A$, according to (\ref{akrTFT})] is assumed to be dependent
 on $k^{}_FR$  
through the
occupation numbers $n_i$ in (\ref{mcran}). Therefore, as mentioned above,
one may apply the same averaging procedure \cite{strut,fuhi,sclbook} 
to the occupation numbers $n_{i}$ in (\ref{mcran}).  
The averaging parameters around $\Delta=4
$ and $\mathcal{M}=3$ can be found from study of
the plateau condition (Appendix B3).

We point out that the ETF inertia, $M_{\rm ETF}$, cannot be considered as the
approximation to the smooth $\widetilde{M}_{\rm IP}(0)$ of the IMP because they
describe physically different systems through the corresponding models, 
like in comparison between the LDM stiffness $C_{\rm{LD}}$ 
(or free energy ${\cal F}_{\rm LD}$) and the $\widetilde{C}_{\rm IP}(0)$ 
(or $\widetilde{\cal F}_{\rm IP}$) of the IPM, respectively. Indeed,
there is the dependence of $M_{\rm ETF}$ (\ref{bcol}) on the energy
surface, $b^{}_{\rm S}$, and the in-compressibility modulus, $K$, 
constants describing
the {\it dense} Fermi-liquid drop through $M_{\rm ETF} \propto b^{}_{\rm S}/K$, 
in contrast to the perfect Fermi gas bounded by the 
sharp-edge potential cavity.
Moreover, the inertia $M^{(1)}_{\rm corr}$ (\ref{bcolmcorr}) depends on both 
the ETF friction $\gamma^{}_{\rm ETF}(0)$ (\ref{gcol}), 
$\gamma^{}_{\rm ETF}(0) \approx \gamma_{\rm wf}$, and the LDM susceptibility 
$\chi^{}_{\rm ETF}(0)$ (\ref{parck}). 

Concerning the friction, a renormalization procedure, similar to 
(\ref{SQM}), for the inertia is assumed to be applied too,  
\bea\l{gSQM}
\gamma &=&\gamma^{}_{\rm ETF} + \delta \gamma\;, \nonumber\\ 
\gamma^{}_{\rm ETF}&=&\left(1 + 
C_{\rm LD}/\chi^{}_{\rm ETF}(0)\right)^2\,\gamma^{}_{\rm ETF}(0)\;,\nonumber\\
\delta \gamma&=&\gamma^{}_{\rm IP}(0)-\widetilde{\gamma}^{}_{\rm IP}(0)\;,
 \eea
where $~\gamma^{}_{\rm ETF}(0)$ is defined by (\ref{gtf}), 
and one has $\gamma=\gamma^{}_{\rm ETF}$ because 
$\gamma^{}_{\rm IP}(0)=\widetilde{\gamma}_{\rm IP}(0)=0$, according 
to (\ref{gtfdef}) 
with the help of the equation 
(\ref{GFdef}) at $\epsilon^{}_0 =+0$. However, we emphasize, in particular
for understanding the meaning of the {\it undamped} inertia component
$M_{\rm ETF}$ (\ref{bcol}) that there is a dramatic discrepancy
between the finite friction $\gamma^{}_{\rm ETF}$ (\ref{gSQM}) in the
{\it statistical } ETF model and {\it zero} IPM friction values in the
{\it perfect Fermi-gas} cranking model. The reason is the same essential
difference in physical properties of these two compared models, as mentioned
above in relation to the smooth free energy, stiffness and inertia. The
friction $~\gamma^{}_{\rm ETF}$ (\ref{gSQM}) is equal approximately to the
wall formula $\gamma_{\rm wf}$ up to both small 
 $C_{\rm LD}/\chi^{}_{\rm ETF}(0)$ 
and small $\overline{T}^2$ corrections, 
$\gamma^{}_{\rm ETF} \approx \gamma_{\rm wf}$, see (\ref{gtf}).
Therefore,  as usual, it is called as the one-body friction, in contrast to the
friction of another nature related, for instance, to the two-body collisional
viscosity  
from a realistic residue interaction between particles.
In the ETF model within the phase-space representation, one assumes 
the {\it undamped one-body} motion inside the nucleus as well as an existence
of the nuclear ES formed by a {\it many-body} strong interaction of 
particles as in the LDM \cite{strmagbr,strmagden}. In this
sense, if we take also into account the dynamical ES, the 
wall friction $\gamma_{\rm wf}$ is not
really of the only one-body nature: 
The origin of the friction $\gamma_{\rm wf}$
is  the collisions of the internal independent particles with the ES
considered as an {\it external} wall of the gas container with respect to
those in the nuclear volume (the macroscopic ``piston'' model 
\cite{wall}). The lost of the energy of particles inside of the
system in 
such a model, $-\gamma_{\rm wf} \dot{q}^2$,
is due the work of the {\it external} force coming from the ES (``piston'').
As well-known, there is no such a 
friction in the completely self-consistent
problems where a splitting into the ``volume'' and ``surface'' motion is
absence at all. The 
ES can be also included self-consistently into the system, for
instance, through the boundary conditions derived for the kinetic Landau-Vlasov
equation of motion inside the nucleus in the FLDM
(Section III, Appendix A2 and 
\cite{strmagbr,magstr,magkohofsh,kolmagsh}). Thus, the friction coefficient,
$\gamma_{\rm wf}$, is not really the {\it nuclear dissipation} of the only
one-body nature in such a self-consistent picture. In
particular, we may describe the {\it undamped} (nondissipative) motion by
using the ETF inertia $M_{\rm ETF}$ [see (\ref{bcol}) and (\ref{btf})] 
depending
on $\gamma_{\rm wf}$ because of applying the consistency condition
\cite{hofbook,ivhofpashyam}.

Following the basic ideas of the SCM for dense Fermi systems 
\cite{migdal,strmagbr} in our renormalization procedure we suggest to replace
correspondingly the averaged IPM free energy ${\widetilde {\cal F}}_{\rm{IP}}$,
stiffness $\widetilde{C}_{\rm IP}(0)$ and inertia $\widetilde{M}_{\rm IP}(0)$
of the IPM, see (\ref{fSQM}), (\ref{stSQM}) and (\ref{SQM}), 
by more relevant phenomenological macroscopic quantities 
 ${\cal F}_{\rm LD}$, $C_{\rm LD}$ and $M_{\rm ETF}$. A similar replacement
of the averaged susceptibility $\widetilde{\chi }_{\rm IP}(0)$ for
the perfect  
Fermi gas, which is {\it divergent}
for a 
box potential [according to (\ref{chiGGdef}] by the
{\it finite} ETF (liquid-drop) isolated susceptibility $\chi^{}_{\rm ETF}(0)$
(\ref{parck}) can be used within the SCM, too. Their significant
difference, can be immediately understood from (\ref{parck}) by looking at
the formal zero limit of the LDM surface energy coefficient 
 $b^{}_{\rm S}$ proportional to the diffuseness parameter
$a$ in the ES approximation \cite{strmagbr,strmagden}, see also
(B4) and (B5) of \cite{MGFyaf2007}. These comments might be also helpful in order
to clarify the meaning of the inertia $M_{\rm ETF}$ given by (\ref{bcol}).

Such a theoretical scheme for the stiffness, the
inertia and the friction looks logically more closed and consequent
for the {\it dense} Fermi systems with the strong interacting particles within
the Migdal theory \cite{migdal} because the  
quasiparticle states near the
Fermi surface are mainly involved in the calculations based on the
one-body Green's function representation of (\ref{chiGGdef}) by
 meaning of the renormalization procedure \cite{strmagbr}.
Other contributions of the s.p.\ states far from the Fermi
surface on distance larger a few major shells in this dynamical version
of the SCM \cite{strut,fuhi} are replaced by
another, macroscopic (almost local) components of a many-body nature
\cite{migdal,strmagbr,khodsap}. The macroscopic terms are 
 available at the present moment with using the
nuclear phenomenological properties known from the experiment, such as the 
particle density of infinite nuclear
matter, surface tension, separation energy per one nucleon, in-compressibility
modulus and so on, similarly to the consistent ETF 
approximation (\ref{bcol}) (Section III). Note also
that other versions of the phenomenological macroscopic components of the
friction and inertia can be also considered for the specific dynamical
problems. However, as well known  
\cite{bormot,hofbook,belyaev}, the 
simultaneous use of the irrotational flow inertia and standard hydrodynamical 
friction disagrees with experimental data on the nuclear collective-excitation
energies and fission in many aspects, and the consistent ETF approach
might be preferable for the macroscopic parts of the SCM.

\bigskip
\subsubsection{Numerical results}
\medskip

We compare now the renormalized SCM (SC) inertia $M$ (\ref{SQM}) 
and its semiclassical ETF component 
$M_{\rm ETF}$ (\ref{bcol}) with the 
corresponding quantum IPM (IP) result (\ref{trcollQF}), 
(\ref{mcran}) as
functions of $k^{}_FR$ in Fig.\ \ref{fig23}. 
 For convenience, this comparison
is performed in the irrotational flow units, $M_{\rm irr}$,
with accounting for the ETF relationship (\ref{akrTFT}) between the
particle number $A$ and $k^{}_FR$. In our SCM approach for calculations of the
IP and 
the SC we use the infinitely deep spherical
square-well potential for the equilibrium mean field. In these
calculations, the averaging parameter
$\Delta=4$ for good plateau condition is larger than that found 
in calculations of Fig.\ \ref{fig22} with the same polynomial correction order
$\mathcal{M}=3$. Note that for the perfect quantum Fermi gas in this 
potential one has $M_{\rm IP}=M_{\rm IP}(0)$  (\ref{trcollQF})
because
 the isolated susceptibility $\chi^{}_{\rm IP}(0)$ is
infinity as mentioned above. 
In Fig.\ \ref{fig23}, the  
pronounced shell effects measured by
the deviation of the SC (\ref{SQM}) from the ETF (\ref{bcol}) 
are seen well at zero temperature $T$ for all  $k^{}_F R$. 
The shell oscillation amplitude of the inertia $M$ (\ref{SQM}) (SC)
decreases significantly with increasing temperature $T$
and practically disappears at $T=3 \div 4$ MeV.
As seen from Fig.\ \ref{fig23}, 
for large 
$k^{}_FR$ 
and temperatures $T$, one finds rather close ETF
 values $M_{\rm ETF}$ (dashed)
of the SC inertia (\ref{SQM}), as compared to that of the IPM 
$M_{\rm IP }(0)$ (solid) (\ref{mcran}), and its
average (AIP) $\widetilde{M}_{\rm IP}(0)$ (dash-dotted). Notice, the total
SC inertia (dots) (\ref{SQM}) approximately coincides with the IP results
near $k^{}_FR \approx 12 \div 14 $. However, the $k^{}_FR$ (or particle number
$A^{1/3}$) dependence of our smooth local TF inertia approach,
$M_{\rm TF}\propto (k^{}_FR)^4$ (in fact, even a little stronger), differs essentially
from behavior of both the irrotational flow inertia, $M_{\rm irr}
 \propto (k^{}_FR)^5$, and the AIP one, as should be expected,
see Section IVB3.

Fig.\ \ref{fig24}   
displays the collective ETF
friction $\gamma^{}_{\rm ETF} $
(\ref{gSQM}), SCM
inertia $M$ (\ref{SQM}) (SC)  [with  
ETF inertia
$M_{\rm ETF}$ (\ref{bcol}),  and
the POT free-energy shell corrections $\delta {\cal F}_{\rm scl}$ 
(\ref{dfdcpot}) versus the corresponding independent particle model 
(IP) results 
[see, e.g.\ , (\ref{mcran}) for $M_{IP}=M_{\rm FF}(0)$ and 
(\ref{fSQM}) for $\delta \mathcal{F}_{IP}$ as functions of temperature $T$ at 
$k^{}_FR=13.36$,
respectively. As shown in Fig.\ \ref{fig22} 
by the arrow, this number of $k^{}_FR$ corresponds to a 
minimum of the free-energy
shell correction $\delta {\cal F}(k^{}_FR)$ (\ref{fSQM}) and (\ref{dfdcpot})
at $T=0$. It is an example of the completely closed shells
related to a large  
magic-particle number $A=254$ in the considered
spherical-box equilibrium potential 
through (\ref{akrTF}) for the relationship $A$
to $k^{}_FR$ at zero temperature. Within the temperature interval restricted by
the small parameter $(T/\vareps^{}_F)^2$ of 
a Sommerfeld 
expansion, one may neglect
a change of $k^{}_FR$ with temperatures, $T \ll \vareps^{}_F$, 
for a given particle number $A$. 

The ETF friction 
$\gamma^{}_{\rm ETF}$ (thin frequent dashed), see 
(\ref{gSQM}), is shown at the top of Fig.\ \ref{fig24}. 
It is a slightly 
increasing function of temperature $T$ due to the
corrections proportional to $T^2$. It is typical for the Fermi liquid system
in the regime of rare (zero-sound) collisions of the quasiparticles,
and therefore, significantly differs from the well-known
(decreasing) hydrodynamic behavior, $\propto 1/T^2$,
from the two-quasiparticle collisional viscosity
\cite{hofbook,magkohofsh,belyaev}.
The ETF0  approximation (thick rare
dash-dotted) for the ``intrinsic'' friction $\gamma_{\rm ETF}(0)
$ (\ref{gtf}) is rather close to the temperature-dependent
wall formula WF (thick solid) of the perfect local approximation  
[TF without $\hbar$ ETF corrections, see (\ref{gcol}) without the 
$\gamma_{\rm corr}^{(1)}$ correction]
\cite{hofivyam,hatkoonran}. The latter is approximately equal,
up to small $T^2$ corrections, to the constant 
$\gamma_{\rm wf}$ in (\ref{gtf}). 
A small difference between the full ETF approach for the friction coefficient
$\gamma_{\rm ETF}$ (\ref{gSQM}), and the ``intrinsic'' 
ETF0 approximation $\gamma^{}_{\rm ETF}(0)$ (\ref{gtf}) 
[or the wall formula $\gamma_{\rm wf}$ in (\ref{gtf}), 
$\gamma_{\rm ETF}(0) \approx \gamma_{\rm{wf}}$],
is due to a small collective consistent correction, $\sim 2
C_{\rm LD}/\kappa_{\rm ETF}$, for large 
particle
numbers as mentioned above \cite{MGFyaf2007}. 

In the bottom of Fig.\ \ref{fig24}, one finds a perfect agreement between the
IP (\ref{fSQM}) and the POT (\ref{dfdcpot}) 
temperature dependencies of the free-energy
shell correction $\delta {\cal F}$, as in Fig.\ \ref{fig22} 
for its $k^{}_FR$ 
dependence.  
A similar sharp decrease of the both quantum shell corrections in 
the free energy (\ref{fSQM}) and 
the inertia (\ref{SQM}) at about the same critical temperature
$T_{\rm cr}$ (\ref{Tcr}) are seen from comparison of the bottom and
middle panels of this Figure. The reason is that the corresponding
PO-components of the POT sums for $\delta M$
(see \cite{magvvhof,MGFyaf2007}) and $\delta {\cal F}$ (\ref{dfdcpot})
decrease exponentially with increasing both the temperature $T$ 
and semiclassical
parameter $k^{}_F{\cal L}_{\rm PO}$ due to the same temperature-dependent factor,
$\Phi(\pi k^{}_F {\cal L}_{\rm PO} T/2 \vareps^{}_F)$, written explicitly in
(\ref{fbeta}) for $\delta {\cal F}_{\rm PO}$.
 
The middle panel of Fig.\ \ref{fig24} 
shows transparently that the
SC inertia (dots) rapidly converges to the 
ETF  asymptotics 
for temperatures $T \simg T_{\rm cr} = 2 \div 3$ MeV. 
The ETF and SC inertias 
[through its ETF part, see (\ref{SQM}) and (\ref{bcol})] 
depend on the in-compressibility modulus $K$. 
For its conventional nuclear value, $K=220$ MeV, we obtain rather \
a good
agreement of the ETF approximation versus IP (also its average AIP, thin
frequent dash-dotted) and SC for particle numbers 
$A \sim 200\div 300$ and temperatures $T \simg T_{\rm cr}$, 
see the middle of Figs.\ \ref{fig23} 
and \ref{fig24} 
nearby $k^{}_FR \sim 12
\div 14$ at $T=2$ and $4$ MeV. This agreement
becomes the better the larger temperature $T$, as shown in 
Fig.\ \ref{fig24}   
for example at $k^{}_FR=13.36$. The magnitudes of IP ( and average AIP),
SC and its ETF asymptotic inertias for high enough temperatures,
$T \simg T_{\rm cr}$, are a factor of about 3
larger than the irrotational flow value $M_{\rm irr}$.

As seen from the difference  
between the SC and 
the ETF inertia in the middle of
Fig.\ \ref{fig24}, the shell effects are rather strong for 
temperatures smaller
than $T_{\rm cr}$. 
In the zero temperature limit  
one finds a
minimum of about $2 M_{\rm irr}$ in the IP and SC inertias $M$. Such a
decrease might be related to the magic particle number ($A=254$)
with closed shells. We
emphasize importance of the term $M_{\rm corr}^{(1)}$ (\ref{bcol}) 
in our collective inertia calculations with accounting for the {\it consistency
relation} \cite{hofbook,ivhofpashyam}.

Fig.\ \ref{fig25}  
displays the temperature dependencies of
the quadrupole collective-excitation energies, $\hbar
\sqrt{C/M}$, reduced friction, $\gamma/M$, 
and effective friction, $\eta=\gamma/2\sqrt{C M}$  
(all in the ${\rm F}$
mode), for the SC model
versus their consistent ETF approximation \cite{MGFyaf2007} and IP results at
$\epsilon^{}_0 \rightarrow +0$, 
except for obvious IP zeros of the quantities proportional to
$\gamma$. All the quantities in Fig.\ \ref{fig25}  
are rather slow functions of temperature $T$ at
large enough values, $T \simg T_{\rm cr}$, see (\ref{Tcr}). 
As expected, the temperature for disappearance of the shell effects
 in all these quantities is near the critical $T_{\rm cr}$ (\ref{Tcr})
as in Fig.\ \ref{fig24}  
for the inertia, $\delta M$, and the free energy,
$\delta {\cal F}$, as well as for the stiffness, $\delta C$,
shell corrections (Fig.\ \ref{fig22}). 
For high temperatures, $~T \simg T_{\rm cr}$,
the SC practically coincides with its almost constant TF asymptotics.
The significant shell effects
are naturally manifested at temperatures $T$ smaller than $~T_{\rm cr}$. 

As shown in the top panel of Fig.\ \ref{fig25}, 
the IP (thin solid) and SC (dots) quadrupole-vibration 
energy parameter, $\hbar \sqrt{C/M}$, 
is approximated well by the ETF curve (thin frequent dashed) at
temperatures $T \simg T_{\rm cr}$, where the shell effects are
exponentially small. Such a parameter at finite temperature 
might be used in analysis of the fission experimental data 
\cite{gonfro,hilgonros,frgon}.
Again, a good agreement between IP and SC energies, 
$\hbar \sqrt{C/M}$, can be found for all temperatures. 
It is similar to the results obtained for the inertia 
(Fig.\ \ref{fig23})
because of the same renormalized stiffness $C$
used in both IP and SC calculations, and the only inertia
$M$ is critical in this comparison. The significant {\it shell enhancement}
in the collective vibration energies, $\hbar \sqrt{C/M}$,
at smaller temperatures improves comparison with the experimental 
results for the quadrupole collective states 
in magic cold nuclei \cite{MGFyaf2007}.
In the small temperature limit, these energies are 
strengthed 
because of the minimum of the inertia, $\delta M$, 
and the maximum of the stiffness, $\delta C$,
associated both with the minimum of the free energy, 
$\delta {\cal F}$, shell correction for a magic particle number. 
A small decrease of the vibration energy maximum at low temperatures
is related mainly to 
a negative Coulomb
stiffness correction $C_{\rm LD}^{\rm coul}$ given explicitly 
in \cite{MGFyaf2007}
[see the IP1 (thick solid) for its IP result, and the SC1 
(full heavy rare squares) for the corresponding SC renormalization with the
smooth ETF1  high-temperature  
asymptotics (thick rare dashed) in this part
of Fig.\ \ref{fig25}, see also \cite{bormot,bormot1}. 
The other surface ($A^{-1/3}$)
corrections (included also in IP1, SC1 and ETF1) are much smaller
 because of a large particle number $A$. It should be noted, however, that 
the comparison of our SC results with the experimental data for magic nuclei, 
discussed in \cite{MGFyaf2007} for instance for ${\rm Pb^{208}}$, 
requires certainly more realistic calculations 
\cite{bormot,hofbook,khodsap,berborbro}. On a qualitative level, 
we may only point
out here that our SC results become more close to the experimental data for
magic nuclei mainly 
because of accounting for
the  shell effects.

As seen in the middle panel of Fig.\ \ref{fig25}   
for higher temperatures, 
the ETF reduced friction, $\gamma/M \approx \gamma^{}_{\rm ETF}/M_{\rm ETF}$, 
[(\ref{bcol}) and (\ref{gSQM})] is of  
order of the
estimation \cite{MGFyaf2007}, which is basically 
comparable within the
same order  
as its evaluation from the experimental fission data in 
\cite{hilgonros} for the nearly spherical shapes of compound nuclei.
More exact SC values are notably enhanced at smaller
temperatures 
because of the shell effects. Note that the maximum
of $\gamma/M$ at zero temperature can be explained by minima of the both free
energy ${\cal F}$ and inertia $M$  (Fig.\ \ref{fig24}) 
because there is no
friction shell corrections at 
$\epsilon^{}_0=+0$, according to
(\ref{gSQM}). In this comparison we should take into
account that the ETF approach is expected to be a good
approximation for significantly heated  
systems for which one may
disregard shell effects. We neglected also the
nuclear equilibrium deformations which influence essentially on
the reduced friction for fission processes
\cite{frgon,ivhofpashyam,gonfro,hilgonros}. 
We should expect also the importance of residue interactions as the
two-body viscosity in this comparison. 

Rather a strong overdamped motion, $\eta>1$,
at temperatures $T \simg T_{\rm cr}$, where the ETF practically
coincides with the SC approximation, is shown in 
the bottom panel of 
Fig.\ \ref{fig25}. As seen from comparison of the ETF versus
the SC, the effective friction $\eta$ is decreasing monotonically to about one
with decrease of temperature because of the shell effects.
These calculations are also roughly in agreement with 
evaluations of the effective damping coefficient, $\eta$, found 
from the experimental
data on fission 
\cite{hilgonros}.

Thus, our results for the collective vibration energies at zero
temperature, as well as the reduced, and the effective-friction coefficients
for larger temperatures are qualitatively in reasonable agreement with the
experimental data  
\cite{frgon,gonfro,hilgonros,MGFyaf2007}.
We suggested in this Section 
the modified SCM for calculations of the nuclear transport
coefficients for a slow collective motion. The consistent ETF approach can be
used for the smooth parts of transport coefficients in this SCM
modified with the collective dynamics. We pointed
out the importance of the renormalization (SCM) procedure for inertia 
as for the
shell correction calculations of the nuclear free energy and stiffness.
The shell structure components $\delta M$ for the inertia and $\delta C$ for
the stiffness are 
significant at small temperatures. According to the
POT, they disappear approximately with increasing temperature by 
an exponential law at about the same
critical temperatures $2 \div 3$ MeV as well as the free-energy 
shell corrections
$\delta {\cal F}$. After the SCM renormalization of the quantum transport
coefficients we obtained somewhat improved
results toward experimental data for the
quadrupole vibration excitation energy at small
temperatures, and for the reduced
friction and effective damping parameters at large temperatures; as compared
to the hydrodynamical model. The quantum and semiclassical SCM calculations of
the transport coefficients might be helpful for understanding and overcoming
some difficulties 
within the linear response theory at the finite two-body
dissipation related to a residue interaction.

\subsubsection{Smooth trajectory corrections}
\l{trcoefslowfpm}

In this Section, we shall deal with nonlinear effects in the 
transport coefficients because of their quadratic expression (\ref{gtfdef})
and (\ref{btfdef}) through the Green's function (\ref{Gsem}). Therefore, the
smooth semiclassical nonlocal corrections to the transport parameters 
[(\ref{mQQ0}) and (\ref{gammaQQ0})] because of the classical trajectories 
with finite actions might yield the significant contributions, in addition
to the wall formula and ETF inertia.

The famous wall formula (WF) 
describing the local
one-body friction, suggested originally by  Swiatecki and his collaborators
\cite{wall},  
was re-derived in many works
based on the semiclassical and quantum arguments (see  
\cite{koonran,MGFyaf2007,GMFprc2007}, for instance). 
Important non-adiabatic 
nonlinear corrections were used in the microscopic collective 
classical and quantum  dynamics
\cite{blocki}, in particular 
for calculations of the excitation 
energy and its time derivatives  
  \cite{blocki,BMYpr,BMYijmpe2012}.
Then, the peculiarities of the excitation energies for 
many periods of  
oscillations of the classical
dynamics were  discussed \cite{BMYpr,BMYijmpe2012}
for several multipole surface shapes (the order
of Legendre 
polynomials). 
However, some 
problems in the analytical study of a multipolarity dependence
of the smooth one-body friction and inertia 
should be still clarified within the EGA of the POT (Section II). 

For slow (small-amplitude) vibrations around the spherical 
equilibrium shape of a nucleus described explicitly through the mean field
in terms, e.g., of the cavity with sharp edges, one can
use (\ref{chiGGdef}) of the linear response theory \cite{hofbook} 
to relate the 
transport coefficients (\ref{trcoll}) through the ``intrinsic'' 
(friction $\gamma(0)$, and inertia $M(0)$) ones 
to the one-body Green's function $G(\r^{}_1,\r^{}_2;\vareps)$ 
\cite{magvvhof}. This relationship is especially useful for the 
semiclassical derivations of 
transport coefficients.
We shall not show sometimes, within this Section, the
argument ``0'' of these transport coefficients for 
simplicity of  
notations
writing comments if necessary to avoid misunderstanding. The tilde
above statistically averaged quantities will be disregarded too
within this Section because only such smooth ones will be considered 
up to the end of this Section IVB4.
Using also the 
spherical symmetry of the equilibrium shape and its expansion
in Legendre polynomials $P_n$, one has \cite{koonran}:
\begin{eqnarray}\label{gamsphgen}
\gamma &=&
\int_0^\pi \d \psi\; \sin\psi\;P^{}_n\left(\cos \psi\right)\;
\Upsilon\left(\psi\right),
\nonumber\\
 M &=&\int_0^\pi {\mbox d}\psi\; \sin\psi\;P^{}_n\left(\cos \psi\right)\;
B\left(\psi\right),
\end{eqnarray}
where  $\psi=\theta_2-\theta_1$ is the polar angle
between the two vectors $\r^{}_1$ and $\r^{}_2$. These vectors are 
arguments of the Green's function $G(\r_1,\r_2;\vareps)$ at the 
equilibrium surface in a major plane crossing any 
symmetry $z$
axis because of  
the 
spherical symmetry (Fig.\ \ref{fig26}). 
The kernels,
 $\Upsilon(\psi)$ and $B(\psi)$, for 
the integrals over $\psi$ are given by [(\ref{gtfdef}) and (\ref{btfdef})]
\bea\label{gaminsphker}
&&\hspace{-0.9cm}\Upsilon\left(\psi\right)=
\frac{d^{}_s     \hbar^5R_0^6}{2 m^2}
\left(\frac{
\partial^{2} \Im G}{\partial r^{}_{1}\partial r^{}_{2}}
\right)^{2}_{r=R_0,\vareps=\vareps^{}_{F}}, \nonumber\\
&&B\left(\psi\right) =
\frac{d^{}_s     \hbar^6R_0^6}{2 m^2} \nonumber\\
&\times&\int_0^{\vareps_F}\d \vareps
\left(\frac{\partial^{2} \Im G}{\partial r^{}_{1}\partial r^{}_{2}}
\frac{\partial^2}{\partial \vareps^2}
\frac{\partial^{2} \Re G}{\partial r^{}_{1}\partial r^{}_{2}}
\right)^{}_{r=R_0}. 
\eea
For billiards, one writes
 $\vareps^{}_{F}=p_F^2/2m$, where $p^{}_F$ is the Fermi momentum.
The effective mean field for motion of the particle 
at equilibrium is 
taken as  
the infinitely deep square-well potential 
depending on time 
$t$ through the multipole variations of a time-dependent surface radius 
$R(\theta,t)$ (\ref{surf}),
\cite{blocki}
keeping also the volume and the position of the center of mass  
conserved. 
The deformation parameter was traditionally introduced
as $\alpha=q \sqrt{4 \pi/5}$ to exclude a difference in constant between
the Legendre polynomial description of the ES quadrupole  
and spheroidal shapes
for small deformations \cite{blocki}.
In (\ref{gaminsphker}), $R_0$ is the radius of the
 equivalent sphere; 
 and $\alpha(t)$ is a
periodic function of time, 
$ \alpha(t)=\alpha\cos\left(\omega t\right) $, where $\alpha$ (without
the argument $t$) is an 
amplitude. Thus, the
rate of the excitation energy in time $t$ can be then written 
in terms of the consistent collective inertia $M$, 
and friction $\gamma$ [see (\ref{trcoll})] 
through the intrinsic
ones $M(0)$ and $\gamma(0)$ 
given in (\ref{gamsphgen}):
\be\l{eqmotde}
\d E/\d t=M \dot{\alpha} \ddot{\alpha} +\gamma \dot{\alpha}^2.
\ee

The friction $\gamma$ and inertia $M$ (intrinsic) coefficients 
[(\ref{gamsphgen})
and (\ref{gaminsphker})] 
 can be found with the help of the semiclassical expansion of the Green's 
function $G$ derived by Gutzwiller 
 \cite{gutz,strumag,sclbook,magvvhof}
for $k^{}_FR \sim A^{1/3} \gg 1 $
[$k^{}_F$ is the wave number at the Fermi energy $\vareps^{}_F$, see 
Section IVA2; and  
(\ref{Gsem}), (\ref{GsemGa}), (\ref{GCT0}),
and (\ref{ampiso})].

Following \cite{koonran}, according to the Green's function expansion
(\ref{Gsem}) over the CTs, one may split the transport 
coefficient kernels $\Upsilon(\psi)$ and 
$B(\psi)$ given by (\ref{gaminsphker}),
both averaged over the phase 
space variables \cite{GMFprc2007}, into the two terms, 
\bea\l{gamm0cav0}
\Upsilon(\psi)&=&\Upsilon_{\rm wf}(\psi)+\Upsilon_{\rm corr}(\psi)\;,
\nonumber\\
B(\psi)&=&B_{\rm ETF}(\psi)+B_{\rm corr}(\psi)\;.
\eea
Here
\bea\l{gwfdef}
&\Upsilon_{\rm wf}(\psi)=
\gamma_{\rm wf} \delta(\psi)/\sin\psi\;  ,\nonumber\\
&\gamma_{\rm wf}\approx d_s\hbar (k_FR^{}_0)^4/(10 \pi)
\eea
is the WF 
friction within the TF approach, according to 
\cite{MGFyaf2007}, and
$~B_{\rm ETF}(\psi)=M_{\rm ETF} \delta(\psi)/\sin \psi~$,  
where $M_{\rm ETF}$
presents the ETF  
(volume, surface and curvature) inertia terms 
[see (\ref{bcol})].
These (local TF and nearly local ETF) terms are related to the 
short CT$_0$ component of the Green's function trajectory 
expansion (\ref{Gsem})
after averaging over the phase space variables to remove their oscillations. 
Note that we have taken into account here the friction-dependent 
correction (\ref{bcolmcorr}) to the intrinsic
inertia $M(0)$ in (\ref{bcol}) 
for the smooth consistent 
collective inertia $M_{\rm ETF}$ \cite{MGFyaf2007}, i.e.\ , 
\be\l{btotcol}
M \approx M_{\rm ETF}(0) + 
M_{\rm corr}^{(1)} + M_{\rm corr}^{(2)}\;,
\ee
where 
$M_{\rm ETF}(0)$ is the ETF component (\ref{btf}),  
$M_{\rm corr}^{(1)}$ is the self-consistent correction  (\ref{bcolmcorr}) 
in the ETF inertia (\ref{bcol}), and $M_{\rm corr}^{(2)}$  
the smooth nonlocal
trajectory correction to the ETF inertia,
\be\l{mcorr2}
 \hspace{-0.5cm}M_{\rm corr}^{(2)}=\int_0^\pi \d \psi\; \sin\psi\;
P^{}_n\left(\cos \psi\right)\;
B_{\rm corr}(\psi)\;,
\ee
and similarly, for the smooth trajectory-friction correction 
$\gamma_{\rm corr}^{(2)}$ in the sum 
$\gamma \approx \gamma^{}_{\rm ETF}(0)+\gamma^{(2)}_{\rm corr}$.
We neglected also a relatively small smooth nonlocal 
correction $\gamma_{\rm corr}^{(1)}$ [see (\ref{gtf})] to the friction
$\gamma^{}_{\rm ETF}(0)$ (\ref{gcol}) in 
calculations of the corresponding 
components of the inertia $M$ and friction $\gamma$
($\gamma_{\rm corr}^{(1)} \ll \gamma_{\rm wf}$). 
The second terms $\Upsilon^{}_{\rm corr}(\psi)$ and
$B_{\rm corr}(\psi)$ in (\ref{gamm0cav0}) 
are the nonlocal corrections to the friction and inertia 
kernels, respectively,
\bea\label{gamsphc}
\Upsilon_{\rm corr}(\psi)&=&\gamma_{\rm wf}\sum_{v,w}
\frac{\sin^3\phi\;\cos\phi}{2 v \sin\psi}\nonumber\\ 
&\times& \mathcal{R}_{\Upsilon}(\psi,v,w)\;,
\eea
\bea\label{msphc}
\hspace{-0.5cm}B_{\rm corr}(\psi)&=&-M_{0}\sum_{v,w}
\frac{\sin^3\phi\;\cos\phi |\sin\phi|}{\sin\psi}\nonumber\\ 
&\times&\mathcal{R}_{B}(\psi,v,w),
\eea
where $M_0=d_smR_0^2 \;(k^{}_FR_0)^3/(16 \pi^2)$. 
The summations over $v$ and $w$ 
run over all  
CT$\neq $CT$_0$ from
 ${\bf r}_1$ to  ${\bf r}_2$ in the spherical box (like T$_1$
shown in Fig.\ \ref{fig26}), 
in accordance with the Green's function 
expansion (\ref{Gsem}).
The number of sides $v \geq 2$
(intermediate vertexes plus one) 
and  the winding number  $w \leq $ [$v/2$]
for clockwise and $w \geq -$[$v/2$] for 
anticlockwise motion specify the CT 
(the square brackets
show the integer number of $v/2$). 
As shown in Fig.\ \ref{fig26},
$\phi$ is a half of the central
angle for any chord of the equivalent sides of the trajectory,
$\phi=\phi^{}_{\rm PO}-\psi/(2 v)$,
$\phi^{}_{\rm PO}=\pi w/v$, 
($2 v\phi +\psi=2\pi w$, see Fig.\ \ref{fig26}). 
In (\ref{gamsphc}) and (\ref{msphc}), we introduced also the
modulation factors
\bea\l{modfactor}
&&\hspace{-0.5cm}\mathcal{R}_{\Upsilon}(\psi,v,w)\!=\!1 \!-\! J_0\left[(
\Delta S(\psi,v,w)/\hbar)^{1/2}\right],\nonumber\\
&&\hspace{-0.5cm}\mathcal{R}_{B}(\psi,v,w)\!=\!1 \!-\! J_0\left[(8 
\Delta S(\psi,v,w)/\hbar)^{1/2}\right]\;,
\eea
where $J_0(x)$ is the cylindrical Bessel function of 
zero order; and
\begin{equation}\label{deltaS}
 \Delta S(\psi,v,w)=\pi p^{}_FR_0\;\cos\phi^{}_{PO}\; \sin\psi\;
\end{equation}
is the action perturbation, $\Delta S_{\rm CT}$, 
depending on the relative angle $\psi$,
and the $v,w$ integers which all specify the CT.
Notice first that the calculation of the radial derivatives of the second terms 
(\ref{GsemGa}) in the Green's 
function expansion $G$ (equation (\ref{Gsem})) 
are mainly (at leading order of $\hbar$) reduced to the 
derivatives of the action, $S_{\rm CT}=p \mathcal{L}_{\rm CT}$, 
where 
$\mathcal{L}_{\rm CT}=2R_0 v \sin\phi$ 
is the length of the CT in (\ref{gaminsphker}) 
for cavity-like potentials. 
These derivatives are
the
normal (radial) components
of the particle momenta $p_{r1}$ and $p_{r2}$ at the spherical surface 
which are related to the initial $\r_1$
and the final $\r_2$ point,
$p_{r\; 1} =p_{r\; 2}= p\;\sin \phi$   
(Fig.\ \ref{fig26}).

In contrast to the  
work \cite{koonran}, we derive  the expressions (\ref{gamsphc}) 
for the friction and
(\ref{msphc}) for the inertia solving the 
symmetry breaking problem.
Indeed, we have to integrate the kernels $\Upsilon(\psi)$ and $B(\psi)$ in 
(\ref{gamsphgen}) over all $\psi$ from $0$ to $\pi$. However, in the limit
$\psi \rightarrow 0$ a nonclose isolated trajectory of the type of 
 T$_1$ (Fig.\ \ref{fig26}) turns into
 the degenerated one-parametric family of the closed periodic orbits.
The parameter of each of these families is the angle
of the rotation of a periodic orbit (PO) around the $z$ axis 
directed from the center of the spherical box to the vertex point $\r_1=\r_2=\r$
with the same action ($\phi=\phi^{}_{\rm PO}$).    
The expression for the Green's function
amplitude ${\cal A}^{}_{\rm CT}$ in (\ref{GsemGa}) for the contribution of the 
PO family 
is in principle enhanced by the semiclassical factor $\hbar^{-1/2}$
 as compared to that for the isolated trajectory 
(\ref{ampiso}) due to
the one-parametric PO degeneracy \cite{strumag,migdalrev} 
(Section II).
With $\psi$ decreasing to zero, one has a sharp increase of the Green's 
function amplitudes
within the angle of order of the wave length 
$1/k^{}_F$ over  the size $R^{}_0$
of the Fermi system at $1/k^{}_FR^{}_0 \ll 1$.
Applying these expressions for the Green's function
amplitudes from 
\cite{strumag,magvvhof}
to the smooth friction and inertia corrections (\ref{gamm0cav0})
at $\psi=0$,  one in fact obtains
zeros for the PO contributions because of the summations over all positive $w$
(for clockwise) and negative $w$ (for anticlockwise) motion 
with  the odd summands in $w$.
However, their analytical behavior 
is important for
a continuous match of the transport coefficients with the 
asymptotic Gutzwiller ones
related to the isolated trajectories T$_1$
at nonzero $\psi \gg 1/k^{}_FR_0$.

The smooth transition between this asymptotics for the friction
and inertia kernels, $\Upsilon(\psi)$ and $B_{\rm corr}(\psi)$
and the limit $\psi \rightarrow 0$
can be found by using the 
uniform approximation \cite{sclbook}.  
This
transition can be considered as a perturbation of the action, 
$\Delta S_{\rm CT}=\Delta S(\psi,v,w)$  (\ref{deltaS}),
due to the symmetry breaking at $\psi=0$. Then, we transform 
a cycle variable, say $\varphi$ at $\psi=0$, into 
another cycle one $\widetilde{\varphi}=\widetilde{\varphi}(\varphi)$  
(not necessary to be specified) with the 
increasing parameter $\psi$ 
such that for the phase integral $\Phi^{}_{\rm CT}$ of the semiclassical
Green's function in (\ref{GsemGa}) %
\cite{strumag,sclbook}, 
one finds by definition, 
\bea\label{PhiCT}
&&\Phi^{}_{\rm CT}(\tilde{\varphi}) \equiv \Phi^{}_{\rm CT}(\varphi)\nonumber\\
&=&\tilde{\Phi}^{0}_{\rm CT} + 
\sqrt{
8\; \Delta S^{}_{\rm CT}/\hbar}\; 
\cos\left(\tilde{\varphi}\right)\;,
\eea
for the inertia calculations and without factor 8 
in front of $\Delta S_{\rm CT}$ 
under the square root for the friction ones. 
The unknown constants 
$\widetilde{\Phi}^{0}_{\rm CT}$ and $\Delta S^{}_{\rm CT}/\hbar$ can be found now 
from the two asymptotic solutions at the $\psi \rightarrow 0$ and 
$\psi \gg 1/k^{}_FR^{}_0$ limits for the friction 
$\Upsilon_{\rm corr}(\psi)$ 
and inertia $B_{\rm corr}(\psi)$  
components of (\ref{gamm0cav0}).
 We emphasize that these transport coefficients are quadratic in $G$ rather than
 linear one for the level density calculations 
in the standard procedure for the uniform approximations \cite{sclbook}.
Therefore, in contrast to the standard transformation which is linear in 
the action perturbation $\Delta S^{}_{\rm CT}$, see \cite{sclbook}, called as 
the ``pendulum transformation'',
one finds a square root 
of $\Delta S_{\rm CT}$ in (\ref{PhiCT}).
Finally, one obtains the uniform approximation (\ref{gamsphc}) and 
(\ref{msphc}) for the smooth nonlocal friction and inertia corrections
in terms of the modulation factor $\mathcal{R}(\psi,v,w)$ (\ref{modfactor}).
These results for 
$k^{}_FR_0 \rightarrow \infty$ ($\mathcal{R}(\psi,v,w) \rightarrow 1$) 
as functions 
of $\phi$ coincide formally
with the smooth trajectory 
corrections found in \cite{koonran}.  
By using a different method, 
one obtains the same results as presented in 
\cite{koonran}  if we ignore, nevertheless, 
the significant symmetry breaking for the transition
from the PO family to the isolated CT contributions.
This symmetry breaking leads to the divergences 
which can be removed for the friction calculations by the artificial
procedure suggested in \cite{koonran}. 
One finds the same reason also for the divergences
of the inertia for which, however, 
this procedure does not help (for even $n$). 
Notice also that  the definition
of the angle itself $\phi$ is different from that of \cite{koonran}. 
The  action perturbation $\Delta S_{\rm CT}(\psi)$ (\ref{deltaS})
in units of $\hbar$ with a small $\psi$ measures the width in the 
variable $\psi$, 
$\psi \siml \hbar/\Delta S_{\rm CT} \sim 1/\pi k^{}_FR^{}_0 << 1$, for 
the uniform transition between the two abovementioned 
limits at a finite $k^{}_FR^{}_0$ , the contribution of the 
PO family, and the Gutzwiller limit for
the isolated CT.

Our results (\ref{gamsphc}) and (\ref{msphc}) 
for $\Upsilon_{\rm corr}$ and $B_{\rm corr}$ associated with 
the smooth trajectory corrections  
at finite values $k^{}_FR_0$ essentially differ from 
their asymptotics $k^{}_FR_0 \rightarrow \infty$ 
by  the modulation factor $\mathcal{R}(\psi,v,w)$ defined in 
(\ref{modfactor})
in terms of the Bessel function $J^{}_0$, and therefore, 
from those of \cite{koonran}.
We obtained the finite $\psi$ dependence of
the nonlocal friction- and inertia-correction kernels,
$\Upsilon^{}_{\rm corr}(\psi)$ (\ref{gamsphc}) and 
$B^{}_{\rm corr}(\psi)$ (\ref{msphc}), respectively, due to 
smooth contributions of the T$_1$-like classical 
trajectories (Fig.\ \ref{fig26}).  
In order to see 
the convergence of summations over the trajectories $v$ and $w$ 
in $\Upsilon_{\rm corr}$ 
(\ref{gamsphc}) [or $B_{\rm corr}$ in (\ref{msphc})] 
with our definition of $\phi$, one can
subtract $\sin^3\phi^{}_{\rm PO}\cos\phi^{}_{\rm PO}$ from 
$\sin^3\phi\cos\phi$, and add the same one in the correction
summand. 
The additional extra component of the friction $\Upsilon_{\rm corr}$ (or inertia
$B_{\rm corr}$) is identically zero because of 
 the symmetry
of the summation over negative (anticlockwise) and positive 
(clockwise) winding numbers $w$ provided that
the summand is 
an odd function of $w$,  
as mentioned above. 
Thus, the summation over positive and negative 
$w$ in (\ref{gamsphc}) for the friction coefficient 
correction 
gives one more factor multiplier approximately
$\propto 1/v$ at large $v$ ($\psi \leq \pi$, $|w| \leq [v/2]$). 
This leads to 
the fast convergence of  
sum over the trajectories specified by $v, w$ 
after the summation
in $w$ as $1/v^2$. 
More slow convergence ($\propto 1/v$) 
over these 
trajectories occurs for the inertia coefficient correction (\ref{msphc}).
A slow 
convergence, $\sim 1/v$, takes nevertheless place 
because of oscillations 
in the numerator of the
inertia summation for large $v$ (alternating series at large $v$) .

The smooth 
friction coefficients $\gamma$   
(\ref{gaminsphker}) and (\ref{gamsphc})  
in units of $\gamma_{wf}$
show their essential difference from
the w.f. for smaller multipolarities $n$, 
especially for the quadrupole friction (Table 3).
With increasing $n$, they decrease 
toward one in agreement with the 
results of \cite{koonran}, see last line taken from \cite{koonran}.
However, the friction coefficient  
$\gamma$ 
depends much on the finite $k^{}_FR^{}_0$ value. The limit 
$k^{}_FR^{}_0 \rightarrow \infty$
shown in Table 3 can be compared with the results 
of \cite{koonran}.  
As seen from Table 3, we found
an essential difference of the friction coefficients
for all multipolarities, especially for smaller even 
$n$, from those
of \cite{koonran} for $\gamma/\gamma_{\rm wf}$ 
which are independent of $k^{}_FR^{}_0$. 
But there is no dramatic difference between the smooth friction
for the quadrupole and octupole
dynamical surface distortions, in contrast to that found in \cite{koonran}.

We present the smooth inertia ratios  $M/M_{\rm irr}$ 
[see (\ref{btotcol})] 
for several 
values of $k^{}_FR_0$ in Table 4. The $k^{}_FR_{0}$ dependence
of these ratios appears mainly through the ETF values $M_{\rm ETF}/M_{\rm irr}$
(numbers shown in the brackets). Their self-consistent  positive 
surface component [see $M_{\rm corr}^{(1)}$ (\ref{bcolmcorr}) 
in (\ref{bcol})]  
is dominating 
at smaller 
$k_FR_0$ and disappears at $k^{}_FR_0\simg 10$
as the surface
correction, proportional relatively to $A^{-1/3}$
with respect   
to the volume irrotational-flow inertia. 
The corrections $M_{\rm corr}^{(2)}/M_{\rm irr}$  are much less sensitive to the
variations of $k^{}_FR_0$  than for the friction  
$\gamma_{\rm corr}^{(2)}/\gamma^{}_{\rm wf}$ (c.f. Tables 3 and 4). 
Note that our results 
for the inertias differ
significantly from the ones obtained in \cite{koonran} 
(last line of the Table 4). The quantities 
$M/M_{\rm irr}$ for even $n$ are finite, in contrast to the divergent results
of \cite{koonran} (even after the renormalization procedure suggested in 
\cite{koonran} to avoid divergence). For the quadrupole
 vibrations, one finds the positive values of $M$ of the order of $M_{\rm irr}$ 
for smaller $k^{}_FR_0 \siml 10$ 
(\ref{msphc}).
The inertia $M$ for larger $k^{}_FR_0$ 
are not shown because of almost the same
smooth trajectory (nonlocal) correction $M_{\rm corr}^{(2)}$, and 
the ETF (nearly local) part $M_{\rm ETF}$ 
[$M_{\rm ETF}(0)$ and $M_{\rm corr}^{(1)}$ components]
are unknown as we do not know
what are the parameters, especially the isoscalar surface-energy constant
$b^{}_{\rm S}$ of the ETF ES 
\cite{MGFyaf2007} for so 
large particle
numbers. In particular,
by this reason, the values of the inertia $M/M_{\rm irr}$ in the limit 
$k^{}_FR_0 \rightarrow \infty$ 
should be improved.
With increasing multipolarity $n$, one obtains much larger inertia $M$
as compared to its irrotational value $M_{\rm irr}$ \cite{bormot}. 
For odd multipolarities
one has significantly larger inertia, in contrast
to much smaller values obtained in \cite{koonran}, even after the 
renormalization procedure which is rather artificial for the inertia.

Concluding 
we have obtained the smooth nonlocal corrections
from longer trajectories to the WF 
friction and ETF inertia 
coefficients within the EGA POT by using the uniform approximation for 
solving the symmetry breaking problem.
The convergence of the friction corrections to the WF 
was found 
with increasing multipolarity of the surface distortions. The
inertia parameters are larger than the irrotational flow value
in the nuclear region of particle numbers.

\subsection{MAJOR-SHELL EFFECTS IN THE DISSIPATIVE\\ NUCLEAR DYNAMICS}
 
As shown semiclassically in the previous Sections,
many dynamical problems, in particular, in nuclear physics can be reduced 
to the collective motion of independent particles in a mean field with a 
relatively sharp time-dependent ES 
within the microscopic-macroscopic 
approaches \cite{myswann69,strut}.
In recent years it became apparent that the collective nuclear
dynamics is very much related to the nature of the nucleonic motion.
Behavior of the nucleonic dynamics is important in physical processes
as fission and heavy ion collisions where a great amount of the
collective energy is dissipated into a chaotic nucleonic motion.
We have to mention here very intensive studies of 
the one-body dissipative phenomena described largely through the 
macroscopic wall formula for the excitation energy rate with 
taking into account 
the non-adiabatic nonlinear corrections (Section IVB4 and
\cite{myswann69,wall,blocki,koonran,GMFprc2007}).  
For instance, it was
extended to the microscopic collective 
classical and quantum dynamics.
We like also to emphasize the 
significance of the transparent classical picture through the Poincar\'e 
sections of surfaces and Lyapunov exponents 
showing the order-chaos 
transitions, and  
quantum results for the average of the
time-dependent excitation energy rate \cite{blocki}.  
Then, the peculiarities of the excitation energies for 
many periods of the oscillations of the classical
dynamics were  discussed for the ES described by several Legendre 
polynomials 
\cite{BMYijmpe2012}. The statistics of the 
 spacing between the neighboring levels and its relation to the shell effects
depending on the specific properties of the s.p.\
spectra, as well as the  multipolarity and deformation of the ES
shapes were studied
\cite{BMstatlevPRC2012}. 
Our main purpose in this Section is to look at the  
time-dependent derivatives of the excitation energies of
the quantum gas of 
particles and focus to their correlations with the shell effects at 
slow and small-amplitude collective motion.

\subsubsection{Time-dependent energy rate}

To study the shell effects in the friction coefficients, we 
start with the time-dependent s.p.\
Schr\"odinger equation for
the wave function $\psi(t)$ \cite{blocki}, 
\begin{equation}\label{eqmot}
i \hbar \partial \psi/ \partial t =\hat{H}(t)\psi\;.
\end{equation}
The Hamiltonian $\hat{H}(t)$ describes a gas of 
independent particles. For  
the deformed Woods-Saxon (WS) potential one has
\begin{equation}\label{wspot}
\hspace{-0.2cm}V\left(r, \theta,t \right)=-
\frac{V_0}{1+ \exp\left\{\left[r-R(\theta,t)\right]/a\right\}}\;,
\end{equation}
depending on time $t$ through a time-dependent radius 
of the effective surface $R(\theta,t)$ \cite{wall,blocki,BMYpr} 
\bea\label{radt}
\hspace{-0.5cm}R(\theta,t)&=&\frac{R_0}{\Lambda(t)}\left[1+\alpha(t)\; 
\sqrt{\frac{4\pi}{5}}\; 
Y_{n0}\left(\theta\right)\;\right. \nonumber\\
&+& \left.
\alpha_1(t) \sqrt{\frac{4\pi}{3}}\; 
Y_{10}\left(\theta\right)\right]\;.
\eea
Here, $\Lambda(t)$ is a normalization factor ensuring a
volume conservation, $\alpha_1(t)$ stands 
 for keeping a fixed position of the center of mass 
for odd multipolarities, and
$R_0$ is the radius of the
 equivalent sphere, $Y_{n0}(\theta)=
\sqrt{(2n+1)/4\pi}\;P_n\left(\cos\theta\right)$.
$P_n\left(\cos\theta\right)$ is the Legendre polynomial,
 and $\alpha(t)$ a
periodic function of time. For the collective multipole vibrations 
(\ref{radt}) near the
spherical shape, one may write 
$ \alpha(t)=\alpha\cos\left(\omega t\right) $ where $\alpha$ is their 
amplitude [$\alpha(t)=q(t)(4 \pi/5)^{1/2}$ where $q(t) $ is the deformation
parameter as 
in Section IVB4].

Starting with oscillations from a maximum displacement of the
$P_n$ deformation equal to $\alpha$, one can traditionally \cite{blocki}
introduce  
the adiabaticity parameter $\eta_{\rm ad}$
being the ratio of the biggest wall speed to the biggest
speed of particles:
\begin{equation}\label{adpar}
\eta_{\rm ad}=\alpha \omega/\Omega,\qquad\qquad \Omega=v^{}_F/R_0\;,
\end{equation}
where 
 $\omega$ and  $\Omega$ are 
the frequencies of the collective and particle 
motions with the Fermi velocity $v_F$. 
Note that a condition
of smallness of the frequency $\omega$ 
with respect to $\Omega$  is used  
often
in the nuclear collective dynamics, in particular, 
within the microscopic-macroscopic 
approaches \cite{myswann69,strut,fuhi}.
Again, $\Omega$ determines the distance between the major 
shells \cite{strumag,migdalrev,sclbook},
$\hbar \Omega\sim \vareps^{}_F/A^{1/3} \approx 7-10$ MeV in heavy nuclei,  
with the Fermi energy $\vareps^{}_F$
and the particle number $A$ in nucleus (Section II).

For small amplitude vibrations around the sphere, one can use an approximate
expression  for the rate of the excitation energy of the gas \cite{wall},
\begin{equation}\label{WF}
\frac{dE}{dt}=m\rho \bar{v} \oint \dot{n}^2 \d S\; ,
\end{equation}
where $\rho$ is the particle density, $\bar{v}$ the mean particle speed, and
$\dot{n}$ the normal speed of the surface element of the wall. The
integral is taken over the entire surface.

Solutions to the dynamical equation (\ref{eqmot}) for 
the wave function $\psi(t)$, 
$\psi(t)=\sum_i C_i(t) \phi_i $, 
is taken as an expansion over the eigenfunctions $\phi_i$ of the 
static eigenvalue problem:
 $\hat{H}_0 \phi_i = \vareps_i \phi_i$ ,
where $\hat{H}_0$ is a static Hamiltonian taken at the spherical shape.
Thus, the problem is 
reduced to calculations of the time-dependent coefficients $C_i(t)$.

\subsubsection{Averaged energy rates and shell effects}

The s.p.\ spectra are calculated for the quadrupole and octupole 
shapes of the WS potential (\ref{wspot}) 
are presented
in \cite{BMYijmpe2012}. 
For the solution of the eigenvalue problem, 
it is convenient to
use the expansion over
the well-known deformed axially-symmetric 
harmonic-oscillator basis \cite{blocki}. 
To study the quantum-classical 
correspondence \cite{BMYpr} 
we deal with the 
WS potential (\ref{wspot}) having 
a sharp edge (diffuseness $a=0.1$ fm) and large
depth
($V_0=200$ MeV), similarly to those of a classical motion of 
particles inside the container with the  infinitely
deep depth and sharp walls. 
The spectra are having the strong shell effects at small
deformations near the spherical shape \cite{BMYijmpe2012}.
The magnitudes
of inhomogeneity of the s.p. levels near the Fermi energy,
i.e.\ , the shell effects, become smaller at 
deformations $\alpha \gtrsim 0.1$ because of the 
symmetry breaking (Section II, \cite{strumag,migdalrev,sclbook}).
However, the shell effects are not decreasing with 
deformations  as the considered ES multipolarities have
the same azimuthal symmetry which leads to a  partial 
integrability of the Hamiltonian $\hat{H}_0$ having the 
deformed potential \cite{migdalrev}.

Figs.\ \ref{fig27} and \ref{fig28} 
show the quadrupole $(P_2)$ and octupole $(P_3)$ 
averages of the time derivatives of the excitation energies, 
$\langle dE/dt \rangle$,
for a slow collective motion, $\omega/\Omega=$
$0.2$, with a small amplitude, 
$\alpha \sim 0.1$, and corresponding 
shell-correction energies
 $\delta E$ as functions
of the particle number parameter $N^{1/3}$, $N$ is the neutron number
in nucleus. These averages are proportional to
the one-body friction coefficient $\gamma$,
\begin{equation}\label{dedtgam}
  \langle dE/dt\rangle = \gamma \omega^2 \alpha^2/2\;.
\end{equation}
For the corresponding WF
(\ref{WF}), derived within
the TF approach, one finds
\bea\label{dedtgamwf}
 \langle dE/dt\rangle_{\rm wf} &=& \gamma_{\rm wf} \omega^2 \alpha^2/2\nonumber\\
&=&9 \vareps^{}_F\eta_{\rm ad}\;\alpha\;\omega A/20\;.
\eea
The quantum numerical results for  $\langle dE/dt\rangle$ are small as 
compared to those of the WF
(\ref{dedtgamwf}). However,
as clearly seen from these Figures, especially in 
Fig.\  \ref{fig28}  for the
octupole vibrations, one can observe rather strong correlations
between the fluctuations of the time-dependent energy-rate average 
$\langle dE/dt \rangle$ depending on the particle number,
$N^{1/3}$, and those of the shell-correction energies, $\delta E$, in good 
agreement with the shell effects in spectra near the spherical 
shape (Fig.\ \ref{fig22}). 
These correlations should be expected from the results of   
\cite{BMYijmpe2012} 
for the 
shell fluctuations of the level 
densities and energies, respectively. Note  that  
for $\alpha \lesssim 0.1$, 
the positions of the minima and maxima of
the  shell correction energies, $\delta E$, become almost
the same, at least for not too large
particle numbers, though the amplitude of $\delta E$ as 
oscillating function of
$N^{1/3}$ increases with decreasing deformation, 
c.f.\ the solid curve for $\alpha=0.1$ and dashed
for $\alpha=0.05$ in the bottom panel of
Fig.\ \ref{fig28}.
For such a small $\alpha$,
one should not expect significant differences between the 
friction coefficients
for the quadrupole and octupole vibrations
because the energy level 
structure becomes almost the same for all multipolarities 
at the small-amplitude collective motion.
The excitation energy should be correlated to the shell
structure of levels around the Fermi surface through an 
inhomogeneous distribution of the
energy levels $\vareps_i$ near the  
Fermi  surface $\vareps^{}_F$, i.e.\ , the shell effects in the level 
density at the energy $\vareps \approx \vareps^{}_{F}$.

Notice that the   
smooth semiclassical friction $\gamma/\gamma^{}_{\rm wf}$   
in the limit of the small amplitudes
($\alpha \rightarrow 0$) and frequencies 
($\omega \rightarrow 0$) of  
vibrations of 
the spherical box
is much larger (5 orders of the magnitude) than
the averaged quantum values 
obtained in 
\cite{BMYijmpe2012}
for slightly multipole deformed WS potential. 
Notice also that the results for friction coefficients depend on 
widths of the Gaussian averaging over the phase space variables 
\cite{GMFprc2007}. With decreasing widths one can find the situation with
alternative contributions of the WF $\gamma_{\rm wf}$ and correction 
$\gamma_{\rm corr}$ in (\ref{gamm0cav0}). In this case, the WF
might be not dominating and one has
another explanation of the small-average quantum friction
found in \cite{BMYijmpe2012}
for the quadrupole and octupole vibrations.

The main period of the friction shell
correction $\delta \gamma $ as function of the particle (neutron) number 
variable $N^{1/3}$ is of order of that of 
the energy shell correction  $\delta E$ (see Section IVB).  According to the 
POT  \cite{strumag,migdalrev,sclbook} it can be evaluated largely as 
$2 \tilde{g}\hbar \Omega/3 N^{2/3} \approx (N/A)^{1/3}/2\approx 0.4$ for
large total particle numbers $A$ in the nucleus. (Here,
$\hbar \Omega \approx 2 \pi\hbar/t^{}_{PO}\approx \vareps^{}_F/A^{1/3}$, where 
$t^{}_{\rm PO}$ is the period
of motion of the particle along the shortest PO and 
$\widetilde{g} \approx 3N/4 \vareps^{}_F$ in the TF approximation
for a smooth level density.) This is mainly in agreement 
with our numerical results presented in Figs.\ \ref{fig27} and 
\ref{fig28}. The derivations of oscillating components
of the friction coefficient (including the shell effects) due to 
other contributions of the CTs (also POs) within the POT
will be presented elsewhere.

Thus, we found thus a fairly strong correlations between fluctuations of the
excitation energies for a slow and small-amplitude vibration 
near the spherical shape
and  shell-correction energies. They are especially 
pronounced for the octupole
case for which our dynamical calculations are far going to smaller 
deformations with a good accuracy.
We obtained the smooth corrections
related to longer trajectories to the wall formula  within the POT and
found convergence to the WF
with increasing multipolarity of the surface
distortions.
The investigations of 
correlations between the shell effects and 
the dissipative character in the one-body 
friction coefficients within the semiclassical 
POT is under the way.
 It might be {\it perspective} to use the
combined macroscopic-microscopic approaches
and response function theory
to clear up the results more systematically and analytically. 
Our quantum and semiclassical results can be 
helpful for 
understanding the 
one-body dissipation and inertia at  
slow and faster collective dynamics with different shapes
like the ones met 
in nuclear fission and heavy-ion collisions.


 \section{NUCLEAR COLLECTIVE ROTATIONS}
\l{semshellmi}

This Section is devoted to one of the most remarkable and traditional
subjects of nuclear physics. As mentioned in Introduction, within the 
cranking model (Section VA), one can use the extension of the Strutinsky SCM 
to the rotational problems by Pashkevich and Frauendorf \cite{fraupash,mix}.
Within the semiclassical EGA POT approach, using also the 
response function theory and extended Gutzwiller expansion 
of the Green's function over the classical trajectories
as for other transport coefficients (Section IV), we 
significantly simplify the rotational many-body problem 
calculating the moments of inertia 
in terms of the ETF component (Section VB) and its shell correction 
(Section VC).

\subsection{THE CRANKING MODEL}
\l{cranmod}

Within the cranking model, the nuclear collective rotation of a
Fermi independent-particle system associated with a many-body Hamiltonian
  $\hat{\mbox{H}}$
can be described, to a good approximation \cite{eisgrei}, in the
restricted subspace of Slater determinants, by the eigenvalue problem
for a s.p.\ Hamiltonian
\be\l{new01}
  \hat{H}^{\boldsymbol\om}=\hat{H} + \hat{H}_{\rm CF}^{\boldsymbol\om} \;,
\ee
usually referred to as the {\it Routhian}.
For this Routhian, in the body-fixed rotating frame
\cite{bormot,mix,fraupash}, one has
\be\l{raussian}
  \hat{H}^{\boldsymbol\om} = \hat{H} + \hat{H}_{\rm CF}^{\boldsymbol\om}
  = \hat{H} - \boldsymbol \om \cdot \left(\hat{\boldsymbol\ell}
                                                   + \hat{{\bf s}}\right),
\ee
where $\hat{H}_{\rm CF}^{\boldsymbol\om}$ is the s.p.\ cranking field
which is approximately equal
  (neglecting a smaller centrifugal term, $\propto \om^2$)
to the Coriolis interaction.
  The rotation frequency $\boldsymbol\om$ of the body-fixed coordinate
  system (relative the laboratory system), and which is the Lagrange
  multiplier of our problem, is defined
through the constraint on the nuclear angular momentum ${\bf I}$,
evaluated through the quantum average
$\langle \hat{\boldsymbol\ell} +
\hat{{\bf s}} \rangle^{\boldsymbol\om}={\bf I} $, of the
total s.p.\ operator, $\hat{\boldsymbol\ell} + \hat{{\bf s}}$,
where $\hat{\boldsymbol\ell}$
is the orbital angular momentum and $\hat{{\bf s}}$
is the spin of the quasiparticle,
thus defining a function
$\boldsymbol\om=\boldsymbol\om({\bf I})$.
The quantum average
 of the total s.p.\ operator $\hat{\boldsymbol\ell}
+ \hat{{\bf s}}$ is obtained by evaluating expectation values of the
many-body Routhian
$\hat{\mbox{H}}_{\rm CF}^{\boldsymbol\om}$ 
in the subspace 
of Slater determinants.
For the specific case of a rotation around the $x$ axis ($\om=\om_x$)
which is perpendicular to the symmetry $z$ axis of the axially-symmetric mean
field $V$, one has 
(dismissing for simplicity spin (spin-isospin) variables),
\bea\l{constraint0}
\langle \hat{\ell}_x \rangle^{\om} &\equiv&
d_s \sum_i n_i^{\om} \nonumber\\
&\times&\int \d \r
\;\psi_i^{\om}\; \left(\r\right) \; \hat{\ell}_x\;
\overline{\psi}_i^{\om}\left(\r\right)=I_x\;.
\eea
The spin (spin-isospin) degeneracy $d_s$ accounts for the symmetry of
  the
mean-field potential $V(\r)$.
The occupation numbers $n_i^{\om}$
for the Fermi system of independent nucleons are given by
\bea\l{ocupnumbi}
n_i^{\om}&\equiv& n\left(\vareps_i^{\om}\right)\nonumber\\
&=&\{1+ \exp\left[(\vareps_i^{\om} - \lambda^{\om})/T\right]\}^{-1}\;. 
\eea
In (\ref{constraint0}),  $\psi_i^{\om}(\r)$ are the eigenfunctions,
and
$\overline{\psi}_i^{\om}(\r)$ their complex conjugate;
$\vareps_i^{\om}$
the eigenvalues of the Routhian $\hat{H}^{\om}$ (\ref{raussian});
$\lambda^{\om}$ is the chemical potential. For relatively small frequencies
$\om$ and temperatures $T$, $\lambda^{\om}$ is,  
in a good approximation
equal to the Fermi energy,
$\lambda^{\om} \approx \vareps_{{}_{\! {\rm F}}} =\hbar^2 k_{\rm F}^2/2 m^*$,
where $k_{{}_{\! {\rm F}}}$ is the Fermi momentum in units of $\hbar$
and $m^*$ is the effective mass.
From (\ref{constraint0}), the rotation frequency $\om$ can be
specifically expressed in terms of a given angular momentum
  $I_x$ of the nucleus: $\om=\om\left(I_x \right)$. Within
the same approach,
one approximately has for the particle number
\bea\l{partconspert}
A&=& d_s\sum_i n_i^{\om}
\int \d \r\; \psi_i^{\om}(\r)\;\overline{\psi}_i^{\om}(\r)\nonumber\\
&\approx& d_s \int \d \vareps \; n(\vareps)\;,
\eea
which determines the chemical potential $\lambda^{\om}$ for a given
   number $A$ of nucleons. Since
we introduce the continuous parameter $\om$ and ignore the uncertainty
relation between the angular momentum and
  the rotation angles
of the body-fixed coordinate system,
the cranking model is semiclassical
in nature \cite{ringschuck,GMBBps2015}.
 One may thus consider the collective MI $\Theta_x$, for a rotation
  around the $x$ axis      (omitting, to simplify the notation, spin
  and isospin variables), as a response
of the quantum average
$\delta \langle \hat{\ell}_x \rangle^{\om}$ (\ref{constraint0}), to the
external cranking field $\hat{H}_{\rm CF}^{\om}$ in (\ref{raussian}).
Similarly to the magnetic or isolated susceptibilities
\cite{richter,fraukolmagsan,magvvhof,MGFyaf2007},
one can write
\be\l{response}
\delta \langle \hat{\ell}_x \rangle^{\om}=
\Theta_x(\om) \delta\om\;,
\ee
where
\bea\l{thetaxdef}
&\!\Theta_{x}(\om)=
\partial \langle \hat{\ell}_x\rangle^\om/\partial \om=
\partial^2 E(\om)/\partial \om^2,
\quad\;\;\;
\nonumber\\
&\!\!\! E(\om)=\langle \hat{H} \rangle^\om
\equiv d_s \sum_i n_i^{\om}\!\int \d \r
\;\psi_i^{\om}\left(\r\right)\! \hat{H} 
\overline{\psi}_i^{\;\om}\left(\r\right)\;.
\eea
  A parallel (alignment) rotation with respect to the symmetry $z$ axis
  \cite{mix,kolmagstr,dfcpuprc2004}
 has also been considered in \cite{mskbPRC2010}.

 For a nuclear rotation around the $x$ axis, one can treat, as was shown
  in \cite{inglis,belyaevzel,valat,bormot,fraupash,mix}, the term
  $-\boldsymbol\om \cdot \hat{\boldsymbol\ell} = -\om \, \hat{\ell}^{}_x$
  as a perturbation.
With the constraint (\ref{constraint0})
and the MI (\ref{thetaxdef}) treated in the 
second order perturbation theory,
one obtains the well-known Inglis cranking formula, 
\be\l{inglismi}
\Theta_x=d_s\; \sum_{ij}^{\;\;\;\;\;\;\;\prime}
\;\frac{(n_i-n_j)\;
\Big|\langle j|\hat{\ell}_x|i \rangle \Big|^2}{\vareps_i-\vareps_j}\;,
\ee
where $\langle j|\hat{\ell}_x|i \rangle$ is the matrix elements of the
angular-momentum projection operator $\hat{\ell}_x$ given in the middle 
of (\ref{constraint0}) at $\om=0$, $|i\rangle=\psi_i$ (the superscript 
prime means again the exclusion of all terms with 
$\varepsilon_i=\varepsilon_j$, as in Eq.\ (\ref{mcran})). 
Instead of carrying out the rather involved calculations
presented above, one could, to obtain the yrast line energies $E(I_x)$
for small enough temperatures $T$ and frequencies $\om$, approximate
the angular frequency by $\om=I_x/\Theta_x$ and write the energy in the form
\be\l{yrast}
E(I_x)= E(0) + \frac{I^{2}_x}{2 \Theta_x}\;.
\ee
As usually done, the rotation term above needs to be quantized
through $I^{2}_x \rightarrow I_x (I_x+1)$ in order to study
rotation bands.

\subsection{SELF-CONSISTENT ETF DESCRIPTION \\ OF NUCLEAR ROTATIONS}
\l{etfmi}

Following references \cite{bartelnp,belyaev}, a microscopic description
   of rotating nuclei was obtained in the Skyrme Hartree--Fock formalism,
   within the ETF approach up to order $\hbar^2$. With a variational space
   restricted to 
 Slater determinant,
 the minimization of the
expectation value of
the nuclear Hamiltonian
leads to the s.p. Routhian 
$\hat{H}^{\boldsymbol\omega}_{\isospin}$ (\ref{raussian})
that is determined by a one-body potential $\hat{V}_{\isospin}({\bf r})$,
a spin-orbit field $\hat{{\bf W}}_{\isospin}({\bf r})$, 
and an effective-mass
form factor $f^{\rm eff}_{\isospin}({\bf r})=m/m^*_{\isospin}$ (see
also \cite{brguehak}). In what follows and throughout this Section, 
and to simplify the
     notation, we shall omit the {\it hat} sign on quantum mechanical
     operators of local quantities within the semiclassical
approximation. 
In the case when the time reversal symmetry is 
broken, the additional
     cranking field $\boldsymbol\alpha_{\isospin}({\bf r})$ and
     spin field form factors ${\bf S}_{\isospin}({\bf r})$ appear.
Here, and throughout this Section, a subscript q refers to the nucleon
      isospin (q = {n, p}), not to be confused with
the deformation parameter or wave number $q$ in other Sections of
     this review. All these fields can be written as functions of the local
     densities and their derivatives, such as the neutron or proton particle
     densities $\rho_{\isospin}({\bf r})$, the kinetic energy densities
     $\tau_{\isospin}({\bf r})$, 
the spin densities (also referred to
as spin-orbit densities)
${\bf J}_{\isospin}({\bf r})$, the current densities
${\bf j}_{\isospin}({\bf r})$, and
the spin-vector densities
$\boldsymbol\rho_{\isospin}({\bf r})$. \
Note that in the present subsection, $\tau_{\isospin}({\bf r})$
stands for the kinetic-energy
density which should not be confused with the neutron skin variable in
Section III and Appendix A.
In principle, two
additional densities appear, a spin-vector kinetic-energy density
$\boldsymbol\tau_{\isospin}({\bf r})$ and a tensor coupling
$J^{}_{i j}({\bf r})$
between spin and gradient vectors, which have, however, been neglected
since their contribution should be small, as suggested by \cite{BFH87}.

The cranking-field form factor $\boldsymbol\alpha_{\isospin}({\bf r})$
     contains two contributions, one of them coming from the orbital part
     of the constraint, $-\boldsymbol\om \cdot \boldsymbol\ell$, which has
     been shown in \cite{BV78} to correspond to the Inglis cranking formula
     \cite{inglis}, while the other, the Thouless--Valatin self-consistency
     contribution \cite{TV62} has its origin
in the self-consistent response
of the mean field to the time-odd part of the density matrix generated by
the cranking term of the Hamiltonian.
The aim is now to find functional relations for the local densities
$\tau_{\isospin}(\r)$, ${\bf J}_{\isospin}(\r)$,
${\bf j}_{\isospin}(\r)$ and $\boldsymbol\rho_{\isospin}(\r)$ in
terms of the particle densities $\rho_{\isospin}({\bf r})$,
in contrast to those given by
Grammaticos and Voros \cite{GV79} in terms of the form factors
$V_{\isospin}$,
$f^{\rm eff}_{\isospin}$, ${\bf W}_{\isospin}$,
$\boldsymbol\alpha_{\isospin}$ and
${\bf S}_{\isospin}$.
Taking advantage of the fact that, at the leading TF order,
the cranking field form factor is given by \cite{bartelnp}
\be\l{ETF02}
  \boldsymbol\alpha^{({\rm TF})}_{\isospin}
                        = f^{\rm eff}_{\isospin}\left( {\bf r}
    \times \boldsymbol\omega \right)\;,
\ee
one simply obtains the rigid-body value for the Thomas--Fermi current density
\be\l{ETF03}
   {\bf j}^{({\rm TF})}_{\isospin} = \displaystyle \frac{m}{\hbar}
   \left( \boldsymbol\om \times {\bf r} \right) \, \rho_{\isospin}\;.
\ee
This result is not that trivial, since it is only through the effect
of the Thouless--Valatin self-consistency terms that such a simple
result is obtained. Notice also that (\ref{ETF03}) corresponds to a
generalization to the case $f^{\rm eff}_{\isospin} \neq 1$ of a result
already found by Bloch \cite{Bl54}.

Equation (\ref{ETF03}) can also be considered as an extension of the
     Landau quasiparticle (generalized TF) theory
     \cite{landau,abrikha,belyaev} presented in Section III and applied now,
     (if particle collisions can be neglected in the kinetic equation 
(\ref{LVeq})),
     to the mean-field case of rotating Fermi-liquid systems
     (cf. (\ref{ETF03}) with the current density as an average of the
     \textit{particle} velocity, $\p_{\rm rot}/m=\boldsymbol\om \times \r$,
     rotating with the frequency $\boldsymbol\om$ \cite{belyaev}).
In particular, the re-normalization of the cranking field form factor
$\boldsymbol\alpha^{({\rm TF})}_{\isospin}=
f_{\rm q}^{\rm eff} \boldsymbol\alpha_{\rm o}$ with
\be\l{alpha0}
\boldsymbol\alpha_{\rm o} =
\left( {\bf r} \times \boldsymbol\omega \right)\;,
\ee
by (\ref{ETF02}) can be also explained
as related to the 
corrections,
$f_{\rm q}^{\rm eff} \neq 1$, obtained
by Landau \cite{landau} with using
both the Galileo principle and the Thouless--Valatin self-consistency
corrections to a particle mass $m$ due to the quasiparticles'
(self-consistent) interaction
through a mean field. They lead in \cite{bartelnp}
to  the self-consistent TF angular momentum
of the \textit{quasiparticle}
$\boldsymbol\ell_{\isospin} =f_{\isospin}^{\rm eff} \boldsymbol\ell_{\rm o}$ with the
classical angular momentum
$
\boldsymbol\ell_{\rm o}= \r \times \p 
$
of the particle,
so that $-\boldsymbol\om\cdot \boldsymbol\ell_{\isospin}=
\boldsymbol\alpha_{\isospin} \cdot \p$. This effect is similar to that
for the kinetic energies of the \textit{quasiparticles},
$\vareps_{\isospin}=p^2/(2 m^*_{\isospin})=f_{\isospin}^{\rm eff}\vareps_{\rm o}$
where $\vareps_{\rm o}=p^2/(2m)$. 
With this transparent connection to the Landau quasiparticle
theory, it is clear that
there is no contradictions with the TF 
$\hbar \rightarrow 0$ limit (\ref{alpha0}) of the
     current densities in the FLDM, accounting for the TF particle
     densities.
In (\ref{ETF03}), $\hbar$
 appears formally due to a traditional
 use of the dimensionless units for the angular momenta in the
quantum-mechanical picture to compare with experimental
nuclear data. Another reason is related to
a consistent treatment of the essentially quantum
spin degrees of freedom, beyond the Landau quasiparticle approach
to the description of Fermi liquids, which have no \textit{straight} classical
correspondence, in contrast 
to the orbital angular momentum
$\boldsymbol\ell$.
For already smooth quantities, the convergence in the TF limit
     ($\hbar \rightarrow 0$) can be realized after a statistical 
(macroscopic) averaging
     over many s.p.\ (or more generally, many-body) quantum states.
This averaging removes the fluctuating (shell) effects  which
appear in the denominators of the exponents within the POT ( 
Section II). 
Finally, the spin paramagnetic  effect
 can be considered within the ETF approach, 
through a macroscopic averaging, along 
with the orbital diamagnetic contribution.
For instance, the spin-vector density does not
have a \textit{straight} classical analogue,
such as the 
orbital angular momentum, and has to be considered as a quantity
     of leading order $\hbar$.

It is now important to notice \cite{bartelnp} that in an ETF expression
     of order $\hbar^2$, it is sufficient to replace quantities, such as
     the cranking field form factor $\boldsymbol\alpha_{\isospin}$, by their
     ETF expressions.
(after the statistical averaging mentioned above).
In the same way, to obtain a semiclassical expression for the
     spin-vector densities $\boldsymbol\rho_{\isospin}$ that is correct up
     to order $\hbar$, only TF expressions are required. One thus obtains
     for $\boldsymbol\rho_n$ and $\boldsymbol\rho_p$ a system of linear
     equations that can be easily resolved \cite{bartelnp}.
One also notices from this system of
equations that the spin-vector densities are proportional to the
angular velocity $\omega$.
Exploiting the well known analogy of the microscopic Routhian problem with
electromagnetism, one may then define {\it spin susceptibilities}
$\chi_{\isospin}$,
\be\l{ETF04}
  \boldsymbol\rho_{\isospin} = \hbar \,
\chi_{\isospin} \, \boldsymbol\omega \; .
\ee

The key question now is to assess the sign of
these susceptibilities and to
decide whether or not the corresponding alignment is of a
``Pauli paramagnetic''
character. The study of \cite{bartelnp} shows that this is the case, i.e., that
the spin polarization is, indeed, of paramagnetic character, thus confirming
the conclusions of the investigation performed
by Dabrowski \cite{Da75} in a simple model of non-interacting
nucleons.

Since the cranking field $\boldsymbol\alpha_{\isospin}$ is, appart from
     its contribution coming from the constraining field
     $\boldsymbol\alpha_{\rm o}$ (\ref{alpha0})
     only determined by the current densities ${\bf j}_{\isospin}$ and the
     spin-vector densities $\boldsymbol\rho_{\isospin}$, one can then write
     down in a simple way \cite{bartelnp} the contributions to the current
     densities ${\bf j}_{\isospin}$ going beyond the TF approach.
The semiclassical corrections
of order $\hbar^2$ can be split into
contributions
$(\delta {\bf j}_{\isospin})_{\ell}$ and
$(\delta {\bf j}_{\isospin})_{s}$ coming
respectively from the orbital motion and the spin degree of freedom. It is
found \cite{bartelnp} that the orbital correction
$(\delta {\bf j}_{\isospin})_{\ell}$
corresponds to a surface-peaked {\it counter-rotation} with respect to the
rigid-body current proportional
to $\left( \boldsymbol\omega \times {\bf r} \right)$,
thus recovering the Landau diamagnetism characteristic of a finite Fermi gas.
With the expressions of the current densities ${\bf j}_{\isospin}$
and the spin-vector
densities $\boldsymbol\rho_{\isospin}$ up to order $\hbar^2$,
one can then write down the
corresponding ETF expressions for the kinetic-energy density
$\tau_{\isospin}(\r)$ and
spin-orbit density ${\bf J}_{\isospin}(\r)$.

Taking the explicit ETF functional expressions up to order $\hbar^2$
      of all the densities entering 
our problem, one is able to write down the energy of the
nucleus in the laboratory frame as a functional of these local densities,
\be\l{endentau}
E = \int \d \r \; \rho \; \mathcal{E}[\rho_{\isospin},
\tau_{\isospin}, {\bf J}_{\isospin}, {\bf j}_{\isospin}\;,
\boldsymbol\rho_{\isospin}],
\ee
where $\rho=\rho_n+\rho_p$. 
Upon some integration by parts, one finds that
$\mathcal{E}$ can be written as a sum of
 the energy density per particle of
the nonrotating system
$\mathcal{E}(0)$ and its rotational part,
in line of (\ref{yrast}).
Within the ETF approach, one has from (\ref{endentau})
\be\l{endentauSCM}
E_{\rm ETF}= \int \d\r \rho \mathcal{E}(0) +
\frac12 \Theta_{\rm ETF}^{({\rm dyn})}\,\om^2\;,
\ee
where $\Theta_{\rm  ETF}^{({\rm dyn})}$ is the ETF dynamical moment of
inertia for the
nuclear rotation with
a frequency $\boldsymbol\om$. This MI is given in the form:
\bea\l{ETF05}
&  \Theta_{\rm ETF}^{({\rm dyn})}
  = \Theta_{\rm orb.}^{({\rm dyn})} + \Theta_{\rm spin}^{({\rm dyn})}
  =\! m \sum_{\isospin} \int \d \r \left\{ r_\perp^2 \,
  \rho_{\isospin} \right. \nonumber\\
& \left. - \!\frac{f^{\rm eff}_{\isospin} \rho^{1/3}_{\isospin}}
  {(3 \pi^2)^{(2/3)}} + \left[\frac{\hbar^2}{2m} \!+\! W^{}_0
  \left(\rho \!+\! \rho_{\isospin}\right) \right] \, \chi_{\isospin}
  \right\}\;,
\eea
where
$ r_\perp$ is the distance of a given point to the rotation axis. 
The Skyrme-force strength parameter of the spin-orbit
interaction $W^{}_0$ is defined in Section III3 \cite{brguehak}.

One notices that the ETF term which comes from the orbital motion
turns out to be the classical rigid-body (TF) MI. Semiclassical corrections
of order $\hbar^2$ come from both the orbital motion
($\Theta_{\rm orb.}^{({\rm dyn})}$) and from the spin degrees of freedom
($\Theta_{\rm spin}^{({\rm dyn})}$). The contribution $\Theta_{\rm orb.}^{({\rm dyn})}$
is
negative corresponding to a surface-peaked counter rotation in the rotating
frame. Such a behavior is to be expected
for a N-particle system bound by
attractive short-range forces  \cite{BJ76}.
The spin contribution
$\Theta_{\rm spin}^{({\rm dyn})}$ turns out to be of the {\it paramagnetic} type,
thus leading to a positive contribution which corresponds to an alignment of
the nuclear spins along the rotation axis. It can
also be shown \cite{BBQ93}
that the ETF kinematic moment of inertia,
\be\l{mikin}
  \Theta_{\rm ETF}^{({\rm kin})} = \displaystyle
\frac{\langle{\boldsymbol\ell}+{\bf s}\rangle^{\om}}{\omega},
\ee
is identical
to the ETF dynamical moment of
inertia presented above.

It is now interesting to study the importance of the Thouless--Valatin
self-consistency terms.  This has accomplished by calculating
the MI
in the ETF approximation but omitting, this time, the
Thouless--Valatin terms. One then finds \cite{bartelnp} the following
expressions for the dynamical MI, in what is simply the
Inglis cranking (IC) limit
\bea\l{ETF06}
  \Theta_{\rm IC}^{({\rm dyn})} 
&=& m \sum_{\isospin} \int \d \r 
\left[\frac{\rho_{\isospin}}{\left(f^{\rm eff}_{\isospin}\right)^2}\right.\nonumber\\ 
  &+& \left.\frac{m B_3}{\hbar^2}  \rho_{\isospin} \rho_{\bar {\isospin}}
          \left(\frac{1}{f^{\rm eff}_{\isospin}} -
\frac{1}{f^{\rm eff}_{\bar {\isospin}}} \right)^{\!\!2} 
                                        \right] r_\perp^2,
\eea
where $\bar{\isospin}$ is the {\it other}
charge state ($\bar{\isospin} \!\!=\!\! p$ when
$\isospin \!\!=\!\! n$ and vice-versa)
and $B_3$ is defined through the Skyrme force parameters
$t^{}_1, t^{}_2, x^{}_1$ and $x^{}_2$ (see \cite{bartelnp}).
Apart from the corrective term in $\rho_{\isospin} \, \rho_{\bar {\isospin}}$,
one notices
that the first term in the expression above, which is the leading term,
 yields, at least for a standard HF-Skyrme
    force where $f^{\rm eff}_{\isospin} \geq 1$, a smaller MI 
than the corresponding term in (\ref{ETF05})
containing the Thouless--Valatin corrections.
It is also worth noting that in
this approximate case, the kinematic MI is given by
\be\l{ETF07}
  \Theta_{\rm IC}^{({\rm kin})}
        = m \sum_{\isospin} \int \d \r \;
\frac{\rho_{\isospin}}{f^{\rm eff}_{\isospin}} \;
r_\perp^2\;,
\ee
which turns out to be quite different from the above given
dynamical MI, (\ref{ETF06}),
obtained in the same limit (ETF limit, omitting the
Thouless--Valatin self-consistency terms).

To investigate the importance of the different contributions to the
     total moment of inertia, self-consistent ETF calculations up to order
     $\hbar^4$ have been performed \cite{bartelpl} for 31 nonrotating
     nuclei, imposing 
a spherical symmetry,
and
using
the SkM$^*$ Skyrme effective nucleon-nucleon interaction \cite{BQB82}. Such
calculations yield variational semiclassical density profiles for neutrons
and protons \cite{brguehak} which are then used to calculate the above given
moments of inertia. 
The nuclei included in this study are
$^{16}$O,
$^{56}$Ni, $^{90}$Zr, $^{140}$Ce, $^{240}$Pu and three isotopic chains for
Ca ($A \!=\! 36-50$), Sn ($A \!=\! 100-132$) and Pb ($A \!=\! 186-216$).
The results of these calculations
are displayed in Figure \ref{fig29}  
taken from
\cite{bartelnp}.

One immediately notices the absence of any significant isovector dependence.
The good reproduction of the total ETF moment of inertia obtained by the
TF (rigid-body) value is also quite striking. 
One finds that the semiclassical orbital and spin corrections are in
     fact not small individually but cancel
each other to a large extent. To illustrate this fact the ETF moments
obtained by omitting only the spin contribution are also shown on the Figure.
One thus obtains a reduction of the Thomas--Fermi result that is about 6\% in
$^{240}$Pu but as large as 43\% in $^{16}$O.

The Inglis cranking approach performed at the TF level underestimates
the kinematic moment of inertia by as much as 25\%, and the dynamical MI by about 50\% in heavy nuclei, demonstrating in this way the importance
of the Thouless--Valatin self-consistency terms.

In \cite{bartelnp}, a crude estimate of the semiclassical corrections due to orbital and spin degrees of freedom has been made by considering the nucleus as a piece of
symmetric nuclear matter (no isovector dependence as already indicated by the
self-consistent results shown in Fig.\ \ref{fig29} 
above).
It turns out that these
semiclassical corrections have an identical $A$ dependence
($A_{}^{-2/3}$ relative to the leading order TF,
i.e.\ , the classical rigid-body term)
\be\l{ETF08}
  \Theta_{\rm  ETF} = \Theta^{(\rm RB)}_{\rm TF} 
     \left[ 1 + \left( \eta_{\ell} +
\eta_s \right) A^{-2/3} \right] \; .
\ee
A fit of the parameters $\eta_{\ell}$ and $\eta_s$ to
the numerical
results displayed
in Fig.\ \ref{fig29} 
yields $\eta_{\ell} = -1.94$ and $\eta_s = 2.63$
giving a total (orbital
+ spin) corrective term of $0.69 \, A^{-2/3}$. For a typical rare-earth
nucleus (A = 170) all this would correspond to a total corrective term equal to
2.2\% of the classical rigid-body value,
$\Theta^{(\rm RB)}_{\rm TF}$, resulting from a -6.3\% correction
for the orbital
motion and a 8.5\% correction for the spin degree of freedom.

Whereas in the calculations that lead to Fig.\ \ref{fig29} 
above,
spherical symmetry
was imposed, fully variational calculations have been performed in
\cite{bartelpl}, imposing however the nuclear shapes to be of spheroidal form.
In this way, the
nuclear rotation clearly impacts on the specific form of the
matter densities $\rho^{}_n$ and $\rho^{}_p$ which, in
turn, in the framework
of the ETF approach determine all the other local densities, as explained
above.

Trying to keep contact to usual shape parametrisation with the
      standard quadrupole parameters $\beta$ and $\gamma$, these are 
chosen in such a way as to
     yield the same semi-axis lengths of the quadrupole drop as those
     obtained for the spheroids. As a result,
Figure \ref{fig30}  
shows the evolution of the equilibrium
solutions (the ones that minimize the energy for given angular momentum) 
as a function of the nuclear spin I. 
One clearly observes that at low
values of
the angular momentum
(I in the range between 0 and 50 $\hbar$) the nuclear drop takes on
an oblate
shape, corresponding to increasing values of the quadrupole 
deformation parameter $\beta$
with increasing I values, but keeping the non-axiality parameter fixed at
$\gamma = 60^{\circ}$.  For larger values of the total angular momentum
(I beyond 55 $\hbar$), one observes a transition into triaxial shapes, where
the nucleus evolves rapidly to more and more elongated shapes. For even
higher values of I (I beyond 70 $\hbar$) the nucleus approaches
the fission
instability. These results are in excellent qualitative agreement with those
obtained by Cohen, Plasil and Swiatecki \cite{CPS74} in a rotating LDM
(see \cite{matmatnak2010_12,hinsatnak2010_11} 
for a more microscopic
     description of the transitions between oblate and prolate shapes in
     nuclear rotational bands).

It is
amusing to observe
here
a {\it backbending} phenomena at the semiclassical
level when one is plotting,
as usual, the MI $\Theta_{\rm ETF}$
versus the rotational angular momentum,
    as displayed in Fig.\ \ref{fig31}.
One should, however,
insist on the fact that this {\it backbending} has strictly nothing to do with
the breaking of a Cooper pair. The rapid increase
of the moment of inertia
at about I$ \,= 60 \hbar$
with a practically constant (or even slightly
decreasing)
rotational frequency $\om$ comes simply
from the fact that at such a value of I
(between I $\approx$ 60 and I $\approx$ 70) the nucleus elongates
substantially
increasing in this way its deformation and at the same time its
MI.

It is therefore interesting to notice that the semiclassical ETF
approach leads to a moment of inertia that is very well approximated
by its TF, i.e., the classical rigid-body value. Thouless--Valatin terms which
arise from the self-consistent response of the mean field to the time-odd
part of the density matrix generated by the cranking piece of the
Hamiltonian are naturally taken care of in this approach. Semiclassical
corrections of order $\hbar^2$ coming from the orbital motion and the spin
degree of freedom are not small individually, but compensate each other
to a large extent.

One has, however, to keep in mind that shell and pairing effects,
      that go beyond the ETF approach, are not included 
in this description. These effects are not only both present, but influence
each other to a large extent, especially for the collective high-spin
rotations of strongly deformed nuclei 
\cite{belyaevhighspin,pomorbartelPRC2011,sfraurev}.

\subsection{SHELL-STRUCTURE MOMENT OF INERTIA}
\l{semshell}

We apply the EGA POT (Section II)
for the derivation of the MI through the rigid-body MI
(with the shell corrections)
in the NLA (Section IVA)
related to the collective statistically equilibrium rotation
with a given frequency $\omega$ \cite{mskbPRC2010}. For simplicity,
we shall discard the spin and isospin degrees of freedom, in particular,
the spin-orbit and asymmetry interaction.
Notice also that from the results presented in
Figs.\ \ref{fig29} and \ref{fig31}  
(with the help of
Fig.\ \ref{fig30}), 
 one may conclude that the main contribution to
the moment of inertia
of the strongly deformed heavy nuclei can be found within the
ETF approach to the rotational problems as a smooth
 rigid-body MI. However, sometimes the MI shell corrections play the
dominating role as traps in the yrast line for the deformed nuclei at high spins,
like in $^{60}$Dy \cite{strutdy152}, recall also the asymmetry of 
nuclear fission fragments because of the shell effects.

\medskip

\subsubsection{
Green's function trajectory expansion for the MI}
\label{greenfun}

For the derivations of shell effects \cite{strut} within the POT
\cite{sclbook,gutz,strumag,bt76,creagh,migdalrev,MKApanSOL1},
it turns out to be helpful
to use the coordinate representation of the MI
through the Green's functions $G\left(\r_1,\r_2;\vareps \right)$
as for the other transport coefficients (Section IV)
\cite{mskbg,mskbPRC2010,magvvhof,MGFyaf2007,GMSnpae2007}.
In the coordinate representation, 
the MI $\Theta_x$ as a susceptibility  
(\ref{thetaxdef}) [or (\ref{response})] which is similar to 
the response function (\ref{chiGGdef}) for
the collective 
vibrations, can be expressed in terms of the s.p.\  
Green's function $G$ by using its spectral
representation (\ref{GFdef}) through the Inglis formula 
(\ref{inglismi}),
\bea\l{micoorrepom} 
&& \Theta_{x}(\omega) = \frac{d_s}{\pi} \int^{\infty}_0 {\rm d}\vareps\;
n({\vareps}) \nonumber\\
&\times& \Re \int {\rm d} \r_1 \int {\rm d} \r_2 \; 
\ell_{x}(\r_1)\; \ell_{x}(\r_2) \nonumber\\ 
&&\times 
  \left[G\left(\r_1,\r_2;\vareps -\hbar \omega \right)
+ \overline{G}\left(\r_1,\r_2;\vareps +\hbar \omega \right) \right]
\nonumber\\ 
&&\times \Im \left[G\left(\r_1,\r_2;\vareps\right)\right]\;.   
\eea
The angular momentum operator $\ell_x$ in (\ref{micoorrepom})
takes a similar role of the s.p.\ operator named $\hat{F}$ (or $\hat{Q}$) 
in Section IV 
\cite{magvvhof,GMFprc2007,MGFyaf2007,belyaev}.
For the adiabatic rotations one can neglect here the 
$\omega$-dependence, 
\bea\label{micoorrep}
\Theta_{x}&=&\frac{2 d_s}{\pi}\int^{\infty}_0 {\rm d}\vareps\;
n({\vareps})\nonumber\\ 
&\times&\int {\rm d} \r_1 \int {\rm d} \r_2 \; 
\ell_{x}(\r_1)\; \ell_{x}(\r_2) \; \nonumber\\
&\times&  \Re \left[G\left(\r_1,\r_2;\vareps \right) \right]\;
\Im \left[G\left(\r_1,\r_2;\vareps\right)\right]\;.    
\eea
This representation is useful within the semiclassical POT, 
weakening the criterium 
of the quantum perturbation approximation: The maximal 
rotational-excitation
energy $\hbar \omega$ for which the approximation is valid 
becomes then significantly larger 
than the nearest neighbor s.p.-level spacing around 
the Fermi surface $\vareps^{}_F$, but still somewhat smaller than the energy 
distance between major shells $\hbar \Omega$ ($\hbar \Omega \!\approx\! 
\vareps^{}_F/A^{1/3}$) as shown by Migdal \cite{Migdal75}. 
This is in contrast to the cranking formula (\ref{inglismi}) derived
with the help of the 
quantum perturbation criterion
of smallness of the excitation energies with respect to
this s.p.\ level spacing. The latter restriction appears
probably because of using
the spectral representation of the Green's
function $G$ in (\ref{micoorrepom}) and (\ref{micoorrep}). Therefore,
one can assume that this restriction is weakened through 
the coordinate
representation of $G$, valid even for a quasi-continuous spectra of 
the semiclassical approximation.
${}$

\subsubsection{Semiclassical Green's function and particle density}

For the s.p.\ Green's function $G\left(\r_{1},\r_{2};\vareps\right)$
in (\ref{micoorrepom}) and (\ref{micoorrep}), we shall use the  
semiclassical Gutzwiller trajectory expansion 
(\ref{Gsem}), (\ref{GsemGa}) and (\ref{ampiso}) as in Section IV \cite{gutz}.
The sum runs over the CT with particle energy 
$\vareps$ from a point $\r^{}_1$ to a point $\r^{}_2$ 
(see Fig.\ \ref{fig1}). 
There are several reasons leading to oscillations of the MI 
obtained from (\ref{micoorrep}), some local  
($\r^{}_2$ is close to $\r^{}_1$) and 
some nonlocal (they are different), where  
the local PO part is related to the 
shell effects, whereas nonlocal (non-PO) contributions 
could have their origin in 
reflections of the particle from the boundary. The NLA
($|\s^{}_{12}| \!=\! | \r^{}_2 \!-\! \r^{}_1 | \!\ll\! R$) 
is valid after a statistical averaging over many microscopic 
quantum states. 
Then, the maximal possible value of the parameter, which is the 
product of the two dimensionless
quantities $\;\om/\Omega$ and $\;S/\hbar$, used in \cite{dfcpuprc2004} 
as a {\it small parameter} of the perturbation approach of Creagh to the 
classical dynamics \cite{sclbook}, in our EGA
can be somewhat larger. This implies that  
$(\om/\Omega) \; (S/\hbar) \simg 1$, under the usual 
semiclassical condition  $S/\hbar \sim k^{}_F R\sim A^{1/3} \gg 1$; 
where $k^{}_F=p^{}_F/\hbar$ is the Fermi momentum in $\hbar$ units, and 
$R$ the mean nuclear radius.
According to (\ref{Gsem}),  one  
can split also the Green's function $G(\r^{}_{1},\r_{2};\vareps)$ 
into a contribution $G^{}_{{\rm CT}_0}$ (\ref{GCT0}) coming from the 
direct path between 
the two points, $\r_1$ and $\r_2$,  and a contribution $G^{}_{{\rm CT}_1}$ that contains the 
contributions from all other trajectories involving reflections 
(\ref{Gsem}) (Fig.\ \ref{fig1}).
For the component $G^{}_{{\rm CT}_0}$ 
related to the trajectory
CT$_0$ for which the action $S^{}_{{\rm CT}_0}$ 
disappears in the limit $\r_2 \to \r_1$, one finds again (\ref{GCT0})
with using $\s=\s_{1,2}$,
particle momentum $\,\, p(\r) \!=\! \sqrt{2m[\vareps - V(\r)]\,}$,
and potential $V(\r)$, where $\;\r=(\r_1+\r_2)/2$.
According to (\ref{micoorrep}), for a semiclassical 
statistical-equilibrium rotation with  
constant frequency $\om$,
one approximately 
obtains \cite{mskbPRC2010,belyaev}  
\bea\l{misplit}
\Theta \approx \Theta_{\rm GRB} &=& 
m \int \d \r \; r_\perp^2 \, \rho_{\rm scl}(\r)\nonumber\\ 
&=& \Theta_{\rm ETF} + \delta \Theta_{\rm scl}\;, 
\eea
where $ \Theta_{\rm GRB}=\Theta_{\rm ETF}^{\rm GRB} + 
\delta \Theta_{\rm scl}^{\rm GRB}$ is 
the generalized rigid-body (GRB) MI
with (Section VB and \cite{bartelnp,belyaev})
\be\l{dthetaETF}
\Theta_{\rm ETF} \approx \Theta_{\rm ETF}^{\rm GRB}=
m \int {\rm d}\r\; r_\perp^2 \, \rho^{}_{ETF}\left(\r\right)\;,
\ee
and $ \delta \Theta_{\rm scl} $ 
its shell correction \cite{mskbPRC2010,belyaev},
\be\l{dtsclsplit}
\delta \Theta_{\rm scl} \approx \delta \Theta_{\rm scl}^{\rm GRB}=
m \int {\rm d}\r \; r_\perp^2 \, \delta \rho_{\rm scl}(\r)
\ee
with $ r_\perp^2 = y^2 + z^2 $. 
Such a splitting (\ref{misplit}) is associated with 
that of the spatial particle density
$\rho(\r)$, 
\be\l{partdensplit} 
\hspace{-0.4cm}\rho(\r) \!=\! 
 - \frac{1}{\pi}\Im \int {\rm d} \vareps n\left(\vareps\right)
\!\left[G\left(\r_1,\r_2;\vareps\right)\right]_{\r_1=\r_2=\r}.
\ee
Substituting (\ref{Gsem}) into (\ref{partdensplit}), one obtains
$\rho(\r)$  
in terms of its ETF particle density
$\rho^{}_{\rm ETF}$ and its shell
correction $\delta \rho_{\rm scl}(\r)$,
\be\l{rhoscl}
\rho^{}_{\rm scl}\left(\r\right)=
\rho^{}_{\rm ETF} + \delta \rho_{\rm scl}(\r)\;,
\ee
where 
\bea\l{rhotf}
\rho^{}_{\rm ETF}\left(\r\right) &=&
- \frac{1}{\pi}\,\Im  \int  {\rm d} \vareps 
\widetilde{n}\left(\vareps\right)\nonumber\\
&\times&
\left[G_{{\rm CT}_0}\left(\r_1,\r_2;\vareps\right)\right]^{}_{\r_1=\r_2=\r}
\eea
and 
\bea\l{drhoscl}
\delta \rho_{\rm scl}\left(\r\right) &=&
-\frac{1}{\pi} \Im  \int \d \vareps \delta n\left(\vareps\right) \nonumber\\
 &\times&\left[G^{}_{{\rm CT}_1} 
\left(\r_1,\r_2;\vareps\right)\right]^{}_{\r_1=\r_2=\r}\;.
\eea
The standard decomposition of the occupation numbers 
\begin{equation}\label{ndecomp}
n=\widetilde{n}+\delta n
\end{equation}
into the smooth and fluctuating parts is used as usually in the SCM 
\cite{fuhi}.
The crossing terms coming from the substitution of (\ref{ndecomp})
and (\ref{Gsem}) into (\ref{partdensplit}) almost do not contribute after
the phase space integration by the SPM.

\subsubsection{MI phase space trace formulas}

Substituting (\ref{partdensplit}) with (\ref{Gsem}) into 
(\ref{misplit}), for the total semiclassical MI $\Theta_x$, one obtains
the phase-space trace formula \cite{GMBBps2015,GMBBprc2016}:
\bea\l{misclps}
 \Theta_{\rm scl} &\approx& d_s \, m \!\int\! \d \vareps\; \vareps\; n(\vareps)
\nonumber\\
&\times&\int \frac{{\rm d} \r_2 {\rm d} \p_1}{(2 \pi \hbar)^3}\;
\frac{r_{2\perp}^2}{\vareps} \, g_{\rm scl}(\r_2,\p_1;\vareps)\; \nonumber\\
&=& \Theta_{\rm ETF} +\delta \Theta_{\rm scl}\;,
\eea
where 
\begin{eqnarray}\label{fscl}
&&g_{\rm scl}(\r_2,\p_1;\vareps) = \frac{\partial f(\r,\p)}{\partial \vareps}
\nonumber\\
&=&G_{\rm scl}(\r_2,\p_1;\vareps)\; 
\exp\left[i \, \p_1\left(\r_1\!-\!\r_2\right)/\hbar\right]\;.
\end{eqnarray}
The usual Wigner distribution function $f(\r,\p)$ in the phase space
$\r,\p$, and now $r_{2\perp}^2=y_2^2+z_2^2$, were introduced 
(to simplify the notations
the subscript 2 in $r_{2\perp}^2$ will be omitted in the following). Here 
$G_{\rm scl}(\r_2,\p_1;\vareps)$ is a 
semiclassical Green's function
in the mixed phase-space representation obtained by the
Fourier transformation of
  $G_{\rm scl}(\r_1,\r_2;\vareps)$ 
(\ref{Gsem}), 
\begin{eqnarray}\label{psgfun} 
&\!G_{\rm scl}\!(\r_2,\p_1;\vareps)\!=\! 
\Re \!\sum_{\rm CT}
  \!\Big|\!J_{\rm CT}\!\left(\p^{}_{1\perp},\p^{}_{2\perp}\!\right)\!\Big|^{1/2}
 \! \delta\left[\vareps\!-\!H(\r_2,\p_2)\right]\!\!\nonumber\\
&\times
\exp\!\left[\frac{i}{\hbar} 
S_{\rm CT}\left(\r_1,\r_2;\vareps\right) \!-\!
\frac{i\pi}{2}\mu^{}_{\rm CT}\right], 
\end{eqnarray}
where $J_{\rm CT}\left(\p^{}_{1\perp},\p^{}_{2\perp}\right)$ is the 
Jacobian for the transformation from the component $\p^{}_{1\perp}$ of the 
momentum $\p_1$ that is perpendicular to the reference CT
to the perpendicular component $\p_{2\perp}$ of the momentum $\p_2$. 
We formally inserted the additional integral over $\r_1$ with  
$\delta(\r_2-\r_1)$ into the middle of (\ref{misplit}) 
and transformed the spatial coordinates $\r_1$ and $\r_2$ to the 
phase space variables $\r_2$ and $\p_1$ \cite{MAFptp2006,belyaev}. 
Using then the
Fourier transformation of this $\delta$ function of the coordinate difference
$\r_2-\r_1$ to a new momentum ${\tilde \p}$ and integrating, by the stationary 
phase method, the MI in such a phase space representation over the component 
of ${\tilde \p}$, perpendicular to the classical trajectories, one arrives at 
(\ref{misclps}) \cite{MKApanSOL1}. 
Note that under the perfect local approach of the NLA ($\r_1 \rightarrow \r_2 
\rightarrow \r)$ and $\p_1 \rightarrow \p_2 \rightarrow \p$
the ETF CT$_0$ component (\ref{GCT0}) of the Green's function 
(\ref{Gsem}) is related to an energy density of the 
TF distribution function (\ref{fscl})
\be\l{distrfuntf}
g_{\rm scl}(\r,\p;\vareps)  
\rightarrow g^{}_{\rm TF}(\r,\p;\vareps) = \delta(\vareps-H(\r,\p)).
\ee

\subsubsection{ETF distribution function density}

As shown in Appendix C, 
the ETF distribution density 
component, $g^{}_{\rm ETF}(\r,\p;\vareps)$, can be derived within
the ETF approach by using the inverse Laplace transformation (\ref{laplac}), 
\be\l{getf}
g^{}_{\rm ETF}(\r,\p;\vareps)=g^{}_{\rm TF} + g^{}_{\rm S}(\r,\p;\vareps)\;,
\ee
where $g^{}_{\rm TF}$ is given by (\ref{distrfuntf}).
Taking also into account the same phase-space integration, one can simplify
more this expression excluding formally the terms which do not contribute
because of the integral over momentum $\p$.
For the ETF surface correction $g^{}_{\rm S}$, one obtains
\bea\l{fetf1}
&&g_{\rm S}(\r,\p;\vareps)=  
\hbar^2
\left\{\left(-\frac{\nabla^2V}{4m}\right)\right.\nonumber\\
&\times& \left.\frac{
\partial^2\delta\left(\vareps-H(\r,\p)\right)}{\partial \vareps^2} 
\right.\nonumber\\
&+&\left.
\left[\frac{\left(\nabla V\right)^2}{6m} +
\frac{p^2 \nabla^2V}{18 m^2}\right]\;
\frac{
\partial^3\delta\left(\vareps-H(\r,\p)\right)}{\partial \vareps^3}\;
\right.\nonumber\\
&-&\left.\frac{p^2 \left(\nabla V\right)^2}{24m^2}\;
\frac{
\partial^4\delta\left(\vareps-H(\r,\p)\right)}{\partial \vareps^4}
\right\}\;.
\eea
The gradients  
of the potential, $(\nabla V)^2$ and $\nabla^2 V$, can be expressed in terms of
the those of the TF particle density within the same $\hbar^2$ precision,
,%
\be\l{tfden}
\rho^{}_{\rm TF}=d_s \frac{p_\lambda^3(\r)}{6 \pi^2 \hbar^3}\;,
\ee
where
\be\l{pf}
p_{\lambda}(\r)=\sqrt{2m \left[\lambda-V(\r)\right]}\;.
\ee
Differentiating (\ref{tfden}) and solving the obtained linear
system of equations with respect to the potential gradient terms, 
one results in (\ref{dv2}) and (\ref{d2v}).
These expressions are more convenient to use in a more general case
including  billiard systems as spheroidal cavity.

\subsubsection{ETF and shell structure energies}

Equation (\ref{misclps}) strongly resemblances 
(except for the additional factor 
$m r_{\perp}^2/\vareps$) with the expression for the semiclassical s.p.\ energy 
\bea\l{escl}  
 E_{\rm scl} &=& d_s \int \d \vareps\; \vareps\; 
n(\vareps)\; g_{\rm scl}(\vareps)\nonumber\\
 &=& d_s \!\int\! \d \vareps\; \vareps\;n(\vareps) 
\int \frac{\d \r_2 \d \p_1}{(2 \pi \hbar)^3}
\nonumber\\
&\times& g_{\rm scl}(\r_2,\p_1;\vareps) 
   \approx  E_{\rm ETF} + \delta E_{\rm scl}\;,
\eea
where $E_{\rm ETF}$ is the ETF energy and $\delta E_{\rm scl}$ the energy 
shell correction. We also used 
the phase-space trace formula for the semiclassical level density 
$g_{\rm scl}(\vareps)$ \cite{gutz,strut,strumag,sclbook,MAFptp2006,migdalrev}
with a similar decomposition,
\bea\l{gscldef}
  g_{\rm scl}(\vareps) &=& \int \frac{\d \r_2 {\rm d} \p_1}{(2 \pi \hbar)^3}\; 
      g_{\rm scl}(\r_2,\p_1;\vareps) \nonumber\\
&\approx& g^{}_{\rm ETF}(\vareps) + \delta g_{\rm scl}(\vareps) \;.
\eea

Substituting (\ref{dv2}) and (\ref{d2v}) into  
 (\ref{fetf1}) for the ETF surface-distribution density, and then, 
to (\ref{escl}) for
the ETF energy, one obtains 
\be\l{etf}
E_{\rm ETF}=E_{\rm TF}+E_{\rm S}\;, 
\ee
where
\be\l{etfdens}
E_{\rm S}=d^{}_s \int \d \vareps\; \vareps\; n(\eps) 
\int \frac{\d \r \d \p}{(2 \pi \hbar)^3}\; g^{}_{\rm S}\approx
\sigma \mathcal{S}
\ee
with
\be\l{sigma0}
\sigma=\frac{\hbar^2}{72 m } \int_{-\infty}^{\infty} \frac{\d \xi}{\rho}
\left(\frac{\partial \rho}{\partial \xi}\right)^2\;.
\ee
We used here the local orthogonal-coordinate
system $\xi,~~\r_{\parallel}$
with the $\xi$ axis perpendicular to the nuclear surface, and the 
two other $\r_{\parallel}$ coordinates tangent to the surface (Section III)
\cite{strtyap,strmagden}.
Integrating first in 
(\ref{etfdens}) over the energy $\vareps$  by parts and 
using the properties of the
$\delta$ functions, we take the integral over $p$ applying the relation
$p=\sqrt{2m (\vareps -V(\r))}$ for $\vareps=\lambda$ 
which then appears from these properties. 
For the last integration, one transforms 
the remaining  integral over the spatial coordinates $\r$ to the local
coordinate system $\d \r=\d \xi \d \r_{\parallel}=\d \xi \d S$. Within 
the ES (leptodermous) approximation 
\cite{strtyap,strmagden,BMRV,BMRps2015},
where $a/R \approx A^{-1/3}\ll 1$ is  a small parameter for a large deformed
nucleus,  at the leading order one can keep the main highest-order
derivatives of the particle density $\rho$. After simple algebraic
transformations, one finally arrives at (\ref{etfdens}) where $\sigma $
is the tension coefficient. Notice that the same result (\ref{etfdens}) 
can be obtained directly
from the surface ($\hbar^2$) of the ETF kinetic energy density functional
(\ref{tau2}) \cite{sclbook},
\be\l{tau2}
\tau^{}_{\rm S}=\frac{\hbar^2}{2m}\left[
\frac{\left(\nabla \rho\right)^2}{36 \rho}
+\frac{\nabla^2\rho}{3}\right]\;,
\ee
by using the same technics of the ES approximation 
(Section III and \cite{strtyap,strmagden,BMRV,BMRps2015}). 
Note that we used also the Gaussian theorem
for the integration of the Laplacians over the infinite 
spatial-coordinate 
volume (the densities with all their derivatives tend to zero in the 
limit far infinitely from the nucleus). This result should be expected
for the IP model for arbitrarily deformed 
nuclear ES. As shown in Section III, taking into account the interaction 
depending on the particle density as in the Skyrme forces, one may
get a more realistic expression for the 
tension coefficient $\sigma$ than given in (\ref{sigma0})
[see (\ref{bsplus})
for the surface energy constant, 
$b^{}_{\rm S}=b^{(+)}_{\rm S}4 \pi r_0^2 \sigma$].
As for perspective, the extension of these calculations to account for 
the spin-orbit interaction and 
neutron-proton asymmetry can be done strictly following 
Section III and 
\cite{BMRV,BMRps2015,BMRprc2015}.

\subsubsection{Surface terms in the MI shell correction}

Multiplying and dividing (\ref{misclps}) identically by the energy 
$E_{\rm scl}$ (\ref{escl}), one finally arrives at
\be\l{MIE} 
\Theta_{\rm scl} \approx  
          m \, \langle \frac{r_{\perp} ^2}{\vareps} \rangle \, E_{\rm scl}\;,
\ee
where brackets mean the average over the phase space variables
$\r_2$, $\p_1$ and energy $\vareps$ with a weight $\vareps$, i.e.\ ,
\be\l{psav}
\langle \frac{r_{\perp} ^2}{\vareps}\rangle =
\frac{\int {\rm d} \vareps\; \vareps\; 
\int {\rm d} \r {\rm d} \p \;(r_{\perp} ^2/\vareps)\; 
g_{\rm scl}(\r,\p;\vareps)}{
\int \d \vareps\; \vareps 
\int \d \r \d \p\; g_{\rm scl}(\r,\p;\vareps)} \; .
\ee
Using now the same subdivision in terms of the ETF and shell components
for the MI (\ref{misclps}) and the s.p.\ energy (\ref{escl}),  
after a statistical averaging for finite temperatures $T$ one 
obtains
\be\l{dmiF}
  \delta \Theta_{\rm scl} \approx  
           m \, \langle r_{\perp} ^2 /\vareps \rangle\;\; \delta \mathcal{F},
\ee
where $\delta \mathcal{F}$ is the free-energy shell correction
(Section IIC1 and IVB1).
Notice that in the spirit of the SCM   
\cite{strut,fuhi}, equation (\ref{MIE}) is used here only for
deriving the relation (\ref{dmiF}) between the leading-order
shell corrections of the MI and the ones of the free-energy, 
which are always due to the inhomogeneity of the s.p.\ (quasiparticle) 
spectrum near the Fermi surface, see Section IVB1.
This does not mean that the total MI $\Theta_{\rm scl}$ can be 
associated with the s.p.\ nature. In line with the standard SCM, one can, 
however, use a renormalization procedure, replacing a smooth both 
s.p.\ energy 
and MI component by the corresponding macroscopic statistically averaged 
quantities, in particular through the ETF method (Section VB). 
In these derivations we used also
the improved stationary phase (periodic orbit) conditions for the evaluation 
of integrals over the phase space variables $\r_2$ and $\p_1$ \cite{MAFptp2006}.
Within the POT, at a given temperature $T$, the PO sum 
(\ref{dfdcpot}), 
presented for the semiclassical 
free-energy shell correction $\delta {\cal F}_{\rm scl}$ 
in 
 billiard potentials, takes  a more general form (\ref{dfscl}) 
for any potential wells
\cite{strut,strumag,kolmagstr,fraukolmagsan,mskbPRC2010,belyaev}. 
The PO components
of the oscillating (free) energy and level-density shell corrections
are taken all 
at the chemical potential $\vareps =\lambda$ for $\om=0$, 
which, at zero temperature, is equal
to the Fermi energy $\vareps^{}_F$. 
see more specific expressions of $\delta g^{}_{\rm PO}$, 
for instance (\ref{dgPO}), in terms of the PO classical degeneracy,
the stability factors, and the action along the PO in Section II 
\cite{gutz,strut,strumag,sclbook,belyaev,migdalrev,MKApanSOL1}. 
In (\ref{gscldef}), $g^{}_{\rm ETF}(\vareps)$ is the smooth ETF component, and 
$ \delta g_{\rm scl}(\vareps)$ the semiclassical oscillating contribution  
\cite{sclbook} where the latter can be expressed in terms of the PO sum, 
as shown in Section II.
POs appear through the stationary phase condition 
(which, in the present context, is equivalent to the PO condition, 
see Section IIB and   
\cite{MAFptp2006})
for the calculation of the integrals over $\r_2$ and 
$\p_1$ in (\ref{gscldef}) by the ISPM
\cite{spheroidptp,MAFptp2006,migdalrev,MKApanSOL1}.
For the phase-space and energy average 
$\langle r_{\perp}^2 / \vareps \rangle$ (\ref{psav}),
one again obtains approximately a decomposition into an ETF and a 
shell-correction contributions, through the distribution function 
$g_{\rm scl}(\r_2,\p_1;\vareps)$ (\ref{fscl}) 
and the Green's function (\ref{psgfun})
with the help of the decomposition (\ref{Gsem}),
\begin{equation}\label{ell2}
  \langle r_{\perp}^2 / \vareps \rangle 
  \approx \langle r_{\perp}^2 / \vareps \rangle^{}_{\rm ETF} + 
  \delta \langle r_{\perp}^2 / \vareps \rangle .
\end{equation}

For this coefficient within the ETF
in the ES approximation, one obtains
\begin{equation}\label{ell2ms0}
  \langle \frac{r_{\perp}^2 }{ \vareps }\rangle_{\rm ETF}
  \approx  \frac{\Theta_{\rm TF} + \Theta_{\rm S}}{E_{\rm TF} +E_{\rm S}}\;,
\end{equation}
where $E_{\rm TF}$ and $\Theta_{\rm TF}$ are the TF components \cite{sclbook} while
$E_{\rm S}$ and $\Theta_{\rm S}$ are the ETF surface energy corrections  given
by (\ref{etfdens}) .  
For spheroid cavity,
one obtains the explicit expressions:
\be\l{entf}
E_{\rm TF}=\frac{d_s p^{5}_F}{15 \pi \hbar^3m}\;a^2 b\;,
\ee
\bea\l{mtf}
\Theta_{\rm TF}&=&\frac{d_s p^{3}_F}{12 \pi^2 \hbar^3 m}\;\int \d \r \;(y^2+z^2)
\nonumber\\
&=&\frac{2d_s p^{3}_F}{45 \pi \hbar^3}\;\pi a^4 b\;
\left(1+\eta^2\right)\;,
\eea
where $a$ and $b$ are the semi-axes of spheroidal cavity,
with $a^2b=R^3$, and $R$ being the radius of the equivalent sphere,
$\eta =b/a$ is the deformation parameter. (This parameter should be not
confused with $\eta$ which denotes other quantities in Section IV and VB)
For the ETF surface corrections $\Theta_{\rm S}$
in the case of the spheroidal
cavity, one explicitly obtains
\be\l{Msfin}
\Theta_{\rm S}= \frac{4 a^4\sigma}{\lambda}
\left[\eta^2 I_0 + \frac{\pi}{4}\left(1-2\eta^2\right)\;I_1\right]\;,
\ee
where
\bea\label{I0}
I_0&=&\int_0^1\d \zeta \;\sqrt{\frac{
1 + \zeta(\eta^2-1)}{1-\zeta}}\;\nonumber\\
&=&1+\frac{
\left[1+(\eta^2-1)\right] \arctan\sqrt{\eta^2-1}}{\sqrt{\eta^2-1}}\;,
\eea
\bea\l{I1}
I_1&=&\int_0^1 \zeta \d \zeta\; \sqrt{\frac{1 + \zeta(\eta^2-1)}{1-\zeta}}
\nonumber\\
&=&\frac{2 \eta}{3 \sqrt{\eta^2-1}}\left\{\left[1+2(\eta^2-1)\right]
E\left(\frac{\sqrt{\eta^2-1}}{\eta}\right)\right. \nonumber\\
&-&\left.
K\left(\frac{\sqrt{\eta^2-1}}{\eta}\right)\right\}\;,
\eea
$E(k)$ and $K(k)$ are the standard complete elliptic integrals \
\cite{byrdbook}. Thus,
for the spheroid cavity, from (\ref{ell2ms0}) with 
(\ref{entf}), (\ref{mtf}), (\ref{Msfin}) and (\ref{etfdens})
one finally arrives at
\begin{equation}\label{ell2ms}
  \langle \frac{r_{\perp}^2 }{ \vareps }\rangle_{\rm ETF}
\approx \frac{a^2+b^2}{3 \lambda}\;
\frac{1+\Theta_{\rm S}/\Theta_{\rm TF}}{1+E_{\rm S}/E_{\rm TF}}\;,
\end{equation}
\be\l{Mstf}
\frac{\Theta_{\rm S}}{\Theta_{\rm TF}}=\frac{
45b^{}_{\rm S}\left[\eta^2I_0+\pi(1-2\eta^2)I_1\right]}{
4\pi \eta^{2/3}(1+\eta^2)\lambda (k^{}_Fr^{}_0)^3\;A^{1/3}}\;,
\ee
\be\l{Estf}
\frac{E_{\rm S}}{E_{\rm TF}}=\frac{
15b^{}_{\rm S}\overline{\mathcal{S}}}{
16 \eta^{2/3}\lambda (k^{}_Fr^{}_0)^3\;A^{1/3}}\;,
\ee
$b^{}_{\rm S}=4\pi r_0^2\sigma$,
$\overline{\mathcal{S}}$ is the spheroid surface area
$\mathcal{S}$ in units of $a^2$, $\mathcal{S}=a^2 \overline{\mathcal{S}} $,
\be\l{spheroidS}
\overline{\mathcal{S}}=2\pi\left(1+\frac{\eta^2}{\sqrt{\eta^2-1}}\right)\;
\arcsin\left(\frac{\sqrt{\eta^2-1}}{\eta}\right)\;,
\ee
$\sigma $ is the surface tension coefficient (\ref{sigma}) \cite{strmagden,magsangzh}.
Using units of the classical rigid-body
(TF) MI, 
\begin{equation}\l{}
\Theta_{\rm TF}=m \left(a^2+b^2\right) A/5\;,
\end{equation}
one  results in 
\be\label{dtxtxtf}
\frac{\delta \Theta_x}{\Theta_{\rm TF}}=
\frac{5\left(1+\Theta_{\rm S}/\Theta_{\rm TF}\right)}{1+ E_{\rm S}/E_{\rm TF}
}\;\frac{\delta \mathcal{F}}{3 A\lambda}\;.
\ee

\subsubsection{Comparison with quantum calculations}

Figs.\ \ref{fig32} and \ref{fig33} 
 show a good comparison between the 
semiclassical ISPM 
MI shell corrections (\ref{dmiF}) obtained 
with (index $+$)  
surface terms and the
quantum-mechanical (QM) result.
In the zero-temperature limit, the  
shell-correction free energy
$\delta \mathcal{F}$ becomes obviously identical to the 
shell-correction  energy
$\delta E$, and according to (\ref{dfscl}) 
\be\l{dmiE}  
  \delta \Theta 
\approx \langle \frac{r^{2}_{\perp}}{\varepsilon}\rangle \, \delta E \;.
\ee
The QM approach is determined 
through the ETF average (\ref{ell2ms0})  
for 
$\langle r_{\perp}^2/\eps \rangle$ with a realistic 
surface-energy constant
$b^{}_{\rm S}\approx 20 $~MeV (Section III)
whereas the energy shell correction $\delta E$   
(equal $\delta F$ at zero
 temperature $T$) is calculated 
by the  SCM 
using the quantum spectrum. 
The relationship (\ref{partconspert}) 
between the chemical potential $\lambda$ and the particle number $A$
in the nucleus is used 
in the semiclassical calculations, e.g.\ , of $\delta \mathcal{F}$ 
(\ref{dfscl}),
with an averaging parameter which is much smaller than the distance between
major shells $\hbar \Om$ for a sake of convenience. 
A large supershell
effect appears in 
$\delta \Theta_x$, especially for larger 
deformations
 in the PO bifurcation region (Fig.\ \ref{fig33}).

The effect of surface 
corrections, (\ref{Mstf}) and (\ref{Estf}), 
is analyzed in
   Figs.\ \ref{fig34} 
and \ref{fig35} 
that show, together with the 
result of the quantum
  calculation, the shell components $\delta \Theta_x/\Theta_{\rm TF}$
   obtained with (ISPM$_{+}$) and without (ISPM$_{-}$) these surface
   corrections. The difference between both curves is seen to be more
   important for small particle numbers, which can be easily understood
   since the surface corrections decrease as $A^{-1/3}$ as seen from 
   (\ref{Mstf}) and (\ref{Estf}).
The contribution of the shorter three-dimensional orbits bifurcated from the 
equatorial ones are dominating in the case of large deformations 
(Fig.\ \ref{fig33}), 
in contrast to the small deformation region where
the meridian orbits are predominant (Fig.\ \ref{fig32}),  
in accordance with \cite{migdalrev,spheroidptp}.
One also observes that
the surface corrections become more significant  
with increasing
deformation of the system. 

 For small temperatures one has 
$\delta F_{\rm scl}\!\approx\!\delta E_{\rm scl}$, 
and therefore, a remarkable  interference of the 
dominant short three-dimensional and meridian 
 orbits is  shown in  
\cite{migdalrev,spheroidptp,GMBBps2015,GMBBprc2016}. 
Their bifurcations in the superdeformed region
give essential contributions to the MI through the
      (free) energy shell corrections.

For heated Fermi systems, we calculate the  
quantum-mechanical and the semiclassical  
shell-correction free energy (Fig.\ \ref{fig36})
for the spheroidal cavity at finite temperatures and 
deformations as a function of the particle number variable $A^{1/3}$  
for temperature $T=1$ MeV as compared to the cold case ($T=0$). 
For simplicity, we neglect in this comparison  the relatively small
 surface corrections of (\ref{ell2ms0})
in the MI shell components (\ref{dmiF}).
Minima of $\delta \mathcal{F}$ are related to magic particle numbers 
at a finite temperature. The
factor $\langle r^{2}_{\perp}/\varepsilon \rangle$, 
(\ref{ell2ms0}), appearing in (\ref{dmiF})  
in particular in (\ref{dmiE}) 
can be simply evaluated  within the cranking model 
(which is, as already pointed 
out in the introduction, semiclassical in nature) by 
using the simplest TF estimate  
for the distribution function (\ref{distrfuntf}) (Sections VC3 and VC4).
Thus, neglecting $\hbar$ corrections 
due both to surface [in (\ref{ell2ms0})] and
curvature terms \cite{GMBBprc2016},
one can also disregard shell effects in this average factor (\ref{ell2})
of (\ref{dmiF}). 
Note that the CT$^{}_0$ component of the 
Green's function (\ref{Gsem}) 
corresponds mainly to the TF approach through 
the leading term of the 
nearly local approximation, i.e.\ , the local approximation.
For the spheroidal cavity potential, one then obtains from 
(\ref{ell2ms0})
\be\l{Eq.33} 
  \langle \frac{r^{2}_{\perp}}{\varepsilon} \rangle
    \approx \langle \frac{r^{2}_{\perp}}{\varepsilon} \rangle^{}_{\rm TF}
                     = \frac{a^2 + b^2}{3 \lambda} \; .
\ee
Using this more simple estimate, one may  evaluate the 
shell-correction MI for
different temperatures $T$, and deformations $\eta$ defined as
$\eta = b/a~$  (Fig.\ \ref{fig37}). In that case of 
small deformations ($\eta=1.2$), 
there is almost no contribution from PO 
bifurcations, whereas for large deformations a  
significant contribution of PO 
bifurcations is observed. 
In addition, Fig.\ \ref{fig37} shows a big 
weakness of this effect with increasing temperature, especially 
for larger both deformations and particle numbers.
As seen from Fig.\ \ref{fig38}, for 
small temperatures 
the orbits, which give the dominating contribution into the 
shell structure, at large deformations are 
the bifurcating POs: 
The shortest (four) three-dimensional (3D) POs 
which appear from the corresponding parent equatorial (EQ) orbits 
at $\eta \! \approx\! 1.6\! \div\! 2.0$; and the 
shortest meridional (two elliptic
and one hyperbolic) POs emerging
at smaller 
deformations ($\eta=1$ and $\sqrt{2}$) \cite{spheroidptp,migdalrev}. 
Therefore, the ISPM amplitudes of
oscillations of the level density from the bifurcating POs
are enhanced as compared to other POs far from the bifurcation
(or symmetry-breaking) deformations. 
Both kinds of PO families mentioned above 
yield the essential contributions 
through 
$\delta \mathcal{F}$ 
at zero temperature (i.e.\ , $\delta E$). 
At finite temperatures, the main contribution to the MI 
shell structure is due to the shortest EQ orbits 
because of the exponential temperature-dependent factor in (\ref{dfscl})
for the 
shell-correction free energy $\delta \mathcal{F}$ (see also Fig.\ \ref{fig36}).
In addition,  the factor 
$1/t_{\rm PO}^2$ in (\ref{dfscl}) 
with the time period 
$t^{}_{\rm PO}$ of the particle motion along the PO enhances
shorter EQ POs at a finite temperature, too. All these properties 
differ significantly from the classical perturbation results of [20] 
where the EQ orbits do not 
contribute and 3D PO contributions are not considered.

Figs.\ \ref{fig38} and \ref{fig39} show the temperature 
dependence of the MI 
shell corrections. With increasing temperature $T$, one observes 
\cite{strumag,kolmagstr,magkolstrizv1979,richter,fraukolmagsan,mskbg,mskbPRC2010,belyaev}
an exponential decrease of the 
shell-correction free energy
as this is obvious from (\ref{dmiF}) and (\ref{dfscl}).  
For larger particle numbers $A^{1/3}$ (larger $k^{}_F R \approx 
2 A^{1/3}$) and temperatures $T$ [$T> T_{\rm cr}$ (\ref{Tcr})], the 
shorter EQ orbits become dominating as compared to more degenerate
but longer 3D and meridional orbits, as seen 
from Fig.\  \ref{fig38}.

The shell corrections to the MI (\ref{dmiF}) 
turn out to be
relatively much smaller than the classical rigid-body (TF) 
component. This is similar to the shell-correction 
(free) energy  $\delta E$ (or $\delta \mathcal{F}$) as compared with the 
corresponding TF term.
However, many important physical effects, such as fission 
isomerism and 
high-spin physics depend basically on the shell effects. 
Our nonperturbation results 
for the MI shell corrections can 
of course be applied for larger rotational frequencies 
and larger deformations 
($\eta \sim 1.5 - 2.0$) where the bifurcations 
play the dominating role like 
in the case of the deformed harmonic oscillator 
\cite{mskbg,mskbPRC2010,belyaev}.

In this Section, we derived the shell components $\delta \Theta$ of the
moment of inertia
in terms of the free-energy shell correction $\delta \mathcal{F}$ within the
nonperturbative extended Gutzwiller POT for any effective mean-field
potential using the phase-space variables.
For the deformed spheroidal cavity, like for  
the harmonic oscillator potentials \cite{mskbPRC2010,belyaev},
we found a good agreement
between the semiclassical POT and quantum results for
$\delta \mathcal{F}$ and $\delta \Theta$
using the Thomas-Fermi approximation for
$\langle r_{\perp}^2/\vareps \rangle$
at several critical deformations and temperatures.
For smaller temperatures a
very interesting
interference of the
dominant short three-dimensional and parent equatorial orbits and
their bifurcations in the superdeformed region
is shown to appear.
For larger temperatures, the shorter EQ orbits are dominant.
An exponential decrease of the shell corrections with
increasing temperature is analytically demonstrated.


\section{Conclusions}
\label{concl}

In this review, we present
the semiclassical extended Gutzwiller approach (EGA)
to the s.p. Green's
function and the periodic orbit theory
for a description
of the level-density and (free) energy shell corrections.
Phase space trace formulas for the level density and energy shell
corrections were introduced for any Hamiltonian 
(Section II).

  Analytical expressions for
  the
  surface symmetry-energy constants were derived within the local-density ETF
approach by using simple isovector
solutions of the particle density in leading order of the 
leptodermous effective-surface approximation,
 taking into account  the derivatives of the 
symmetry energy, the spin-orbit interaction and 
the isovector surface
gradient terms, as demonstrated in Section III. We used the
  surface symmetry-energy constants for
  several Skyrme-force parametrizations to
calculate the energies and sum rules of the isovector dipole resonance
(IVDR) strength and
the transition densities within the Fermi liquid-drop model (FLDM).
It turns out that
the surface symmetry-energy constants 
  are quite sensitive to
the parameters of the Skyrme force,
  in particular through the $\mathcal{C}$ coefficients in  
(\ref{Cpm}) appearing
in the density gradient terms of the isovector part of the energy
density. The values of these isovector constants are found to
  have also a strong influence on the spin-orbit interaction. 
The IVDR strength is shown to split into a main and several
satellite peaks.  This IVDR splitting and 
the mean isovector giant dipole-resonance (IVGDR) 
energies and energy weighted sum rules  are found,
within the  FLDM to be
in good agreement with the experimental data.

The transport coefficients for the
low-lying collective vibrational states were derived by using the
ETF components of the semiclassical periodic orbit theory. We
  also suggested (Section IV) to use
the shell correction method (SCM) to determine
the transport coefficients within the response function theory in close
analogy with the Strutinsky shell correction method for the (free) energy. 
We found an
enhancement of the inertia for the low-lying vibration states as
compared to the irrotational flow inertia. We thus obtain
a good agreement of the ETF semiclassical trajectory EGA for the main 
averaged characteristics, such as the energies,
transition probabilities, and EWSRs for the low-lying quadrupole and octupole
collective states. Taking  
also shell corrections to the smooth ETF
transport coefficients into account, one obtains an improved comparison with
experimental data, in particular near magic nuclei 
where these corrections play an
important role. Smooth nonlocal trajectory corrections were derived
for the inertia and friction parameters
  of low-energy collective vibrations
by solving the symmetry breaking problem within the uniform
approximation. We confirm that with increasing multipolarity of the
nuclear shape vibrations, the smooth friction coefficients tend to the
famous wall formula. In particular
 for the quadrupole and octupole collective modes, we found a
strong correlation,  as functions of the particle number, 
between the average dissipative time-dependent energy
rate and the energy shell correction.

 Semiclassical functional expressions for the moments of inertia were derived
 in the framework of the ETF approach (Section V). We used
 these analytical expressions to obtain a self-consistent description
 of rotating nuclei where the rotation velocity impacts on the structure
 of the nucleus. It has been shown that such a treatment leads, indeed,
 to the Jacobi phase transition to triaxial shapes as already predicted
 in \cite{CPS74} within the rotating LDM. 
We emphasize that the simple
rigid-body moment of inertia gives a quite accurate
approximation for the full ETF value.
Being aware of the mutual interplay
 between rotation and pairing correlations 
\cite{belyaevhighspin,sfraurev,pomorbartelPRC2011}, 
it would be
 especially  interesting to work out an approach that is able to
determine  the nuclear structure depending on its angular velocity,
as we have  done here in the ETF approach, but taking pairing correlations
and their rotational quenching explicitly into account.

We also
derived, within the nonperturbative extended POT, the shell corrections
of the MI in terms of the
free-energy shell corrections,
through those of the rigid-body MI of the
equilibrium rotations,
which is exact for the HO and 
quite accurate for the spheroidal potential.
Phase-space trace formulas for
the MI shell corrections
were obtained for any mean field potential accounting for the surface
corrections to the ratio of the MI and
free-energy shell corrections. An
exponential decrease of all shell corrections
with increasing temperature is observed as expected.
We also observe an enhancement of the amplitude of the
MI shell corrections due to the bifurcation catastrophe phenomenon.

As for further perspectives, it would be certainly
worth to apply our results to
calculations of the structure of the IVDR
within the Fermi-liquid droplet model in order
to determine the value of the fundamental surface and volume
(including derivative terms) symmetry-energy
constants from a
comparison with experimental data for the satellite resonances
\cite{ponomarev,adrich,wieland} and neutron skins \cite{vinas5},
as well as
  with other
theoretical calculations
\cite{vretenar1,vretenar2,vretenar3,nester1,nester2,BMRPhysScr2014,
BMRprc2015}.
For a further extension of
the description of  the low-lying isovector collective states, one has
to use the POT for including the shell effects semiclassically
\cite{sclbook,strumag,GMFprc2007,BMYijmpe2012,BMps2013}.

It would be also worth to apply this semiclassical theory
to the shell corrections of the MI and other transport 
coefficients, such as the inertia and friction parameters, 
for more realistic edge-like
 potentials with surface diffuseness \cite{MKApanSOL1}  within
the nuclear collective dynamics involving magic nuclei
\cite{dfcpuprc2004,magvvhof,MGFyaf2007,GMFprc2007}. One of the most attractive
applications of the semiclassical periodic-orbit theory, in line with one
of the main activities of V.G.\ Solovjov,
is its extension to the spin-orbit and pairing interactions 
\cite{brackquenNPA1981,brackrocciaIJMPE2010},
and their influence on the collective
vibrational and rotational excitations in heavy deformed neutron-rich
nuclei
\cite{belyaevhighspin,sfraurev,pomorbartelPRC2011,abrpairing}.

In nuclear physics, the spin of the nucleons plays an important role in the
MI calculations \cite{ringschuck,bartelnp,belyaev,sfraurev}.
As shown in \cite{bartelnp,belyaev}, it leads in particular to the
essential paramagnetic effects in the MI through its smooth ETF part.
It would be valuable to also include the spin degrees of freedom into the
semiclassical MI shell correction since it leads to the well-known spin-orbit
splitting which significantly changes the nuclear shell structure.
The analytical expressions for the MI obtained at the 
present stage have therefore
only a somewhat restricted value for the use in real nuclei, but could be
directly applied for metallic clusters and quantum dots.
The extension of the POT \cite{AB_jphys2002,BAPZ_ijmpe2004} to the MI
shell correction calculations including the spin degree of freedom
would therefore, constitute an essential progress in the understanding
of the semiclassical relation between the nuclear MI 
and free-energy shell corrections.
On the way to a more realistic study of this relation, let us mention,
in addition, the application of the POT for
  deformed and diffuse-surface power-law 
potentials \cite{MKApanSOL1} to our calculations of the MI shell corrections.
  In this connection, one has to recall
also the inclusion of pairing correlations, especially far from deformed
magic nuclei \cite{sfraurev}.
In this respect, the idea of applying the Strutinsky SCM to the
  calculation of nuclear energies in a Hartree-Fock-Bogoliubov type
approach at finite temperatures \cite{brackquenNPA1981} 
through the level density shell corrections for the spectra might
  also be useful
to develop a more realistic semiclassical theory.
Another interesting application of our semiclassical nonperturbative POT
would consist in taking into account the non-adiabatic effects, in
particular the $\omega$ dependence of the classical trajectories.
The most important more long-term future problem might be to 
perform comparison between the experimental data for the nuclear 
rotational bands with the POT results
for the MI, including the smooth ETF and PO-shell corrections. 
We also point out that extensions of these POT results would be 
extremely interesting for the inertia and friction coefficients
for low-lying collective states using more realistic Hamiltonians.
%


\bigskip
\centerline{{\bf ACKNOWLEDGMENTS}}
\medskip
%
%

The authors gratefully acknowledge
  many fruitful discussions with
S.\ Aberg, V.\ I.\ Abrosimov, K.\ Arita, 
J.\ P.\ Blocki, R.\ K.\ Bhaduri, M.\ Brack,
V.\ Yu.\ Denisov, S.\ N.\ Fedotkin, S.\ Frauendorf,
F.\ A.\ Ivanyuk,
V.\ M.\ Kolomietz, M.\ Kowal, M.\ Matsuo,
K.\ Matsuyanagi,
J.\ Meyer, V.\ O.\ Nesterenko, V.\ V.\ Pashkevich,
M.\ Pearson, V.\ A.\ Plujko, P-G.\ Reinhard, 
P.\ Ring, E.\ E.\ Saperstein,
A.\ I.\ Sanzhur, S.\ Siem, J.\ Skalski,
X.\ Vinas, and V.\ G.\ Zelevinsky,
One of us (A.G.M.) is also very
  grateful for the nice hospitality extended to him
during his working visits of the
Technical Munich University in Garching,
the University
of Regensburg in Germany,
  the Hubert Curien Institute of the University of Strassburg, the
National Centre for Nuclear Research in 
Otwock-Swierk and Warsaw of Poland,
and
  the Physics
Department of the Nagoya Institute of Technology.
  Many thanks go also to
the Japanese Society of
  the
Promotion of Sciences
for
  their
financial support, Grant No. S-14130.

\appendix

\setcounter{equation}{0}

\renewcommand{\theequation}{A.\arabic{equation}}
\renewcommand{\thesubsection}{A\arabic{subsection}}

\section{ 
TO THE EFFECTIVE SURFACE \\ APPROXIMATION}
\l{appA1}

\subsection{Solutions of the isovector \\ Lagrange equation}
\l{appA11}

The Lagrange equation for the variations of the isovector particle
density $\rho_{-}$ is given in the local coordinate system
where   $\xi$
is the distance of a given spatial point $\r$ to the ES, and 
the other orthogonal tangent-to-ES variables can be 
taken for instance
as the cylindrical coordinate of the projection of a point $\r$
to the ES and the azimuthal angle around the symmetry $z$ axis  by
\cite{BMRV,magsangzh}
\bea\l{lagrangeqminus}
&&2 \mathcal{C}_{-}\frac{\partial^2 \rho_{-}}{\partial \xi^2 } + 2
\mathcal{C}_{-}\mathcal{H} \frac{\partial \rho_{-}}{\partial \xi}\;\nonumber\\
&-& \frac{\d}{\d \rho_{-}}\left[\rho_{+}
\vareps_{-}\left(\rho_{+},\rho_{-}\right)\right]
 + \Lambda_{-}=0\;. 
\eea
Here $\mathcal{H}$ is the mean curvature of the ES 
($H=1/R$ for the spherical ES). The
isovector chemical-potential correction $\Lambda_{-}$ was
introduced \cite{magsangzh}
like the isoscalar one $\Lambda_{+}$, worked out in detail in
\cite{strmagbr,strmagden}.
Up to the leading terms in a small parameter $a/R$, one obtains
from
(\ref{lagrangeqminus})
\be\l{lagrangeqminus0}
2\mathcal{C}_{-} \; \frac{\partial^2 \rho_{-}}{\partial \xi^2} -
\frac{\d}{\d \rho_{-}} \left[\rho_{+} \vareps_{-}(\rho_{+},\rho_{-})\right]= 0\;.
\ee
We neglected here the higher order terms proportional 
to the first derivatives of the
particle density $\rho_{-}$ with respect to $\xi$ and
the surface correction to the isovector chemical potential in
(\ref{lagrangeqminus}) (\cite{strmagbr,strmagden} for
the isoscalar case).
For the dimensionless isovector density
$w_{-}=\rho_{-}/(\overline{\rho} I)$ one finds after simple transformations the
following equation and the boundary condition in the form
\bea\l{yeq0minus}
&&\hspace{-0.5cm}\frac{\d w_{-}}{\d w} \!=\!c_{\rm sym}
\sqrt{\frac{\overline{\mathcal{S}}_{\rm sym}(\eps)(1\!+\! \beta w)}{e[\eps(w)]}}
\sqrt{\Big|1 \!-\!\frac{w_{-}^2}{w^2}\Big|}\;,\nonumber\\
&&
\quad w_{-}(w=1) =1\;,
\eea
where $\beta$ is the SO parameter defined below (\ref{ysolplus}),
$\overline{\mathcal{S}}_{\rm sym}=\mathcal{S}_{\rm sym}/J$, $c_{\rm sym}$
is given by (\ref{csymwt}) and
$\mathcal{S}_{\rm sym}(\eps)$ in (\ref{symen}).
The above equation determines the isovector density $w_{-}$ as a function of
the isoscalar one $w(x)$ (\ref{ysolplus}). In the
quadratic approximation for $e[\eps(w)]$ [up to a small asymmetry correction
proportional to $I^2$
in (\ref{epsilonplus})], one finds an explicit analytical expression
in terms of elementary functions \cite{BMRV}.
Substituting
$w_{-}=w\;\cos \psi$ into (\ref{yeq0minus}),
and taking the approximation $e=(1-w)^2$,
one has the following
first order differential equation for a new function $\psi(w)$:
\bea\l{ueqminus}
&&\hspace{-2.7cm}- \frac{w(1-w)}{c_{\rm sym}}\;\sin \psi\;\frac{\d \psi}{\d w} 
\nonumber\\
&&\hspace{-2.2cm}=\sqrt{\overline{\mathcal{S}}_{\rm sym}(\eps)(1+\beta w)}\; \sin\psi 
\;-\frac{1-w}{c_{\rm sym}}\;\cos\psi\;,\nonumber\\
\,\,\psi(w=1)=0\;. 
\eea
The boundary condition for this equation is related to that
of (\ref{yeq0minus}) for $w_{-}(w)$.
This equation looks more complicated because of the
trigonometric nonlinear terms.
However, it allows to obtain the simple approximate
analytical solutions within standard perturbation theory.
Indeed, according to (\ref{yeq0minus}) and (\ref{ysolplus}),
where we do not have an explicit $x$-dependence,
we note that $w_{-}$ is mainly a sharply decreasing function of
$x$ through $w(x)$ within a small diffuseness region of the
order of one in dimensionless units (Figs.\ 
\ref{fig2} and \ref{fig3}). Thus,
 we may find  approximate solutions to equation (\ref{ueqminus})
with its boundary condition in terms
of a power expansion of a new function
$\widetilde{\psi}(\gamma)$ in powers of a  new small
argument $\gamma$ (\ref{csymwt}),
\be\l{powser}
\widetilde{\psi}(\gamma)\equiv
\psi(w)=\sum_{n=0}^{\infty} c_n\;\gamma^n(w)\;,
\ee
where the coefficients $c_n$ and $\gamma$ are defined in
(\ref{csymwt}). Substituting the power series
(\ref{powser}) into (\ref{ueqminus}), one expands first
the trigonometric functions into a power series of $\gamma$,
according to the boundary condition in (\ref{ueqminus}).
As usual, using standard perturbation theory,
we obtain a system of algebraic equations
for the coefficients $c_n$ (\ref{powser}) by
equating coefficients from both sides of
(\ref{ueqminus}) with the same powers of $\gamma$. This
simple procedure leads to a system of algebraic recurrence relations
which determine the coefficients $c_n$ as functions of the parameters
$\beta$ and $c_{\rm sym}$ of (\ref{ueqminus}),
\bea\l{cn}
&&\!\!c_0=0,\qquad c_1=\frac{1}{\sqrt{1+\beta}}\;,\nonumber\\
&&c_2=\frac{c_1}{2 c_{\rm sym}(1\!+\!\beta)}\left(
\beta c_{\rm sym}^2 \!+\! 2 \!+\! \frac{L}{3J} c_{\rm sym}^2(1\!+\!\beta)\right),
\nonumber\\
&&\!\!c_3=-c_1\left\{
\frac{4}{3} c_1^2 \!-\!3 \frac{c_1 c_2}{c_{\rm sym}} \!-\!
\frac{c_2 c_{\rm sym}}{2 c_1}\left(\beta c_1^2 \!+\! 
\frac{L}{3 J}\right)\;\right. 
\nonumber\\
&&\!\!- \left.\frac18 \beta^2 c_{\rm sym}^2 c_1^4
\!+\! \frac{K_{-}c_{\rm sym}^2}{36 J}
\!+\!\frac{c_{\rm sym}^2L}{12 J} \left(\beta c_1^2  \!-\!
\frac{L}{6 J} \right)\right\},
\eea
and so on. In particular, up to second order in $\gamma$,
we derive analytical solutions as functions of $\beta$, $c_{\rm sym}$, $J$
and $L$ in an explicitly closed form:
\bea\l{wsol3}
&&\widetilde{\psi}(\gamma) = \gamma\left(c_1 +
c_2 \gamma\right),\quad
c_1=\frac{1}{\sqrt{1+\beta}}\;,\\
&&c_2= \frac{\beta c_{\rm sym}^2 +2
+  L c_{\rm sym}^2 (1+\beta)/(3J)}{2 (1+\beta)^{3/2} c_{\rm sym}}\;.
\eea
Thus,  using the standard
perturbation expansion method of solving
$\widetilde{\psi}(\gamma)$ in terms of the power series of 
$\gamma$ up to $\gamma^2$, one obtains
the quadratic expansion of $\psi(w)$, (\ref{ysolminus}),
with $\widetilde{c}=c_2/c_1$.
Notice that one finds a good convergence of the power expansion
of $\widetilde{\psi}(\gamma(w))$ (\ref{wsol3}) in $\gamma(w)$ for
$w_{-}(x)$ at the second order in $\gamma(w)$ because of  
values of $c_{\rm sym}$ larger one for all Skyrme forces presented in Table 1
[(\ref{csymwt}) for $c_{\rm sym}$].

\subsection{The macroscopic boundary \\ conditions}
\label{appA12}

For the derivation of the expression for surface tension coefficients
$\sigma_{\pm}$, we first write the system of the Lagrange equations
by using variations of the energy density $\epsi(\rho_{+},\rho_{-})$
with respect to the isoscalar and isovector densities $\rho_{+}$
and $\rho_{-}$. Then, we substitute the solution of the first Lagrange equation
for variations of the
isoscalar density $\rho=\rho_{+}$ in the energy density
(\ref{enerdenin}) \cite{strmagbr,strmagden} into the second
Lagrange equation (\ref{lagrangeqminus}) 
for the isovector density $\rho_{-}$. Using
the Laplacian in  the variables $\xi$ and 
other cylindrical coordinates introduced above 
\cite{strmagbr} we keep the major terms in this second equation
within the improved precision in small parameter
$a/R$. The improved precision means that we take into account
the next terms proportional to
the first  derivatives of particle densities [along with
the second ones of (\ref{lagrangeqminus0})], and small surface
corrections $\Lambda_{\pm}$ to the
isoscalar and isovector Lagrange multipliers $\lambda_{\pm}$.
Within this improved precision,
one finds the second Lagrange equation (\ref{lagrangeqminus}) 
by the variations of the energy
density $\epsi(\rho_{+},\rho_{-})$ (\ref{enerdenin}) with respect to
 the isovector
particle density $\rho_{-}$.
Multiplying (\ref{lagrangeqminus})
by $\partial \rho_{-}/\partial \xi$ we integrate over
the  coordinate $\xi$,
normal-to-ES direction, from a spatial point $\xi_{in}$
inside the volume (at $\xi_{in} \siml -a$ )
to $\infty$ term by term.
Using also integration by parts, within the ES approximation
this results in the macroscopic boundary conditions (together with the 
isoscalar boundary
condition from
\cite{strmagbr,strmagden,magsangzh,kolmagsh,magstr,magboundcond})
\bea\l{macboundcond}
\left(\overline{\rho}\;I\;
\Lambda_{-}\right)_{ES} &=&  P_{s}^{(-)} \equiv 2 \sigma_{-} \mathcal{H}\;, 
\nonumber\\
 \left(\overline{\rho}\;
\Lambda_{+}\right)_{ES} &=&  P_{s}^{(+)} \equiv 2 \sigma_{+} \mathcal{H}\;.
\eea
Here, $P_s^{(\pm)}$ are the
isovector and isoscalar surface-tension (capillary) pressures and
 $\sigma_{\pm}$ are the corresponding tension coefficients;
see their expressions in (\ref{sigma}). We point out that the lower limit
$\xi_{in}$ can be approximately extended to $-\infty$ as in (\ref{sigma})
for $\sigma_\pm$ because of a fast convergence of the integral over $\xi$
 within the surface layer, at $a/R \ll 1~$. 
The integrands contain, indeed, the square of the first derivatives, 
$(\partial \rho_{\pm}/\partial \xi)^2 \propto (R/a)^2$, 
and therefore, the integral over $\xi$ 
converges exponentially rapidly within the ES layer $|\xi| \leq a$. This leads
to the aditional small factor $a/R$ in (\ref{sigma}), 
$\sigma_{\pm} \propto R/a$. Therefore, at this higher order of the improved
ES approximation, one may neglect high order corrections 
in the calculation
of derivatives of $\rho_{\pm}$ themselves by using the analytical universal
density distributions $w_{\pm}(x)$ [(\ref{ysolplus}) 
and (\ref{ysolminus})] within the ES layer which do not depend on the specific
properties of the nucleus as mentioned in the main text. 
(These mean-curvature corrections are small terms proportional to the first derivative
$\partial \rho_{-}/\partial \xi$ and $\Lambda_{-}$ in
(\ref{lagrangeqminus}),
as for the isoscalar case 
\cite{strmagbr,strmagden,magstr,magboundcond}).
In these derivations,  the obvious boundary
conditions of disappearance of the particle densities and all their
derivatives with respect to $\xi$ outside of the ES for $\xi \rightarrow \infty$
($\xi \gg a$) were taken into account too.

The Lagrange multipliers $\Lambda_{\pm}~$, multiplied by
$\overline{\rho}I$ and $\overline{\rho}~$, in the parentheses
on the left-hand sides of 
 (\ref{macboundcond})
are the volume isovector ($\overline{\rho}I \Lambda_{-}$) and isoscalar
($\overline{\rho} \Lambda_{+}$) pressure excesses, respectively
\cite{magsangzh}.
These pressures due to the surface curvature can be derived by
using the
volume solutions of
the Lagrange equations for the particle densities [obtained by 
variating 
the energy density $\epsi$ and neglecting
all the derivatives of the particle densities in (\ref{enerdenin})],
\bea\l{rhovol2}
\!\!\!\rho_{-} &\approx& \overline{\rho}\; \left[I\left(1 +
\frac{9 \Lambda_{+}}{K}\right) +
\frac{\Lambda_{-}}{2J} \right], \nonumber\\
 \!\!\!\rho_{+}&\approx&
\overline{\rho}\left[1+\frac{9 \Lambda_{+}}{K}
\left(1 - \frac{81\Lambda_{+}}{2K}\right) -
\frac{18 J}{2K}\;I^2\right]\;. 
\eea
Inserting $\Lambda_{+}$ and $\Lambda_{-}$ from (\ref{macboundcond})
into (\ref{rhovol2}), one finds
\be\l{rhovolbsminus2}
\hspace{-0.4cm}\rho_{-}\!=\!\overline{\rho}\;I\;
\left[1 \!+\! \frac{6 b_{\rm S}^{(+)}\;\mathcal{H}\;r_0}{K} \!+\!
\frac{2 b_{\rm S}^{(-)}\;\mathcal{H}\;r_0}{6 J\;I^2}\right]\;.
\ee
As seen from (\ref{rhovolbsminus2}),
 the isovector density correction to the volume density
$\rho_{-}$ because of
a finiteness
of the coupled system of the two Lagrange equations depends on both
isoscalar and isovector
surface energy constants $b_S^{(\pm)}$ in the first-order expansion of the small
parameter $a/R$. If we are not too far from the valley of stability, $I$ is an
additional small parameter, and the isovector corrections are small compared
with the isoscalar values
[$b_S^{(-)} \propto I^2$, $\Lambda_{-}\propto I$;
see \ (\ref{rhovol2}), (\ref{rhovolbsminus2}), and (\ref{bsminus})].
Thus,  (\ref{macboundcond}) has  a clear
physical meaning as the macroscopic boundary conditions
for equilibrium of the isovector and isoscalar
forces (volume and surface pressures) acting on the ES \cite{bormot,kolmagsh}.
Equations in (\ref{macboundcond}) can be used as the boundary conditions to the
volume Lagrange equations obtained by neglecting  derivatives of the
particle densities $\rho_{\pm}$ over $\xi$.
Note that the isovector tension coefficient  $\sigma_{-}$
is much smaller than the isoscalar one $\sigma_{+}$
[see (\ref{sigma})] as
$\sigma_{-} \propto  I^2$ because of $\rho_{-} \propto I$ and $I \ll 1$ 
near the nuclear beta-stability line. Another reason is
the smallness of ${\cal C}_{-}$
as compared to ${\cal C}_{+}$ for the realistic Skyrme forces
(Table 1)
\cite{chaban,reinhard}.
From comparison of (\ref{rhovol2})  and
(\ref{rhovolbsminus2}) for $\rho_{-}$ [see also (\ref{sigma})],
one may also evaluate
\be\l{lambdatotminus}
\Lambda_{-} =
\frac{2 \sigma_{-} \mathcal{H}}{\overline{\rho}\;I} \approx
\frac{2 b_{\rm S}^{(-)}}{3 I A^{1/3}}
\sim k_{\rm S} I \frac{a}{R}\;.
\ee
which is consistent with (\ref{lagrangeqminus})
($r_0 \mathcal{H} \sim a/R$ in these estimations, see
corresponding ones in \cite{strmagbr,strmagden}).

\subsection{ 
Derivations of the surface energy \\ and
its coefficients}
\l{appA13}

For calculations of the surface energy components
$E_S^{(\pm)}$ of the energy $E$ in  (\ref{energy})
within the same improved ES approximation as described above
in Appendix A we first
separate the volume terms related
to  the first two terms of (\ref{enerdenin}) for the energy density
$\epsi$ per particle. Other terms of the energy density
 $\rho \epsi(\rho_{+},\rho_{-})$ in
(\ref{enerdenin}) lead to the surface components $E_{\rm S}^{\pm}$
(\ref{Espm}), as they
are concentrated near the ES.
Integrating the energy density $\rho \epsi$
per unit of the volume [see (\ref{enerdenin})]
over the spatial coordinates
$\r$ in the local coordinate system defined above
within the ES approximation, one finds
\bea\l{Espm1}
E_S^{\pm} &=&
\oint  \d S \int\limits_{\xi_{\rm in}}^{\infty} \d \xi
\left[\mathcal{C}_{\pm}\left(\nabla \rho_{\pm}\right)^2
\right.\nonumber\\
&+& \left.\rho_{+}\vareps_{\pm}\left(\rho_{+},\rho_{-}\right)\right]\;\approx 
\sigma_{\pm}\;\mathcal{S}\;, 
\eea
where $\xi_{\rm in} \siml -a$
\cite{strmagbr,strmagden,magsangzh}.
Local coordinates 
were used because the integral
over $\xi$ converges rapidly within the ES layer which is effectively taken
for $|\xi|\siml a$. Therefore again, we may extend
formally $\xi_{in}$ to $-\infty$ in the first (internal) integral
taken over the ES in the normal-to-ES direction $\xi$ in
(\ref{Espm1}).
Then, the second integration is performed over the closed surface
of the ES. The integrand over $\xi$ contains terms of the order of
$(\overline{\rho}/a)^2 \propto (R/a)^2~$. 
However, the integration is effectively performed over the edge region
of the order of $a$ that leads to
the additional smallness proportional to $a/R$ as in Appendix A.
At this leading order the 
dependence of the internal
integrand on orthogonal-to-$\xi$ coordinates can be neglected.
Moreover, from the Lagrange equations [(\ref{lagrangeqminus0})
for the isovector case]
at this main order, one can realize that the terms without 
particle density gradients in (\ref{Espm1}) are equivalent to the gradient
terms. Therefore, for the calculation
of the internal integral we may approximately reduce the integrand
over $\xi$ to
derivatives of the universal particle densities
of the leading order $\rho_{\pm}(\xi)$ in $\xi$
using
\be\l{derequiv}
\hspace{-0.2cm}\mathcal{C}_{\pm} \!\left(\nabla \rho_{\pm}\!\right)^2 \!+\!
\rho_{+}\vareps_{\pm}\!\left(\rho_{+},\rho_{-}\!\right) \!\approx\!
2\mathcal{C}_{\pm}(\partial \rho_{\pm}/\partial \xi)^2
\ee
[(\ref{ysolplus}) and (\ref{ysolminus}) for $w_{\pm}(x)$].
We emphasize that the isovector gradient terms are obviously
 important for these calculations.
Taking approximately
the integral over $\xi$ within the infinite
integration region ($-\infty <\xi <\infty$) out of the integral over
the ES ($\d S$) we are left with the
integral over the ES itself that is the
surface area $\mathcal{S}$. Thus,
we arrive finally at the right hand side of (\ref{Espm1})
with the surface tension coefficient $\sigma_{\pm}=b^{(\pm)}_S/(4\pi r_0^2)$
[(\ref{sigma}) for $b^{(\pm)}_S$].

Using now the quadratic approximation $e[\eps(w)]=(1-w)^2$
in (\ref{sigma}) for $b_{\rm S}^{\pm}$
($\mathcal{D}_{-} = 0$),
one obtains (for $\beta<0$, see Table 1)
\be\l{bsJpm}
b_{\rm S}^{(\pm)}= 6 \overline{\rho}\; \mathcal{C}_{\pm}\;
\frac{\mathcal{J}_{\pm}}{r_0 a}\;,
\ee
where
\bea\l{Jp}
&&\hspace{-0.8cm}\mathcal{J}_{+}=
\int\limits_0^1 \d w\; \sqrt{w(1+\beta w)}\;(1-w)\;\nonumber\\
&=&\frac{1}{24(-\beta)^{5/2}} 
\left[\mathcal{J}_{+}^{(1)}\; \sqrt{-\beta(1+\beta)}\;
\right.\nonumber\\ 
&+&\left.\mathcal{J}_{+}^{(2)}\; \arcsin\sqrt{-\beta}\right],
\eea
with
\be\l{jplusA13}
\mathcal{J}_{+}^{(1)}=3 + 4 \beta(1+\beta),\quad
\mathcal{J}_{+}^{(2)}=-3-6\beta\;.
\ee
For the isovector energy constant $\mathcal{J}_{-}$, one finds
\bea\l{Jm}
&&\hspace{-0.5cm}\mathcal{J}_{-}\!=\!\frac{-1}{1\!+\!\beta}
\!\int\limits_0^1\! \d w \sqrt{w(1\!+\!\beta w)}\!\nonumber\\
&\times& (1-w)(1+\widetilde{c} \gamma(w))^2\!=\!
\frac{\widetilde{c}^2}{1920 (1+\beta) (-\beta)^{9/2}}\nonumber\\
&&\times
\left[\mathcal{J}_{-}^{(1)}
\left(\frac{c_{\rm sym}}{\widetilde{c}}\right)
\sqrt{-\beta(1+\beta)}\right.\nonumber\\
&&+\left.{\cal J}_{-}^{(2)}\left(\frac{c_{\rm sym}}{\widetilde{c}}\right)\arcsin\sqrt{-\beta}\right],
\eea
with
\bea\l{jminusa13}
&&\hspace{-0.1cm}\mathcal{J}_{-}^{(1)}(\zeta)= 
105- 4 \beta\left\{95 +75 \zeta 
+\beta \left[119+10\zeta (19+6\zeta)\right.\;
\right.\nonumber\\
 &&\!\!+ \left.\left. 8 \beta^2
\left(1+ 10\zeta(1+\zeta)\right) + 8 \zeta 
\left(5 \zeta (3 +2 \zeta) -6\right)\right]\right\},
\eea
\bea\l{jminus2a13}
&&\hspace{-1.5cm}\mathcal{J}_{-}^{(2)}(\zeta)=15 \left\{7+2\beta \left[
5 (3 + 2 \zeta)
+
8 \beta (1+\zeta)\;\right.\right.\nonumber\\
&\times&\left.\left.\left(3 +\zeta +2 \beta (1+\zeta)\right)\right]\right\}.
\eea
These equations determine explicitly the analytical expressions for the
isoscalar ($b_{\rm S}^{(+)}$) and isovector ($b_{\rm S}^{(-)}$) 
energy constants in terms of
the Skyrme force parameters, see
(\ref{ctilde}) for $\widetilde{c}$ and (\ref{csymwt}) for $c_{\rm sym}$
and $\gamma(w)$.
For the limit $\beta \rightarrow 0$ one has
from (\ref{Jp}) and (\ref{Jm})
$\mathcal{J}_{\pm} \rightarrow 4/15$. With 
(\ref{skin}) and (\ref{fw}) one arrives also
at the explicit analytical expression  for the isovector stiffness $Q$
as a function of $\mathcal{C}_{-}$ and $\beta$. In the limit
$\mathcal{C}_{-} \rightarrow 0$  one obtains $k^{}_S \rightarrow 0$ and
$Q \rightarrow \infty$ because of the finite limit of
the argument 
$c_{\rm sym}/\widetilde{c}\rightarrow 2(1+\beta)/[\beta +(1+\beta) L/(3J)]$
of the function
$\mathcal{J}_{-}$ in (\ref{Jm})
[see also (\ref{ysolminus}) for
$\widetilde{c}$ and (\ref{csymwt})
for $c_{\rm sym}$].

\subsection{Simple case of symmetrical \\ nuclei}
\l{appA14}

For the simplest case of the symmetric nuclei, $N=Z$ ($I=0$),
one has from (\ref{enerdenin}) (omitting the subscripts ``plus'')
\be\l{enerdenins} 
\epsi\left(\rho\right)=-b^{}_{\rm V} + \vareps(\rho) +
\left(\mathcal{C}+\frac{\Gamma}{4 \rho}\right)
\frac{\left(\nabla \rho\right)^2}{\rho}\;.
\ee
For simplicity we neglect the spin-orbit terms along with the asymmetry ones.
Variating the energy functional (\ref{energy}) with this energy density 
per particle, we obtain the Lagrange equation \cite{strmagden}:
\be\l{lagreqs}
2 \left(\mathcal{C} +\frac{\Gamma}{4\rho}\right) \triangle \rho -
\frac{\Gamma}{4 \rho^2} \left(\nabla \rho\right)^2 +\Lambda=0\;,
\ee
where $\Lambda=\lambda +  b^{}_{\rm V}$ is the correction to the separation
energy per particle $-b^{}_{\rm V}$ in the chemical potential $\lambda$.
Introducing a local orthogonal coordinate system with the 
normal-to-ES coordinate $\xi$, one gets for 
the particle density $\rho_0$ at leading order
in the leptodermous parameter $a/R $ a simple ordinary differential
equation:
\be\l{rhoeq0} 
\frac{\d \rho_0}{\d \xi} =-\frac{2 \rho \vareps^{1/2}(\rho)}{
\sqrt{4 \mathcal{C}\rho +\Gamma}}, \quad \Gamma=\frac{\hbar^2}{18 m}\;.
\ee
This equation can be solved analytically at arbitrary surface-interaction
constant $\mathcal{C}$ for the quadratic approximation
$\vareps(\rho)=[K/18 \rho^2_{\infty} ] \left(\rho-\rho_{\infty}\right)^2~$,
where $K$ is the in-compressibility modulus of  
infinite symmetric matter.
Transforming (\ref{rhoeq0}) to that for the dimensionless 
particle density, 
$w(x)=\rho(\xi)/\overline{\rho}~$, $~x=\xi/a$, for $\mathcal{C}=0$ 
(symmetric gas of independent nucleons), one finds
\be\l{weq0}
\hspace{-0.4cm}w'(x)\!=\!-\zeta w \sqrt{\eps(w)},\,\,\, \zeta\!=\!2 a \sqrt{K/(18\Gamma)}\;.
\ee
Differentiating once more term by term over $x$ and using the ES definition
$w''(x)=0$ at the ES, $x=0$, one arrives at
the boundary condition:
\be\l{boundconds0}
2 \eps(w^{}_0) + w^{}_0\eps^\prime(w^{}_0)=0\;.
\ee
For the quadratic $\eps(w)$, one finds the solution $w_0=1/2$.
Integrating (\ref{weq0}) with $\eps(w)=(w-1)^2$ and using the 
boundary condition (\ref{boundconds0}), one obtains the 
explicit Fermi-function solution:
\be\l{wsols0}
w(x)=\left[1+\exp\left(\zeta x\right)\right]^{-1}\;.
\ee
For large $x$, one has from (\ref{wsols0}) asymptotically, 
$w(x)\rightarrow \exp(-\zeta x)$ for $x \rightarrow \infty$. 
Therefore, one can define the diffuseness
parameter $a$ from the usual condition, $\zeta=1$ so that the particle density
$w(x)$ will be decreased at large $x$ in $e$ times, i.e., 
\be\l{difs0}
a=\sqrt{\frac{18 \Gamma}{4K}}=\sqrt{\frac{\hbar^2}{4mK}}\;.
\ee
Another limit case of $\mathcal{C}\neq 0$ but neglecting $\Gamma$ was 
considered in Section III  
(see \cite{magsangzh} for a more general case of 
$\mathcal{C}\neq 0$ and $\Gamma \neq 0$, simultaneously).

For the energy (\ref{energy}) with (\ref{enerdenins}), one has
\bea\l{energys}
E&=&-b^{}_{\rm V}A +\int\d \r \left[
\left(\mathcal{C}+\frac{\Gamma}{4 \rho}\right)\left(\nabla \rho\right)^2 +
\rho \vareps(\rho)\right]\nonumber\\
&=&E_{\rm V}+E_{\rm S}\;,
\eea
where $E_{\rm V}=-b^{}_{\rm V}A$ is the volume and 
$E_{\rm S}=\sigma \mathcal{S}$ is the surface
components with the tension coefficient
\be\l{sigmas}
\sigma= \int_{-\infty}^{\infty} \d \xi \left(\mathcal{C} +
\frac{\Gamma}{4 \rho}\right)\left(\frac{\partial \rho}{\partial \xi}\right)^2\;.
\ee
For calculations of the surface energy component $E_{\rm S}$ from
the second integral in (\ref{energys}), one notes that 
we need the particle density $\rho \approx \rho_0$ at leading order
in  small parameter $a/R$ by the same reasons as mentioned in 
Appendix A3. Therefore, according to the Lagrange equation
at this order (\ref{rhoeq0}), the two terms in the 
square brackets of the integral in (\ref{energys}) are identical, see 
(\ref{derequiv}).
Using (\ref{sigmas}) for the tension coefficient $\sigma$ 
and (\ref{rhoeq0}) at $\mathcal{C}=0$ 
one finds analytically (after transforming to the dimensionless
quantities and changing the integration variable from 
$x$ to $w$),
\be\l{sigmas0}
\sigma=\frac{\hbar \overline{\rho}}{36}\;\sqrt{{K}{m}}\;.
\ee
Other limit cases are considered in Section III, this Appendix A
and in \cite{magsangzh}.

\setcounter{equation}{0}

\renewcommand{\theequation}{B.\arabic{equation}}
\renewcommand{\thesubsection}{B\arabic{subsection}}
\section{
TO CALCULATIONS OF TRANSPORT \\  COEFFICIENTS}
\l{appA2}
\subsection{Coupling constants}
\l{appA21}
The consistency condition for the single-particle operator 
$\rm{\hat{F}}$ (\ref{foperdV}) 
of the external field writes \cite{bormot,hofbook}
\bea\l{consisteqF}
&\delta \langle {\rm\hat{F}}\rangle_\omega = \kappa_{\rm{FF}}\,\delta
q_\omega\;, \nonumber\\ 
& \kappa_{\rm{FF}}= - \chi_{\rm{FF}}(0) -
C_{\rm{FF}}(0)\;,
\eea
where $\kappa_{\rm{FF}}$ is the coupling constant, and 
$\chi_{\rm{FF}}(0)$ the  
isolated susceptibility in the $\rm{F}$ mode. 
For a quasi-static process, the first consistency
relation in (\ref{consisteqF}),
\bea\l{consisteqFF}
&&\hspace{-0.5cm}\delta \langle \hat{\rm F}\rangle 
= \int \d \r \hat{\rm F}(\r)\, \delta
\rho(\r,q) \;\nonumber\\
&=& \int \d \r \hat{\rm F}(\r)\, \left(\frac{\partial
\rho(\r,q)}{\partial q}\right)_{q=0} \delta q\;,
\eea
determines the coupling constant $\kappa^{}_{\rm{FF}}$ by
\begin{equation}\label{kappaFF}
\kappa^{}_{\rm{FF}} = \int \d \r \, \hat{\rm F}(\r)\, \left(\frac{\partial
\rho (\r,q)}{\partial q}\right)_{q=0}.
\end{equation}
Within the considered macroscopic model, the particle density variation (transition density) can be presented as a sum of the ``volume'' and ``surface'' 
parts in the ES  
approximation \cite{strmagbr,strmagden},
\be\label{transden}
\hspace{-0.4cm}\delta \rho\left(\r,q\right) \!=\! \delta \rho_{\rm{vol}}(\r,q)y(\xi) 
\!-\!\rho_{\rm in} \frac{R}{a}\frac{\partial y(\xi)}{\partial
\xi}Y_{\lambda 0}(\theta).
\ee
Here $\rho_{\rm in}$ is the particle density inside of the system far from 
the ES, see (\ref{kappaQres}) (for simplicity the
low index in $\rho_{\rm in}$ 
is omitted in the main text). 
The ES is defined as the spatial points of maximal 
particle-density gradient $\nabla \rho(\r)$.
The radial coordinate dependence of the particle density is approximated via $\rho(r,\theta,q)= \rho_{\rm{in}}\, y(\xi)$, $~\xi=[r-R(\theta,q)]/a~$ where $y(\xi)$ is a gradual step-like profile function with approximately
a sharp change from 0 to 1 near the nuclear surface, $r=R(\theta,q)$, within a small transition region of the order of a diffuseness parameter, 
$a=(4 \mathcal{C} \overline{\rho}/b_{\rm V})^{1/2}$. For the
coefficient $\mathcal{C}$ in front of the 
$[\nabla \rho(r)]^2$ term of the effective nuclear Skyrme forces, one has
\begin{eqnarray}\label{betajy} 
&&\hspace{-0.5cm}\mathcal{C}\!=\!\frac{4 \pi r_0^5
b_{\rm S}^2}{27 b_{\rm V} \mathcal{J}^2}\,,\quad
\mathcal{J}\!=\!\int\limits_{-\infty}^{\xi_0} {\d \xi}
\left(\frac{\d y(\xi)}{\d \xi}\right)^2 \!\nonumber\\
&&\approx \frac{8}{15},\qquad
y(\xi) = \frac{\rho(\xi)}{\overline{\rho}} 
\approx \tanh ^2(\xi - \xi_0). 
\end{eqnarray}
The profile function $y(\xi)$ in (\ref{transden}) and
 (\ref{betajy}) with
$\xi_0/R=\rm{ArcTanh}(\sqrt{y_0})=0.658...$ for the value $y=y_0=1/3$
at the ES was approximated in (\ref{betajy}) as the analytical
solution ``Par'' derived in \cite{strmagden}. It is the simplest
parabolic approximation, $-b^{}_{\rm V}+K(1-y)^2/18$, for the energy density
per particle inside of the nucleus up to a small relatively kinetic
energy $[\nabla \rho(r)]^2$ correction in the ES layer of the width
$a$. As shown in \cite{strmagden}, this ``Par'' solution
$y(\xi)$ for the particle density is in good agreement with the
Hartree-Fock calculations of the averaged particle densities and
nuclear energies based on several Skyrme force parameters, except for
small quantum (in particular, shell) effects outside of the narrow ES layer.
From (\ref{betajy}) one has the approximate relationship between
the surface energy constant $b^{}_{\rm S}$, and the diffuseness parameter $a$,
$b^{}_{\rm S} \approx 4 b^{}_{\rm V} a/(5 r_0)$ \cite{strmagden}.

In the framework of the ES approximation, at leading order of
expansion in parameter $a/R\sim A^{-1/3}$, for the operator
$\hat{F}(\r)$ of (\ref{colresp}) for $\lambda \geq 2$ one has
\begin{eqnarray}\label{operesa}
\hat{F}(\r) &=&\left(\frac{\delta V}{\delta \rho}\, \frac{\partial
\rho(\r,q)}{\partial q}\right)_{q=0} \;\nonumber\\
&\approx& -R
Y_{\lambda 0}(\hat{\r})\,\left(\frac{\delta V}{\delta \rho}\, \frac{\partial
\rho}{\partial r}\right)_{q=0} \nonumber\\
&=&\frac{RK}{9
\rho}\,Y_{\lambda 0}(\hat{\r})\,\left( \frac{\partial \rho}{\partial
r}\right)_{q=0},
\end{eqnarray}
up to small relatively corrections of the order of $6 b^{}_{\rm S}/(K A^{1/3})$
 to the ``volume'' particle density variations in
(\ref{transden}), see (\ref{kappaQres}). In order to evaluate
the variational derivative $\delta V/\delta \rho$ in
(\ref{operesa}), we used now the thermodynamical relation
(energy conservation equation), 
$\d \lambda = -S\,\d T + {\mbox d}\mathcal{P}/\rho + \d V,$
where $S$ is the nuclear entropy, $\mathcal{P}$ the pressure, and
$V$ is a quasi-static external field \cite{hofbook}. Then, the
conservation of particle number at constant temperature $T$ (constant
chemical potential $\lambda$ and $T=0$ in this specific case) and the
definition of in-compressibility, $K= 9 (\partial {\cal P}/\partial
\rho)_{q=0}$, were taken into account in the third equation of
(\ref{operesa}). Substituting (\ref{operesa}) and
(\ref{transden}) into (\ref{kappaFF}) and taking smooth
$r$-dependent quantities, as compared to the sharp radial derivatives
of the particle density, $~\rho(r) \propto y((r-R)/a)$, at $~q=0~$ off
the integral over $~r~$, we may use the ES approximation for a
surface tension, $b^{}_{\rm S}/(4\pi r_0^2)$ (\ref{sigma}) 
\cite{strmagbr,strmagden}.
With this expression for $b^{}_{\rm S}$, up to 
small terms of the high relatively order in $A^{-1/3}$, 
in particular, those of (\ref{cchi}), and small 
 particle-density corrections, 
$\sim [6 b^{}_{\rm S}/(K
A^{1/3})]^2$, in the nuclear volume,
 one approximately obtains
\begin{eqnarray}\label{kld}
\hspace{-0.7cm}\kappa_{\rm FF} &\approx& - R\, \int \d \r \, \hat{\rm F}(\r)\,
Y_{\lambda 0}(\hat{\r})
\left(\frac{\partial \rho}{\partial r}\right)_{q=0}\!\nonumber\\
& =& -\frac{K\, b^{}_{\rm S}
 \, R^4}{72 \pi \,\rho\, \mathcal{C}\; r_0^2}\;, 
\end{eqnarray}
see (\ref{kappaQres}) for the particle density $\rho$ 
and (\ref{betajy}) for $\mathcal{C}$. 
Using (\ref{betajy}), from (\ref{kld}) for 
the coupling constant $\kappa_{\rm FF}~$,
one arrives at (\ref{kappaF}).

Similarly, from the consistency condition (\ref{consisteq}), one may write
\begin{equation}\label{kappaQ}
\hspace{-0.4cm}\delta \langle \hat{Q} \rangle = \int \d \r
\;\hat{\rm Q}(\r)\;\delta \rho(\r,t) =\kappa_{\rm QQ}\;\delta
q(t)\;,
\end{equation}
where $\delta \rho(\r,q)$ is the particle density variation 
(\ref{transden}) with the same edge-like function $y(\xi)$ described above. Up to small negligibly corrections in expansion over parameter $a/R$, 
from (\ref{kappaQ}) and (\ref{transden}), one finds
\begin{eqnarray}\label{coupconsurfcor}
&\kappa_{\rm QQ}\!=\!\rho R^{\lambda+3}\;\left[1+
\left(\frac{\xi_0}{R}-1\right)(\lambda+2)\frac{a}{R}\right.\nonumber\\
&+\left.(\lambda+1)(\lambda+2)
\left(\frac{\xi_0^2}{2R^2}-\frac{\xi_0}{R}+ {\rm log
2}\right)\frac{a^2}{R^2}\right]\;.
\end{eqnarray}
In these derivations we neglected the contributions of the ``volume''
part of dynamical particle-density variations of (\ref{transden}).
The boundary condition for pressures of the ES 
approach \cite{strmagbr,strmagden} was
used to relate the slow ``volume'' and ``surface'' vibration
amplitudes in (\ref{transden}). The radial ``volume'' 
particle-density dependence in
(\ref{transden}), $\delta \rho_{\rm vol}(\r) \propto j_{\lambda}(qr)
Y_{\lambda 0}(\theta)$, is evaluated like in the macroscopic zero-sound
Fermi-liquid models 
 ($q$ is the wave number,
$\omega \approx q p^{}_F/m$, $q R \siml 1$ for nuclear low-lying
excitations, see Section III and \cite{strmagden,kolmagsh}) 
which leads to a small relatively factor,
$3(\lambda - 1)(\lambda + 2)b^{}_{\rm S}/[(2\lambda + 3)K A^{1/3}]$,
 as compared to the leading
``surface'' term of the sum (\ref{transden}). Other corrections were
obtained from expansion of the integral taking from the ``surface'' part of
the particle density variation in (\ref{transden}) with respect to a 
small parameter,
$a/R \approx 5 b^{}_{\rm S} r_0/(4 b^{}_{\rm V} R) \approx 1.4/A^{1/3}$, at second
(curvature) order. The analytical solution (\ref{betajy}) for $y(\xi)$
was explicitly used for the integrations over the radial variable in
(\ref{kappaQ}). Up to small relatively ``volume'' corrections,
$\sim b^{}_{\rm S}/(K A^{1/3})$, and those of (\ref{coupconsurfcor}), one
obtains (\ref{kappaQres}).

\subsection{Calculations of the Jacobian}
\l{appA22}

For calculations of the Jacobian, 
$~{\cal J}_{\rm CT}\left(\p_1,t^{}_{\rm CT};\r_2,\vareps\right)~$, 
in expansion (\ref{Gsem}) with
(\ref{GsemGa}) and (\ref{ampiso}) let us specify the CT with the
momentum $\p_1$ at the initial $\r_1$, and the final $\r_2$
point
for a given energy $\vareps$, see Fig.\ \ref{fig26}. 
The time $t^{}_{\rm CT}$ for a particle motion along the path CT is
determined by its length ${\cal L}_{\rm CT} $, $~t^{}_{\rm CT}=m{\cal
L}_{\rm CT}/p$, $p=\vert\p_1\vert$ in the edge-like (billiard-like) 
potentials.  
It is convenient to transform the Jacobian ${\cal
J}_{\rm CT}$ to the cylindrical coordinate system $\rho,z,\varphi$
($x=\rho \,\cos{\varphi},~ y=\rho \,\sin{\varphi}$), as shown in
Fig.\ \ref{fig26}.  
For transformation of the momentum variables, one can
 use a similar cylindrical system $p_\rho, p_z, \Phi_p$
($p_x=p_\rho\,\cos\Phi_p, p_y=p_\rho \,\sin\Phi_p$) to take into
account the azimuthal symmetry  
\cite{strumag}. By making use
of the usual properties of Jacobian transformations, one writes
\bea\l{jac} 
&\hspace{-2.0cm}{\cal J}_{\rm CT }\left(\p_1,t^{}_{\rm CT};\r_2,\varepsilon\right)
 = (p_{\rho 1}/\rho_2) \nonumber\\ 
 &\hspace{-1.0cm}\times 
{\cal J}_{\rm CT }\left(p_{\rho 1},p_{z 1},t_{\rm CT};\rho_2,z_2,\varepsilon\right)
=\frac{m^2}{p}\,\frac{p_{\rho
1}}{\rho_2}\, \left(\frac{\partial \theta_{p_y 1}}{\partial
y_2}\right)_{\rm CT }, \nonumber\\
 &\hspace{-1.0cm}\left(\frac{\partial \theta_{p_y 1}}{\partial y_2}\right)_{\rm CT}
= \frac{1}{{\cal L}_{\rm CT}}\;.
\eea
Here, we introduced the Cartesian coordinate system with the axis $x$
along the CT, and the perpendicular axis $y$ with the
center moving along the CT (Fig.\ \ref{fig1} and
 \cite{sclbook,gutz,strumag}).
From simple geometrical
relationships, the stability factor of $~\partial \theta_{p_y
1}/\partial y_2~$ in (\ref{jac}) for the central planar
CT in this $x,y$ coordinate system was obtained
through its invariant length, $~\mathcal{L}_{\rm CT} = 
2 R v^{}_{\rm CT} \,\sin \phi$,
$~\phi=(\psi + 2 \pi w^{}_{\rm CT})/2 v^{}_{\rm CT}$, 
$~\psi=\theta_2-\theta_1$, $v^{}_{\rm CT}$ and $w^{}_{\rm CT}$
are the numbers of chords and rotations around the symmetry center
along the CT, respectively (Section IVB4). 
Using obvious geometrical
relations, for cylindrical $\rho$-components of the initial momentum, $p_{\rho 1}$, and the
final spatial coordinate, $\rho_2$, one may find rather lengthy
analytical expressions as functions of the initial and finite
spherical coordinates of the CT. 
However, we may transform the integration
variables $\r_1$ and $\r_2$ to the specific Wigner's $\r,\s$
(\ref{transform}) which are related, in the nearly local
approximation $s/R << 1$, to the special spherical-coordinate system
with the $z'$ axis 
directed to the initial point $\r_1$, as displayed
in Fig.\  \ref{fig26} 
. In these new coordinates, the ratio of the
momentum $p_{\rho\;1}$ to the coordinate $\rho_2$ is 
given by $p_{\rho 1}/\rho_2=p \,\cos \phi/(R\,\sin \psi)$. After substitution of
this ratio into (\ref{jac}), one obtains 
\begin{eqnarray}\label{jac0} 
&\mathcal{J}_{\rm CT}\left(\p_1,t_1;\r_2,\vareps\right)= 
\frac{m^2\, \cos \phi}{2 v^{}_{\rm CT}
R^2\, \sin \phi\, \sin \psi}\;,\nonumber\\ 
&\mathcal{J}_{{\rm CT}_0}=\frac{m^2}{s^2}\;, \quad 
s=\mathcal{L}_{{\rm CT}_0}=\vert \r_1 -
\r_2 \vert\;. 
\end{eqnarray}
With this Jacobian 
$\mathcal{J}_{{\rm CT}_0}$ at $~v^{}_{{\rm CT}_0}=1$ and
$~w^{}_{{\rm CT}_0}=0$,
neglecting the Maslov phase, related 
to the caustic and turning points,
for small enough $s/R$ we arrive approximately at (\ref{GCT0}).

\subsection{Calculations of the inertia}
\l{appA23}

Within the nearly local approximation (i), the expression (\ref{GCT0}) 
for the Green's function component $G_{{\rm CT}_0}$ can be
applied in (\ref{mQQ0}) for the inertia $M_{\rm QQ}(0)$. For the
integration over $\s$ in (\ref{mQQ0}) we may use the spherical
coordinate system with the center  
at the point $\r \approx \r_1$ for
a given $\r$ and the polar $z$ axis along $\r^{}_1$ (Fig.\ \ref{fig1}). The
integration over $\r$ can be performed in the usual 
spherical-coordinate system with the symmetry center 
of the spherical box and  $z$ axis. The NLA
(i) and this coordinate system simplify
much the integration limits. We may subtract and add identically the
same local part with its unlocal surface correction,
$\tilde{M}_{\rm QQ}^{(\rm{0})}(0)$ (\ref{mQQ}), separating the
correlation-like terms in the integrand of (\ref{mQQ0}).
Introducing also dimensionless variables, $\wp=r/R$, $\sigma_s=s/R$
(for simplicity, the subscript $s$ in $\sigma_s$ 
will be omitted sometimes 
in this Appendix), $u=kR$,  one obtains
\bea\label{mQQ1}
\hspace{-1.0cm}\tilde{M}_{\rm{QQ}}(0)&=&
\tilde{M}_{\rm{QQ}}^{(\rm{0})}(0)+\tilde{M}_{\rm{QQ}}^{(1)}(0)\nonumber\\
&&+\tilde{M}_{\rm{QQ}}^{(2)}(0)+\tilde{M}_{\rm{QQ}}^{(3)}(0)\;, 
\eea
where
\begin{eqnarray}\label{mQQ00}
\hspace{-1.0cm}\tilde{M}_{\rm QQ}^{(0)}(0)&=&\frac{d_s m^3 R^{2\lambda+6}}{4 \pi
\hbar^4}\langle \int \limits_{0}^{1} \d \wp \;\wp^{2(\lambda+1)}\nonumber\\
&\times& \int
\limits_{0}^{1 + \wp} \d \sigma \;\sigma^2\; \tilde{B}\left(u^{}_{F}
\sigma\right) \; \rangle_{\rm av}\;, 
\end{eqnarray}
\bea\label{mQQ10}
\hspace{-0.4cm}\tilde{M}_{\rm QQ}^{(1)}(0)&=&\frac{d_s m^3 R^{2\lambda+6}}{\pi^2
\hbar^4}\langle \int \limits_{0}^{1} \d \wp \;\wp^{2(\lambda+1)} \nonumber\\ 
&\times&\int
\limits_{0}^{1 + \wp} \d \sigma \;\sigma^2 \;\Delta_{Q}(\wp,\sigma)\;
\tilde{B}\left(u^{}_{F} \sigma\right)\rangle_{\rm av}\;,
\eea
\begin{eqnarray}\label{mQQ01}
\hspace{-0.6cm}\tilde{M}_{\rm QQ}^{(2)}(0)&=&\frac{d_s m^3 R^{2\lambda+6}}{\pi^2
\hbar^4}\langle \int \limits_{0}^{1} \d \wp \;\wp^{2(\lambda+1)} \nonumber\\
&\times&\int
\limits_{0}^{1 + \wp} \d \sigma \;\sigma^2 \; \Delta_{B}\left(u^{}_{F}
\sigma\right)\rangle_{\rm av}\;,
\end{eqnarray}
\begin{eqnarray}\label{mQQ11}
\tilde{M}_{\rm QQ}^{(3)}(0)&=&\frac{d_s m^3 R^{2\lambda+6}}{\pi^2
\hbar^4}\langle \int \limits_{0}^{1} \d \wp \;\wp^{2(\lambda+1)} \nonumber\\
&\times&\!\!\!\!\!\int
\limits_{0}^{1 + \wp} \d \sigma \;\sigma^2 \;\Delta_{Q}(\wp,\sigma)\;
\Delta_{B}\left(u^{}_{F} \sigma\right)\rangle_{\rm av}\;. 
\end{eqnarray}
We introduce $w=k^{}_{F}R \sigma_s~$, $j_1(x)$ as the spherical Bessel
function,  and 
$\mbox{Si}(x)$ as the integral sine. 
The correlation-like functions are denoted by
 $\Delta$ in (\ref{mQQ10})--(\ref{mQQ11}). One of them is defined
 by $\Delta_{B}=B-\tilde{B}$, with 
\begin{eqnarray}\label{Jwf}
B(w) &=&\int \limits_{0}^{w} \d x\; \sin(x)\;j_1(x) \nonumber\\
&=&
\frac{1}{2}\;\mbox{Si}\left(2 w\right) -\frac{1}{2 w} \left[1 -
\cos\left(2 w\right)\right]  
\nonumber\\ &\rightarrow& \frac{\pi}{4} -
\frac{1}{2 w} +\frac{1}{4 w} \nonumber\\
&\times&\cos\left(2 w\right) - \frac{1}{8 w^2}
\sin \left(2 w\right) + \cdots\;.
\end{eqnarray}
The SCM energy spectrum averaging of $B$
 over $k^{}_{F}R$ is denoted by $\tilde{B}(u^{}_{F} \sigma_s)$. 
The other correlation-like function,
 $\Delta_{Q}$, is given by
\begin{eqnarray}\label{QQint}
&&\Delta_{Q}\left(\frac{\sigma}{\wp}\right)= \frac{1}{4 \pi
r^{2\lambda}}\int \d \Omega \int \d \Omega_{s}\times\nonumber\\
&\times& \left[Q\left(\r +
\frac{\s}{2}\right) Q\left(\r -
\frac{\s}{2}\right)-r^{2\lambda}Y_{\lambda 0}^{2}(\cos \theta)\right]\nonumber\\
&=& c_{\lambda}^{(2)}\left(\frac{\sigma_s}{\wp}\right)^2+
c_{\lambda}^{(4)}\left(\frac{\sigma_s}{\wp}\right)^4 \nonumber\\ 
&+&
c_{\lambda}^{(6)}\left(\frac{\sigma_s}{\wp}\right)^6+ \cdots, 
\end{eqnarray}
 where $c_{2}^{(2)}=-5/6$, $c_{2}^{(4)}=1/16$, $c_{2}^{(n \geq 6)}=0$
at $\lambda=2$, and $c_{3}^{(2)}=-7/4$, $c_{3}^{(4)}=7/16$,
$c_{3}^{(6)}=-1/64$, $c_{3}^{(n\geq 8)}=0$ at $\lambda=3$ etc. The
integrals (\ref{QQint}) were evaluated over all possible
spherical angles of the vectors $\r$ and $\s$ in the considered nearly
local approximation (i), where the only small values $s/R$ of 
order of a
few relative wave lengths, $1/k^{}_FR$, give the leading contributions;
$\d \Omega= \sin \theta \d \theta \d \varphi$, $\d \Omega_{s}= \sin
\theta_{s} \d \theta_{s} \d \varphi_{s}$. The integration over the
modulus of vector $\s$ was extended approximately to the maximal one
for a given $r \approx r_1$. Then, we integrated over all such modules
of vector $\r$ within the approximation mentioned above.

The phase-space averaging in (\ref{mQQ00})--(\ref{mQQ11}) can be
exchanged with the integrations over the spatial coordinates. For
calculations of the inertia $\widetilde{M}_{\rm QQ}(0)$ (\ref{mQQ0}), the
function $B(k^{}_{F}R \sigma_s)$ (\ref{Jwf})  
can be expanded in small
semiclassical parameter $1/k^{}_{F}R$, see the asymptotics in (\ref{Jwf})
for large arguments. As seen from this asymptotics, its 
oscillating terms are removed by
Strutinsky averaging over
$u^{}_{F}=k^{}_{F}R$
\cite{strut,fuhi,sclbook,MGFnpae2005,MGFyaf2007,MGFnpae2008},
\begin{eqnarray}\label{Jwfavdef}
M_{\Gamma }\left(u^{}_{F}\sigma_s\right)&=& \int \limits_{-\infty}^\infty
{\mbox d} x\, B\left[\left(u^{}_{F}+x\Gamma \right)\sigma_s\right] 
\nonumber\\
&\times&\left(1+
x \Gamma/u^{}_{F} \right)^{2(\lambda+3)}\,f_{\rm av}^{(2 \mathcal{M})}(x)\;, \nonumber\\
f_{\rm av}^{(2 \mathcal{M})}(x)&=&\frac{1}{\sqrt{\pi}}\,e^{-x^2}\, 
\mbox{P}_{2\mathcal{M}}(x)\;. 
\end{eqnarray}
The correction polynomial of the order of $2 \mathcal{M}$, 
$~\mbox{P}_{2 \mathcal{M}}(x)= \sum \limits_{\tau=0,2,...}^{2 \mathcal{M} }
v_{\tau}\,H_{\tau}(x)$, is defined through the recurrence
relations, $~v_{\tau}=-v_{\tau-2}/2\tau,~ v_0=1$,
$~H_{\tau}(x)$ is the standard Hermite polynomial. The second
multiplier in the integrand of (\ref{Jwfavdef}) takes into account
that we average 
over $R$ [or really over the particle number A, according to
(\ref{akrTF})] from the variable $k^{}_{F}R$ for a fixed $k^{}_{F}$ related to
the well-known value of infinite matter particle density, see the
main text after (\ref{mfmirrA}).

As shown in \cite{GMFprc2007}, using the Strutinsky averaging over 
$k^{}_FR$
which removes oscillations,  one asymptotically obtains a smooth quantity:
\begin{equation}\label{Jwfav}
\widetilde{B}\left(u^{}_{F} \sigma_s\right) = \frac{\pi}{4}-\frac{1}{2 u^{}_{F}
\sigma_s}\;,
\end{equation}

According to (\ref{mQQ01}), (\ref{mQQ11}) and (\ref{Jwfav}), the
SCM average of the correlation-like terms
$\widetilde{M}_{\rm QQ}^{(2)}(0)$ (\ref{mQQ01}) and
$\widetilde{M}_{\rm QQ}^{(3)}(0)$ (\ref{mQQ11}) are zeros because
such an averaging is performed in $k^{}_{F}R~$, 
and these quantities are
linear in $\Delta_{B}$, i.e. by definition, $\widetilde{\Delta}_{B}=0$.
The part of $M_{\rm QQ}^{(1)}(0)$, see (\ref{mQQ10}), related to
the constant $\pi/4$ in $\widetilde{B}(u^{}_{F}\sigma_s)$ (\ref{Jwfav}) can be
neglected as expressed through the linear correlation function,
$\langle Q(\r+\s/2)Q(\r-\s/2) - Q^{2}(\r)\rangle_{\rm av}$, averaged
in phase-space variables 
\cite{kadbaym,kolmagpl} 
(Section \ref{semicl}).

Integrating now analytically the remaining integrals over 
$\sigma(=\sigma_s)$ and
$\wp$ in both the equation (\ref{mQQ00}) for $M_{\rm QQ}^{(0)}(0)$
and the nonzero component of (\ref{mQQ10}) for
$M_{\rm QQ}^{(1)}(0)$, corresponding to the second term in
asymptotics (\ref{Jwfav}) of $\widetilde{B}$, with the help of
(\ref{Jwfav}) and (\ref{QQint}), one arrives at
\begin{eqnarray}\label{mQQVSsplit}
M_{\rm QQ}^{(0)}(0) &\approx& \widetilde{M}_{\rm QQ}^{({\rm vol})}(0) +
\tilde{M}_{\rm QQ}^{({\rm S}1)}(0)\;,\nonumber\\ 
M_{\rm QQ}^{(1)}(0)
&\approx& \tilde{M}_{\rm QQ}^{(\rm{S}2)}(0)\;,
\end{eqnarray}
where $\tilde{M}_{\rm QQ}^{({\rm vol})}(0)$ is the local volume part
(\ref{mQQ}) of the inertia related to the first constant term 
($\pi/4$) in asymptotics (\ref{Jwfav}) in (\ref{mQQ00}) [for simplicity, the
upper index $\rm{vol}$ is omitted in (\ref{mQQ})].
Formally, in the macroscopic limit, $k_{F}R \to \infty$, 
$B\left(u^{}_{F} \sigma\right)$ (\ref{Jwf}) before and after SCM
averaging [see the definition (\ref{Jwfavdef})] tends to the edge-like
function with the constant asymptotic value of $\widetilde{B} \rightarrow 
\widetilde{B}(\infty)=\pi/4$ at all $\sigma_s$, see (\ref{Jwfav}), 
corresponding exactly to the volume local approximation (\ref{mQQ}) to
(\ref{mQQ0}). For its two surface corrections one finds
\begin{eqnarray}\label{mQQ1cor12}
\widetilde{M}_{\rm QQ}^{(S1)}(0)&=&\frac{d_s m^3 R^{2\lambda+6}
\zeta_{\lambda}^{(1)}}{12\pi^2 \hbar^4\;u^{}_{F}}\;,\nonumber\\
\widetilde{M}_{\rm QQ}^{(S2)}(0)&=&
\frac{d_s m^3 R^{2\lambda+6}\zeta_{\lambda}^{(2)}}{12\pi^2 \hbar^4 u^{}_{F}}\;, 
\end{eqnarray}
where $\zeta_{\lambda}^{(1)}$ and $\zeta_{\lambda}^{(2)}$ are number constants
given immediately after (\ref{mFmirrsurfcor}).

Collecting all the volume (local) (\ref{mQQ}), the surface
[relatively $\propto 1/k^{}_{F}R$, (\ref{mQQ1cor12}) of
(\ref{mQQVSsplit})], and curvature [ $\sim
1/(k^{}_{F}R)^2$] corrections from (\ref{kld}) for $\kappa^{}_{\rm{FF}}$,
(\ref{coupconsurfcor}) for $\kappa^{}_{\rm QQ}$, as well as originated by
means of (\ref{g2chm}), as consistency corrections [ $\propto
\gamma_{\rm QQ}^2(0)$] in (\ref{trcoll}), one finally arrives at the
complete inertia $M_{\rm FF}$, see (\ref{mFmirrsurfcor}).

\subsection{The friction coefficient}
\label{appA424}

Using the approximation (\ref{GCT0}) for the Green's function component
$G_{{\rm CT}_0}$ in the friction $\widetilde{\gamma}_{\rm QQ}(0)$ (\ref{gammaQQ0})
and the same coordinate systems, as in the derivations 
of the inertia $\widetilde{M}_{\rm QQ}(0)$, in the nearly
local case (i) one obtains
\begin{equation}\label{gQQ1}
\widetilde{\gamma}_{\rm QQ}(0)=
\widetilde{\gamma}_{\rm QQ}^{(0)}(0)+\widetilde{\gamma}_{\rm QQ}^{(1)}(0)+
\tilde{\gamma}_{\rm QQ}^{(2)}(0)\;,
\end{equation}
where $\widetilde{\gamma}_{\rm QQ}^{(0)}(0)$ is the volume 
local part (\ref{gammaQQ}),
\begin{eqnarray}\label{gQQ01}
\hspace{-0.5cm}\widetilde{\gamma}_{\rm QQ}^{(1)}(0)&=&\frac{d_s m^2 R^{2\lambda+4}}{\pi^2 \hbar^3}\langle
\int \limits_{0}^{1} \d \wp \;\wp^{2(\lambda+1)} \nonumber\\
&\times&\int \limits_{0}^{1 + \wp} \d \sigma \; \left[
\sin^{2}\left(u^{}_{F} \sigma\right) - \frac{1}{2}\right]
\rangle_{\rm av}\;, 
\end{eqnarray}
\begin{eqnarray}\label{gQQ11}
&&\tilde{\gamma}_{\rm QQ}^{(2)}(0)=\frac{d_s m^2 R^{2\lambda+4}}{\pi^2 \hbar^3}\langle
\int \limits_{0}^{1} \d \wp \;\wp^{2(\lambda+1)}\nonumber\\ 
&&\times\int \limits_{0}^{1 + \wp} \d \sigma \;\Delta_{Q}(\sigma/\wp) 
\left[
\sin^{2}\left(u^{}_{F} \sigma\right) - \frac{1}{2}\right]
\rangle_{\rm av}\;. 
\end{eqnarray}
We neglected the linear correlation function, 
$<Q(\r+\s/2)Q(\r-\s/2) - Q^{2}(\r)>_{\rm av}~$, 
averaged over
phase-space variables, as in the derivations of the inertia 
\cite{kadbaym}.

The phase-space averaging in (\ref{gQQ01}) and (\ref{gQQ11}) 
can be exchanged with the integrations over the spatial coordinates. 
As $\langle \sin^{2}\left(u^{}_{F} \sigma\right) - \frac{1}{2}\rangle_{\rm av}=0$,
 the corrections (\ref{gQQ01}) for $\widetilde{\gamma}_{\rm QQ}^{(1)}(0)$ and 
(\ref{gQQ11}) for $\widetilde{\gamma}_{\rm QQ}^{(2)}(0)$ are zeros. Therefore, 
we are left with the single local term (\ref{gammaQQ})  for 
$\widetilde{\gamma}_{\rm QQ}(0)$ within the considered NLA
(i).

\subsection{The isolated susceptibility}
\label{appA234}
\bigskip

Similarly, like in the case of the inertia and friction derivations, 
for the averaged isolated susceptibility, 
\begin{eqnarray}\label{chiQQ0}
\widetilde{\chi}_{\rm{QQ}}(0) &=& \frac{2 d_s}{\pi}\,
\langle \int \rm{d}\r
\int \d \s \;\nonumber\\
&\times&\hat{\rm{Q}}\left(\r+\frac{\s}{2}\right)\;
\hat {\rm Q}\left(\r-\frac{\s}{2}\right)
\int \limits_0^\infty \d \vareps \,n(\vareps)
\nonumber\\
&\times&
\Im G_{{\rm CT}_0}\left(\r+\frac{\s}{2},\r-\frac{\s}{2};\vareps\right)
\nonumber\\
&\times&\Re G_{{\rm CT}_0}\left(\r+\frac{\s}{2},\r-\frac{\s}{2};\vareps\right)
\rangle_{\rm av}\;, 
\end{eqnarray}
one finds 
\begin{equation}\label{cQQ1}
\widetilde{\chi}_{\rm QQ}(0)=
\widetilde{\chi}_{\rm QQ}^{(0)}(0)+\widetilde{\chi}_{\rm QQ}^{(1)}(0)+
\widetilde{\chi}_{\rm QQ}^{(2)}(0)\;.
\end{equation}
Here $\widetilde{\chi}_{\rm QQ}^{(0)}(0)$ is the volume local part (\ref{chiQQ}),
\bea\label{cQQ01}
&\hspace{-0.8cm}\widetilde{\chi}_{\rm QQ}^{(1)}(0)=
-\frac{d_s m y^{}_{F}R^{2\lambda+2}}{2 \pi^2 \hbar^2}\langle
\int \limits_{0}^{1} \d \wp \;\wp^{2(\lambda+1)}\nonumber\\
&\times j_{0}\left(u^{}_{F}(1+\wp)\right)
\rangle_{\rm av}\;, 
\eea
\begin{eqnarray}\label{cQQ11}
&\hspace{-0.8cm}\widetilde{\chi}_{\rm QQ}^{(2)}(0)=
\frac{d_s m y_F^2 R^{2\lambda+4}}{\pi^2 \hbar^2}\langle
\int \limits_{0}^{1} \d \wp \;\wp^{2(\lambda+1)}\nonumber\\ 
&\times\int \limits_{0}^{1 + \wp} \d \sigma \;Q_{\rm Q}(\sigma/\wp)\; 
j_{1}\left(u^{}_{F} \sigma\right)\rangle_{\rm av}\;, 
\end{eqnarray}
$j_n(x)$ is the spherical Bessel function. We neglected again the linear correlation function, $<Q(\r+\s/2)Q(\r-\s/2) - Q^{2}(\r)>_{\rm av}$, averaged in phase-space variables, as above. 
Note that the splitting into the two terms, the local 
$\widetilde{\chi}_{\rm QQ}^{(0)}(0)$ (\ref{chiQQ})
and its correction $\widetilde{\chi}_{\rm QQ}^{(1)}(0)$ was found after the integration of $j_1(w)$, 
over its argument, $w=2 u^{}_{F}\sigma_s$ in (\ref{chiQQ0}), as the
values in lower and upper limits
of the integrand. 

The phase-space averaging in (\ref{cQQ01}) and (\ref{cQQ11}) can be again exchanged with the integration over the spatial radial coordinate $\wp$. 
Therefore, the leading term of the average $<j_1(u^{}_{F}\sigma_s>_{\rm av}$ 
at large $k^{}_{F}R$ is vanishing because of its SCM part, 
and the corrections (\ref{cQQ01}) for $\widetilde{\chi}_{\rm QQ}^{(1)}(0)$ and
(\ref{cQQ11}) for $\widetilde{\chi}_{\rm QQ}^{(2)}(0)$ 
are approximately zeros. Thus, in the NLA
case (i) we are left with the only local term $\chi_{\rm QQ}^{(0)}(0)$ (\ref{chiQQ}).

\section{
WIGNER-KIRKWOOD METHOD  \\ FOR DISTRIBUTION FUNCTION DENSITIES}
\l{appA3}

The Wigner-Kirkwood method starts with the Gibbs operator \cite{sclbook},
$\hat{C}_\beta\!=\!\exp(-\beta \hat{H})$,
where $\hat{H}$ is the s.p.\ quantum-mechanical Hamiltonian.  
In the case that
$\hat{H}$ is time independent, the coordinate-space  
representation of the Gibbs operator, the so-called Bloch density matrix,
  is given by
\be\l{gibbscoor}
\hspace{-0.3cm}C(\r^{}_1,\r^{}_2;\beta)\!=\!
\sum_i\overline{\psi}_i(\r^{}_1)\exp(\!-\beta\;\varepsilon_i)\psi_i(\r^{}_2),
\ee
where $\psi_i$ and $\varepsilon_i$ are the eigenfunctions and eigenvalues
of the Hamiltonian ($\hat{H} \psi_i\!=\!\varepsilon_i \psi_i$). 
Therefore, after  formally replacing $\beta\!=\!it/\hbar$, the  
Bloch density matrix $C(\r^{}_1,\r^{}_2;\beta)$
is seen to be nothing but  the s.p.\   
time-dependent 
propagator (Green's function)  
$K(\r^{}_1,\r^{}_2;t)$ and one can use
the corresponding Schr{\"o}dinger
equation for the calculation of $C(\r^{}_1,\r^{}_2;\beta)$ \cite{sclbook}.
Note that the POT in the extended Gutzwiller version 
starts with the solution of this equation for the
propagator $K(\r^{}_1,\r^{}_2;t)$ in terms of the 
Feynman path integral. Its calculation by the stationary phase method 
leads to the semiclassical expression for $K(\r^{}_1,\r^{}_2;t)$, 
and then, one can get
the semiclassical expansion of the Green's function, 
$G(\r^{}_1,\r^{}_2; \varepsilon)$,
and its traces,  
namely the level density, $g(\varepsilon)$, and the 
particle density $\rho(\r)$ 
(at $\r^{}_1\!\rightarrow\! \r^{}_2\!=\!\r$, see Section II).
The shell components of these densities can be
 expressed in terms of the closed trajectories (see the main text
for the case of the oscillating level-density part written in terms
of POs). 
Thus, the POT can be developed for the Bloch density matrix 
$C(\r^{}_1,\r^{}_2;\beta)$ itself.

In order to solve semiclassically the Schr{\"o}dinger equation for 
the Bloch function
$C(\r^{}_1,\r^{}_2;\beta)$, one can make a transformation, first from 
$\r^{}_1$ and $\r^{}_2$ to the center-of-mass and relative coordinates, 
$\r\!=\!(\r^{}_1\!+\!\r^{}_2)/2$ 
and $\s\!=\!\r^{}_2\!-\!\r^{}_1$, and then, by the Fourier transformation to the
phase-space variables, $\{\r,\p\}$,  
what corresponds to a Wigner 
transformation from 
$C(\r^{}_1,\r^{}_2;\beta)$ to $C_W(\r,\p;\beta)$,
\bea\l{wigner}
&&C_W(\r,\p;\beta)=
\int \frac{\d \s}{(2 \pi \hbar)^3}\nonumber\\ 
&\times& C(\r-\s/2,\r+\s/2;\beta)\;
\exp\left(i\p \s/\hbar\right)\;.
\eea
This reduces one complicated Schr\"odinger equation
to an infinite system of much simpler first-order ordinary
differential equations
(at each power of $\hbar$,
which can be analytically integrated \cite{sclbook})  .

The advantage of the Wigner-Kirkwood method is obviously to generate
   smooth quantities averaged over many quantum states to smooth out quantum
   oscillations as 
shell effects.
The POT on the contrary is aimed at 
the derivation
of analytical expressions for the shell components of the partition 
function, and thereby of the level and  particle densities. 
In the Wigner-Kirkwood method, the main term of the expansion
of $C_W(\r,\p;\beta)$ is proportional to the classical distribution function 
$f_{\rm cl}(\r,\p)$, and $\hbar$ corrections can be obtained 
by solving
a simple system of differential equations at each power of $\hbar$.
 Strictly speaking, there is no convergence of this asymptotical 
expansion because
of presence of the $\hbar$ in the rapidly oscillating exponents.  
Therefore, to get the convergent series in $\hbar$
of the ETF approach, 
one first has to use local averaging in the phase 
space variables, and then, 
expand smooth quantities in a $\hbar$ series,  
in contrast to the shell-structure
POT. In this way, the simple 
ETF $\hbar$ expansions of    local quantities, such as the
   particle density   $\rho(\r)$, kinetic energy density   $\tau(\r)$, and
   level density   $g(\varepsilon)$ are obtained.

The canonic partition function $\mathcal{Z}(\beta)$  is derived
by integrating over the whole space the diagonal Bloch matrix 
$C(\r,\r;\beta)\!=\!C(\r;\beta)$,
\be\l{partfun}
\mathcal{Z}(\beta)=\int \d \r\; C(\r;\beta)=
\sum_i \exp (-\beta \varepsilon_i)\;.
\ee
The trace, $\mathcal{Z}\!=\!\mbox{Tr} [\exp(-\beta \hat{H})]$, 
can be taken for any complete set of states. 
For the semiclassical expansion involving an integral 
over the phase
space, it is more convenient
to take plane waves as the complete set. We may then write
\be\l{partfunps}
\mathcal{Z}(\beta)=
\int \frac{\d \r\; \d \p}{(2 \pi \hbar)^3} e^{-i \p \r/\hbar}\;
e^{-\beta \hat{H}}\;e^{i \p \r/\hbar}\;.
\ee
As the kinetic operator in $\hat{H}$ does not commute with the potential
$V(\r)$, it is convenient to use the following representation \cite{sclbook}:
\be\l{wudef}
e^{-\beta \hat{H}}e^{i \p \r/\hbar}=e^{-\beta H_{\rm cl}}\;
e^{i \p \r/\hbar}\;w(\r,\p;\beta)\;,
\ee
where $H_{\rm cl}$ is the classical Hamiltonian that appears in
 (\ref{distrfuntf}) and (\ref{fetf1}). 
(Subscript in $H_{\rm cl}(\r,\p)$ was omitted in the main text.)
Solving
the Schr{\"o}dinger equation for the function $w$,
\bea\l{weq}
&\hspace{-0.9cm}\frac{\partial w}{\partial \beta}=-i\hbar\left[
\frac{\beta}{m} \left(\p\cdot \nabla V\right)w -\frac{1}{m}
\left(\p\cdot \nabla w\right)\right]\nonumber\\
&+ \hspace{-0.5cm}\frac{\hbar^2}{2m}\left[\beta^2 \left(\nabla V\right)^2w -
\beta \left(\nabla^2 V\right) w \right.\nonumber\\
 &+\left.\nabla^2w -
2\beta \left(\nabla V \cdot \nabla w\right)\right]\;,
\eea
with the boundary condition $\mbox{lim}^{}_{\beta \to 0}w(\r,\p;\beta)\!=\!1$,
one assumes that $w(\r,\p;\beta)$ can be expanded in  
a power series in 
$\hbar$:
\be\l{wexp}
w=1+\hbar w_1 + \hbar^2 w_2 + \cdots \;.
\ee
Equating
terms of the same power in $\hbar$ from both sides of this differential
equation, one obtains 
the $\hbar$ corrections:
\begin{equation}\label{w1}
w_1=-\frac{i\beta^2}{2m}\;\p\cdot \nabla V\;, \hspace{1.8cm}  
\end{equation}
and
\bea\label{w2}
&&w_2=-\frac{\beta^2}{4m} \nabla^2V+ \frac{\beta^3}{6m} 
\left(\nabla V\right)^2\;\nonumber\\
&-&\frac{\beta^4}{8m^2} \left(\p\cdot \nabla V\right)^2
+\frac{\beta^3}{6m^2} \left(\p\cdot \nabla\right)^2 V\;.
\eea
The semiclassical series for the  partition function takes then the form:
\bea\l{partfunpsexp}
\mathcal{Z}(\beta)&=&
\int \frac{\d \r\; \d \p}{(2 \pi \hbar)^3} e^{-i \p \r/\hbar}\;
e^{-\beta H_{\rm cl}}\nonumber\\
&\times&\left(1+\hbar w_1 + \hbar^2 w_2 + \cdots \right)\;.
\eea
Differentiating the TF particle density $\rho^{}_{\rm TF}$ (\ref{tfden})
and solving the obtained linear
system of equations for the gradients 
of the potential, 
one finds 
\be\l{dv2}
\left(\nabla V\right)^2= 
\left(\frac{\pi^2\hbar^2}{m (3\pi^2 \rho)^{1/3}}\right)^2
\left(\nabla \rho\right)^2,
\ee
\be\l{d2v}
\nabla^2V=\frac{\pi^2\hbar^2}{m (3\pi^2\rho)^{1/3}}
\left[
\frac{\left(\nabla \rho\right)^2}{3\rho}
-\nabla^2 \rho\right]\;,
\ee
where the subscript TF on the density has been omitted. 
These expressions are more convenient to use in the more general case,
   including billiard systems, in particular, the spheroidal cavity.

For calculations of the semiclassical distribution function 
$g(\r,\p;\varepsilon)$, one can apply the inverse 
Laplace transformation:
\bea\l{laplac}
& \hspace{-0.5cm}g(\r,\p;\varepsilon) \!=\! \frac{\partial f(\r,\p)}{\partial \varepsilon}
    \!=\! \frac{1}{2 \pi i}\; \int_{\beta_r - i \infty}^{\beta_r + i \infty} d\beta \; 
\nonumber\\
&\times\exp\left[\beta \left(\varepsilon - H_{\rm cl}\right)
\right]
 \left(1+\hbar w_1+\hbar^2 w_2\right)\;,  
\eea
 where $w_1$ and $w_2$ are the semiclassical corrections 
of (\ref{w1}) and (\ref{w2}).
 The integration in the complex $\beta$ plane in (\ref{laplac}) has to be
 taken along the imaginary axis, at a distance $\beta_r$ such that all
 singularities are located at its left. 
The linear term in $\hbar$, i.e.\ , the term $w_1$ that is linear in $\p$,
does not
 contribute to the phase-space (momentum) integral for the energy $\vareps$ and
 for the MI $\Theta$ in (\ref{misclps}). 
 Calculating the integral in (\ref{laplac}) using (\ref{w2}),
 one arrives, after some simple algebraic transformations, at 
 (\ref{fetf1}).
%

\clearpage
\widetext
\vspace{0.2cm}
\noindent
\hspace{-1.0cm}
\begin{table*}[pt]
\ms
\begin{center}
\begin{tabular}{|c|c|c|c|c|c|c|c|c|c|c|c|}
\hline
 & SkM$^*$ & SGII & SLy5 &SLy5$^*$
& SLy6 & SLy7 & SVs28 & SVs32 &
SVm08 & SVK226 & SVk02\\
\hline
$\overline{\rho} $  & 0.16 &  0.16 & 0.16 & 0.16
  & 0.17 & 0.16 & 0.16 & 0.16 &
 0.16 & 0.16 & 0.16  \\
fm$^{-3}$ &  &   &  &   &  &  &  &  & &  & \\
$b^{}_{V}$  & 15.8&   15.6 & 16.0 & 16.0
   & 17.0 & 15.9 & 15.9 & 15.9 &
 15.9 & 15.9 & 15.9  \\
MeV &  &   &  &   &  &  &  &  & &  & \\
$K$  & 217&   215& 230 & 230
   & 245 & 230 & 234 & 234 &
 234 & 226 & 234   \\
MeV &  &   &  &   &  &  &  &  & &  & \\
$J$   & 30.0 & 26.8 & 32.0& 32.0
& 32.0 & 32.0 & 28.0 & 32.0 &
 30.0& 30.0 & 30.0  \\
MeV &  &   &  &   &  &  &  &  & &  & \\
$L$   & 47.5 & 37.7 & 48.3& 45.9
& 47.4 & 47.2 & 7.5 & 59.5 &
 42.0& 35.5 & 37.0  \\
MeV &  &   &  &   &  &  &  &  & &  & \\
$\mathcal{C}_{+}$  & 57.6 &   43.9 & 59.3 &60.1
 & 54.1 & 52.7& 49.6 & 51.8 &
 50.9& 51.4 & 50.7  \\
MeV$\cdot$fm$^{5}$ &  &   &  &   &  &  &  &  & &  & \\
$\mathcal{C}_{-}$  & -4.79 & -0.94 & -22.8  &-24.2
 & -15.6 & -13.4 & 19.6 & 26.0 &
 36.9 & 30.6 & 21.9 \\
MeV$\cdot$fm$^{5}$ &  &   &  &   &  &  &  &  & &  & \\
$c_{\rm sym}$ & 3.24 & 6.07& 1.58 & 1.54
   & 1.77& 1.95 & 1.48 & 1.40 &
 1.13 & 1.22 & 1.46 \\
$\beta$ & -0.64 & -0.54 & -0.58 & -0.52

   &-0.62 & -0.65 & -0.48 & -0.47 &
 -0.51 & -0.48 & -0.48 \\
\hline
\end{tabular}
\end{center}
\vspace{0.5cm}
\caption{ Nuclear-matter parameters of different Skyrme-force parametrizations.}
\label{table1}
\end{table*}
\noindent
\hspace{-1.0cm}
\begin{table*}[pt]
\ms
\begin{center}
\begin{tabular}{|c|c|c|c|c|c|c|c|c|c|c|c|}
\hline
 & SkM$^*$ & SGII & SLy5 &SLy5$^*$ &
SLy6 & SLy7 & SVs28 & SVs32 &
 SVm08 & SVK226 & SVk02\\
\hline
$k_{S,0}$ & -2.47 & -0.53 & -12.6 & -13.1 &
  -9.03 & -7.09 & 11.4 & 15.6 &
 37.1  & 23.7 & 12.7    \\
MeV & & & & &  & & & & & &    \\
$k^{}_S$  & -2.48 &-0.46 & -14.6 & -15.0 &
  -10.1 & -7.61 & 13.3  & 18.2 &
 46.7  & 29.5 & 14.8  \\
MeV & & & & &  & & & & & &    \\
$\nu^{}_0$  & 163 & 21.9 & 0.59 & 0.92 &
  1.21 & 1.99 & 0.90  & 0.84 &
 0.89  & 0.79 & 0.89   \\
$\nu$  & 2.27 & 1.89 & 0.28 & 0.60 &
  0.62 & 0.73& 0.58  & 0.61 &
 0.86  & 0.70 & 0.59    \\
$Q_0$  &  59642 & 29908 & 73 & 72 &
 137 & 287 & -62 & -55 & -62  & -30 &
 -63     \\
MeV & & & & &  & & & & & &    \\
$Q$  &   823 & 2570 & 42 & 41 &
  63 & 98 & -34 & -34 & -34  & -21
 & -36   \\
MeV & & & & &  & & & & & &    \\
$\tau^{}_0/I$ & 0.006 & 0.004 & 0.41 & 0.43
& 0.26 & 0.16 & 0.43 & 0.53 &
 0.040 & 0.89 & 0.45   \\
$\tau/I$ & 0.055 & 0.014 & 0.59 & 0.60 &
  0.40 & 0.28 & 0.62 & 0.73 &
 1.68 & 1.18 & 0.64  \\
$D_0$,MeV &   &  &
   &   &  & 
   &  &  &
  &  &  \\
 $^{132}$Sn & 89 & 91 & 101 & 89 & 104 & 102 & 78 & 79 & 81 & 77 &  84    \\
$D$,MeV &    &   &
 &  &   & 
   &   &   &
   &    &   \\
 $^{68}$Ni & 91 & 92 & 100 & 88 & 104  & 95  & 79 & 80 & 83 & 78 & 85   \\
~~$^{132}$Sn & 89  & 91 &
100  & 89  &  103  & 95
   & 77 &  78  &
  81  & 76   &  83   \\
~~$^{208}$Pb &  90  &  91 &
109 & 88  & 102 & 93
   & 77 & 78 &
 81 & 76  & 82  \\
\hline
\end{tabular}
\end{center}

\vspace{0.5cm}
\caption{ Different symmetry-energy coefficients for the 
Skyrme-forces.}
\label{table2}
\end{table*}

\begin{table*}[pt]
\bs
\begin{center}
\begin{tabular}{|c|c|c|c|c|c|c|c|c|c|}
\hline
$k^{}_FR_0$\textbackslash $n$ & 2 & 3 & 
 4 & 5 & 6 & 7 & 8 & 9 & 10\\
\hline
5.0 & 0.92 & 1.07 & 0.98 
& 1.02 &  0.99 & 1.01 & 1.00 & 1.00 & 1.00\\
10.0 & 0.73 & 1.15 & 0.92 
& 1.05& 0.97 & 1.02 & 0.99 & 1.01 & 0.99\\
15.0 & 0.67 & 1.19 & 0.86 
& 1.09& 0.88 & 1.04 & 0.97  & 1.02 & 0.99\\
20.0 & 0.69 & 1.20 & 0.82 
& 1.12& 0.82 & 1.07 & 0.95 & 1.04 & 0.97\\
25.0 & 0.76 & 1.19 & 0.80 
& 1.14& 0.77 & 1.08 & 0.94 & 1.05 & 0.97\\
30.0 & 0.84 & 1.16 & 0.80 
& 1.14& 0.87 & 1.10 & 0.91 & 1.07 & 0.95\\
35.0 & 0.91 & 1.13 & 0.83 
& 1.14& 0.86 & 1.11 & 0.90 & 1.07 & 0.94\\
40.0 & 0.95 & 1.10 & 0.86 
& 1.13& 0.86 & 1.11 & 0.89 & 1.08 & 0.93\\
45.0 & 0.97 & 1.08 & 0.89 
& 1.11& 0.87 & 1.11 & 0.89 & 1.09 & 0.92\\
50.0 & 0.97 & 1.06 & 0.93 
& 1.09& 0.88 & 1.10 & 0.89 & 1.09 & 0.91\\
$\infty$ & 1.11 & 0.92 & 1.06 
& 0.95 & 1.08 & 0.96 & 1.03 & 0.97 & 1.03\\
\cite{koonran} & 0.00 &  0.85 & 0.45
& 0.90 & 0.62 & 0.93 & 0.71 & 0.94 & 0.76\\
\hline
\end{tabular}
\end{center}
\caption{ Smooth friction coefficients.}
\label{table4}
\end{table*}
\begin{table*}[pt]
\bs
\begin{center}
\begin{tabular}{|c|c|c|c|c|c|c|c|c|c|}
\hline
$k^{}_FR_0$ \textbackslash $n$ &  2 & 3 & 
 4 & 5 & 6 & 7 & 8 & 9 & 10\\
\hline
5  &  1.43  & 1.98 & 2.54 & 3.62
& 3.90  & 5.00 & 5.26  & 6.37 & 6.63
\\
 &  (1.37)  & (2.05) & (2.74) &(3.42)
& (4.10)  & (4.79) & (5.47)  & (6.16) & (6.84)
\\
6  &  1.09  & 1.49 & 1.90 & 2.76
& 2.93  & 3.80 & 3.96  & 4.83 & 4.99
\\
 &   (1.03)  & (1.55) & (2.07) & (2.59)
&  (3.10)  & (3.62) & (4.14)  & (4.66) & (5.17)
\\
7  &  0.88  & 1.20 & 1.53 & 2.24
& 2.36  & 3.08 & 3.19  & 3.92 & 4.02
\\
 &  (0.84)  & (1.25) & (1.67) & (2.09)
& (2.51)  & (2.92) & (3.35)  & (3.76) & (4.18)
\\
8  &  0.75  & 1.01 & 1.27 & 1.90
& 1.97  & 2.61 & 2.67  & 3.31 & 3.37
\\
 &  (0.70)  & (1.06) & (1.41) & (1.76)
& (2.11)  & (2.46) & (2.82)  & (3.17) & (3.52)
\\
9  &  0.65  & 0.87 & 1.09 & 1.66
& 1.70  & 2.27 & 2.30  & 2.88 & 2.91
\\
 &  (0.61)  & (0.91) & (1.22) & (1.52)
& (1.83)  & (2.13) & (2.44)  & (2.74) & (3.05)
\\
10  & 0.41  & 0.77 & 0.96 &1.47
& 1.49  & 2.02 & 2.03  & 2.55 & 2.56
\\
 &  (0.54)  & (0.81) & (1.08) 
& (1.35) & (1.62)  &  (1.89) &  (2.16)  & (2.42) & (2.69)
\\
 \cite{koonran} &   & 0.050  & 
& 0.025  &   & 0.020 &  & 0.015 &  \\
\hline 
\end{tabular}
\end{center}
\caption{ Smooth inertia coefficients.}
\label{table3}
\end{table*}
\clearpage
\begin{figure}
\begin{center}
\includegraphics[width=0.6\textwidth,clip=true]{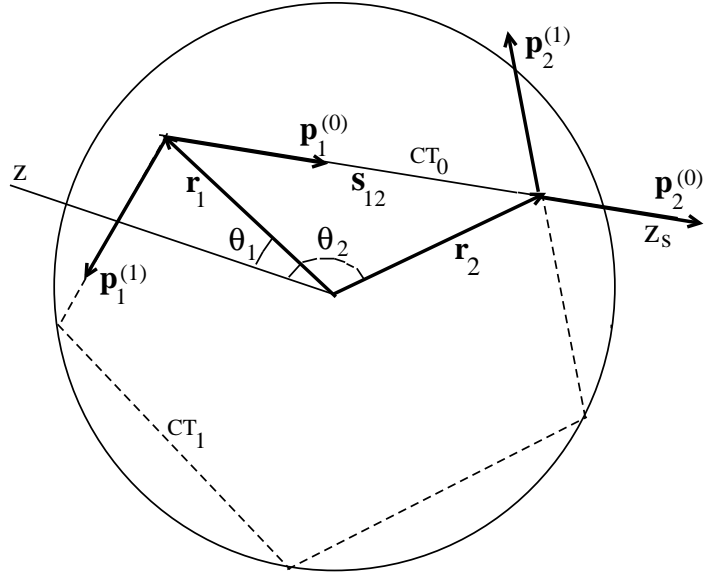}
\end{center}
\caption{Trajectories connecting points $\r_1$ and $\r_2$ without (CT$_0$; solid
  line) and with reflections (CT$_1$; dashed line). Initial 
($\p_1^{(0)}$ and $\p_1^{(1)}$) and final momenta ($\p_2^{(0)}$ and $\p_2^{(1)}$)
  of a particle at these points are also shown, together with the polar
  axises $z$ and $z_s$ and the corresponding angles $\theta_1$ and
  $\theta_2$.}
\label{fig1}
\end{figure}
\vspace{-2.0cm}
\begin{figure*}
\begin{center}
\includegraphics[width=0.75\textwidth,clip=true]{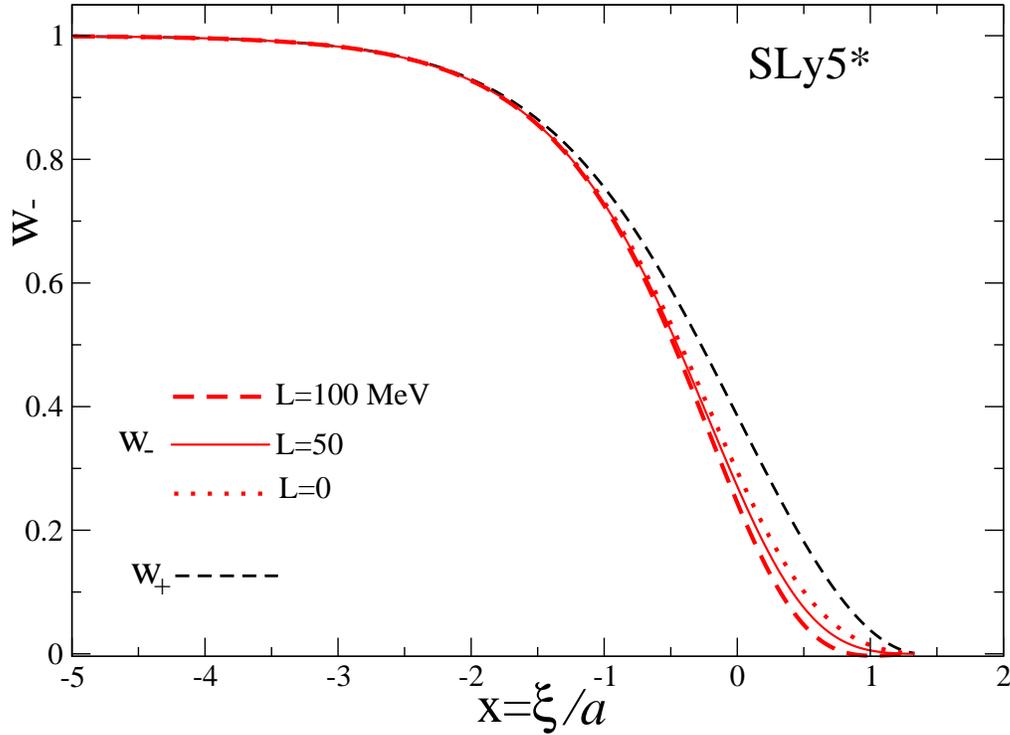}
\end{center}

\vspace{-0.8cm}
\caption{Dimensionless isoscalar and isovector particle densities, $w_{+}$ and
    $w_{-}$ respectively, as function 
of the (dimensionless) distance to
    the ES $x = \xi/a$ for the SLy5 Skyrme interaction (thin dashed and
    solid lines, respectively). To investigate the sensitivity with respect
    to variations of $L$, the isovector density is displayed for three
    different values of the droplet model parameter $L$, keeping all other
    parameters unchanged.}
\label{fig2}
\end{figure*}
\begin{figure*}
\begin{center}
\includegraphics[width=0.8\textwidth,clip=true]{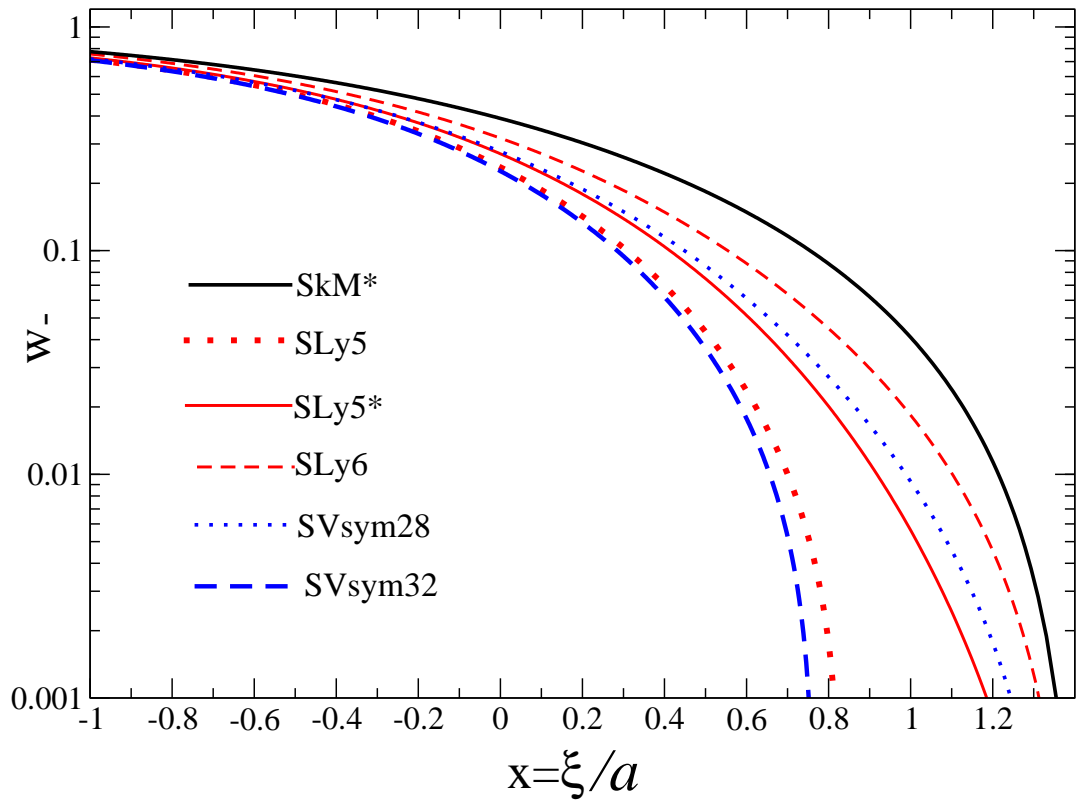}
\end{center}

\vspace{-0.2cm}
\caption{ Dimensionless isovector density $w_{-}$ 
(on a logarithmic scale) as
  function of the (dimensionless) distance to the ES $x = \xi/a$ obtained
  for several Skyrme forces within the quadratic approximation to e+[(w)].}
\label{fig3}
\end{figure*}

\vspace*{1.0cm}
\begin{figure*}
\begin{center}
\includegraphics[width=0.8\textwidth,clip=true]{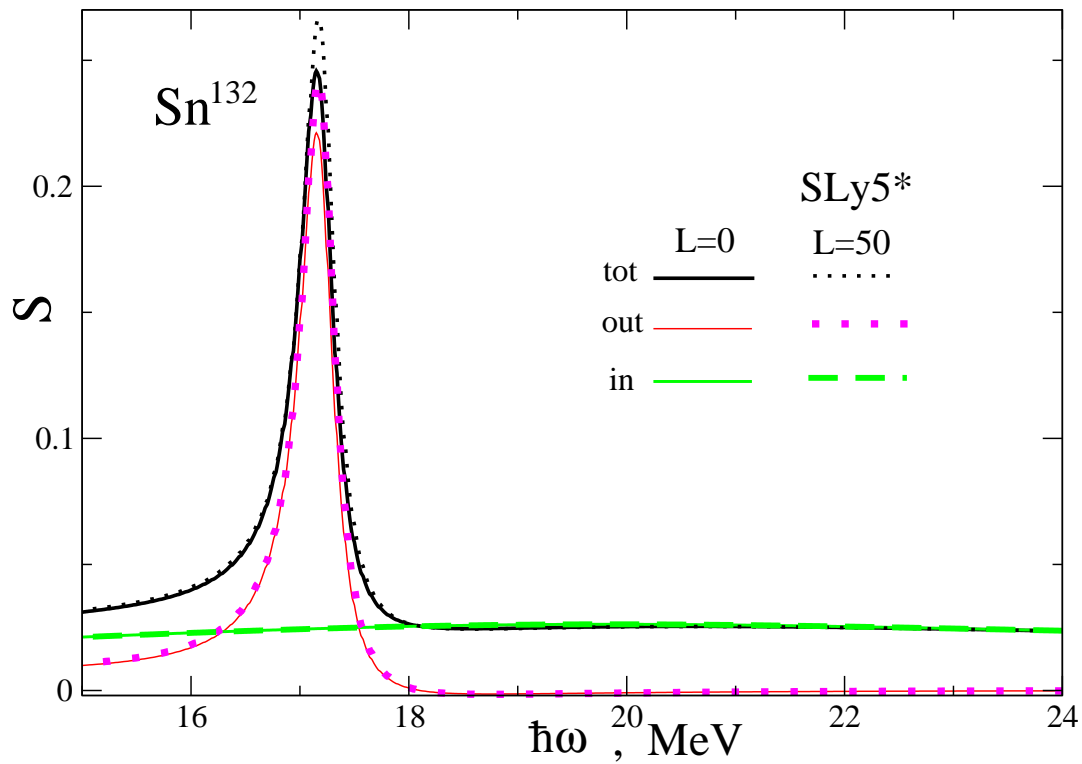}
\end{center}

\vspace{0.02cm}
\caption{ IVDR strength functions $S$ for vibrations of the 
nucleus $^{132}$Sn as
    function of the excitation energy $\hbar \omega$ obtained for the Skyrme
    force SLy5$^*$ \cite{pastore,jmeyer} with a value of $L=50$ MeV (dashed
    and dotted line) and with a zero value for $L$ (solid line). 
 Out-of-phase  and in-phase curves are 
shown separately for the main and the
    satellite excitation mode, respectively.}
\label{fig4}
\end{figure*}

\vspace*{1.0cm}
\begin{figure*}
\begin{center}
\includegraphics[width=0.7\textwidth,clip=true]{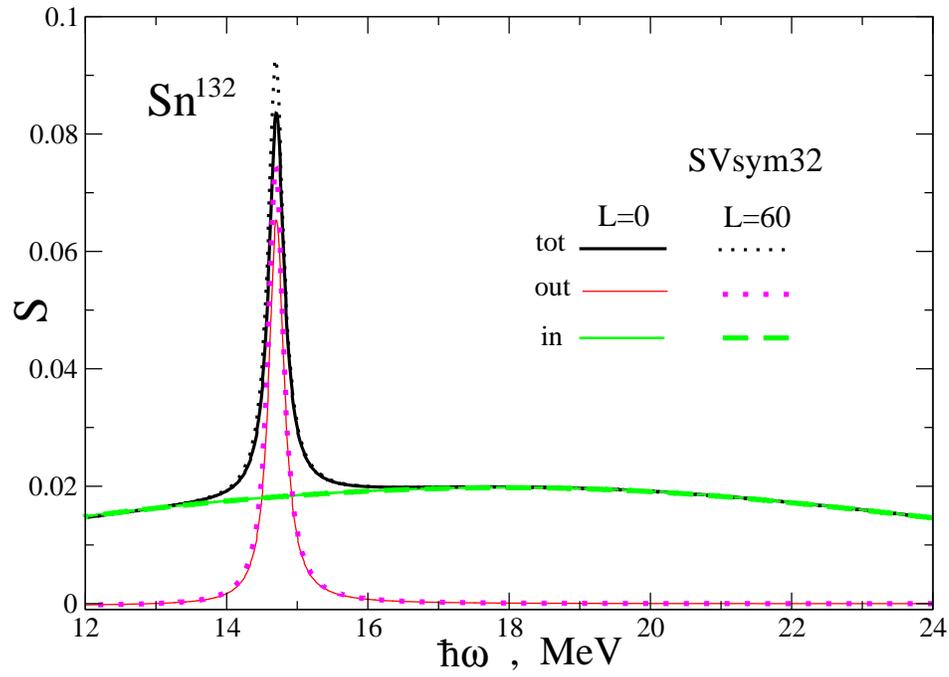}
\end{center}

\vspace{-0.1cm}
\caption{ Same as Fig.\ 4 for the Skyrme force SVsym32 
\cite{reinhardSV} for
   a value of $L \approx 60$ MeV (dashed and dotted curves). For
    comparison the curves for $L=0$ are also shown.  }
\label{fig5}
\end{figure*}
\begin{figure*}
\begin{center}
\includegraphics[width=0.9\textwidth,clip=true]{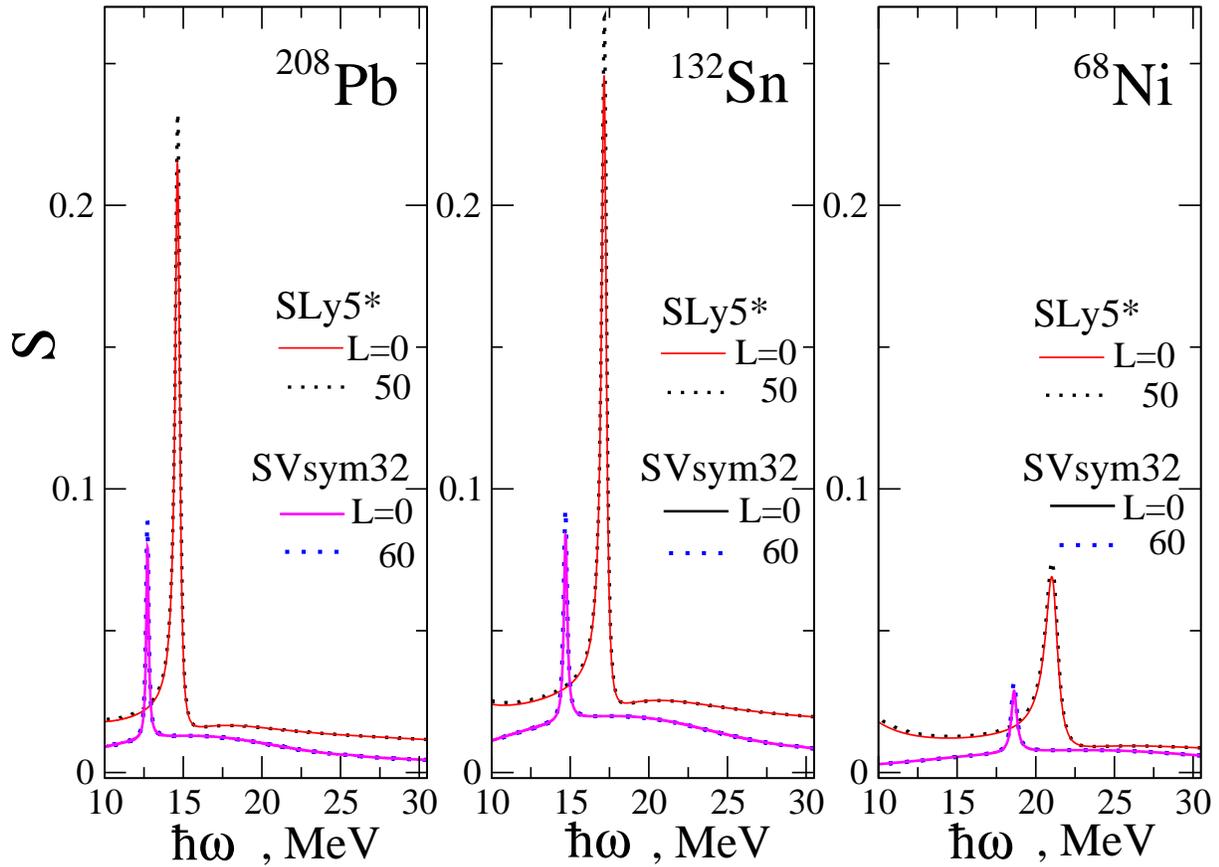}
\end{center}

\caption{ Total IVDR strength functions $S$ 
as function of the excitation energy
$\hbar \omega$ obtained for different double magic nuclei with the
    Skyrme SLy5* (upper curve) and SVsym32 (lower curve) forces. The
    sensivity of the IVDR strength on the slope parameter $L$ at the main
    peak is seen to be very weak.}
\label{fig6}
\end{figure*}
\vspace{0.5cm}
\begin{figure*}
\begin{center}
\includegraphics[width=0.98\textwidth]{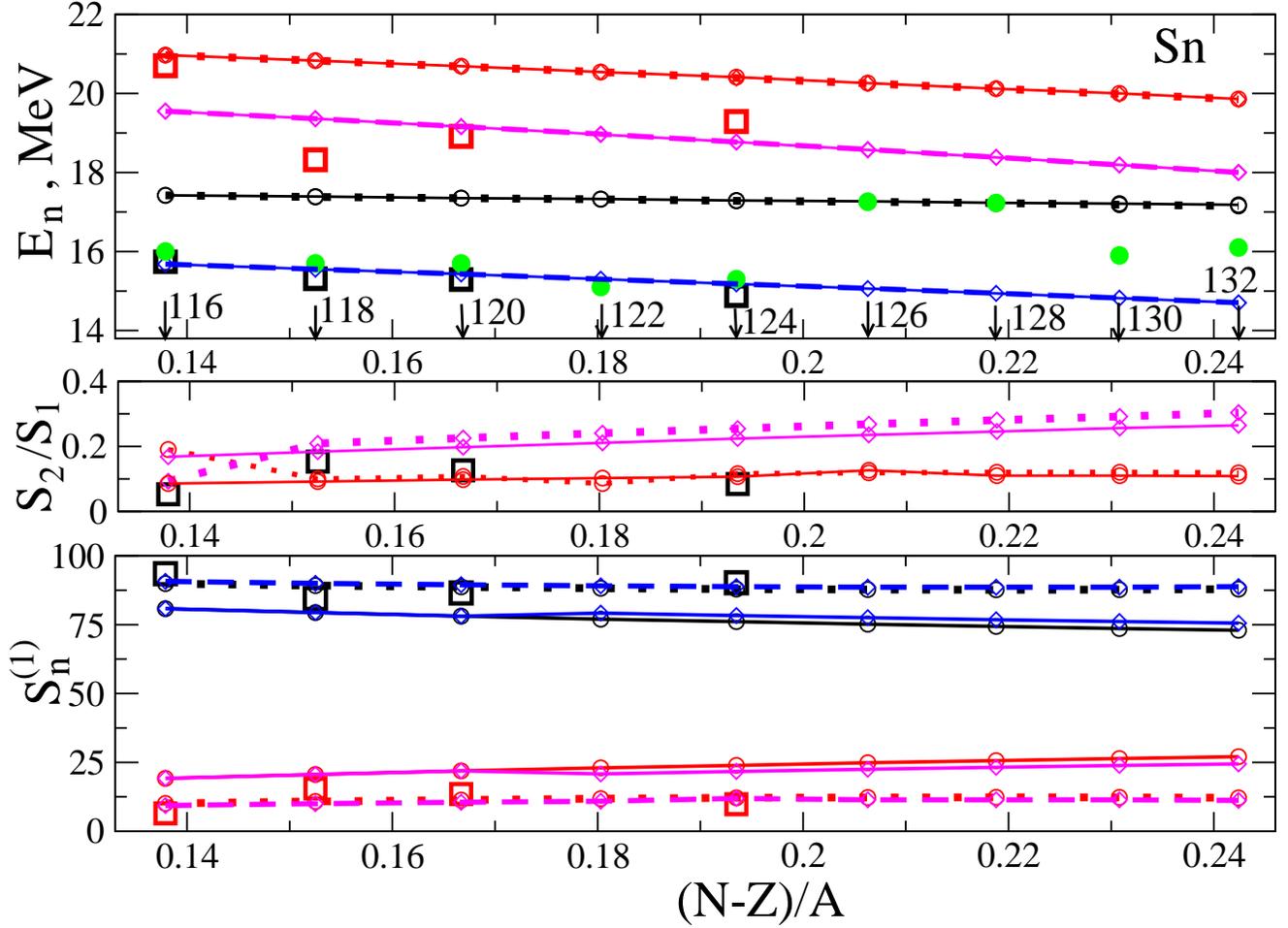}
\end{center}

\vspace{0.5cm}
\caption{ Energies and strength of the IVDR as function of the isospin asymmetry
   parameter $I=(N-Z)/A$ along the Sn isotopic chain from $^{116}$Sn to
   $^{132}$Sn.
   {\it Top:} Energies $E_1$ and $E_2$ of the main and the satellite peak
   respectively. Open squares correspond to experimental data from integral
   cross sections (see the text) for the main peak (1, lower squares) and the
   satellite peak (2, upper squares). Large full circles represent
   experimental IVGDR data. Results obtained in theoretical calculations
   with the Skryme forces SLy5* (open circles and dotted curve) and the
   SVsym32 (open diamonds and dashed curve) are also shown, where the lower
   curves correspond to the main and the upper ones to the satellite peak.
   {\it Middle:} Ratio of stength of satellite versus main peak with the
   experimental IVDR data given by open squares. Theoretical results for
   SLy5* are shown by the solid curve with open circles (for $L=50$ MeV)
   and by tiny solid squares with open circles (for L=0), and for SVsym32
   by the solid curve with open diamonds (for L=60 MeV) and the small solid
   squares with open diamonds (for $L=0$).
   {\it Bottom:} EWSR contributions of the main and the satellite peaks,
   normalized to 100\%, as explained in the text, with the same notation as
   in the top of the figure (from \cite{BMRprc2015}).}
\label{fig7}
\end{figure*}

\vspace{0.2cm}
\vspace{1.0cm}
\begin{figure*}
\begin{center}
\includegraphics[width=0.7\textwidth,clip=true]{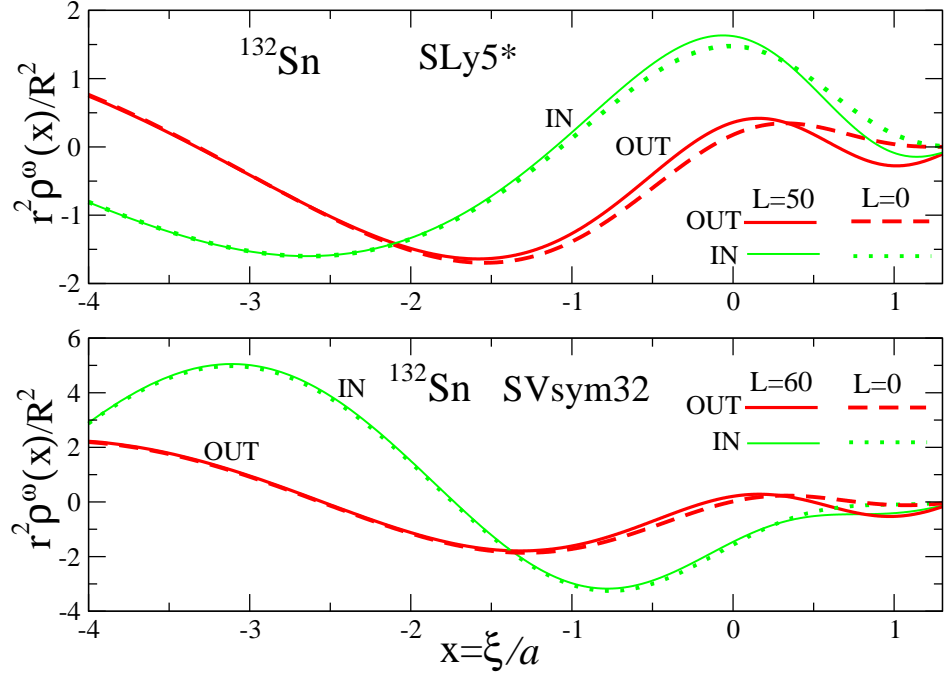}
\end{center}

\vspace{-0.2cm}
\caption{ Main IVDR {\it in-phase} ($\delta \rho_{+}$, 
dots and thin solid line)
    and {\it out-of-phase} ($\delta \rho_{-}$, dashed and thick solid line)
    transition densities (multiplied by $(r/R)^2$) as function of the
    dimensionless distance coordinate $x=\xi/a \approx (r-R)/a$ for the
    satellite peak in the nucleus $^{132}$Sn as obtained with the Skyrme
    interactions SLy5$^*$ (upper part) and SVsym32 (lower part).}
\label{fig8}
\end{figure*}

\vspace{1.0cm}
\begin{figure*}
\begin{center}
\includegraphics[width=0.8\textwidth,clip=true]{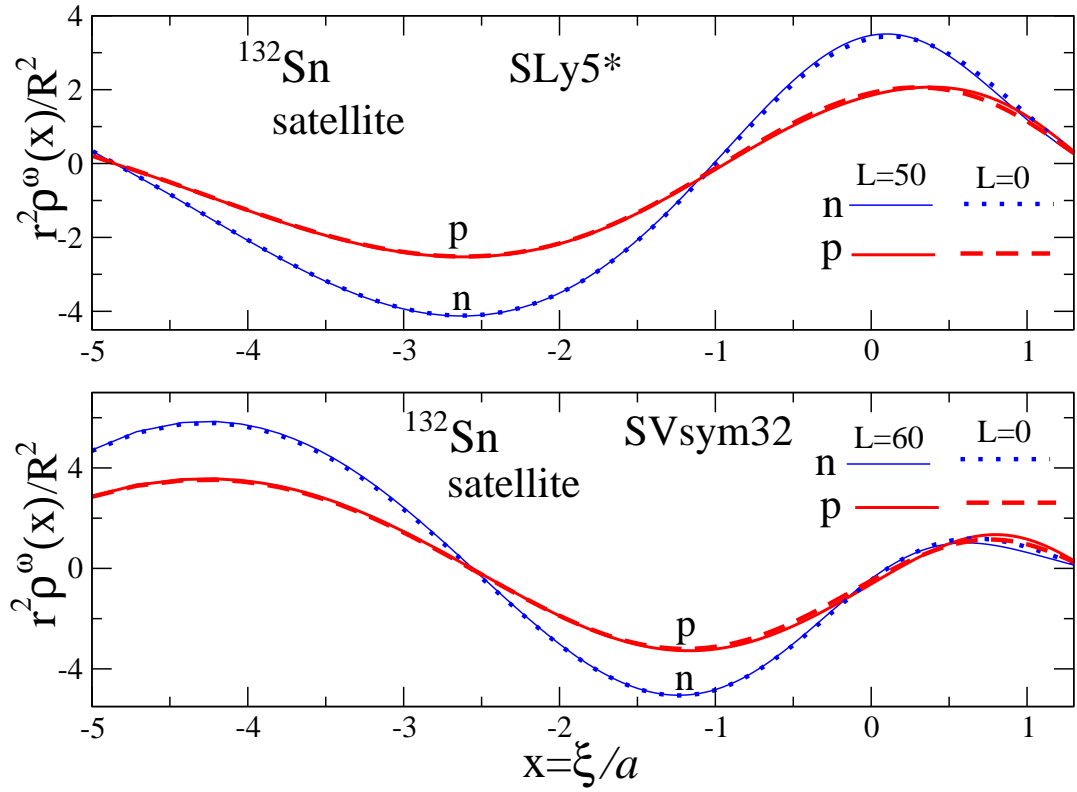}
\end{center}
\vspace{-0.2cm}
\caption{ Same as Fig.\ 8, but for the IVDR 
neutron ($n$) and proton ($p$)
    transition densities (\ref{rhonp}) for the satellite energy $E=E_2$.}
\label{fig9}
\end{figure*}

\begin{figure*}
\begin{center}
\includegraphics[width=0.8\textwidth,clip=true]{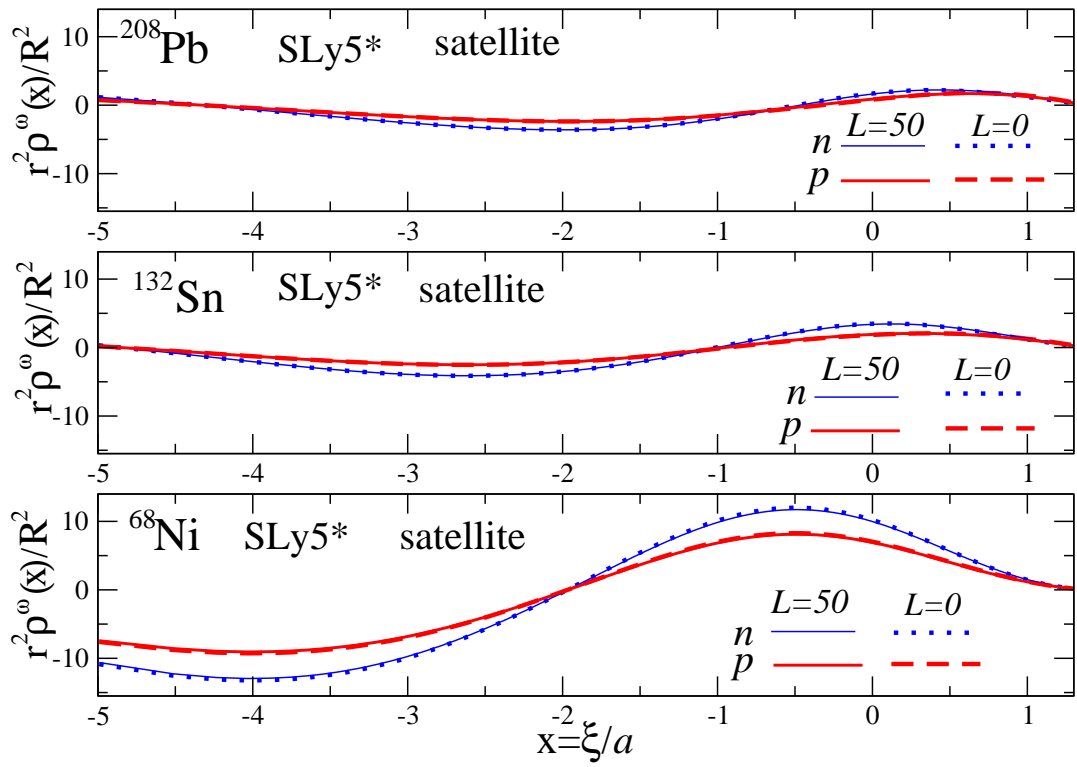}
\end{center}

\vspace{0.2cm}
\caption{IVDR n-p transition densities $\rho^\omega(x)$, as in Fig.\ 9, but
    obtained with the Skyrme interaction SLy5* for three different nuclei,
    $^{208}$Pb (top), $^{132}$Sn (middle) and $^{68}$Ni (bottom), for the
    satellite peak. The sensitivity of our results with respect to the value
    of the derivative constant $L$ (in MeV) is shown to be small.}
\label{fig10}
\end{figure*}
\begin{figure*}
\begin{center}
\includegraphics[width=0.8\textwidth,clip=true]{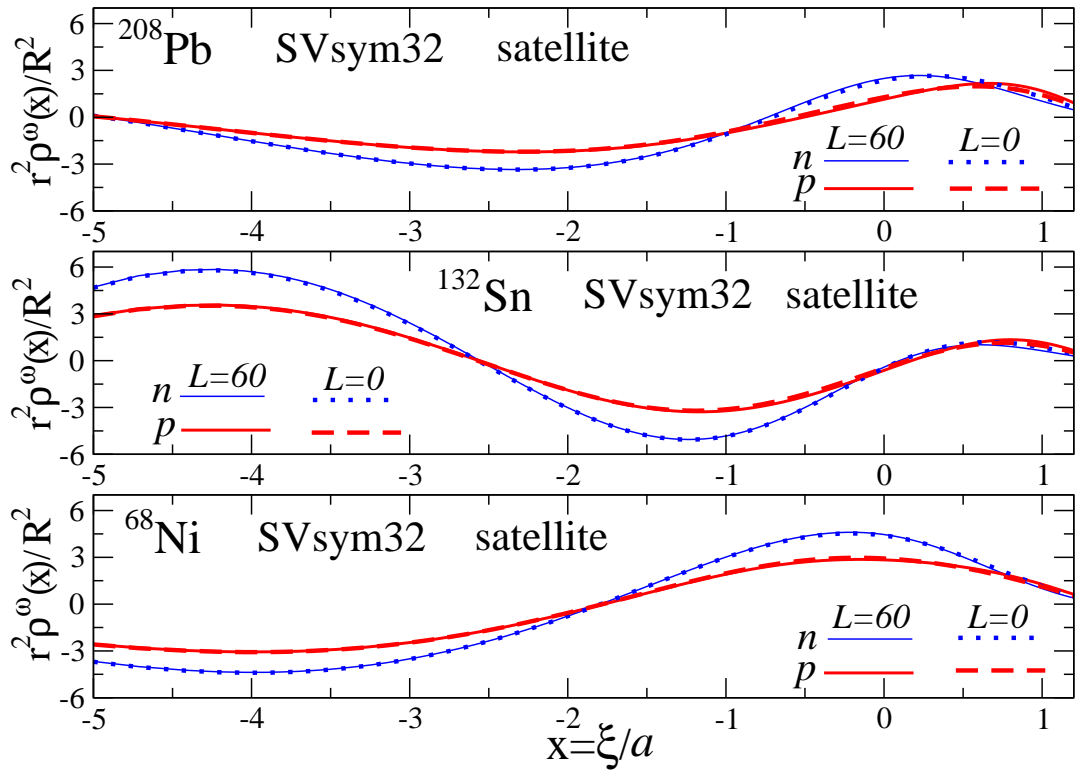}
\end{center}

\vspace{0.2cm}
\caption{ Same as Fig.\ \ref{fig10} but for the Skyrme force SVsym32.}
\label{fig11}
\end{figure*}
\begin{figure*}
\begin{center}
\includegraphics[width=0.7\textwidth,clip=true]{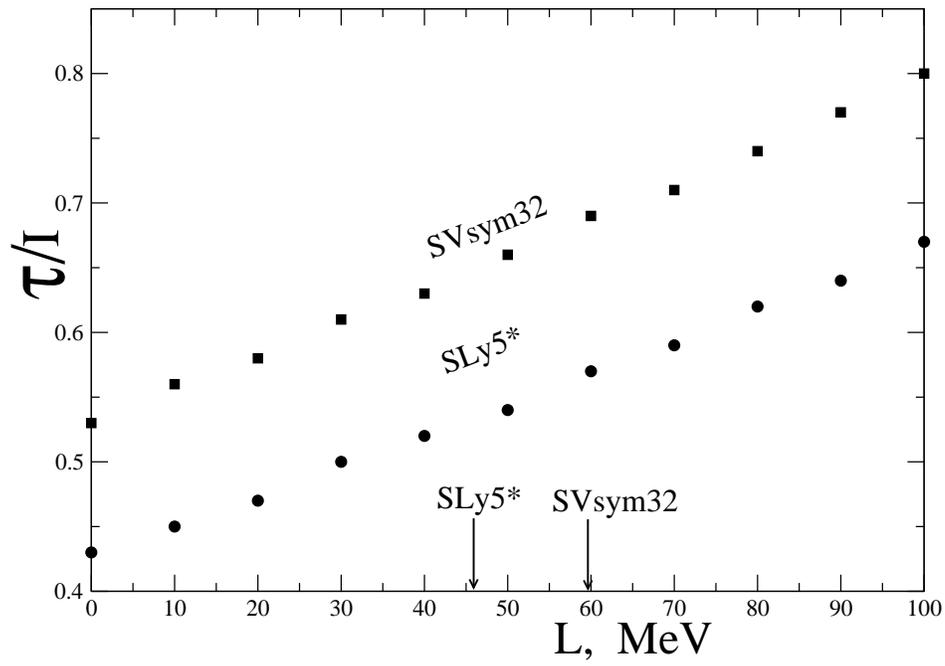}
\end{center}

\vspace{1.0cm}
\caption{Dimensionless neutron skin thickness $\tau$ (\ref{skin}) in units of the
  asymmetry parameter $I$ as function of the derivative constant $L$ for
  the SLy5* and SVsym32 forces; arrows indicate the $L$ values
  corresponding to the two different Skyrme forces.}
\label{fig12}
\end{figure*}

\vspace{1.0cm}
\begin{figure*}
\begin{center}
\includegraphics[width=0.8\textwidth,clip=true]{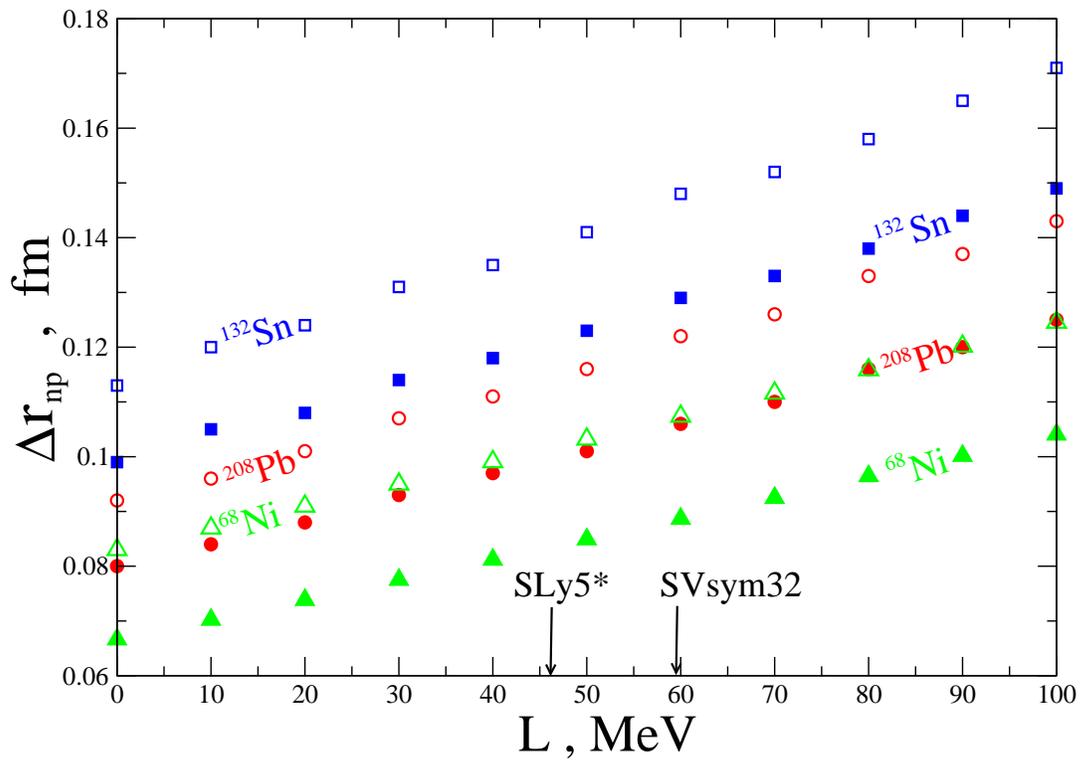}
\end{center}

\vspace{-0.2cm}
\caption{ Neutron skin thickness, 
$\Delta r_{np} =\sqrt{3/5}(R_n-R_p)$, as function of 
the asymmetry
    parameter $I$ with a comparison between experimental data 
obtained for
    the indicated nuclei from antiprotonic atoms with a droplet model fit
    taken from Ref.\ \cite{vinas5}.
}
\label{fig13}
\end{figure*}
\vspace{-0.5cm}
\begin{figure*}
\begin{center}
\includegraphics[width=0.8\textwidth,clip=true]{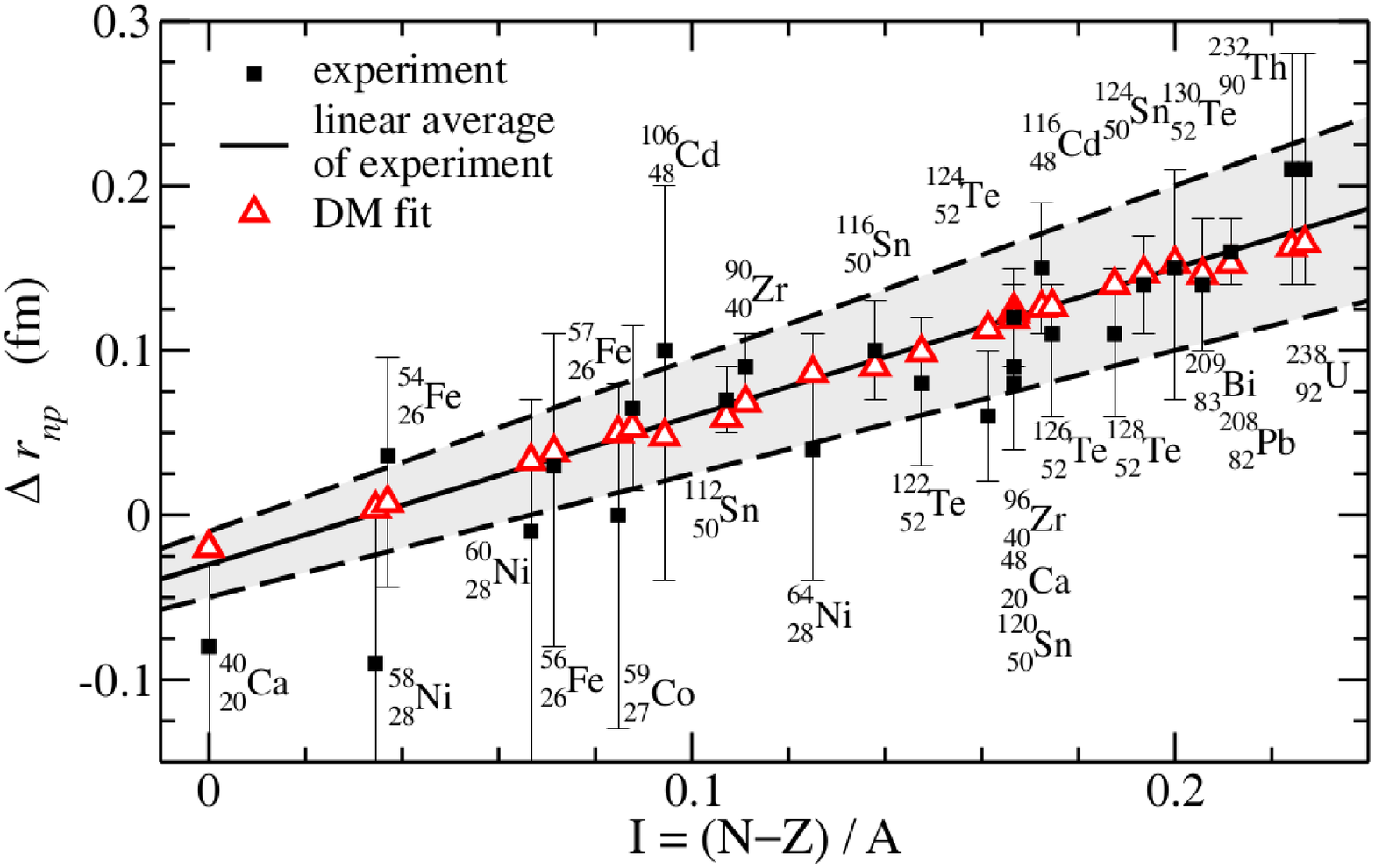}
\end{center}

\vspace{-0.3cm}
\caption{ Neutron skin thickness, 
$\Delta r_{np} =\sqrt{3/5}(R_n-R_p)$, as function of 
the asymmetry
    parameter $I$ with a comparison between experimental data 
obtained for
    the indicated nuclei from antiprotonic atoms with a droplet model fit
    taken from Ref.\ \cite{vinas5}.}
\label{fig14}
\end{figure*}
\begin{figure*}
\begin{center}
\includegraphics[width=0.8\textwidth,clip=true]{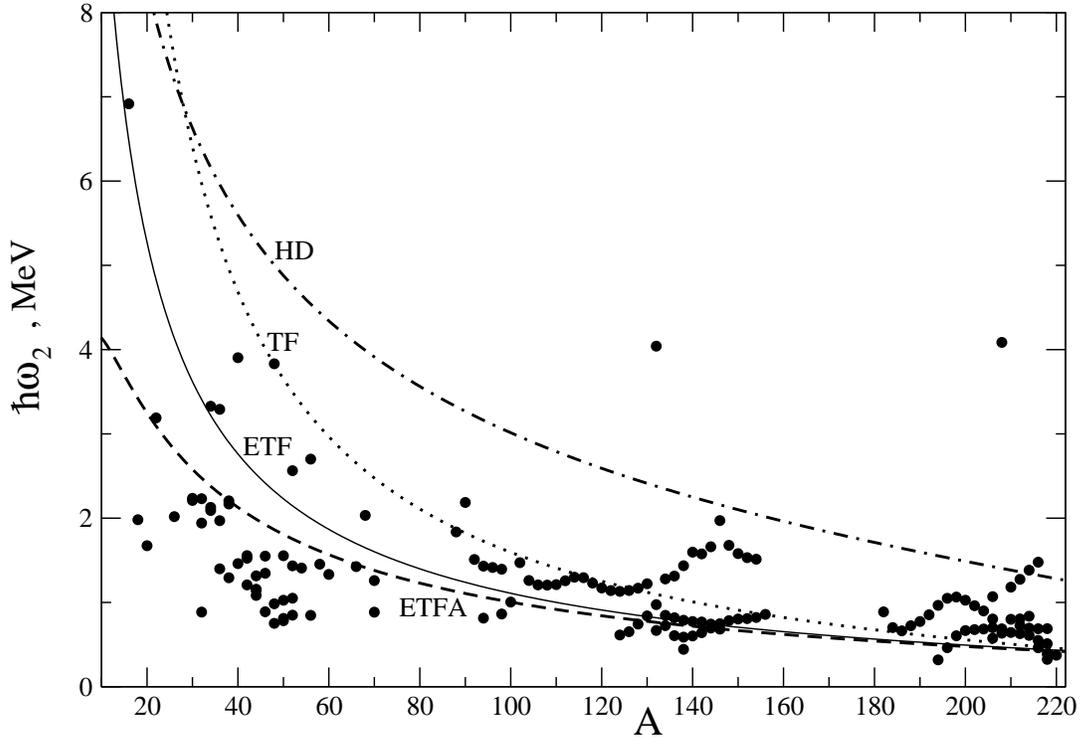}
\end{center}
\caption{ Low-lying quadrupole vibration energies 
$\hbar \omega_{2}$ versus
    particle number $A$ with a comparison between experimental data (full
    dots) for nearly spherical nuclei \cite{table,raman} with 
quadrupole    deformations 
$q_2 < 0.05$ \cite{table,plugor,plugorkad}
    and different theoretical models: the TF approach (dotted line), the
    ETF approach (solid line) that accounts for surface and curvature
    corrections, the asymptotic ETFA formula (\ref{hwsurfcoras})
    (dashed line) and the standard hydrodynamical model \cite{bormot}
    (dash-dotted line).} 
\label{fig15}
\end{figure*}
\begin{figure*}
\begin{center}
\includegraphics[width=0.8\textwidth,clip=true]{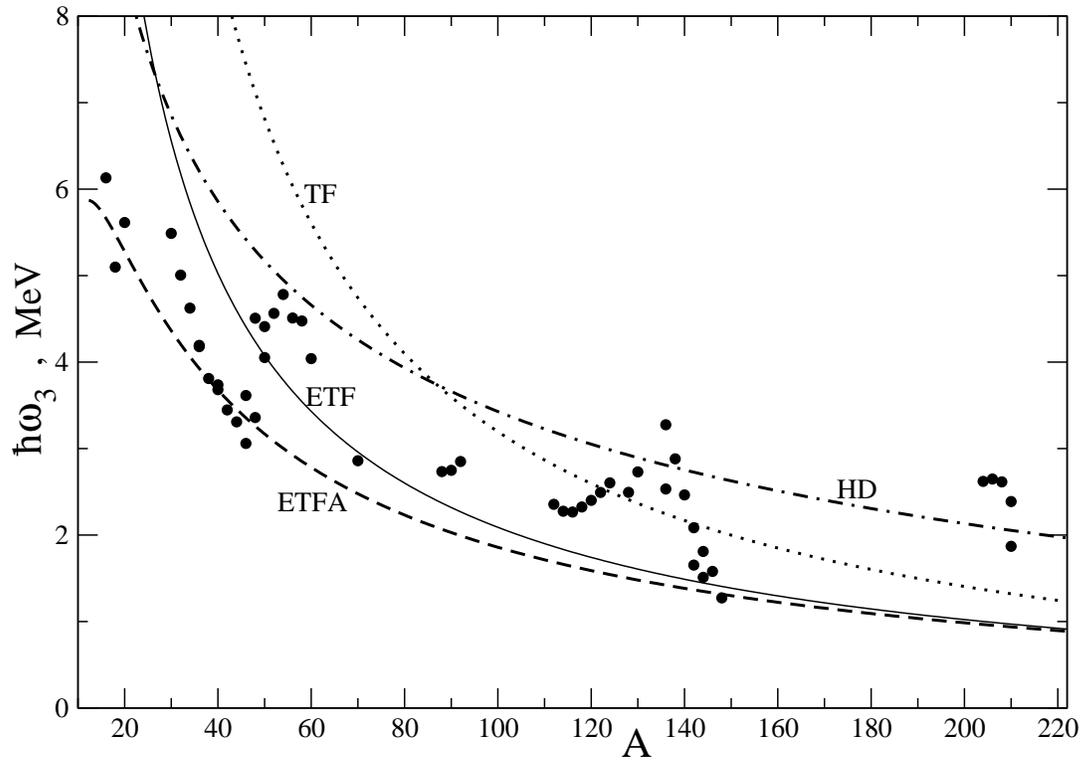}
\end{center}
\caption{ Same as Fig.\ \ref{fig15} but for octupole vibrational 
states; the
 experimental data are taken from 
\cite{table,kibedi}.} 
\label{fig16}
\end{figure*}
\begin{figure*}
\begin{center}
\includegraphics[width=0.8\textwidth,clip=true]{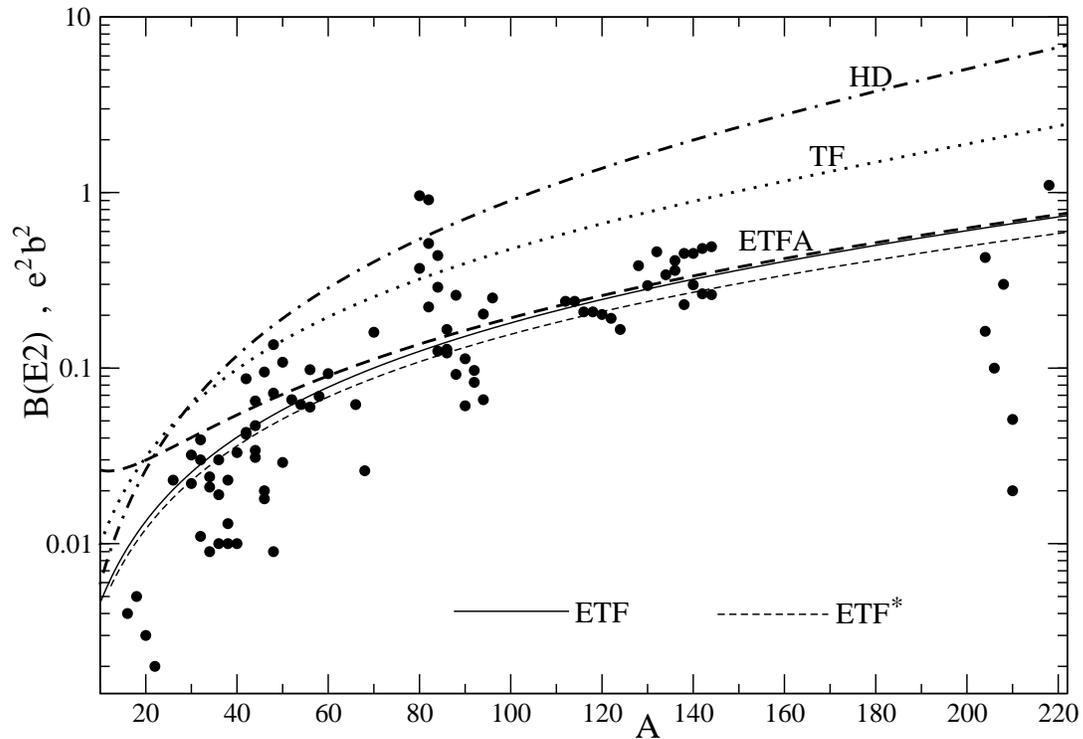}
\end{center}
\caption{ Reduced B(E2) transition probabilities, in units 
e$^2\,$b$^2$, on a
   logarithmic scale, for the quadrupole transition $0^{+} \rightarrow
   2^{+}$ over a very wide mass region, with experimental data \cite{table,
   raman} given by solid circles. 
Different semiclassical models, the TF
   approach (dotted line), the full ETF approach (solid line) that accounts
   for surface and curvature corrections  
with, and the ETF* approach 
(short-dashed
   line)  without $\eta$ corrections of (\ref{semiBLint}) in 
(\ref{semiBLappr}); 
and the analytical asymptotics ETFA (large-dashed line), see
   (\ref{hwsurfcoras}),
 are compared with the hydrodynamical model
   (dash-dotted line).}
\label{fig17}
\end{figure*}
\begin{figure*}
\begin{center}
\includegraphics[width=0.75\textwidth,clip=true]{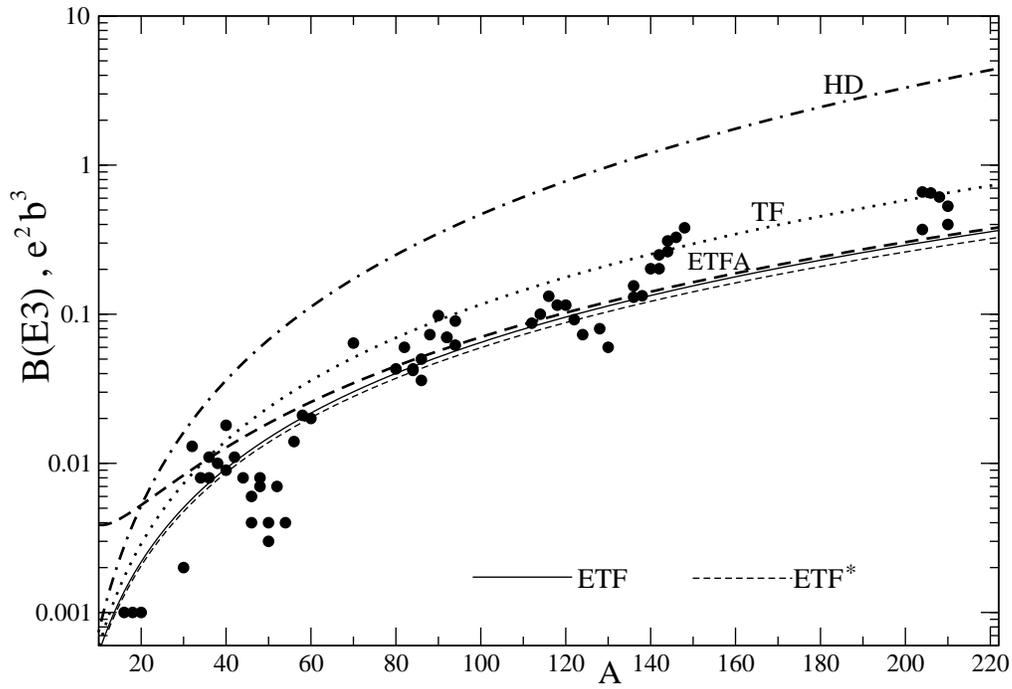}
\end{center}
\caption{Same as Fig.\ \ref{fig17} but for octupole 
transition $0^{+} \rightarrow
   3^{-}$, in units e$^2\,$b$^3$, with experimental data (solid 
circles)
   from 
\cite{table,kibedi}.}
 \label{fig18}
\end{figure*}
\begin{figure*}
\begin{center}
\includegraphics[width=0.75\textwidth,clip=true]{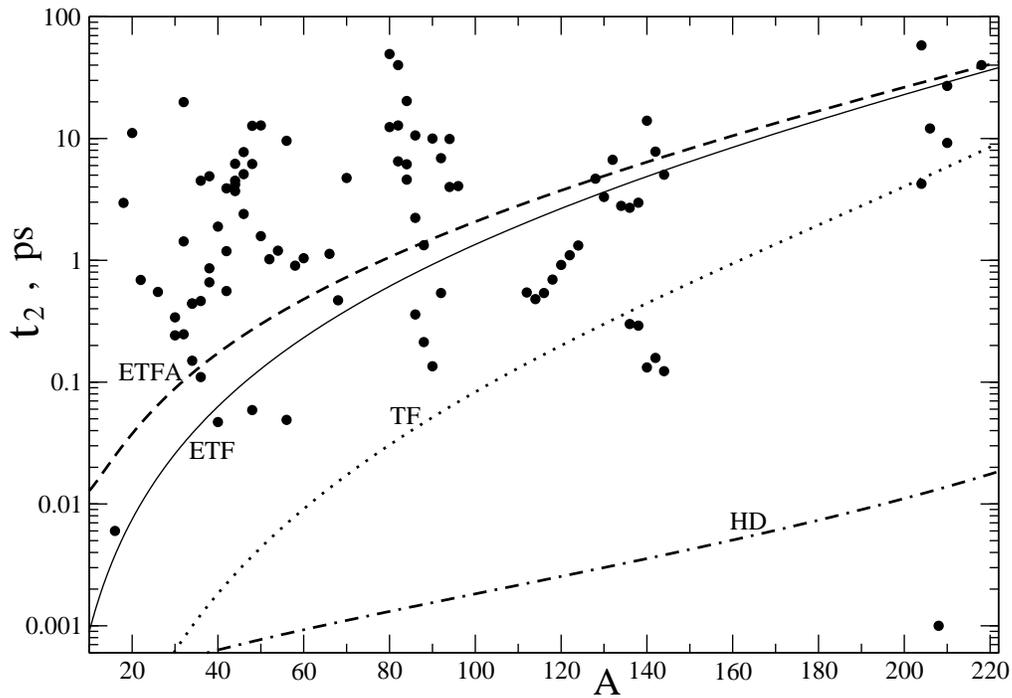}
\end{center}
\caption{Half lives $t^{}_{2}$ (in ps units) for the 
quadrupole transition
   $2^{+} \to 0^{+}$, with experimental data (solid circles) 
from 
   \cite{raman,table}; notations are the same as in Fig.\ \ref{fig17}.} 
\label{fig19}
\end{figure*}
\begin{figure*}
\begin{center}
\includegraphics[width=0.8\textwidth,clip=true]{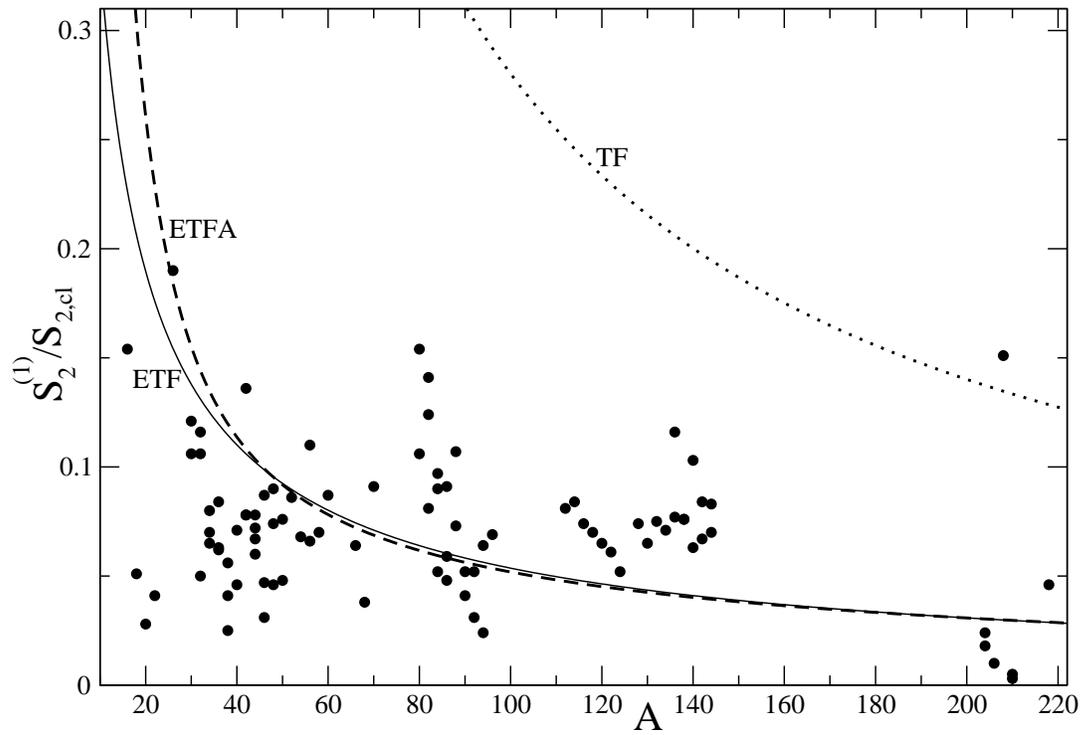}
\end{center}
\caption{Quadrupole EWSR $S_{2}^{(1)}$
(\ref{momstren})
in units of
   $S_{2,{\rm cl}}$
(\ref{EWSR1})
with solid circles representing
   $\hbar \omega_{2} B(E2)$ with the experimental vibration energies
   $\hbar \omega_{2}$ and reduced transition probabilities $B(E2)$ of
   \cite{raman,table} (see Figs.\ \ref{fig15} and \ref{fig17}). The
   semiclassical $S_{2}^{(1)}$ are given by (\ref{sumrules}), and the
   other notation is the same as in Fig.\ \ref{fig17}.}
\label{fig20}
\end{figure*}
\begin{figure*}
\begin{center}
\includegraphics[width=0.8\textwidth,clip=true]{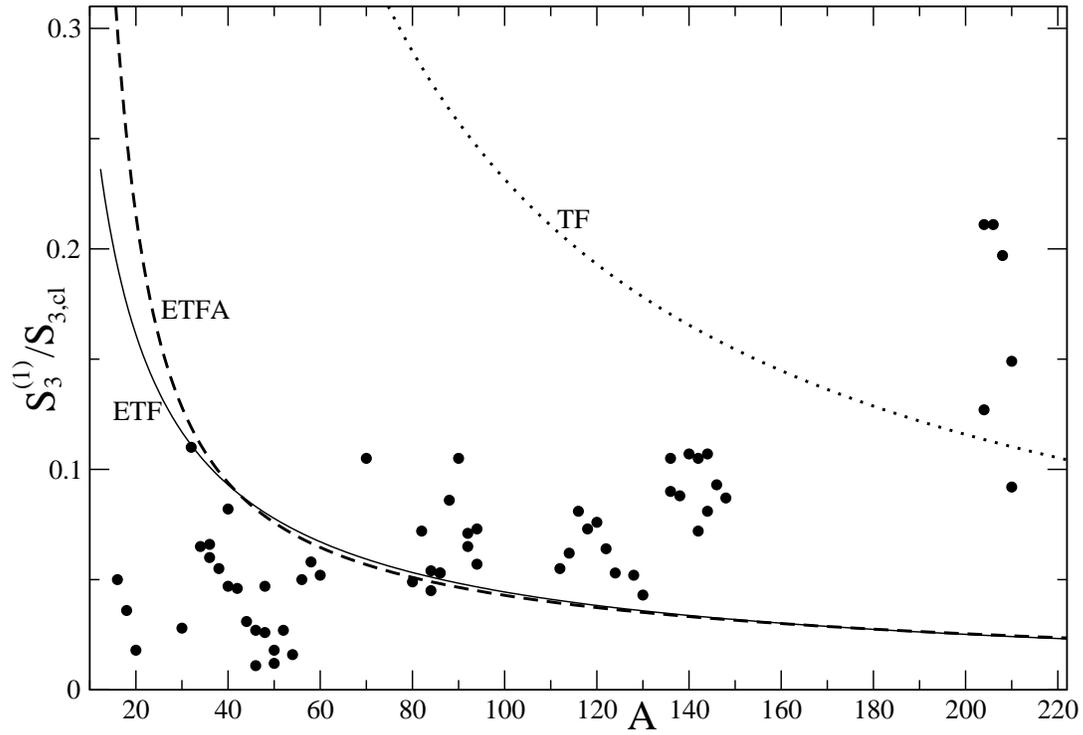}
\end{center}
\caption{   Same as Fig.\ \ref{fig20} but for the octupole 
EWSR $S_{3}^{(1)}$ in units of $S_{3,{\rm cl}}$, with solid circles showing 
$\hbar\omega_{3} B(E3)$ with the experimental $\hbar \omega_{3}$ and $B(E3)$
   from \cite{kibedi,table}.}
\label{fig21}
\end{figure*}
\begin{figure*}
\begin{center}
\includegraphics[width=0.8\textwidth,clip=true]{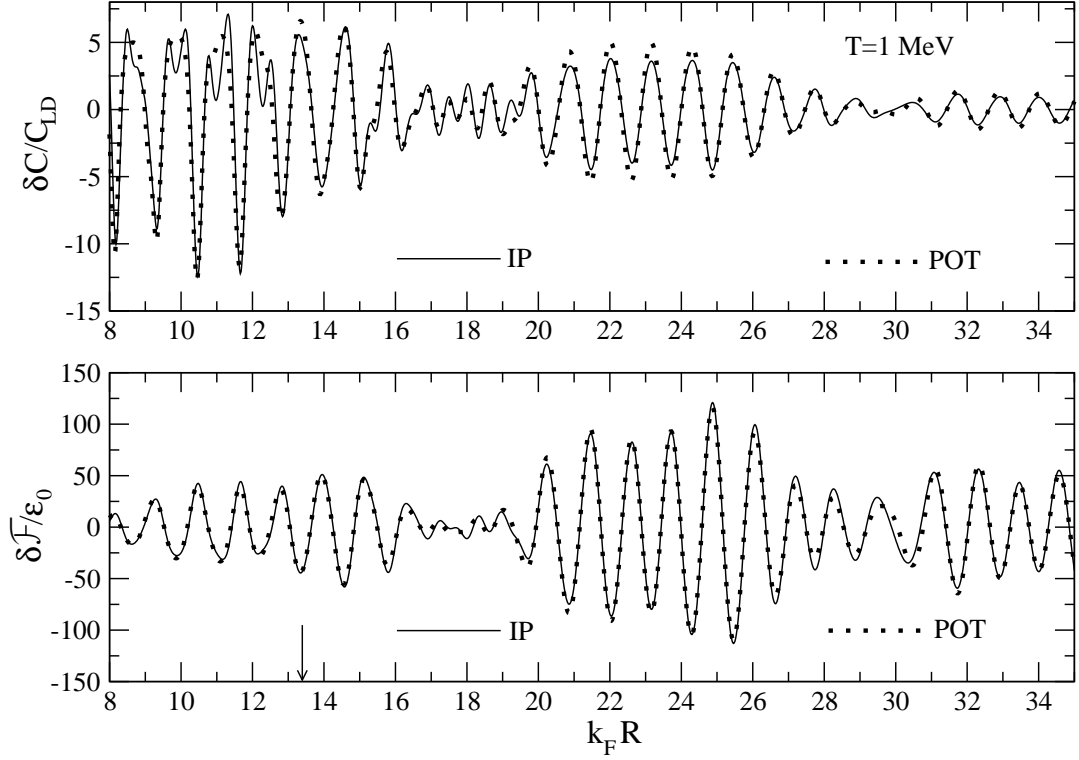}
\end{center}
\caption{ Quadrupole ($\lambda=2$) stiffness, $\delta C$ (top), 
and free energy,
   $\delta {\cal F}$ (bottom), shell corrections, respectively in units of
   $C_{{\rm LD}}$ and $\vareps_0=\hbar^2/(2mR^2)$, as a function of $k_F R$,
   at a temperature of $T=1$ MeV. IP model
, see (\ref{fSQM}) and
   (\ref{stSQM}), and POT curves 
[(\ref{dfdcpot})] are given
   respectively by solid and dotted lines.}
\label{fig22}
\end{figure*}
\begin{figure*}
\begin{center}
\includegraphics[width=0.8\columnwidth,clip=true]{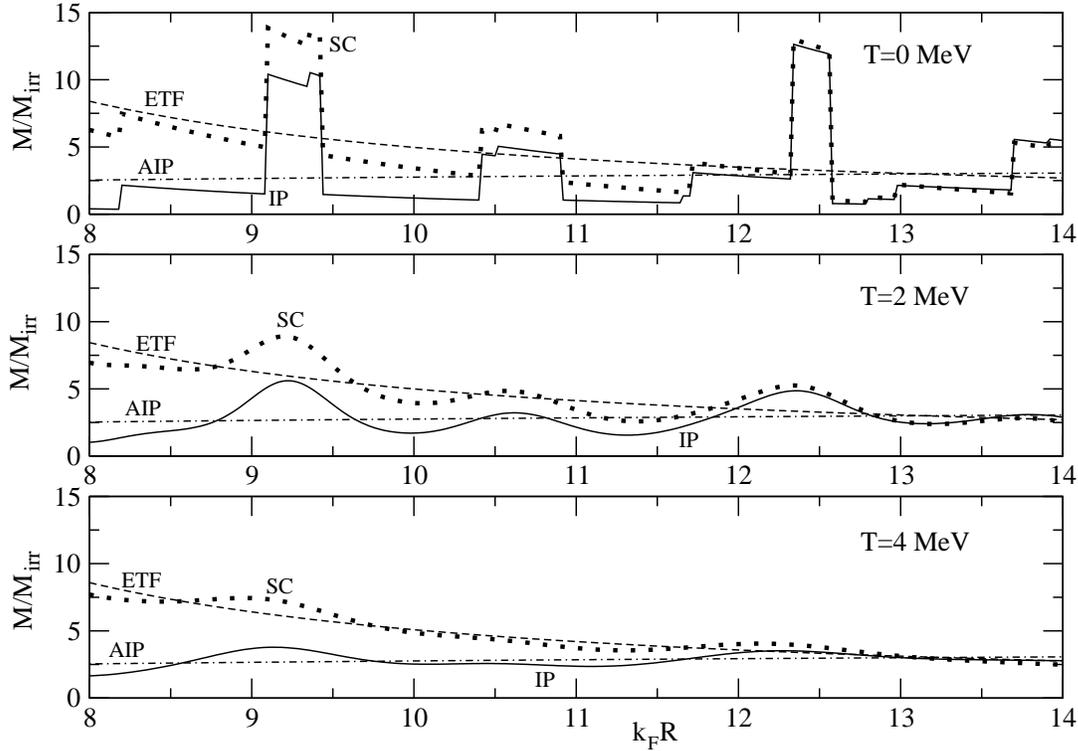}
\end{center}
\caption{ Inertia $M~$ in units of the irrotational flow 
value $M_{\rm irr}$ as
   function of $k_F R$ for temperatures T$\,=\, 0, 2, 4\,$MeV. Quantum
   cranking-model results (IP, solid curve), 
see (\ref{mcran}), and its
   average (AIP,  dash-dotted line) are compared with the renormalized
   inertia (\ref{SQM}) (SC, dotted curve) and the Extended Thomas-Fermi
   inertia $M_{\rm ETF}$
(\ref{bcol}), ETF). The
   parameters used are the same as in Fig.\ \ref{fig22}, besides
of the averaging parameters (see main text). }
\label{fig23}
\end{figure*}
\clearpage
\begin{figure*}
\begin{center}
\includegraphics[width=0.9\textwidth,clip=true]{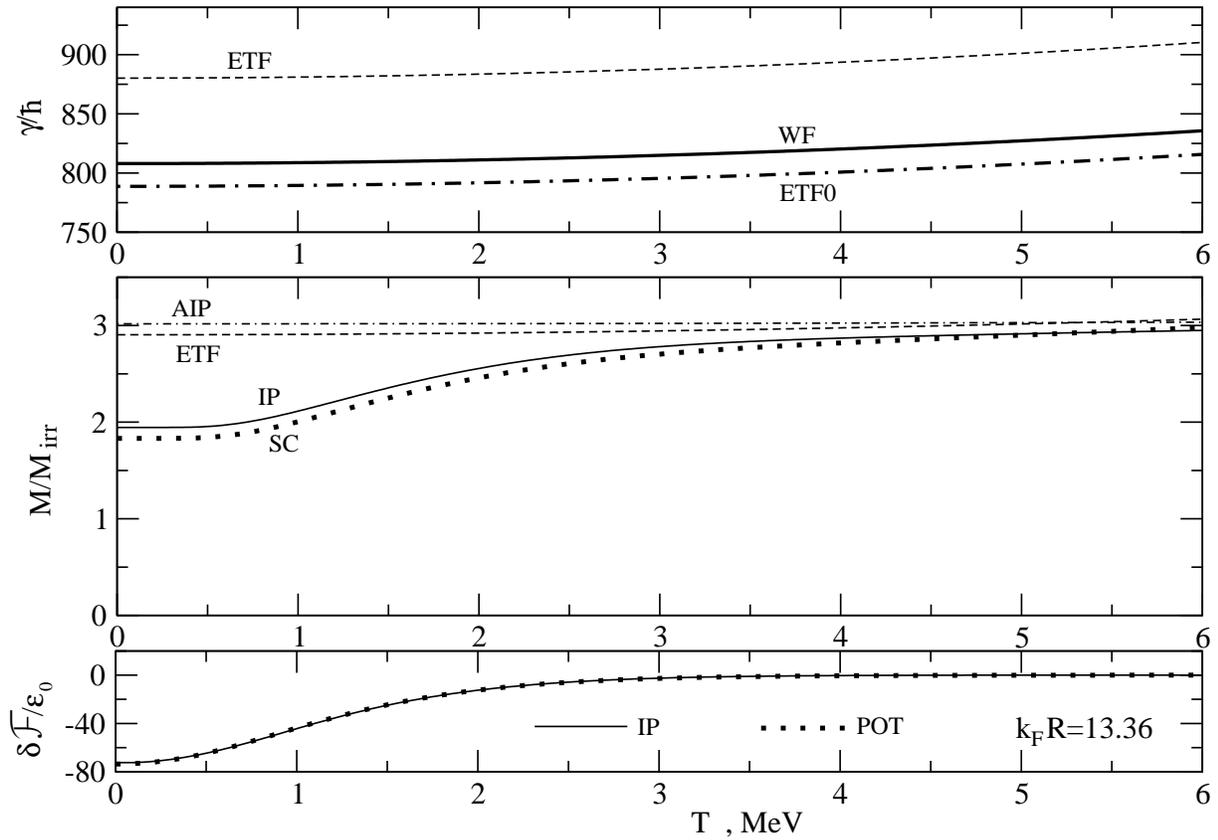}
\end{center}
\caption{ Friction $\gamma$, inertia $M$, and free energy 
shell corrections 
$\delta \mathcal{F}$, in units of $\hbar$, the irrotational flow value
   $M_{{\rm irr}}$, and $\vareps_0$ respectively, determined at a value of
   $k_F R=13.36$, corresponding through (\ref{akrTF}) to $A \approx 254$,
   as function of the temperature T (in MeV).
   {\it top}:  Comparison between ETF friction 
(\ref{gSQM}) 
(ETF,   dashed line), local part of wall formula 
of (\ref{gtf}) (WF, solid),
 and the
   approach of (\ref{gcol}) for $\gamma_{\rm ETF}(0)$ 
(ETF0, dash-dotted
  line)
   {\it middle}: comparison between cranking model inertia $M_{\rm{IP}}(0)$,
   (\ref{mcran}), (IP, solid line), its average ${\tilde M}_{\rm{IP}}(0)$,
   AIP (dash-dotted line), the renormalized (SCM) inertia (\ref{SQM}) (SC,
   dots) and the ETF approach of (\ref{bcol}) for $M_{\rm ETF}$ 
   {\it bottom}: Free-energy shell corrections $\delta {\cal F}$ from the
   IP, 
see (\ref{fSQM}), and the POT model, equation
(\ref{dfdcpot}), are
   presented; the parameters used are the same as in 
Figs.\ \ref{fig22} and \ref{fig23}.
}
\label{fig24}
\end{figure*}
\clearpage
\begin{figure*}
\begin{center}
\includegraphics[width=0.9\textwidth,clip=true]{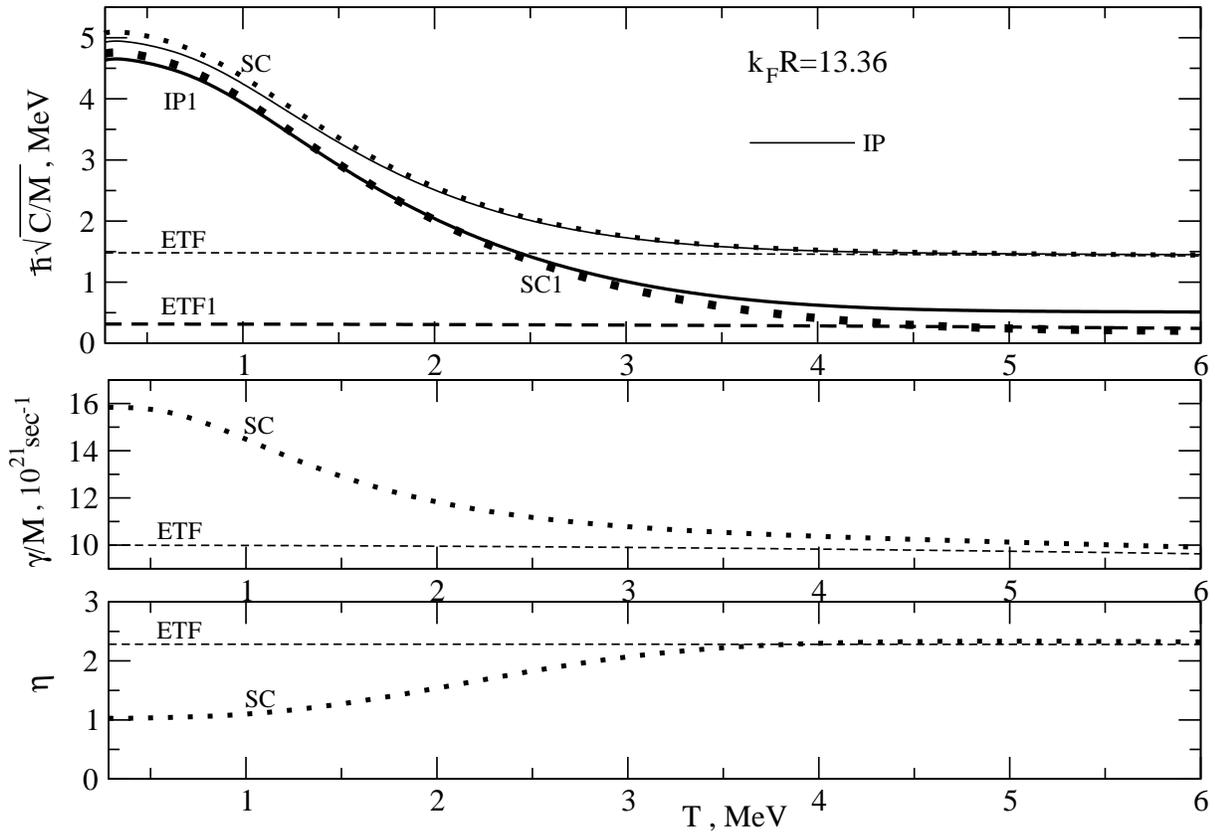}
\end{center}
\caption{Collective vibration energy $\hbar \sqrt{C /M}$, reduced friction
 $\gamma/M$, and effective frictions $\eta=\gamma/2\sqrt{M C}$,
 determined at the same $k^{}_F R$ value as in Fig.\ \ref{fig24}, versus
 the temperature $T$.
 {\it top}: The ETF curve (thin dashed) and the SC curve (dots) are
 determined respectively through the stiffness $C_{\rm{LD}}$ (\ref{stSQM})
 and inertia $M_{\rm ETF}$ (\ref{bcol}) for the first and through the SCM
 stiffness $C$ (\ref{stSQM}) and the inertia $M$ (\ref{SQM}) for the
 latter. The curves denoted by IP1 (thick solid line) and ETF1 (thick
 dashed line) are obtained in the IP and ETF approach, but including the
 Coulomb and surface corrections as explained in the main text, and SC1
 (heavy squares) shows the corresponding SCM quantity;
 {\it middle}: ETF and SC curves show the reduced friction defined with
 the inertia $M_{\rm ETF}$ and the renormalized SCM inertia $M$
 (\ref{SQM}) respectively, with the friction $\gamma = \gamma_{\rm ETF}$
 given by (\ref{gSQM});
 {\it bottom}: Effective friction $\eta=\gamma/2\sqrt{M C}$ obtained with
 a value of $J=30$ MeV in the ETF and SC approaches; 
all other parameters
 are the same as in Figs.\ \ref{fig22} and \ref{fig23}.
}
\label{fig25}
\end{figure*}
\clearpage
\begin{figure*}
\begin{center}
\includegraphics[width=0.75\textwidth,clip=true]{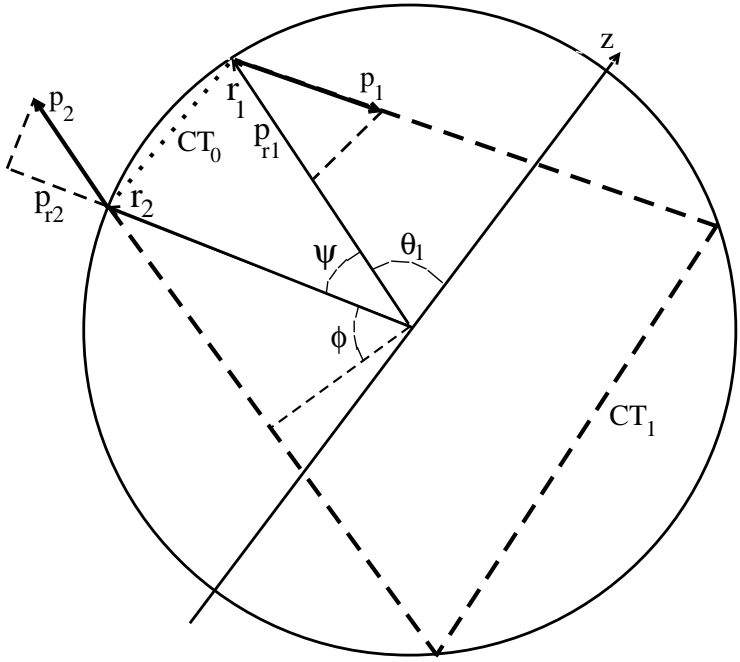}
\end{center}
\caption{ Classical trajectories CT$_0$ 
(dotted line) and CT$_1$ (dashed line)
   from initial point ($\r^{}_1, \p^{}_1$) in phase space to final point
   ($\r^{}_2, \p^{}_2$), as in Fig.\ \ref{fig1}, but at the boundary, with
   radial momentum components $p_{r1}$ and $p_{r2}$ respectively; $\psi$ is
   the angle between the $\r^{}_1$ and $\r^{}_2$ vectors, while
   $\theta^{}_1$ (of $\r^{}_1$) and $\phi$ are the angles in a spherical
   coordinate system with the polar axis $z$.
}
\label{fig26}
\end{figure*}
\begin{figure*}
\begin{center}
\includegraphics[width=0.8\textwidth,clip=true]{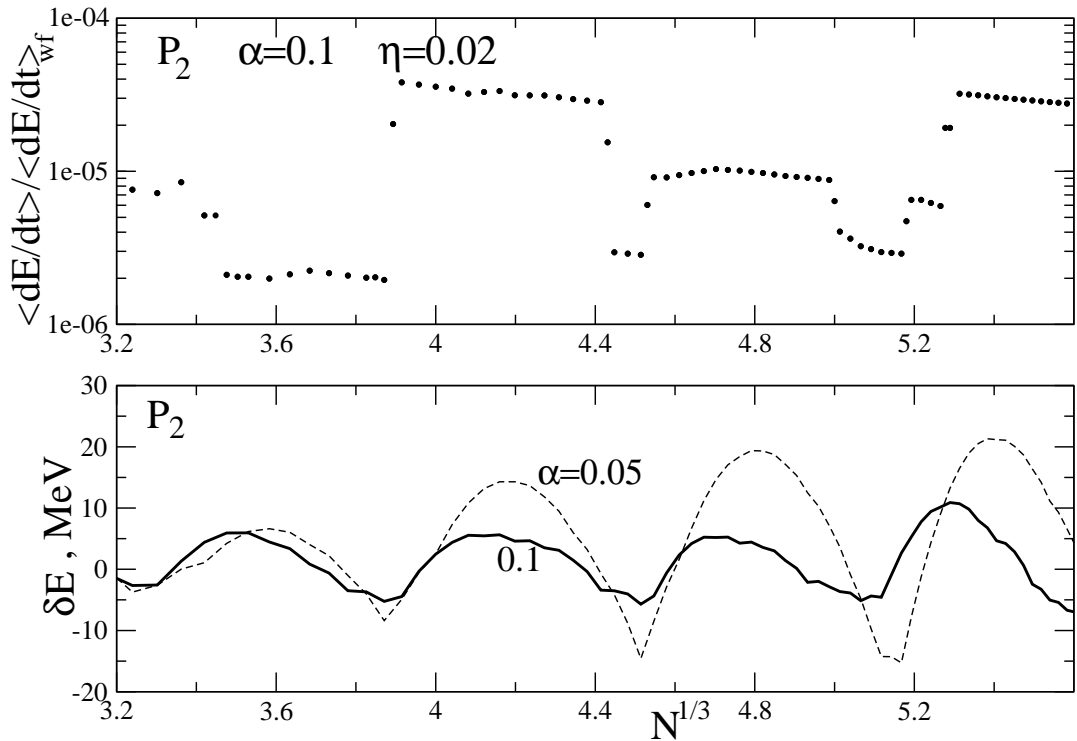}
\end{center}

\vspace{-3.0ex}
 \caption{ Quadrupole vibration
   mean-time derivatives $\langle \d E/\d t\rangle$ (\ref{dedtgam}) in WF
   units (\ref{dedtgamwf}) (top) and shell correction energy 
$\delta E$
   (bottom) as functions of the particle (neutron) number $N^{1/3}$ for two
   different values of the $\alpha$ parameter.}
\label{fig27}
\end{figure*}
\clearpage
\begin{figure*} 
\begin{center}
\includegraphics[width=0.9\textwidth,clip=true]{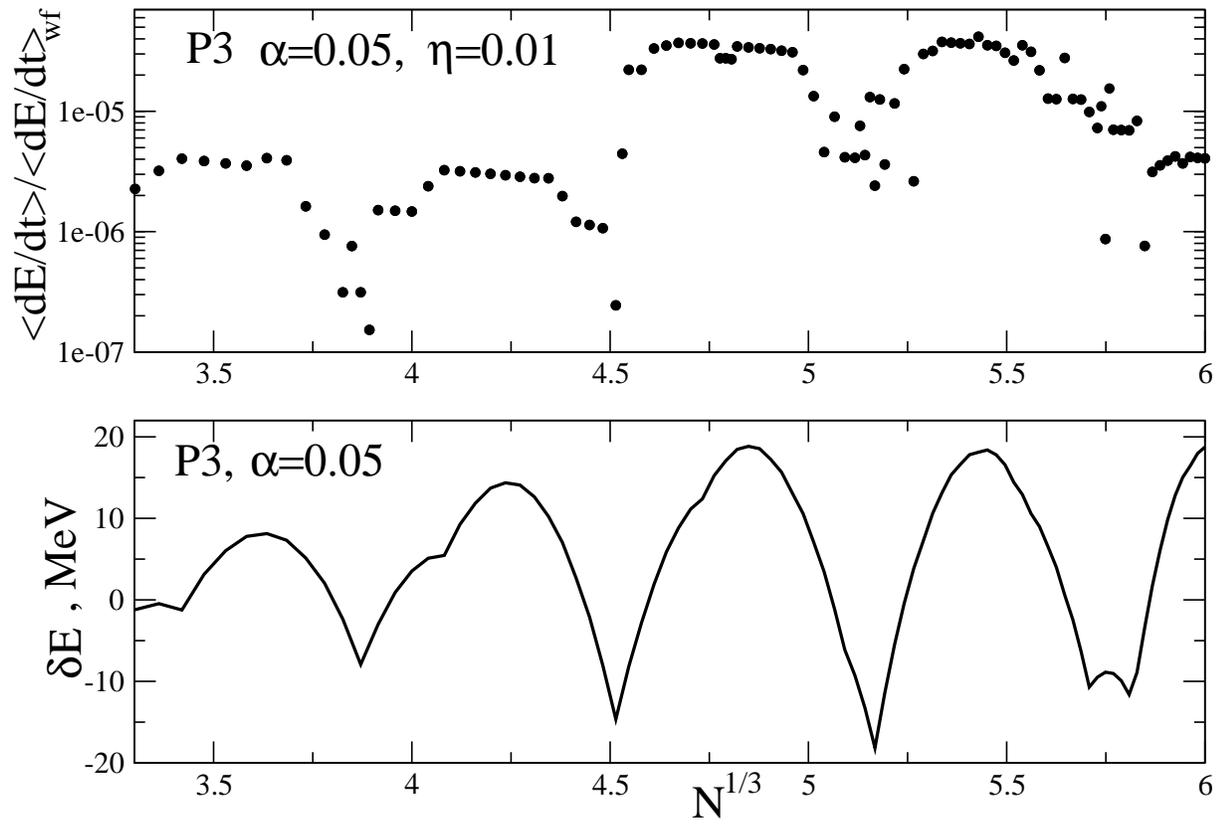}
\end{center}

\vspace{-1ex}
 \caption{ Same as Fig.\ \ref{fig27} but for octupole 
vibrations.}
\label{fig28}
\end{figure*}
\begin{figure*}
\begin{center}
\includegraphics[width=0.7\textwidth,clip=true]{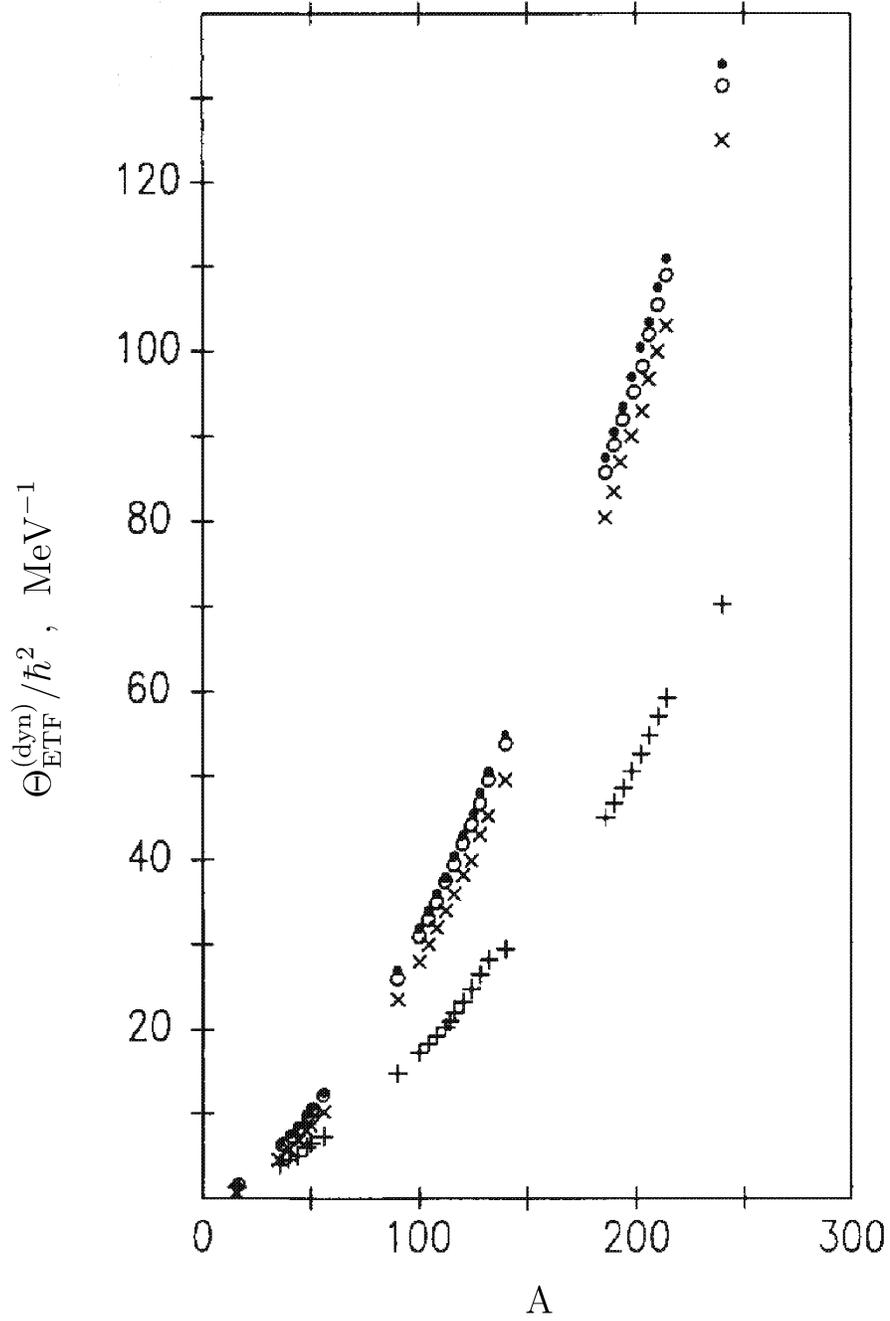}
\end{center}
\vspace{-0.5cm}
\caption{ Semiclassical moments of inertia $\Theta_x$ 
(divided by $\hbar^2$ and
   expressed in MeV$^{-1}$) as functions of the mass  number $A$. ETF
   results correspond to full dots and crosses (x) are obtained upon
   neglecting the  spin degrees of freedom. TF MIs are plotted as open
   circles. Plus signs (+), finally, refer to the dynamical Inglis cranking
   MI (after \cite{bartelnp}).
}
\label{fig29}
\end{figure*}
\begin{figure*}
\begin{center}
\includegraphics[width=0.8\textwidth,clip=true]{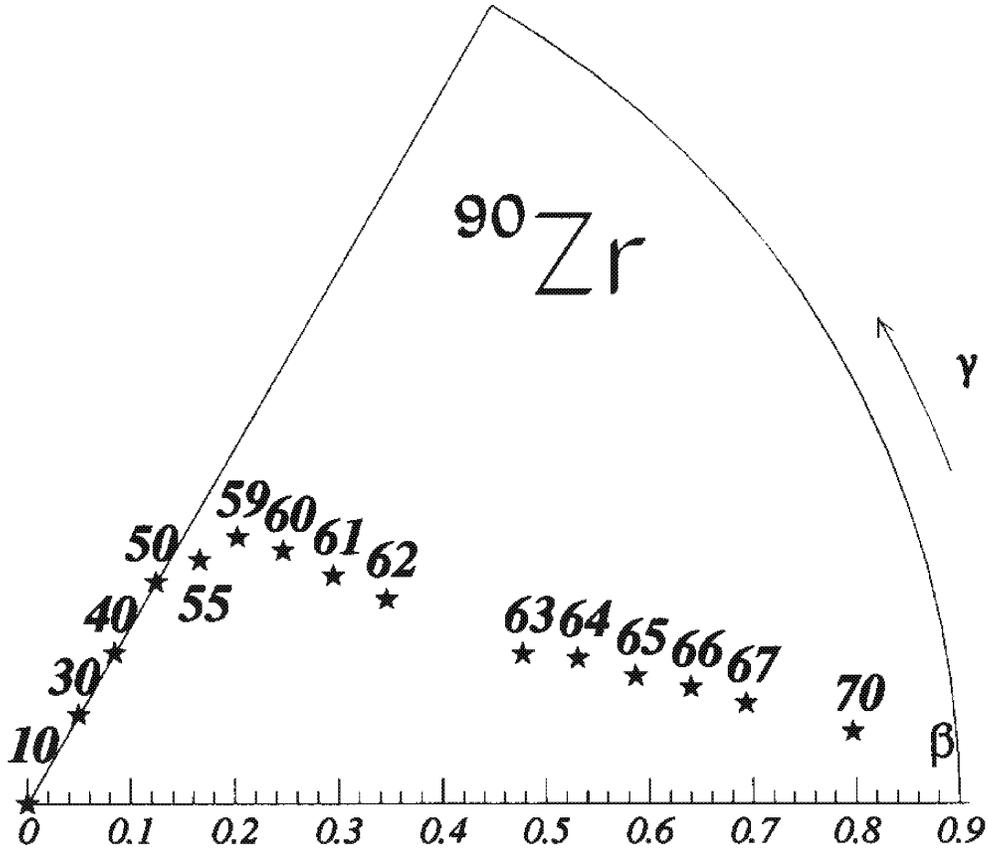}
\end{center}
\vspace{-0.5cm}
\caption{ Equilibrium deformations, in the ($\beta,\gamma$) plane, 
obtained for the
   nucleus $^{90}$Zr for different values $I$ of the total angular momentum
   (after \cite{bartelpl}).}
\label{fig30}
\end{figure*}
\begin{figure*}
\begin{center}
\includegraphics[width=0.8\textwidth,clip=true]{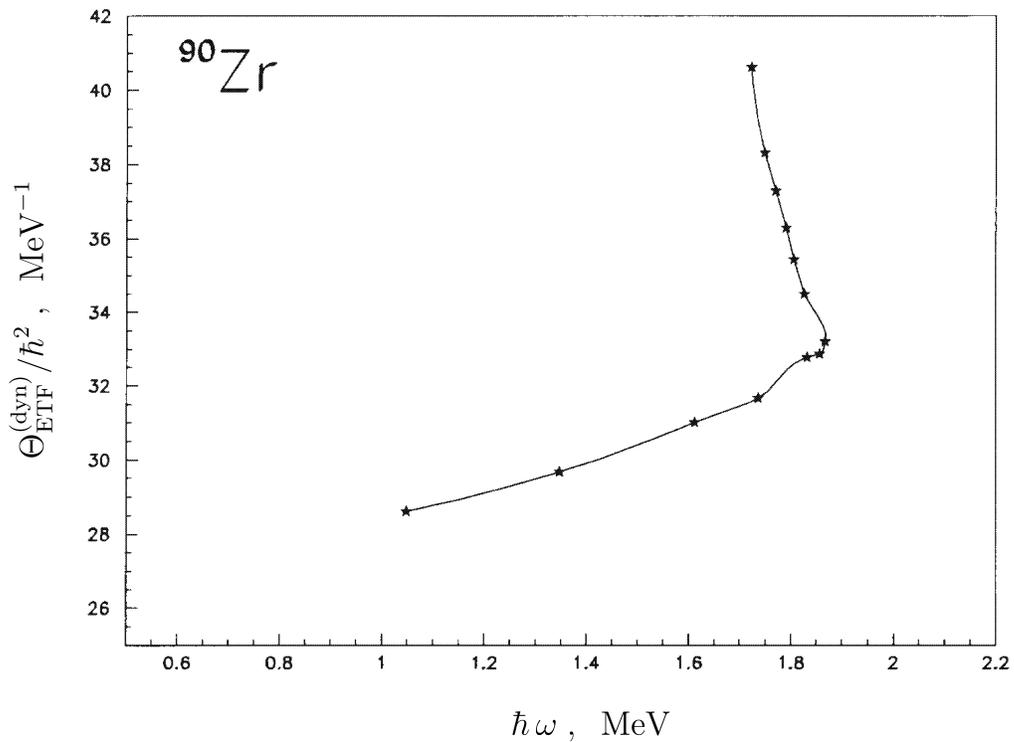}
\end{center}
\vspace{-0.5cm}
\caption{  Variational ETF moment of inertia $\Theta_{ETF}$ 
(in $\hbar^2$ MeV$^{-1}$
   units) for $^{90}$Zr as function of the rotational energy $\hbar \om$
   (after \cite{bartelpl}).  Stars correspond 
to given values of the angular momentum $I$, as done in 
Fig. 30, starting with $I=40\hbar$.
}
\label{fig31}
\end{figure*}
\begin{figure*}
\begin{center}
\includegraphics[width=0.9\textwidth,clip=true]{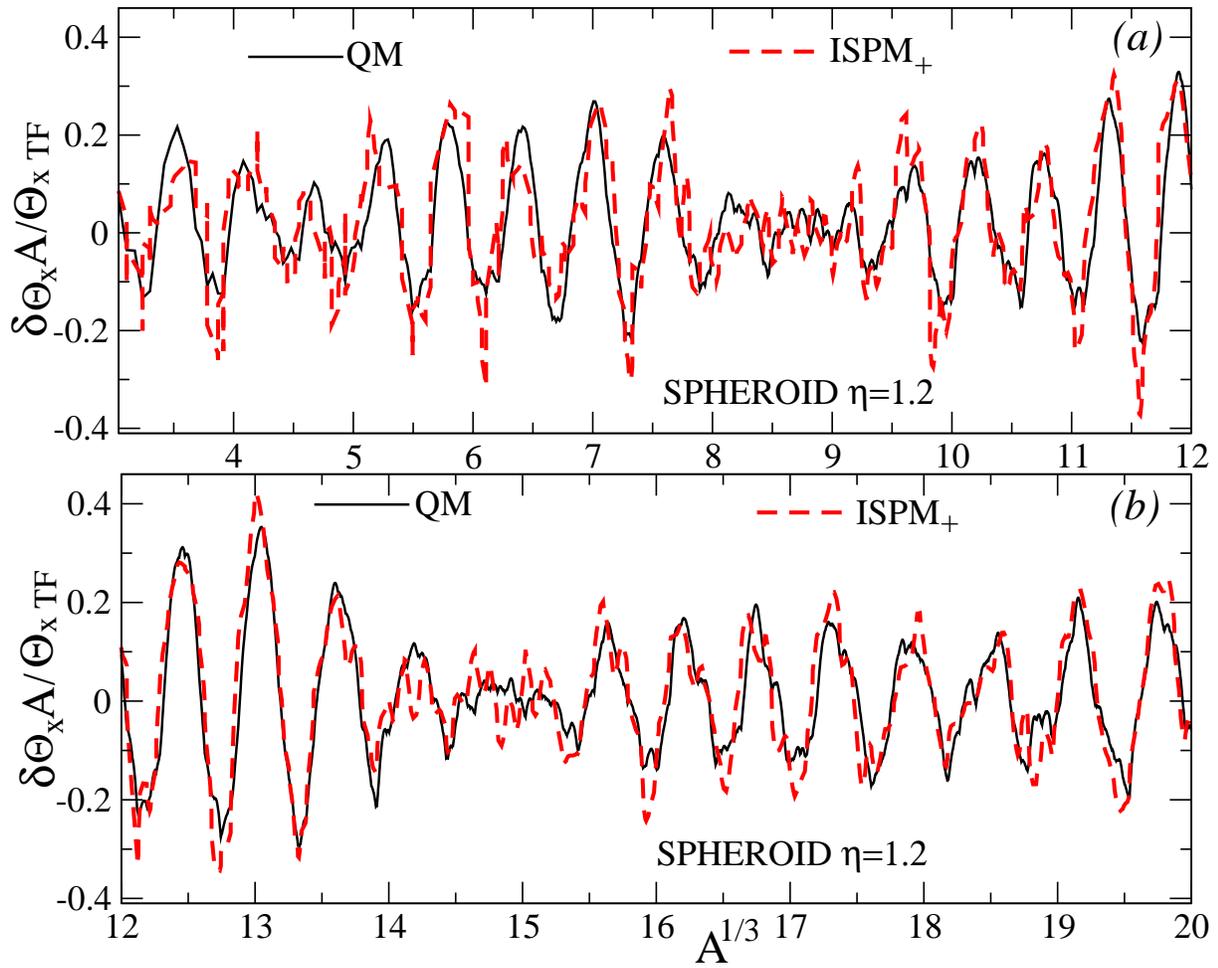}
\end{center}
\vspace{-0.4cm}
\caption{ MI shell components $\delta \Theta_x$ (in TF units) 
as function of
   $A^{1/3}$ at a deformation of $\eta=b/a=1.2$ obtained in a
   quantum-mechanical (QM, full line) and a semiclassical calculation,
   including surface corrections (ISPM$_+$, dashed line) for smaller (upper
   part) and larger particle numbers (lower part).
}
\label{fig32}
\end{figure*}
\begin{figure*}
\begin{center}
\includegraphics[width=0.8\textwidth,clip=true]{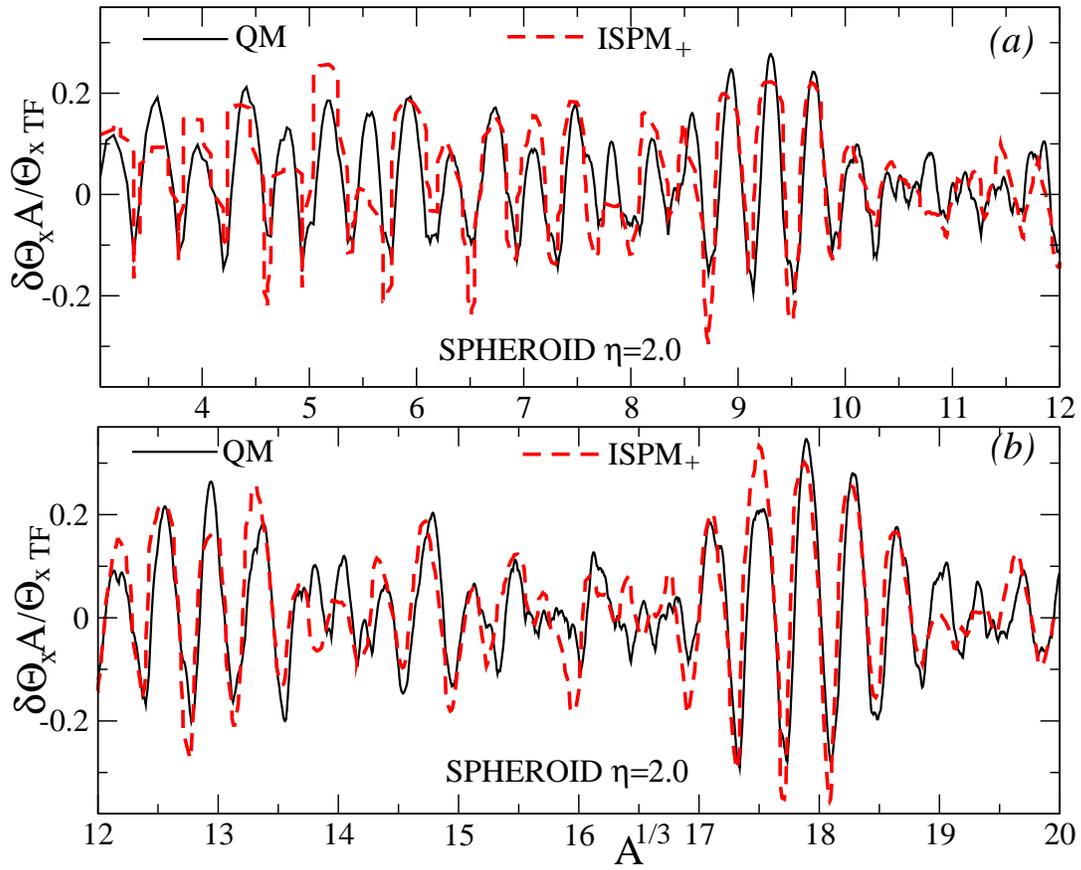}
\end{center}
\vspace{-0.4cm}
\caption{ Same as Fig.\ \ref{fig32} but for a deformation of $\eta =2.0$.
}
\label{fig33}
\end{figure*}
\begin{figure*}
\begin{center}
\includegraphics[width=0.8\textwidth,clip=true]{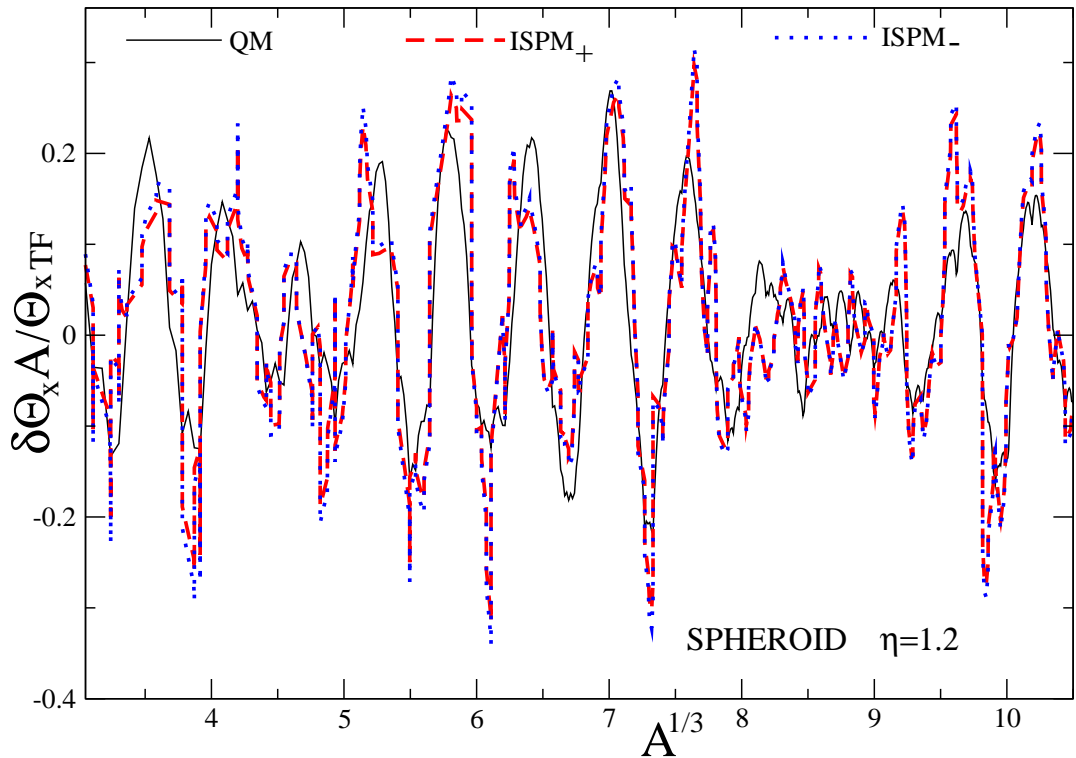}
\end{center}
\vspace{-0.2cm}
\caption{ Comparison between the shell components $\delta \Theta_x$ of the MI
   (in TF units) obtained with (ISPM$_+$, dashed line) and without
   (ISPM$_-$, dotted line) surface corrections as function of $A^{1/3}$.
   For comparison the quantum result (solid line) is also shown.}
\label{fig34}
\end{figure*}
\begin{figure*}
\begin{center}
\includegraphics[width=0.9\textwidth,clip=true]{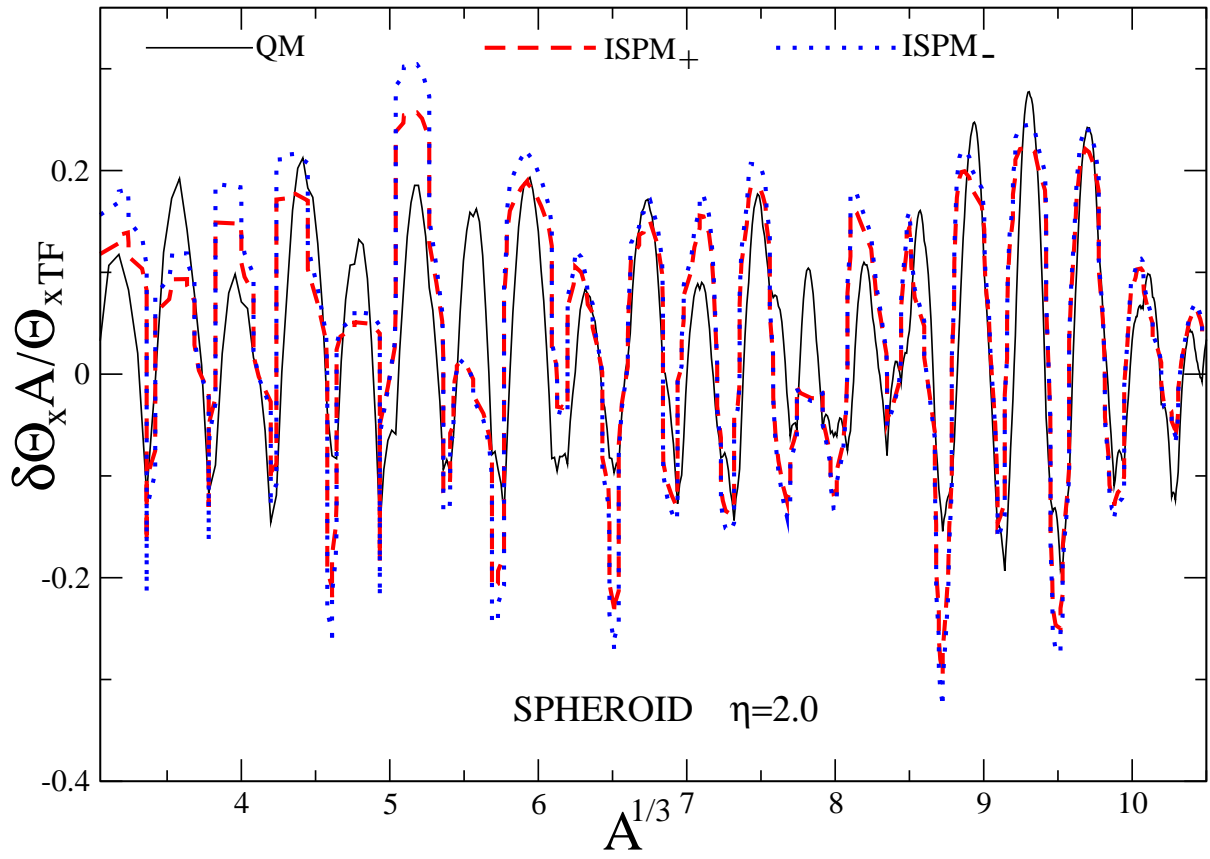}
\end{center}
\caption{ Same as Fig.\ \ref{fig34} but for a 
deformation of $\eta =2.0$.}
\label{fig35}
\end{figure*}

\clearpage
\begin{figure*}
\begin{center}
\includegraphics[width=0.8\textwidth,clip=true]{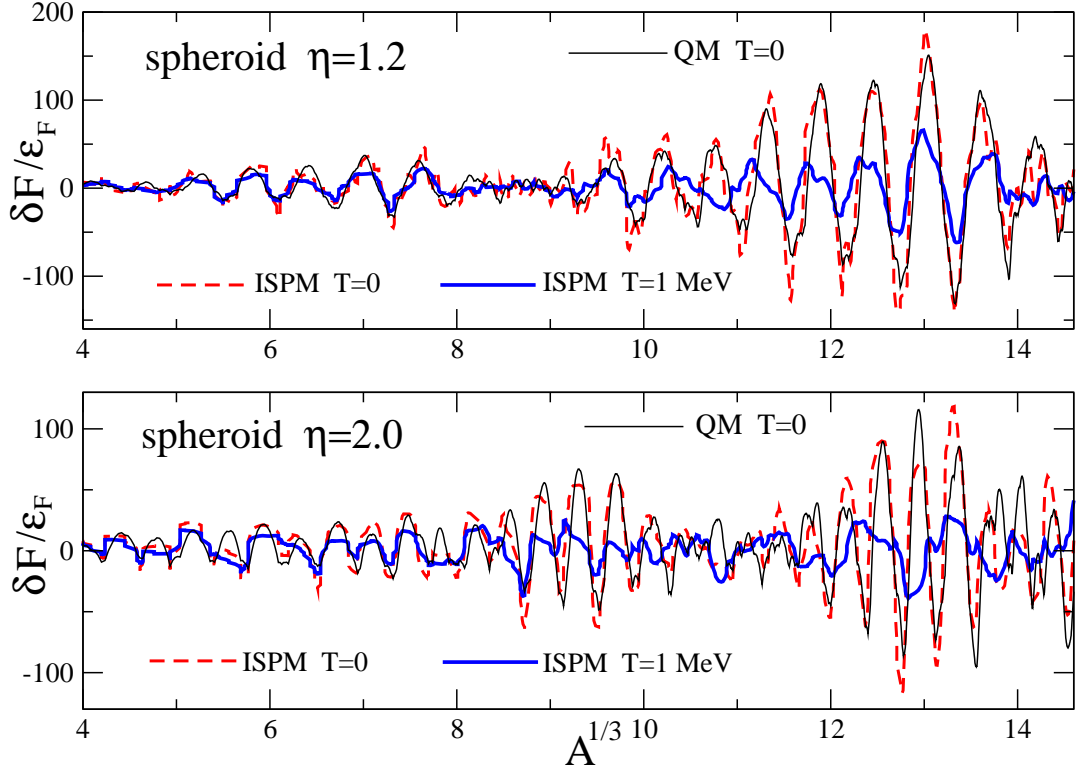}
\end{center}
\caption{ Free energy shell-correction $\delta \mathcal{F}$ (in units of
 $\varepsilon^{}_F$) as function of the particle number variable
 $A^{1/3}$ for spheroidal deformations $\eta=1.2$ (top) and $\eta=2.0$
 (bottom); comparison between the results of a quantum-mechanical
 calculation at zero temperature (thin solid line) with semiclassical
 ISPM calculations at $T=0$ (dashed line) and $T=1$ MeV (solid line).} 
\label{fig36}
\end{figure*}
\begin{figure*}
\begin{center}
\includegraphics[width=0.8\textwidth,clip=true]{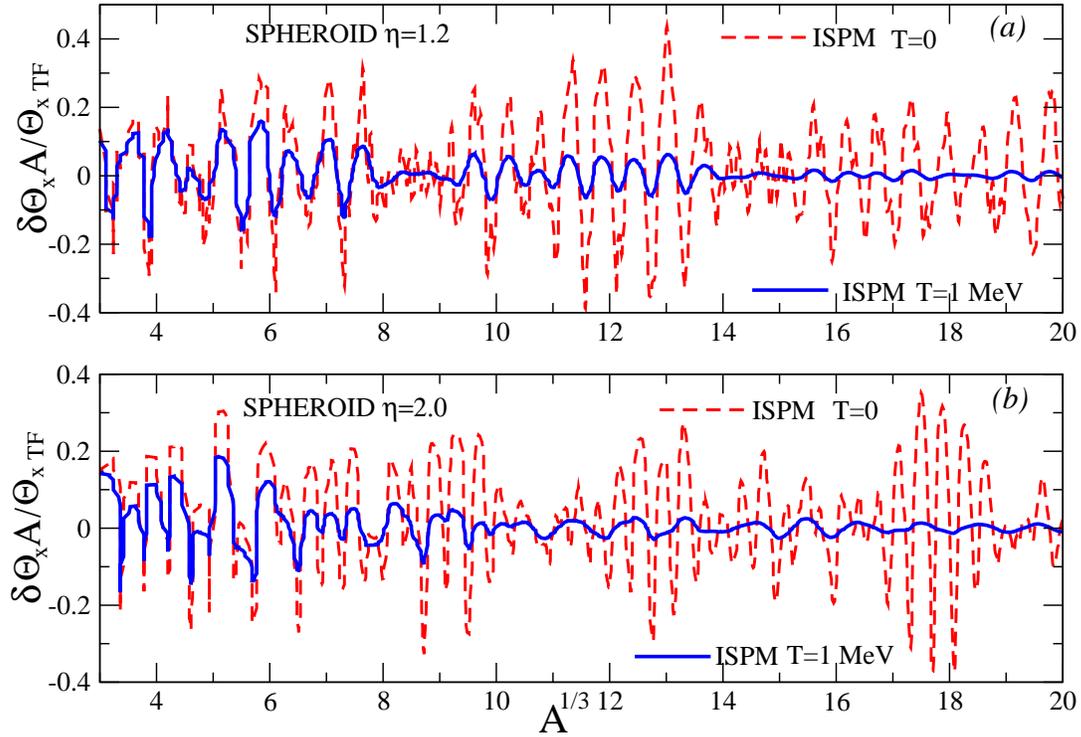}
\end{center}
\caption{ Semiclassical ISPM MI shell components 
$\delta \Theta^{}_x$ (in TF units)
 at zero (dashed line) and $T=1$ MeV temperature (solid line) as function
 of $A^{1/3}$ at deformations $\eta=1.2$ (upper part) and $\eta=2.0$
 (lower part).} 
\label{fig37}
\end{figure*}
\begin{figure*}
\begin{center}
\includegraphics[width=0.8\textwidth,clip=true]{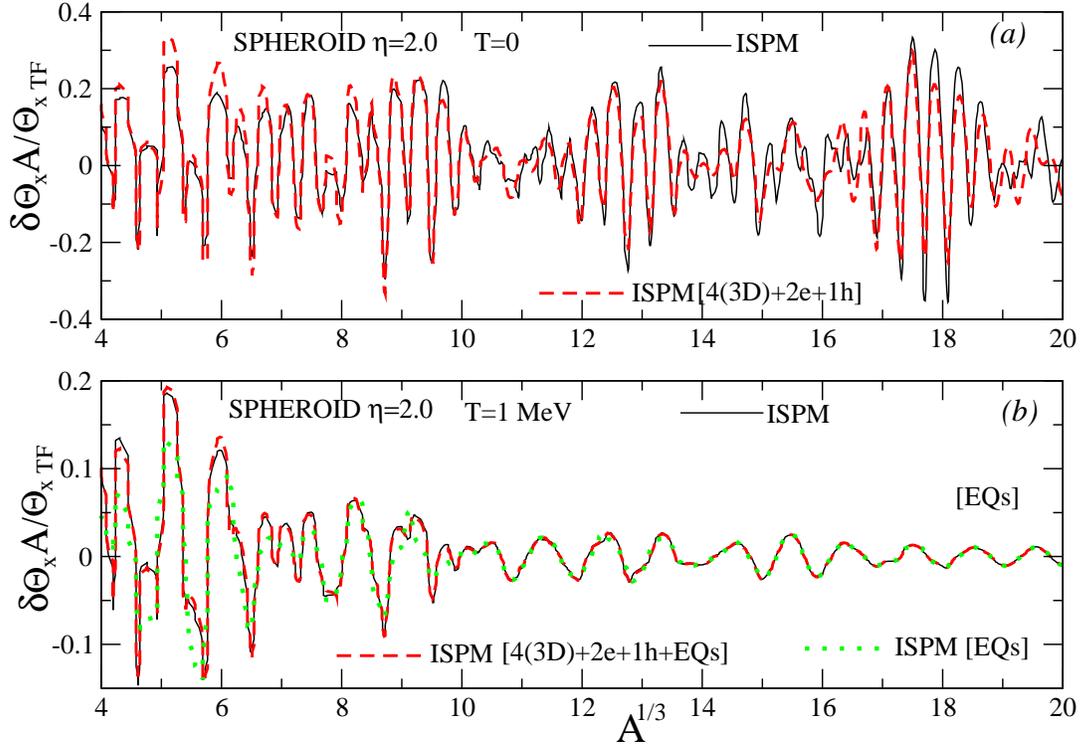}
\end{center}
\vspace{-0.1cm}
\caption{ PO contributions to the moment of inertia shell component
 $\delta \Theta^{}_x$ (in TF units) at temperature $T=0$ (upper part) and
 $T=1$ MeV (lower part) as function of $A^{1/3}$. The
 dashed line gives the contribution coming from the 4 shortest
 three-dimensional (3D) POs (which appear from the corresponding parent
 equatorial (EQ) orbits at $\eta \! \approx\! 1.6\! \div\! 2.0$) and from
 the shortest meridional POs (two elliptic and one hyperbolic) emerging at
 smaller deformations.} 
\label{fig38}
\end{figure*}
\begin{figure*}
\begin{center}
\includegraphics[width=0.8\textwidth,clip=true]{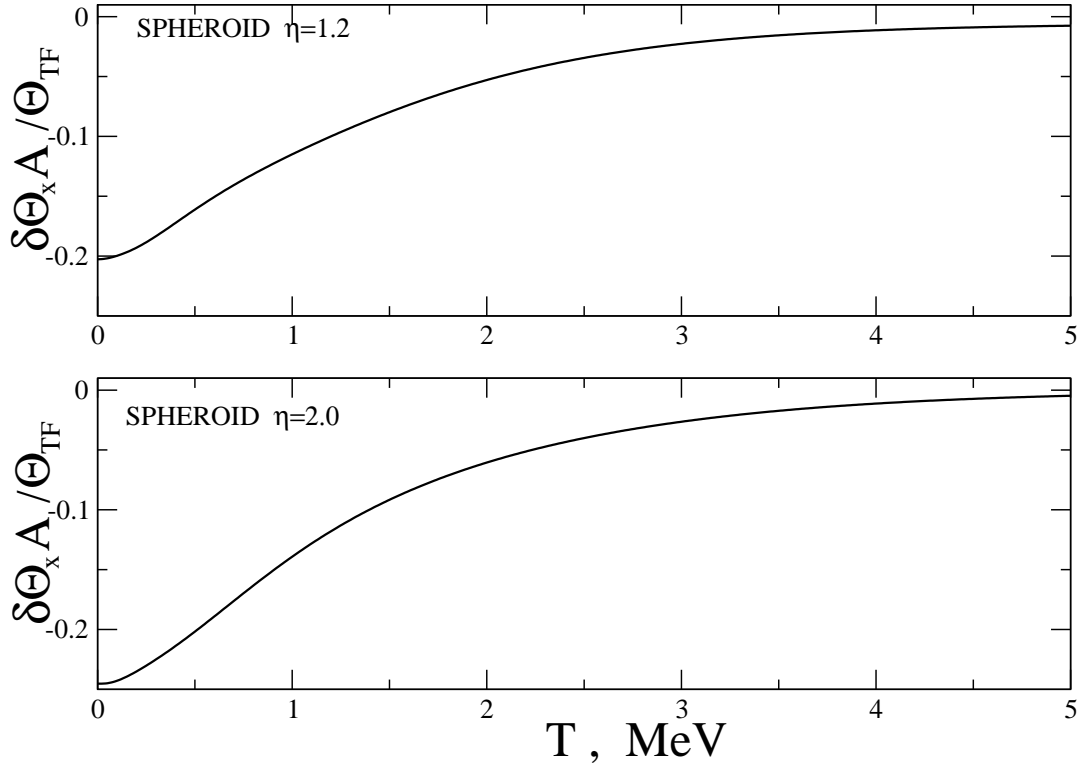}
\end{center}
\caption{ ISPM MI shell components as functions of the 
temperature $T$ for the
 same two deformations studied in Figs.\ \ref{fig32}--\ref{fig35} for
 particle numbers ($A \approx$ 166 and 186) at, respectively, one of
 the minima $k^{}_F R \approx$ 11.70 and 12.25.}
\label{fig39}
\end{figure*}

\end{document}